%% file: dissertation.tex
\documentclass[fleqn,b5paper,10pt,twoside,openright]{book}
\input{mysettings}

\input{mystyle}
\usepackage{slashed}

\begin{document}
\renewcommand{\contentsname}{Contents}

\newsavebox{\rotbox}

\pagestyle{empty}
\include{front}

\include{front2}

\pagestyle{fancyplain}
\renewcommand{\thepage}{\arabic{page}}

\newpage
\pagestyle{empty}

\renewcommand\figurename{Figure}
\tableofcontents
\include{chapter_1/chapter_1}  \markboth{}{}
\include{chapter_2/chapter_2}  \markboth{}{}
\include{chapter_3/chapter_3}  \markboth{}{}
\include{chapter_4/chapter_4}  \markboth{}{}
\include{chapter_5/chapter_5}  \markboth{}{}
\include{chapter_6/chapter_6}  \markboth{}{}
\include{chapter_7/chapter_7}  \markboth{}{}
\include{chapter_8/chapter_8}  \markboth{}{}
\include{chapter_9/chapter_9}  \markboth{}{}
\pagestyle{empty}

\appendix
\include{appendix_a/appendix_a} \markboth{}{} 
\pagestyle{empty}
\include{appendix_b/appendix_b} \markboth{}{}
\pagestyle{empty}
\include{publications}  \markboth{}{}

{\footnotesize
\renewcommand\bibname{References}
\addcontentsline{toc}{chapter}{References}

\bibliography{dissertation}
\bibliographystyle{toine}
}

\pagestyle{empty} \markboth{}{}
\pagestyle{empty}
\include{dankwoord/dankwoord}  \markboth{}{}

\end{document}

%% file: mysettings.tex
\usepackage{color}
\usepackage{palatino}
\usepackage{mathpazo}
\usepackage{rotating}
\usepackage{graphicx}
\usepackage{epsfig}
\usepackage{setspace,section,relsize}
\usepackage[Chris]{style/fncychap_rense}
\usepackage[cmex10]{amsmath}
\usepackage[shortcuts]{extdash}
\usepackage[rigidchapters]{titlesec}
\usepackage[british,dutch]{babel}
\usepackage{subfigure}
\usepackage{amssymb}
\usepackage[innercaption]{sidecap}
\usepackage{multirow}
\usepackage{bigdelim}
\usepackage{bigstrut}
\usepackage{afterpage}
\usepackage{longtable}
\usepackage{enumerate}
\usepackage{hyperref}
\AtBeginDocument{}
\renewcommand\bibname{References}

\usepackage{fancyhdr}
\pdfoutput=1

\usepackage[inner=30mm,outer=20mm]{geometry}
\newcounter{savesection}
  {\setcounter{savesection}{\value{section}}%
   \setcounter{section}{0}%

    \setcounter{equation}{0}
  
   \setcounter{figure}{0}
  
  }%
  {\setcounter{section}{\value{savesection}}}
\usepackage{style/tweaklist}
\renewcommand{\itemhook}{\setlength{\topsep}{-0pt}%
  \setlength{\itemsep}{0pt}}

\renewcommand{\enumhook}{\setlength{\topsep}{-0pt}%
  \setlength{\itemsep}{0pt}}

\titlespacing*{\subsubsection}{0pt}{1\baselineskip}{0.1\baselineskip}
\titlespacing*{\subsection}{0pt}{1.2\baselineskip}{0.1\baselineskip}
\titlespacing*{\section}{0pt}{1.4\baselineskip}{0.4\baselineskip}

\usepackage{verbatim}
\usepackage{cite}
\usepackage{afterpage}
\newcommand{\PB}[2]{\{ #1, #2\}_{\rm P.B.}}
\newcommand{\cG}{\ensuremath{{\mathcal G}}}
\newcommand{\cL}{\ensuremath{{\mathcal L}}}
\newcommand{\cH}{\ensuremath{{\mathcal H}}}
\newcommand{\cD}{\ensuremath{{\mathcal D}}}
\newcommand{\cM}{\ensuremath{{\mathcal M}}}
\newcommand{\cO}{\ensuremath{{\mathcal O}}}
\newcommand{\cP}{\ensuremath{{\mathcal P}}}

\newcommand{\ve}{\varepsilon}
\newcommand{\ed}{\mathrm{d}}
\newcommand{\tr}{\mathrm{tr}}

\newcommand{\bBox}{\bar{\square}}
\newcommand{\bnabla}{\bar{\nabla}}
\newcommand{\bg}{\bar{g}}
\newcommand{\be}{\bar{e}}
\newcommand{\bo}{\bar{\omega}}

\newcommand{\bR}{{\mathbb R}}
\newcommand{\vev}[1]{\langle#1\rangle}


%% file: mystyle.tex
\newcommand{\ml}[1]{}

%% file: front.tex
\pagestyle{empty}

\begin{large}
\begin{center}

\begin{figure}[!ht]
\includegraphics{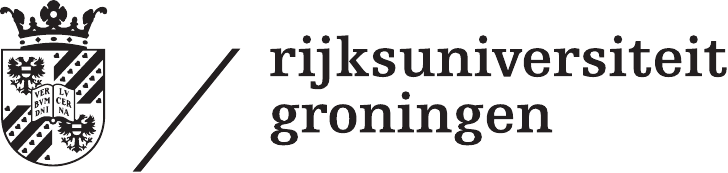}
\end{figure}
\vspace{2.0cm}
{\hspace{-0.0cm} \bf \huge
Chern-Simons--like Theories\\
\vspace{0.3cm}
of Gravity}\strut\\
\vspace{2cm}
{\bf  Proefschrift\strut}\\
\vspace{1.4cm}
ter verkrijging van de graad van doctor aan de\strut\\
Rijksuniversiteit Groningen\strut\\
op gezag van de\strut\\
rector magnificus prof. dr. E. Sterken\strut\\
en volgens besluit van het College voor Promoties.\strut\\
\bigskip
De openbare verdediging zal plaatsvinden op\strut\\
\bigskip
vrijdag 26 september 2014 om 11.00 uur\strut\\
\vspace{1.4cm}
door\\
\vspace{1.4cm}
\textbf{Wouter Merbis}\\
\bigskip
geboren op 10 augustus 1985\\
\smallskip
te Amsterdam\\
\end{center}

\vfill
\newpage

\noindent%
\makebox[4.5cm][l]{\bf Promotor} \strut\\
Prof.~dr.~E.~A.~Bergshoeff\strut\\

\vspace{1cm}
\noindent
\makebox[4.5cm][l]{\bf Beoordelingscommissie}\strut\\ Prof.~dr.~M.~Ba\~nados\\
Prof.~dr.~D.~Grumiller\\
Prof.~dr.~M.~de Roo\\

\vfill

\end{large}

%% file: front2.tex
\renewcommand{\thepage}{\arabic{page}}

\begin{flushright}
\end{flushright}
\vspace{5cm}

\newpage

\begin{flushright}
\end{flushright}

\vspace{15.0cm}
\noindent
This work is part of the research program of the Foundation for Fundamental Research on Matter (FOM), which is part of the Netherlands Organization for Scientific Research (NWO).

\vspace{0.5cm}
\noindent
ISBN: 978-90-367-7274-7 (printed)\\
ISBN: 978-90-367-7273-0 (digital)

%% file: chapter_1/chapter_1.tex
\setcounter{chapter}{0}
\selectlanguage{british}

\chapter[Introduction]{Introduction}
\label{chapter:introduction}

\begin{quote}\em

\end{quote}
\newpage
\pagestyle{headings}

\section{Introduction}
Gravity may be the most well-known force of nature to the majority of people. For everyone on earth, the effects of the gravitational force are so familiar, that its consequences have formed a part of our natural intuition and habits. In contrast, for the theoretical physicist gravity remains perhaps the most enigmatic force of nature. Of course, our everyday experience of the gravitational force, as well as the orbital motion of the planets in our solar system was known since the work of Isaac Newton. Also the underlying principle of gravitational attraction through the curvature of spacetime itself is well understood after Einstein's seminal work in 1915. The problem with gravity most troublesome for theoretical physicists nowadays is that Einstein's theory of gravity is incompatible with the principles which underlay all of the other known forces of nature: the quantization of physical properties. 

A consistent theory of quantum gravity could provide valuable insights into fundamental questions which predate written history; where does the world around us come from? How did it all start? Due to the technological advances of the last centuries, we have now entered an era of high precision experiments both on the nature of fundamental interactions and on the history and evolution of our universe. These experiments have become the guiding principles in the formation of two ``Standard Models'': the Standard Model of particle physics (SM) describing the dynamics of subatomic particles and the Standard Model of Big Bang cosmology ($\Lambda$CDM), describing the evolution of the universe. 

Both these Standard Models are a phenomenological success; they describe the experiments to an astounding level of accuracy (especially the standard model of particle physics). They describe the physics of two completely different regimes; the SM of particle physics governs the very small, while the $\Lambda$CDM model describes the largest scales imaginable. However, we know that the combination of the two is not the full story. The Standard Model of particle physics does not include gravitational interactions and a canonical quantization of gravity leads to a non-renormalizable theory. The $\Lambda$CDM model demands the presence of a dark sector of matter and energy. The dark matter is non-baryonic and hence not described by the SM, while the SM predictions for the dark energy contributions are off by many orders of magnitude. 

Both of these problems are intrinsically related to gravity, since it is through its interactions with gravity that we have come to know about the presence of dark energy and dark matter in the universe. Einstein's theory of general relativity (GR), explains the gravitational force as the interplay between matter and the curvature of spacetime; matter tells the spacetime how to curve and the curved spacetime tells matter how to move. To explain the present experimental tests, like galactic rotation curves, the acoustic peaks in the CMB and the accelerating expansion of the universe, the presence of dark matter is needed as well as a cosmological constant which is intrinsically related to dark energy. In spite of many searches, as of yet there are no direct experimental indications for the existence of dark matter and no known mechanism which can explain the correct value of the cosmological constant. Since their presence is currently solely detected through gravity, an alternative possibility may be that our present understanding of the gravitational force is only an approximate one. Perhaps spacetime curves a bit differently on galactic and cosmological length scales. Hence it is wise to supplement inquiries into the nature of the dark sector of the universe with explorations into the nature of GR itself. 

This leads us to ask some fundamental questions about general relativity. Is it even possible to modify GR in a mathematically consistent way? What are the essential ingredients that are needed to explain gravity as the interplay between matter and a dynamical spacetime and is there room for modification? In light of these questions it is instructive to investigate the defining principles of the theory. Einstein's discovery of GR is in some sense based on a leap of insight; he set out to find a theory invariant under general coordinate transformations satisfying the equivalence principle (which states that inertial mass and gravitational mass are one and the same thing, implying the universality of gravitational attraction). What he found was a fully non-linear theory of Riemannian geometry describing the dynamical nature of spacetime itself, but this answer is not unique; it is possible to construct different theories with general coordinate invariance which satisfy the equivalence principle. 

Here we will consider another way to understand general relativity, which differs from its original formulation. We consider GR as the unique, classical theory for interacting massless spin-2 particles (gravitons) invariant under linearized diffeomorphisms. A resummation of all interaction terms consistent with this symmetry will essentially reproduce General Relativity \cite{Gupta:1954zz,Weinberg:1965rz,Deser:1969wk}. From this construction, general coordinate invariance and the equivalence principle follow as a consequence. This approach is natural from the point of view of field theories for massless bosonic particles. 
Furthermore, the problem of finding a consistent theory of quantum gravity in this light states that we should look for a quantum theory of massless interacting spin-2 particles. Indeed, most of the proposals for a quantum theory of gravity, such as (super)string theory, M-theory and loop quantum gravity include a massless graviton in their spectrum. 

These considerations motivate investigations into the nature of interacting spin-2 particles. Combined with the earlier remarks concerning possible modifications of GR, a natural question may be, is it possible for the graviton to have a small mass? These type of modifications of GR are called ``massive gravity''. Although this idea has been around since the 1930s \cite{Fierz:1939ix}, massive gravity theories have seen a recent resurgence in interest mainly because of the resolution of some theoretical difficulties (see \cite{Hinterbichler:2011tt,deRham:2014zqa} for recent reviews). Investigations into massive gravity theories can broadly be categorized into two approaches: one is to assume that the graviton responsible for the gravitational interactions is not truly massless, but rather has a small mass. The other possibility is that the massless graviton of GR interacts with a sector of massive spin-2 particles.\footnote{Note that massive spin-2 modes naturally arise in theories with multiple interacting spin-2 modes, as there is a no-go theorem forbidding massless spin-2 interactions \cite{Boulanger:2000rq}. Hence if there are multiple gravitons, then they must be massive.}

Both approaches may lead to alternatives to the $\Lambda$CDM model of cosmology. If the graviton of GR itself is massive, then this will introduce a Yukawa-like gravitational potential with a characteristic length scale set by the graviton mass. Above this scale the gravitational force will appear weaker than in the GR description. This causes matter and energy to be less sensitive to the gravitational force on large distance scales. Thus to explain the observed acceleration of the universe, we would need a parametrically larger cosmological constant than in the GR description. This process of `degravitation' leads to an effective screening of the cosmological constant, which may contribute to solving the old cosmological constant problem\footnote{The old cosmological constant problem is the large discrepancy between the measured value of the cosmological constant from astronomical observations, and the contribution to the cosmological constant from a calculation of the energy of the vacuum.} \cite{deRham:2010tw}. If GR can be consistently coupled to a sector of massive spin-2 particles, then this may be a candidate for the dark matter content of the universe. In addition, the massive spin-2 sector also contributes to the cosmological constant, leading to a variety of new possibilities for cosmological applications.

This is all assuming it is even possible to find a consistent interacting theory for massive spin-2 particles. This turns out to be a non-trivial task, even at the classical level. The free theory for massive spin-2 particles was constructed by Fierz and Pauli in 1939 \cite{Fierz:1939ix}. Only some time later, after it was realized that interacting spin-2 particles and gravitation were intrinsically related, this construction was ruled out as a phenomenological theory. The problem is that the massive Fierz-Pauli theory does not reproduce GR when taking the massless limit. This is known as the van Dam-Veltman-Zakharov (vDVZ) discontinuity after the independent work of van Dam and Veltman \cite{vanDam:1970vg} and Zakharov \cite{Zakharov:1970cc}. The massive spin-2 carries 5 degrees of freedom in contrast to the two degrees of freedom for a massless spin-2. These local degrees of freedom can be decomposed into different helicity states: a helicity 0 mode (or a scalar), the helicity $\pm1$ modes of a vector and the helicity $\pm2$ states of a massless graviton. When taking the massless limit in Fierz-Pauli theory, the vector modes decouple and the helicity $\pm2$ modes reduce to linearized GR. The scalar mode, however, will couple to the trace of the stress tensor, which describes the matter content. This coupling will produces different physical predictions and hence give a different theory than linearized GR.

The trick to resolving the vDVZ-discontinuity is to take non-linear effects into account. As is well known, GR is a non-linear theory and at a distance scale set by the Schwarzschild radius, non-linear effects will start to play a role. In Fierz-Pauli theory this scale is set by the Vainshtein radius \cite{Vainshtein:1972sx}, which diverges when the graviton mass is taken to zero. This implies that nowhere in this limit the linearized approximation may be trusted and non-linear effects must be taken into account. Vainshtein showed that the non-linear effects produce an effective screening of the scalar mode, restoring the limit to GR. This is now known as the Vainshtein mechanism. However, soon afterwards it was realized that the most general non-linear completions of massive Fierz-Pauli theory propagates six degrees of freedom instead of five. The additional mode being a scalar-ghost, which now carries the name Boulware-Deser ghost, after the authors of \cite{Boulware:1973my}. The mechanism responsible for the correct massless limit hence introduced a new pathology, the presence of ghosts.

The scalar ghost in the non-linear extensions of Fierz-Pauli theory has caused a loss of interest in these models until the turn of the millennium. The matter changed when it was found that adding particular higher-order interactions could remove the scalar ghost from the spectrum of the theory. This is the approach of de Rham, Gabadaze and Tolley (dRGT) \cite{deRham:2010ik,deRham:2010kj}, they first considered a decoupling limit and added interaction terms in a specific combination to remove the higher-derivative terms responsible for the Boulware-Deser ghost. 
In order to study the ghost problem beyond the decoupling limit, the authors of \cite{deRham:2010kj} resum the new interaction terms into a fully non-linear theory of massive gravity involving elementary symmetric polynomials of the square root matrix $\sqrt{g^{-1} f}$. Here $f_{\mu\nu}$ is a reference metric which is needed to contract the indices of the metric (see also \cite{Hassan:2011vm}). This fully non-linear theory was shown to be ghost free in \cite{Hassan:2011hr}. We will return to this construction briefly in chapter \ref{chapter:two_company}.

The dRGT massive gravity theory describes solely a massive spin-2. The idea is that at short distances, the graviton will appear massless and the Vainshtein mechanism restores the limit to GR. At long distances, the mass term is responsible for modified cosmology and solutions which are self-accelerating can be found. A peculiarity of these theories is that we need a (a priori arbitrary) reference metric $f_{\mu\nu}$. The introduction of this reference metric defines a fixed absolute frame which in some sense goes against our intuition from relativity theory. One of the founding principles of GR is that the physics should not depend on our choice of coordinates. However, since the reference metric is not dynamical, the physics in dRGT massive gravity will depend on our choice of reference frame. In an effort to tackle this problem, the reference metric of dRGT massive gravity can be upgraded to a dynamical metric by introducing a kinetic (Einstein-Hilbert) term for it, resulting in a bimetric theory of gravity\footnote{The first introduction of bimetric theories of gravity date back to the work of Isham, Salam and Strathdee in 1971 \cite{Isham:1971gm}.} \cite{Hassan:2011zd}. These models describe the dynamics of a massive spin-2 particle together with a massless one. However, adding the kinetic term for the fixed reference metric complicates the degree of freedom analysis of the theory. At the moment of writing this thesis, there is an active debate in the literature whether bimetric gravity theories are really free of the Boulware-Deser ghost. Compare for instance the results of \cite{Hassan:2011ea,Kluson:2013cy,Kluson:2013lza,Nomura:2012xr,Soloviev:2012wr,Soloviev:2013mia,Deser:2013gpa}

With the above considerations in mind, we can now state the main theme of this thesis: an exploration into the possibility of consistently describing a `physical' massive spin-2 mode in the presence of gravity in three spacetime dimensions. By physical here is meant that there are at least no mathematical inconsistencies, such as ghost particles or tachyons. We have already discussed the importance of understanding interacting theories of spin-2 particles, let us now focus on another aspect in the main theme of this thesis: gravity in three spacetime dimensions.

\section{Moving to Three Dimensions}
Restricting ourselves to three spacetime dimensions (3D) is somewhat removed from the cosmological motivations mentioned above; since we live in a four dimensional world, it is obvious that no realistic cosmological models can be constructed from three dimensional gravity models. However, there are several reasons why we make the jump to three dimensions.

First of all, the three dimensional case provides a relatively easy playing ground to test ideas and perform exact computations in (quantum) gravitational models. For instance, the Hamiltonian analysis of three dimensional bimetric gravity is more tractable and, as we will see in chapter \ref{chapter:three_crowd}, its results suggests a reason why there is a disagreement in the present literature on the degrees of freedom in bimetric gravity in four dimensions. Of course, if a certain model works in three dimensions, then this is no guarantee that it will work in four dimensions as well, but the results can be useful in guiding us in the right direction.

Secondly, general relativity in three dimensions is special, since in this case the massless spin-2  does not have any local degrees of freedom. The theory can be interpreted as a topological gauge theory and has an equivalent formulation as a Chern-Simons (CS) theory \cite{Achucarro:1987vz,Witten:1988hc}. The absence of local degrees of freedom tells us that the theory only depends on global effects. For instance, there are black holes in three dimensional gravity with a negative cosmological constant, the Ba\~nados-Teitelboim-Zanelli (BTZ) black holes \cite{Banados:1992wn}. Locally these black holes look like Anti-de Sitter (AdS) spacetime, but globally they satisfy the properties of a black hole; they have an event horizon and are characterized classically by their mass and angular momentum. These global charges, which can be computed on the boundary of AdS, are different than in the AdS vacuum.

The relation of three dimensional gravity to a Chern-Simons gauge theory is another central theme of this thesis. Since our aim will be to modify 3D GR and this inevitably entails the addition of local degrees of freedom, in 3D this translates to modifying CS gauge theories to describe local degrees of freedom. There is a way to do so, while keeping some of the desirable properties of CS-theories. All of the 3D gravity models in this thesis fall into a class of models which can be called ``Chern-Simons--like''. What we precisely mean by CS--like will be defined in chapter \ref{sec:introductionc3}. It suffices to say here that, like CS gauge theories, they are defined by a first order Lagrangian three-form constructed solely from one-form fields, but we give up the requirement that all fields are Lie algebra valued connections and hence the resulting theories are no longer topological gauge theories. 

Another advantage of three dimensional gravity is that, on manifolds with a boundary, a two dimensional conformal field theory can be found on this boundary. Adopting asymptotically AdS boundary conditions leads to an asymptotic symmetry algebra relating different physical states with the same boundary conditions. This asymptotic symmetry algebra of global charges is two copies of the Virasoro algebra with a classical central extension, as was shown by Brown and Henneaux in 1986 \cite{Brown:1986nw}. In this sense, three dimensional gravity can be thought of as a two dimensional Conformal Field Theory (CFT) and the real dynamics of the 3D massless spin-2 takes place on the boundary, where its excitations fall into representations of the (conformally invariant) Virasoro algebra. The connection between 3D gravity and a two dimensional CFT discovered by Brown and Henneaux is a precursor of what is now known as holography \cite{'tHooft:1993gx,Susskind:1994vu}, or the AdS/CFT correspondence\cite{Maldacena:1997re,Witten:1998zw,Gubser:1998bc}. 

The philosophy behind the holographic principle is that a theory of gravity in $d$ dimensions is dual to a quantum field theory in $d-1$ dimensions. This implies that some gravitational problem can be rephrased as a field theoretical problem and vice versa. The most well-known example is the duality between type IIB string theory on $\rm AdS_5 \times S_5$ and the conformal $\mathcal N =4$ $\rm SU(N)$ super Yang-Mills theory, although many other dualities have been conjectured.\footnote{Many examples of the AdS/CFT-correspondence have passed several non-trivial checks, but a full proof of the duality has not been given. This is due to the fact that usually, when the gravitational side is strongly coupled, the field theory side has weak coupling and vice versa. This makes the duality hard to proof, however, this feature is also responsible for its usefulness, since strongly coupled systems can now be accessed via their weakly coupled dual theory.} Some physicists would support the idea that these type of dualities are a general property of gravitational theories. This sheds new light on the relation between gravitational theories and quantum field theories and on the search for a theory of quantum gravity; it doesn't have to be a gravity theory! It may be that our quantum gravity problem translates to an easier problem in the dual quantum field theory. Solving quantum gravity in this light may be rephrased as finding the dual field theory. 

Holography for three dimensional gravity is particularly interesting, since dual field theory is two dimensional, which are the most well-studied examples of CFTs. Still, despite many efforts (see e.g. \cite{Banados:1998gg,Carlip:2005zn} for reviews and references), the precise structure of the CFT dual to pure 3D gravity is unknown. But even without the precise structure of the dual theory, some very promising results have been obtained. Most notably, in  \cite{Strominger:1997eq,Birmingham:1998jt} the Bekenstein-Hawking entropy of the BTZ black hole was related to the entropy associated to the asymptotic density of states of a 2D CFT using the Cardy formula \cite{Cardy:1986ie,Bloete:1986qm}. 

Finally, in three spacetime dimensions there are other ways to arrive at a ghost-free theory of massive spin-2 modes. Instead of adding explicit mass terms which break some of the gauge symmetries of the theory, one can add higher derivatives of the metric. One way is to supplement Einstein-Hilbert gravity with a gravitational Chern-Simons term, leading to Topologically Massive Gravity (TMG) \cite{Deser:1981wh,Deser:1982vy}. The resulting theory has third order field equations and describes a single helicity 2 mode. This mode has positive energy and mass when the Einstein-Hilbert term is considered with the `wrong' sign; when the sign is opposite to the one in pure 3D gravity.

A parity even higher-derivative theory of massive gravity describing the two helicity $\pm2$ modes of a massive graviton was developed by Bergshoeff, Hohm and Townsend \cite{Bergshoeff:2009hq,Bergshoeff:2009aq} and it is called New Massive Gravity (NMG). Here one adds higher derivatives of the metric in a generally covariant way by adding squares of the curvature invariants: $R^{\mu\nu}R_{\mu\nu}$ and $R^2$. The theory is free of the Boulware-Deser ghost, provided that the curvature squared terms come in a specific combination. In flat space, the massive spin-2 mode has positive energy when one takes the `wrong' sign for the Einstein-Hilbert term. Both for TMG and NMG the situation is more subtle in anti-de Sitter spacetimes. Due to the higher-derivative equations of motion, an Ostrogradski-type instability arises. This indicates that either the massless of the massive graviton is a ghost. In flat space, we may choose the sign in front of the Einstein-Hilbert term such that the massless mode is a ghost, which is the origin of the 'wrong' sign. This leaves a theory of a healthy massive spin-2, since the massless mode is pure gauge in three dimensions. However, in AdS spacetimes, there are BTZ black holes, which are characterized by the global charges associated to the massless mode. Then a wrong sign for the massless modes leads to negative mass for the BTZ black hole and, in addition, the central charge for the boundary CFT is negative. This implies that the representations corresponding to the boundary graviton modes are non-unitary.

As we will discuss in detail in this thesis, this is a property shared by all higher-derivative extensions of 3D general relativity. In this thesis, we will present novel theories which also modify 3D GR but resolve this problem. This was made possible by the observation that TMG and NMG have a description in terms of a Chern-Simons--like theory \cite{Hohm:2012vh}, however, the full set of CS--like theories is larger. It contains models which describe the same number of degrees of freedom as TMG and NMG, but with improved behavior in the context of the AdS/CFT correspondence. The construction of such models is the most important result of the work described here. These novel models do not have an equivalent formulation in terms of an action including higher-derivative corrections to GR, but instead their action contains auxiliary fields. These can be solved for upon using the field equations and we can write these equations in terms of a single metric plus higher derivative corrections, but this is not valid at the level of the action. This property is novel and allows for new possibilities.

%

\section{Outline of this Thesis}
The rest of this thesis will build on the ideas presented in the previous section. Chapter \ref{chapter:GRin3D} contains a review of general relativity in three spacetime dimensions and its relation to Chern-Simons gauge theories. We will explicitly perform a Hamiltonian analysis of 3D GR and review the Brown-Henneaux procedure of calculating the asymptotic symmetry group. This is done not only for the sake of being self contained, but we will also need some of the results in later chapters, when we will study modifications of three dimensional gravity and their asymptotic symmetry group.

In chapter \ref{chapter:two_company} we will define what we mean by Chern-Simons--like theories of gravity and discuss different approaches of introducing the 2 degrees of freedom of a massive spin-2 in three dimensions. The free Fierz-Pauli theory in three dimensions is discussed, as well as its non-linear extension into dRGT massive gravity. We then discuss CS--like theories with auxiliary fields which may be solved for to obtain higher derivative theories of massive gravity, such a TMG and NMG. We conclude the chapter by introducing a CS--like theory inspired by bimetric gravity, called Zwei-Dreibein Gravity (ZDG). We will show that the field equations of ZDG can be written in terms of a single metric and an infinite sum of higher derivative terms, while this is not possible at the level of the action. 

Chapter \ref{chapter:three_crowd} investigates the Hamiltonian form of the general CS--like theory. We find that the presence of secondary constraints is intrinsically linked to the presence of invertible fields in the theory. We then analyze a number of CS--like models using the general formalism and confirm the absence of a third degree of freedom, which could correspond to a Boulware-Deser ghost, on a case by case basis. In the case of ZDG, the Hamiltonian analysis naturally suggests a defining assumption which makes the theory scalar ghost-free.

After having found the Hamiltonian form for the general CS--like model, it becomes relatively easy to identify the first class constraints which generate the gauge symmetries. This is the subject of chapter \ref{chapter:asymptotic_symm}, where we apply the Brown-Henneaux procedure to various 3D massive gravity theories. We find that the higher derivative theories have a negative central charge whenever the bulk massive mode has positive energy. This implies that these theories are non-unitary in AdS. We then show that in ZDG, this clash between bulk and boundary unitarity does not apply, as both the central charge and the energy of the massive mode are positive in some regions of its parameter space.

ZDG has the same linear spectrum as NMG, but improved behavior in light of the AdS/CFT correspondence, in chapter \ref{chapter:MMG} we investigate a CS--like model with the same minimal bulk properties as TMG, but with improved bulk-boundary behavior. We use the techniques discussed in previous chapters to investigate this `Minimal Massive Gravity' theory, which propagates a single helicity 2 mode. We find that the theory has a simple description in terms of a CS--like theory, leading to a novel structure of the field equations in a metric form. We comment on the problem of coupling such a model to matter.

Chapter \ref{chapter:ENMG_VDG} deals with CS--like models which lead to extensions of New Massive Gravity in two ways. One class of models leads to higher-derivative actions which include more than four derivatives of the metric. We present a systematic way of constructing such theories, which are free of scalar-ghosts by construction. The construction is furthermore consistent with the presence of a holographic $c$-theorem. However, the presence of massive spin-2 ghosts or tachyons cannot be avoided and this limits the applicability of these models to a number of special, critical points in the parameter space where the theories become dual to a logarithmic conformal field theory (LCFT). We conclude chapter \ref{chapter:ENMG_VDG} with a construction of extensions of Zwei-Dreibein Gravity. These models propagate multiple massive spin-2 modes and their parameter space contains regions where all bulk modes are physical, while the boundary central charge in AdS$_3$ is positive. 

Chapter \ref{chapter:AdS_LCFT} will discuss the conjecture that at a continuous range of critical points in its parameter space, ZDG is dual to a logarithmic conformal field theory (LCFT). We will provide evidence for this conjecture both at the linearized and the non-linear level. The chapter concludes with the relation between a massive gravity theory with two massive modes at a special tricritical point where both massive modes become massless and the dual LCFT has rank-3. In chapter \ref{chapter:conclusions} we conclude and give some possible directions for future research.

%% file: chapter_2/chapter_2.tex
\pagestyle{empty}
\setcounter{chapter}{1}

\chapter[Gravity in Three Dimensions]{Gravity in Three Dimensions}
\label{chapter:GRin3D}

\begin{quote}\em
As was emphasized in the introduction, three dimensional gravity provides an attractive and simple playing ground for models of quantum gravity. This chapter serves to give an overview of three dimensional gravity and its peculiarities. In particular, special attention is devoted to the formulation of three dimensional general relativity as a Chern-Simons gauge theory. A Hamiltonian analysis of three dimensional gravity is reviewed and it is shown that there are no local, propagating, degrees of freedom. The last part of this chapter is devoted to 3D gravity on manifolds with a boundary, in particular in Anti-de Sitter spacetime. We show explicitly how the algebra of global charges on the boundary of an asymptotically AdS spacetime gives rise to an infinite dimensional conformal algebra.   
\end{quote}
\newpage
\pagestyle{headings}

\section{General Relativity in Three Dimensions}

General Relativity (GR) is the theory describing the interplay between the local distribution of matter and energy and the curvature of spacetime. Much like how in Maxwell's theory of electromagnetism electric and magnetic fields are sourced by static or moving electric charges, in GR the geometry of spacetime, described by a Riemannian metric $g_{\mu\nu}$, is sourced by the presence of static or moving matter and energy. The geometry of spacetime, in its turn, determines how free particles move. 

The field equations describing this intricate relation between the curvature of spacetime and the matter and energy distribution are Einstein's equations:
\begin{equation}\label{einsteineqn}
G_{\mu\nu}+ \Lambda g_{\mu\nu} = 8 \pi G T_{\mu\nu}\,.
\end{equation}
Here $T_{\mu\nu}$ is the stress-energy tensor, defining the local distribution of matter and energy, $G$ is Newton's constant and we use units where the speed of light is set to unity: $c=1$. $\Lambda$ is the cosmological constant and the Einstein tensor $G_{\mu\nu}$ is defined as
\begin{equation}\label{einsteintens}
G_{\mu\nu} = R_{\mu\nu} - \frac12 g_{\mu\nu} R \,,
\end{equation}
where the Ricci tensor $R_{\mu\nu}$ and the Ricci scalar $R$ are contractions of the Riemann curvature tensor $R^{\alpha}{}_{\beta\gamma\delta}$
\begin{equation}\label{riccidef}
R_{\mu\nu} = R^{\alpha}{}_{\mu\alpha\nu} \,, \qquad R = g^{\mu\nu} R_{\mu\nu}\,.
\end{equation}
The Riemann curvature tensor is second order in derivatives of the metric and completely encodes the curvature of spacetime.  

Under certain assumptions, Einstein's equation \eqref{einsteineqn} is the unique field equation relating geometry to a local matter distribution.  Let us  assume that a Riemannian metric $g_{\mu\nu}$ is determined by the field equations
\begin{equation}\label{fieldeqn}
F_{\mu\nu}[g] = \kappa T_{\mu\nu}\,,
\end{equation}
where $\kappa$ is a dimensionful coupling constant and $F_{\mu\nu}$ is a tensor function of the metric and its derivatives. Then Einstein's equation \eqref{einsteineqn} is unique, regardless of the dimensionality of spacetime, if we further assume the function $F_{\mu\nu}$ satisfies these three requirements \footnote{This uniqueness theorem is due to the work of Cartan \cite{Cartan}, Vermeil \cite{Vermeil} and Weyl \cite{Weyl}. A review can be found in \cite{Giddings:1983es}}:
\begin{itemize}
\item $F_{\mu\nu}[g] $ contains derivatives of the metric up to second order.
\item $F_{\mu\nu}[g] $ is linear in second order derivatives
\item Conservation of energy and momentum ($\nabla^{\mu} T_{\mu\nu}=0$) is an automatic consequence of the field equations \eqref{fieldeqn}.
\end{itemize}
This argument is valid in all dimensions, so let us investigate what the consequences in three spacetime dimensions are. 

The symmetry properties of the Riemann tensor ensure that it has $\frac{1}{12}d^2(d^2-1)$ algebraically independent components, where $d$ is the number of spacetime dimensions. The Einstein tensor $G_{\mu\nu}$ is symmetric, and hence it has $\frac12 d(d+1)$ components. For $d=3$, they both have 6 independent components and, since they are related through \eqref{einsteintens} and \eqref{riccidef}, the Riemann tensor is completely determined in terms of the Einstein tensor.
\begin{equation}
R_{\alpha\beta\gamma\delta} = 2(g_{\alpha[\gamma} G_{\beta]\delta} - g_{\beta[\gamma} G_{\delta]\alpha} ) + 2 G g_{\alpha[\gamma} g_{\delta]\beta}\,.
\end{equation}
By the Einstein equation \eqref{einsteineqn}, this implies that the curvature of spacetime is completely determined in terms of the local matter distribution and the cosmological constant. Outside local sources ($T_{\mu\nu} = 0$) spacetime is locally flat ($\Lambda = 0$) \cite{Giddings:1983es,Deser:1983tn} or of constant curvature ($\Lambda \neq 0$)\cite{Deser:1983nh}. Hence, in three dimensions, there are no gravitational waves and no dynamical gravitational degrees of freedom: i.e. there are no massless gravitons in three dimensions.

\section{3D Gravity as a Chern-Simons Gauge Theory}

The absence of local degrees of freedom suggests that three dimensional gravity is trivial. However, global effects of the manifold are also important; in three dimensions the dynamics is topology \cite{Deser:1988qn}. There are black hole solutions in three dimensional gravity with a negative cosmological constant, found by Ba\~nados, Teitelboim and Zanelli in \cite{Banados:1992wn}. These black holes are locally AdS, but globally they are characterized by conserved charges at the boundary of the AdS spacetime, which differ from the AdS vacuum. In fact, as was shown by Brown and Henneaux \cite{Brown:1986nw}, these global charges fall into representations of the Virasoro algebra.\footnote{The mass and angular momentum of the BTZ black hole are the zero modes of the global conserved charges, as we will discuss below in section \ref{sec:ECASG}} This led to the conjecture that three dimensional Anti-de Sitter gravity can equivalently be described by a two dimensional conformal field theory on the boundary of AdS$_3$ \cite{Witten:2007kt}.

In the remainder of this chapter we will review some known properties of pure gravity in three dimensions, its relation to gauge theory and the asymptotic symmetries of global charges in the case of AdS$_3$ gravity. But before we go into this, we expand a bit on GR in a first-order formalism, by choosing a ``noncoordinate basis'' for the tangent space. This is very well known, however, most of the thesis will use this notation and many graduate courses on GR choose to omit a proper discussion of the first-order formulation. So for the sake of being pedagogical and self-contained, we proceed with a brief discussion on the noncoordinate basis, following appendix J of \cite{carrol_book}.

\subsection{Noncoordinate basis}
As Einstein's equations describe the relation between the curvature of spacetime and the local matter and energy distribution, GR is intrinsically related to the Riemannian geometry of manifolds. All vectors at a point $p$ on a manifold $\cM$ define the tangent space $T_p$. It is customary in GR to choose as a basis  $\{\hat{e}_{\mu}\}$ for the tangent space the directional derivatives with respect to some (arbitrarily chosen) coordinates $x^{\mu}$. This particular basis for the tangent space, $\hat{e}_{\mu} = \partial_{\mu}$, is called a coordinate basis and in general it is not normalized to unity or orthogonal (``orthonormal''). The corresponding basis $\{ \hat{\theta}^{\mu}\}$ for the cotangent space $T_p^*$, the space of all linear maps from $T_p$ to the real numbers, then is given by the gradients $d  x^{\mu}$, since $\hat{\theta}^{\mu} \hat{e}_{\nu} = \delta^{\mu}_{\nu}$.

To clarify the relation between General Relativity and gauge theories, it is convenient to instead choose an orthonormal basis for the tangent space $T_p$, which is not related to a choice of coordinates (noncoordinate basis). Let us denote the set of orthonormal basis vectors as $\{ \hat{e}_a\}$ and demand that the inner product of these basis vectors takes the canonical form:
\begin{equation}\label{innerproduct}
g(\hat{e}_{a}, \hat{e}_b) = \eta_{ab}\,.
\end{equation}
Here $\eta_{ab}$ is the Minkowski metric in the `mostly plus' signature convention and $g(\; , \; )$ denotes the usual spacetime metric tensor. The orthonormal set of basis vectors $\{\hat{e}_a \}$ is called the `vielbein' (German for `many legs') in any dimension, which invites the nomenclature zweibein, dreibein, vierbein, etcetera to denote the vielbein in two, three and four dimensions respectively. The old coordinate basis, spanned by $\hat{e}_{\mu}$, can now be expressed in terms of our new orthonormal basis as
\begin{equation}\label{vielbein}
\hat{e}_{\mu} = e_{\mu}{}^a \hat{e}_{a}\,,
\end{equation}
where the components $e_{\mu}{}^a$ form an invertible matrix. It has become standard practice to confuse tensors with their components, and hence we will continue to call $e_{\mu}{}^a$ the vielbein (and very soon: the dreibein, since we are going to work mostly in three dimensions). The inverse vielbein $e^{\mu}{}_a$ are the components of the orthonormal basis vectors $\hat{e}_a$ in the coordinate basis $\{ \hat{e}_{\mu} \}$, but they serve double duty, since they also relate the coordinate basis one-forms $\{\hat{\theta}^{\mu}\} $ of the cotangent space $T^*_p$ in terms of an orthonormal basis of one-forms $\{ \hat{\theta}^a \}$, satisfying $\hat{\theta}^a \hat{e}_b = \delta^a_b$.
\begin{equation}
\hat{\theta}^{\mu} = e^{\mu}{}_a \hat{\theta}^{a}\,.
\end{equation}
In this respect, also the vielbein itself has two duties, next to \eqref{vielbein} they relate the orthonormal basis one-forms to the coordinate basis one-forms ($\hat \theta^a = e_{\mu}{}^a \hat \theta^{\mu}$).

The (inverse) vielbein is a map between the orthonormal basis of the (co)tangent space and the coordinate basis and for all practical purposes, we can think of them as the identity map which can convert Greek (coordinate basis; curved space) indices to Latin (orthonormal basis; flat space) indices and back. Another way to think of the vielbein is as the square root of the metric, since \eqref{innerproduct} implies
\begin{equation}
\label{metric}
g_{\mu\nu} e^{\mu}{}_a e^{\nu}{}_b = \eta_{ab} \,, \quad \text{and hence:} \quad  g_{\mu\nu} = e_{\mu}{}^a e_{\nu}{}^b \eta_{ab}\,.
\end{equation}
The noncoordinate basis introduces an extra arbitrariness; we can change basis vectors independently of our choice of coordinates, as long as the orthonormality condition \eqref{innerproduct} is preserved. To be more precise, if we choose a new set of basis vectors
\begin{equation}
\hat{e}_a \to \hat{e}_{a'} = \Lambda^a{}_{a'} (x) \hat{e}_a\,,
\end{equation}
such that
\begin{equation}
\Lambda^{a}{}_{a'} \Lambda^{b}{}_{b'} \eta_{ab} = \eta_{a'b'}\,,
\end{equation}
then our new basis is equally good. The transformation which preserve the flat metric are, of course, the Lorentz transformations (or orthogonal transformations for a Euclidean-signature metric). We have this freedom at every point $p$ on the manifold and hence they are called local Lorentz transformations (LLT). Besides LLTs, there are the usual general coordinate transformations (GCT). The vielbein $e_{\mu}{}^a$ transforms as a one-form under general coordinate transformations and is thus a Lorentz vector-valued one-form.

To define covariant derivatives on objects with Latin (Lorentz) indices, we need a connection in order to obtain an expression which transforms as a tensor. In the noncoordinate basis, this is the spin connection $\omega_{\mu}{}^a{}_b$. The covariant derivative is then defined in the usual way\footnote{If the tensor X in \eqref{covder_def} carries Greek indices, then also terms involving the Christoffel connection $\Gamma^{\rho}_{\mu\nu}$ are present on the r.h.s.}
\begin{equation}\label{covder_def}
\nabla_{\mu} X^{a}{}_b = \partial_{\mu} X^a{}_b + \omega_{\mu}{}^a{}_c X^c{}_b - \omega_{\mu}{}^c{}_b X^a{}_c\,.
\end{equation}
Demanding that this expression obeys the usual tensor transformation law fixes how the spin connection transforms under LLTs. 
\begin{equation}
\omega_{\mu}{}^{a'}{}_{b'} = \Lambda^{a'}{}_{a} \Lambda^{b}{}_{b'} \omega_{\mu}{}^a{}_b - \Lambda^c{}_{b'} \partial_{\mu} \Lambda^{a'}{}_{c}\,.
\end{equation}
The covariant derivative of a vector is then a tensor itself and since tensors are independent of the basis we use, we can find a relation between the Christoffel connections $\Gamma^{\rho}_{\mu\nu}$ and the spin connection
\begin{equation}\label{christoffel}
\Gamma^{\rho}_{\mu\nu} = e^{\rho}{}_a \partial_{\mu} e_{\nu}{}^a + e^{\rho}{}_a \omega_{\mu}{}^a{}_b e_{\nu}{}^ b\,.
\end{equation}
The torsion tensor, defined as $T^{\rho}_{\mu\nu} = 2 \Gamma^{\rho}_{[\mu\nu]}$, can thus be written as
\begin{equation}\label{torsion}
T_{\mu\nu}{}^a = 2\left( \partial_{[\mu} e_{\nu]}{}^a + \omega_{[\mu}{}^a{}_b e_{\nu]}{}^ b \right)\,,
\end{equation}
where for later convenience the upper index is converted to a Lorentz index using the vielbein. The Riemann curvature tensor can be found in terms of the spin connection by considering the commutator of covariant derivatives
\begin{equation}
[\nabla_{\mu}, \nabla_{\nu}] X^a = R_{\mu\nu}{}^a{}_b X^b + T_{\mu\nu}^{\rho} \nabla_{\rho} X^a\,.
\end{equation}
Explicit computation using \eqref{covder_def} gives:
\begin{equation}\label{Riemann}
R_{\mu\nu}{}^a{}_b = 2 \left( \partial_{[\mu} \omega_{\nu]}{}^a{}_b + \omega_{[\mu}{}^a{}_c\omega_{\nu]}{}^c{}_b\right) \,.
\end{equation}
The expressions for the curvature and torsion tensors illustrate a useful property of this formalism. Since they are both two-forms and strictly anti-symmetric in Greek indices, we can express them in terms of the one-forms
\begin{equation}
e^a = e_{\mu}{}^a d  x^{\mu}\,,
\end{equation}
and the spin connection one-forms
\begin{equation}
\omega^a{}_b = \omega_{\mu}{}^a{}_b d x^{\mu}\,.
\end{equation}
Using exterior derivatives and wedge products and suppressing the Greek indices, we may write:\footnote{The funny looking factor of two comes from a choice of conventions. Some texts prefer to write $(d X)_{\mu\nu} = 2 \partial_{[\mu} X_{\nu]} = \partial_{\mu}X_{\nu} - \partial_{\nu}X_{\mu}$, while I will use $(d X)_{\mu\nu} = \partial_{[\mu} X_{\nu]} = \frac12(\partial_{\mu} X_{\nu} - \partial_{\nu} X_{\mu}  )$, and similarly for the wedge product.}
\begin{equation}
\begin{split}
& R^a{}_b = 2 ( d \omega^a{}_b + \omega^{a}{}_c \wedge \omega^c{}_b )\,, \\
& T^a = 2 ( d e^a + \omega^a{}_b \wedge e^b )\,.
\end{split}
\end{equation}
The curvature and torsion two-forms satisfy a set of Bianchi identities, which can be written as
\begin{equation}
\begin{split}
& \cD T^a \equiv d T^a + \omega^a{}_b \wedge T^b = R^a{}_b \wedge e^b \,, \\
& \cD R^a{}_b \equiv d R^a{}_b + \omega^a{}_c \wedge R^c{}_b - \omega^c{}_b \wedge R^a{}_c = 0\,,
\end{split}
\end{equation}
where we have defined $\cD$ as the covariant exterior derivative.

As a final remark on the general noncoordinate basis formalism before moving to three dimensions, we derive the antisymmetry of the spin connection by demanding metric compatibility $\nabla_{\rho} g_{\mu\nu} =0$. In the noncoordinate basis, the metric is simply the flat Minkowski metric $\eta_{ab}$ and hence
\begin{equation}
\nabla_{\rho} \eta_{ab} = \partial_{\rho} \eta_{ab} - \omega_{\mu}{}^c{}_a \eta_{cb}  - \omega_{\mu}{}^c{}_b \eta_{ac} = 0 \,,
\end{equation}
implies
\begin{equation}
\omega_{\mu\,ab} = - \omega_{\mu\, ba}\,.
\end{equation}
This property, together with vanishing torsion $T^a = 0$, allows one to solve the spin connection explicitly in terms of first-order derivatives acting on the vielbein $e^a$.

\subsection{Chern-Simons gauge theory}
Now that we have established a noncoordinate basis it is time to move to three dimensions and show the relation between General Relativity and Chern-Simons gauge theories \cite{Achucarro:1987vz,Witten:1988hc}.

In three spacetime dimensions, we can use the totally antisymmetric epsilon symbol $\ve^{abc}$ (where $\ve^{012} =1$) to write the spin connection one-form with a single Lorentz index, defining the (Hodge) dualised spin connection
\begin{equation}
\omega^a = \frac12 \ve^{abc} \omega_{bc}\,.
\end{equation}
The dualised Riemann curvature can then be defined such that
\begin{equation}
R^a = d\omega^a + \frac12 \ve^{abc} \omega^b \wedge \omega^c\,,
\end{equation}
from which the useful identity
\begin{equation}\label{EinsteinID}
e^{\mu}{}_a \ve^{\nu\rho\sigma}  R_{\rho\sigma}{}^a = \det (e) G^{\mu\nu} \,.
\end{equation}
follows. This identity will be used frequently in this thesis.

The Einstein equations \eqref{einsteineqn} outside local sources can be written in the noncoordinate basis as:
\begin{equation}
R^a - \frac{\Lambda}{2}\ve^{abc} e_b \wedge e_c = 0\,.
\end{equation}
Together with the torsion constraint $T^a =0$, they are the equations of motion obtained by varying the Einstein-Cartan action
 \begin{equation}\label{SEC}
S_{EC} = - \frac{1}{8 \pi G} \int_{\cM} \left\{ e_a \wedge d\omega^a + \frac{1}{2} \varepsilon_{abc} e^a \wedge \omega^b \wedge \omega^c - \frac{\Lambda}{6} \varepsilon_{abc} e^a \wedge e^b \wedge e^c \right\} \,,
\end{equation}
with respect to $e^a$ and $\omega^a$. 

We already discussed in the last section that outside local sources there are no dynamical gravitational degrees of freedom and locally spacetime is flat or of constant curvature, depending on the value of the cosmological constant. The absence of local dynamics means that gravity in three dimensions is completely determined by global effects, hence it is a topological theory. There are, however, gauge symmetries in the theory, the LLTs and the GCTs (or diffeomorphisms). The generators of these gauge symmetries form a Lie algebra, which again depends on the value of the cosmological constant. For flat space ($\Lambda = 0$) these are precisely the isometries of Minkowski spacetime, or the Poincar\'e algebra ($ISO(2,1)$). Negative cosmological constant gives the isometries of Anti-de Sitter (AdS) spacetime ($SO(2,2)$), while for positive $\Lambda$ the Lie algebra is given by the isometries of de-Sitter (dS) spacetime ($SO(3,1)$). Gravity in three dimensions can thus be understood as a topological gauge theory where the gauge group corresponds to the isometry group of the vacuum spacetime. 

Chern-Simons (CS) theories in three dimensions are also topological gauge theories and indeed the Einstein-Cartan action can be written as a CS theory \cite{Achucarro:1987vz,Witten:1988hc}. The CS action is defined as
\begin{equation}\label{SCS}
S_{CS}(A) = \frac{k}{4\pi} \int_{\cM} \tr \left( A \wedge d A + \frac{2}{3} A \wedge A \wedge A \right)\,,
\end{equation}
where $A$ is a Lie algebra valued connection and the trace denotes a non-degenerate bilinear form on the algebra. Consider the connections
\begin{equation}\label{CSconnection}
A = e^a P_a + \omega^a J_a\,,
\end{equation}
where $P_a$ and $J_a$ have the commutators 
\begin{equation}\label{Lievacuum}
[P_a ,P_b] = - \Lambda \ve_{abc} J^c\,, \qquad [P_a,J_b] = \ve_{abc}P^c\,, \qquad [J_a, J_b] = \ve_{abc} J^c\,.
\end{equation}
They correspond to the Lie algebra $SO(2,2)$  for $\Lambda < 0$, $SO(3,1)$ for $\Lambda > 0$ and $ISO(2,1)$ for $\Lambda = 0$. Together with the bilinear form
\begin{equation}
\tr(J_aP_b) = \eta_{ab}\,,
\end{equation}
we can verify that the Chern-Simons action \eqref{SCS} for the connections \eqref{CSconnection} reduces to the three-dimensional Einstein-Cartan action \eqref{SEC} when the CS-level $k$ is given by
\begin{equation}
k = \frac{1}{4G}\,.
\end{equation}

\subsubsection{CS theory in Anti-de Sitter spacetime}
When the cosmological constant is negative, the maximally symmetric constant cur\-vature background is Anti-de Sitter spacetime, for which $\Lambda = - \frac{1}{\ell^2}$ with $\ell$  the AdS length. The isometry group is $SO(2,2)$ which is isomorphic to $SL(2,\mathbb R) \times SL(2, \mathbb R)$. In this case, the EC action can equivalently be written as a Chern-Simons action \eqref{SCS} for the connections
\begin{equation}
A = A^{+\,a} J^+_a + A^{-\,a} J^-_a = A^+ + A^-\,.
\end{equation}
Here $J^{\pm}_a= \frac12 (J_a \pm \ell P_a)$ are mutually commuting generators of $SL(2,\mathbb R)$ with commutators
\begin{equation}
[J^{\pm}_a ,J^{\pm}_b] = \ve_{abc}J^{\pm\,c}\,, \qquad [J^+_a , J^-_a] = 0\,,
\end{equation}
and the bilinear forms are
\begin{equation}
\tr(J^+_a J^+_b) = \frac12 \eta_{ab}\,, \qquad \tr(J^-_aJ^-_b) = - \frac12 \eta_{ab}\,, \qquad \tr(J^+_a J^-_b) = 0\,.
\end{equation}
Now the Chern-Simons action \eqref{SCS} splits into two parts
\begin{equation}\label{CSpm}
\begin{split}
S_{CS} (A) & = S_{CS} (A^+) + S_{CS}(A^-)\,, \\
& = \frac{k}{4\pi} \int_{\cM} \left\{ A^+_a \wedge d A^{+\,a} +  \frac13 \ve_{abc} A^{+\,a} \wedge A^{+\,b} \wedge A^{+\,c} \right\} \\
& \;\; -  \frac{k}{4\pi} \int_{\cM} \left\{ A^-_a \wedge d A^{-\,a} +  \frac13 \ve_{abc} A^{-\,a} \wedge A^{-\,b} \wedge A^{-\,c} \right\} \,.
\end{split}
\end{equation}
Note that this is the difference of two Chern-Simons actions. If the connections $A^{\pm\,a}$ are given in terms of the dreibein and the spin connection as
\begin{equation}\label{AdSconnections}
A^{\pm\,a} = \omega^a \pm \frac{1}{\ell} e^a\,,
\end{equation}
then \eqref{CSpm} also reduces to the Einstein-Cartan action \eqref{SEC}, now with the CS-level
\begin{equation}
k = \frac{\ell}{4G}\,.
\end{equation}
Assuming an invertible dreibein, the Einstein equation, together with the vanishing of the torsion two-form, can now be written as zero field-strength conditions for the connection $A^{\pm}$.

This formulation of three dimensional AdS gravity is more convenient when studying the asymptotic symmetries (the topic of section \ref{sec:ECASG}). Furthermore, this theory can be extended to gauge theories of $SL(N,\mathbb R) \times SL(N, \mathbb R)$ and eventually $hs[\lambda] \oplus hs[\lambda]$ to describe massless bosonic higher spin fields coupled to gravity in AdS$_3$, see for instance \cite{Aragone:1983sz,Blencowe:1988gj,Bergshoeff:1989ns,Vasiliev:2003ev,Henneaux:2010xg,Campoleoni:2010zq,Campoleoni:2011hg,Gaberdiel:2010pz}.

\section{Hamiltonian Analysis}
\label{sec:ECHamiltonian}
We continue our venture into three dimensional GR by studying the theory following Dirac's procedure for constraint Hamiltonian systems. The fact that the theory is essentially a Chern-Simons gauge theory makes its Hamiltonian formulation particularly simple and allows us to go through the procedure relatively fast (as compared to the higher-dimensional Hamiltonian analysis of GR). Still, some subtleties arise when we consider the theory on manifolds with a boundary, which will be the topic of the next section. The main purpose of this section is to verify explicitly that the theory contains no local degrees of freedom and to see the generators of the gauge symmetries arise in a very natural way. 

We take as a starting point the Einstein-Cartan Lagrangian three-form \eqref{SEC}, dropping the overall factor of $\frac{1}{8\pi G}$ for the moment.\footnote{Here and in the rest of this thesis we will define the Lagrangian three-form $L$ as $S = \frac{1}{8\pi G}\int L$ while the Lagrangian density $\cL$ is defined as $S = \frac{1}{8\pi G} \int \det(e) d^3 x \, \cL$.}
\begin{equation}\label{LEC}
L_{EC} = - \left\{ e \cdot d\omega + \frac{1}{2} e \cdot \omega \times \omega - \frac{\Lambda}{6}  e \cdot e \times e \right\} \,.
\end{equation}
Here and in the rest of this thesis, the wedge products are implicit and the Lorentz indices $a,b,c,\ldots$ are suppressed by using a 3D vector notation in which contractions with $\eta_{ab}$ and $\ve_{abc}$ are represented by dots and crosses respectively. So, for instance, $\ve_{abc} e^a \wedge e^b \wedge e^c$ is denoted as $e \cdot e \times e$. We decompose the dreibein $e{}^a$ and spin-connection $\omega^a$ into its space and time components:
\begin{equation}
e^a = e_0{}^a dt + e_i{}^a dx^i\,, \qquad 
\omega^a = \omega_0{}^a dt + \omega_i{}^a dx^i\,.
\end{equation}
The Lagrangian density becomes:
\begin{equation}\label{LECdecomp}
\cL = \varepsilon^{ij}  e_{i} \cdot \dot{\omega}_{j} + e_{0} \cdot \mathcal P + \omega_{0} \cdot \mathcal J  \,,
\end{equation}
where $\ve^{ij} = \ve^{0ij}$ and $\mathcal P^a$ and $\mathcal J^a$ are defined as:
\begin{equation}
\begin{split}\label{ECpc}
\mathcal P^a & = - \ve^{ij} \left( \partial_i \omega_{j}{}^a + \frac12 (\omega_i \times \omega_j)^a - \frac12\Lambda (e_i \times e_j)^a \right) \,, \\
\mathcal J^a & = - \ve^{ij} \left( \partial_i e_{j}{}^a + ( \omega_i \times e_j)^a \right)  \,,
\end{split}
\end{equation}
The Lagrangian density \eqref{LECdecomp} is first order in time derivatives. Therefore, the canonical momenta do not involve $\dot{q}$'s ($q$ being $(e^a,\omega^a)$ in this case). In fact, the canonical momenta belonging to the spatial parts $e_i^a$ and $\omega_i^a$ are proportional to $\omega_i^a$ and $e_i^a$ respectively. This implies that introducing canonical momenta for the spatial variables is redundant, since an equally good way to describe them is using the spatial parts of the fields themselves. Furthermore, the canonical momenta of the time components are zero. This implies that the time components of the fields are not dynamical and hence do not contribute to the physical phase-space of the theory. We may interpret them as Lagrange multipliers enforcing the constraints $\mathcal P_a =0$ and $\mathcal J_a=0$ and consider the spatial part of the fields $(e_i{}^a, \omega_i{}^a)$ to be the canonical variables of the theory.

The Poisson brackets of the canonical variables read:
\begin{equation}
 \PB{e_{i}{}^a(x)}{\omega_{j}{}^{b}(y)} = - \varepsilon_{ij} \eta^{ab} \delta^{(2)}(x-y)\,.
\end{equation}
The Hamiltonian is just the sum of the primary constraints \eqref{ECpc} enforced by a set of Lagrange multipliers
\begin{equation}
\cH = -  e_{0} \cdot \mathcal P - \omega_{0} \cdot \mathcal J\,.
\end{equation}
The Dirac procedure for constraint Hamiltonian system dictates that we check the consistency of the primary constraints under time evolution. This is done by computing the Poisson brackets of the primary constraints with the Hamiltonian. Since our Hamiltonian is just the set of constraints, this amounts to computing the Poisson brackets of the constraints with themselves. It is convenient to define the smeared operators $\mathcal J[\xi]$ and $\mathcal P[\xi]$ by integrating the constraint functions with an arbitrary Lorentz vector $\xi^a(x)$, 
\begin{equation}\label{ECsmeared}
\mathcal J[\xi] = \int_{\Sigma} d^2x \; \xi(x)\cdot \mathcal J(x)\,, \qquad 
\mathcal P[\xi] = \int_{\Sigma} d^2x \; \xi(x) \cdot \mathcal P(x)\,,
\end{equation}
where $\Sigma$ is a space-like hypersurface. For the moment we will assume that $\xi^a$ will vanish at spatial infinity such that we can ignore boundary terms when we integrate by parts. In the next section we will relax this assumption and generalize the discussion to manifolds with a boundary and non-trivial boundary conditions.

The Poisson brackets of the smeared operators \eqref{ECsmeared} form the algebra
\begin{equation}
\begin{split} \label{PBalgebra}
\PB{\mathcal J[\xi]}{\mathcal J[\eta]} & =  \mathcal J[[\xi,\eta]]  \,, & \PB{\mathcal P[\xi]}{\mathcal J[\eta]} & =  \mathcal P[[\xi,\eta]]  \,, \\
\PB{\mathcal P[\xi]}{\mathcal P[\eta]} & = - \Lambda \, \mathcal J[[\xi,\eta]]   \,,
\end{split}
\end{equation}
where $[\xi,\eta] = \xi \times \eta$.  This is, of course, exactly the Lie algebra of the isometries of the vacuum spacetime \eqref{Lievacuum}, or the $SO(2,2)$ group for AdS ($\Lambda < 0$), $SO(3,1)$ for dS ($\Lambda > 0$) and $ISO(2,1)$ for flat space ($\Lambda = 0$).

The Poisson bracket algebra \eqref{PBalgebra} implies that both the constraint functions $J_a$ and $P_a$ are (primary) first-class and hence generate gauge symmetries (up to possible boundary terms). There are no secondary or second-class constraints. The counting of the dimension of the physical phase space is then
\begin{equation}
2 \times 6 \left\{ \begin{array}{c} \text{canonical} \\ \text{variables} \end{array} \right\} 
- 6 \left\{ \begin{array}{c} \text{primary} \\ \text{constraints} \end{array} \right\} 
- 6 \left\{ \begin{array}{c} \text{gauge} \\ \text{symmetries} \end{array} \right\} 
= 0\,.
\end{equation}
So there are no local degees of freedom.
Since the constraint functions are first-class, they generate gauge transformations. Indeed, when we calculate the Poisson brackets of the smeared constraints with the fields we find that $\mathcal J[\xi]$ generates local Lorentz transformations with gauge parameter $\xi^a$
\begin{equation}
\begin{split}
\delta_{\mathcal J} e_i^a & = \PB{e_i{}^a}{\mathcal J[\xi]} = - (e_{i} \times \xi)^a \,,\\
\delta_{\mathcal J} \omega_i^a & = \PB{\omega_i{}^a}{\mathcal J[\xi]} = - \partial_i \xi^a - (\omega_{i} \times \xi)^a \,,\\
\end{split}
\end{equation}
And $\mathcal P[\xi]$ generates the local translations
\begin{equation}
\begin{split}
\delta_{\mathcal P} e_i^a & = \PB{e_i{}^a}{\mathcal P[\xi]} = - \partial_i \xi^a - (\omega_{i} \times \xi)^a  \,,\\
\delta_{\mathcal P} \omega_i^a & = \PB{\omega_i{}^a}{\mathcal P[\xi]} = - \Lambda (e_{i} \times \xi)^a \,,\\
\end{split}
\end{equation}
When the gauge parameter $\xi^a$ is taken to be proportional to the Dreibein $\xi^a = e_{\mu}{}^a \zeta^{\mu}$, then these transformations reduce to the standard transformation rule for diffeomorphisms on-shell, up to a local Lorentz transformation with gauge parameter $\omega_{\mu}{}^a \zeta^{\mu}$ \cite{Witten:1988hc}. So if we define a new constraint function $\phi[\zeta]$ as
\begin{equation}
\phi[\zeta] =\mathcal P[e_{\mu}{}^{a}\zeta^{\mu}] + \mathcal J[\omega_{\mu}{}^a \zeta^{\mu}]\,.
\end{equation}
Then $\phi[\zeta]$ generates diffeomorphisms on-shell, as it can be verified that
\begin{equation}
\begin{split}
\PB{\phi[\zeta]}{e_i{}^a} & = \cL_{\zeta} e_{i}{}^a + \text{equations of motion}\,,\\
\PB{\phi[\zeta]}{\omega_i{}^a} & = \cL_{\zeta} \omega_{i}{}^a + \text{equations of motion}\,.
\end{split}
\end{equation}
Here $\cL_\zeta$ denotes the Lie derivatives with respect to $\zeta^\mu$: $\cL_\zeta X_{\nu} = \zeta^{\mu} \partial_{\mu}X_{\nu} + X_{\mu} \partial_{\nu} \zeta^{\mu}$.

We have now analyzed the bulk theory and saw how the generators of the gauge symmetries arise as first-class constraints in the Hamiltonian analysis. In the analysis so far, we have ignored boundary terms. We will now consider these terms and show that the presence of a boundary introduces boundary degrees of freedom, which fall into representations of a conformal algebra on the boundary. The main point is that on manifolds with a boundary, the generators of gauge transformations need to be improved with a boundary term. This implies that at the boundary, the generators are no longer first-class and represent a symmetry transformation, connecting physically distinct configurations. 

\section{Asymptotic Symmetries}
\label{sec:ECASG}

It was shown in \cite{Brown:1986nw} by Brown and Henneaux that gravity in AdS$_3$ with non-trivial boundary conditions contains excitations at the conformal boundary. These boundary degrees of freedom fall into representations of the asymptotic symmetry algebra which is generated by global symmetry transformations which preserve the AdS$_3$ (or Brown-Henneaux) boundary conditions. In this section we will review the Brown-Henneaux argument and show that the asymptotic symmetry algebra for three dimensional AdS gravity consists of two copies of the Virasoro algebra with a classical central extension.

Of course, there are many ways to arrive at the expression for the central charge in three dimensional AdS gravity. Starting from the second-order formulation, one could look at the anomalous transformation of the renormalized boundary stress-tensor, as in \cite{Kraus:2005zm}. Or we could first write 3D gravity as a pure Chern-Simons theory and employ the methods described in \cite{Banados:1994tn}. Here we prefer to stay close to the description of three dimensional gravity in terms of the dreibein $e^a$ and the spin connection $\omega^a$, since we will later investigate the asymptotic symmetries of modified theories of gravity in three dimensions, which do contain propagating degrees of freedom in the bulk. The analysis here will loosely follow the lines of Carlip's analysis of Topological Massive Gravity in \cite{Carlip:2008qh}, applied to the Einstein-Cartan theory defined by \eqref{SEC} and supplemented with results obtained and reviewed in \cite{Banados:1994tn,Banados:1998gg}. This will aid us a great deal in chapter \ref{chapter:asymptotic_symm}, where we calculate the asymptotic symmetries and central charges of various theories of massive gravity in three dimensions. For more details on asymptotic symmetries in three dimensional gravity and the relation to a conformal field theory, see  \cite{Carlip:1994hq,Coussaert:1995zp,Banados:1998pi,Banados:1998ta,Carlip:2005zn,Carlip:2005tz,Witten:2007kt,Maloney:2007ud}.

When we consider the theory in AdS$_3$ with $\Lambda = - \frac{1}{\ell^2}$, we can define the new smeared operators $L_{\pm}[\xi]$ as
\begin{equation}
L_{\pm}[\xi] = \mathcal P[\xi] \pm \frac{1}{\ell} \mathcal J[\xi]\,,
\end{equation}
whose algebra splits into two mutually commuting sectors, up to boundary terms
\begin{equation} \label{sl2sl2}
\begin{split}
\{ L_{\pm}[\xi], L_{\pm} [\eta]\} & = \pm \frac{2}{\ell} L_{\pm}[[\xi, \eta]]\,, \\
\{ L_{+}[\xi], L_{-} [\eta]\} & = 0\,. 
\end{split}
\end{equation}
Of course, if had made the field redefinitions \eqref{AdSconnections} in the Einstein-Cartan action \eqref{SEC} and then derive the first-class constraint functions, we have would arrived at this expression for the algebra of Poisson brackets immediately. This is due to the fact that the isometry group of AdS$_3$ is isomorphic to $SL(2,\mathbb R) \times SL(2,\mathbb R)$ and the absence of local gravitational degrees of freedom in the bulk allow us to make this split into mutually commuting sectors everywhere. Later in this thesis, we will modify the CS theory in the bulk to a theory which does contain local degrees of freedom and it will no longer be possible to perform this split everywhere in the bulk spacetime. However, we will see that with the appropriate boundary conditions, this redefinition will still work at the boundary of AdS$_3$.

In computing the Poisson brackets of the primary constraints \eqref{sl2sl2} we have ignored any boundary term and implicitly assumed that the smeared operators are differentiable, i.e. that their derivative with respect to the canonical variables is well-defined. On manifolds with a boundary a boundary term may arise when we vary the smeared operators with respect to the fields
\begin{equation}
\delta L_{\pm} [\xi] = \int_{\Sigma} d^2x \; \xi_a \frac{\partial L_{\pm}^a}{\partial a_i^b}  \delta a^b_i + \int_{\partial \Sigma} d\phi B_{\pm}[\xi,a,\delta a]\,.
\end{equation}
Here $a_i^a = ( e_i{}^a, \omega_i{}^a)$ and $\partial \Sigma$ denotes the boundary of the space-like hypersurface $\Sigma$, parametrised by the coordinate $\phi$. The presence of a non-zero $B_{\pm}[\xi,a,\delta a]$ could lead to delta-function singularities in the Poisson brackets. To remove these, we can define the improved operator $\tilde{L}_{\pm}[\xi]$ as
\begin{equation}\label{Limproved}
\tilde{L}_{\pm} [\xi] = L_{\pm} [\xi] + Q_{\pm}\,,
\end{equation}
where $Q_{\pm}$ is defined such that its variation cancels the boundary term coming from the variation of $L_{\pm}[\xi]$
\begin{equation}
\delta Q_{\pm} = - \int_{\partial \Sigma} d\phi \, B_{\pm}[\xi,a,\delta a]\,.
\end{equation}
Explicit computation of $B_{\pm}[\xi,a,\delta a]$ gives
\begin{equation}\label{delQ}
\delta Q_{\pm} =   \int_{\partial \Sigma} d\phi \; \xi \cdot \delta \left( \omega_{\phi} \pm \frac{1}{\ell} e_{\phi} \right) \,.
\end{equation}
After we specify the boundary conditions and the symmetry transformations which preserve these boundary conditions, it is possible to integrate this expression to obtain the boundary charges $Q_{\pm}$. 

Now that the improved operators have a well-defined variation, we must go back to the Poisson brackets \eqref{sl2sl2} and keep track of the boundary terms. They are:
\begin{equation}\label{ECcentralext}
\{ \tilde{L}_{\pm}[\xi], \tilde{L}_{\pm}[\eta] \} = \ldots \pm \frac{2}{\ell} \int d\phi\, \xi \cdot \left[ \partial_{\phi} \eta + \left( \omega_{\phi} \pm \frac{1}{\ell} e_{\phi}\right) \times \eta \right]\,,
\end{equation}
where the dots denote the bulk part. The boundary term in the Poisson brackets will in general provide a term proportional to $Q_{\pm}(\xi,\eta)$, which is required to improve the bulk part, and a term $K(\xi,\eta)$ which is independent of the fields $e$ and $\omega$ and provides a central extension term in the algebra of Poisson brackets \cite{Banados:1994tn}.

The above considerations show that the generators of gauge symmetries pick up boundary terms $Q$. In general, this implies that the generators of gauge symmetries become second-class at the boundary, as their Poisson brackets do not vanish on the constraint surface due to the boundary term in \eqref{ECcentralext}. The boundary terms $Q$ do not generate gauge symmetries, but they generate (global) symmetry transformation on the space of physical states; they are conserved global charges which take different values for each state. We will now continue the analysis by considering the algebra of global charges which preserve the asymptotic structure of the AdS$_3$ vacuum, i.e. the boundary terms related to diffeomorphisms which do not change leading behavior towards the boundary of AdS$_3$.  

As we saw at the end of the last section, the gauge parameters for diffeomorphisms are proportional to the dreibein $\xi^a = e_{\mu}{}^a \zeta^{\mu}$. In general the algebra \eqref{sl2sl2} will then pick up extra terms proportional to $\{L_{\pm}[\xi], e_{i}^a\}$. 
\begin{equation}
\begin{split}\label{PBdiffs}
\{ L_{+}[\xi], L_{+} [\eta]\} & =  L_{+}\left[ \{L_{+}[\xi], e_{i}^a\}\chi^i - \{L_{+}[\eta], e_{i}^a\}\zeta^i  + \frac{2}{\ell} [\xi, \eta]^a \right]\,, \\
\{ L_{-}[\bar{\xi}], L_{-} [\bar{\eta}]\} & =  L_{-}\left[ \{L_{-}[\bar{\xi}], e_{i}^a\}\bar{\chi}^i - \{L_{-}[\bar{\eta}], e_{i}^a\}\bar{\zeta}^i  - \frac{2}{\ell} [\bar{\xi}, \bar{\eta}]^a \right]\,, \\
\{ L_{+}[\xi], L_{-} [\bar{\eta}]\} & = L_- \left[ \{L_{+} [\xi], e_{i}^a\}\bar{\chi}^i \right] - L_+ \left[\{L_{-}[\bar{\eta}], e_{i}^a\}\zeta^i \right] \,,
\end{split}
\end{equation}
where $\eta^a = e_{\mu}{}^a \chi^\mu$ and similarly with the barred quantities.
These terms will not vanish in the bulk, however it is possible to find a set of symmetry transformations which preserves the asymptotic structure of the AdS space, i.e. for which the symmetry transformation $\delta e_{i}{}^a = \{L_{\pm}[\xi], e_{i}^a\}$ does not introduce any new contributions to the boundary conditions. In addition, the terms appearing in the last line of \eqref{PBdiffs} should vanish asymptotically.

\subsection{Asymptotically Anti-de Sitter Boundary Conditions}
In 3D GR, spacetimes with asymptotic AdS$_3$ boundary conditions can be parametrized in terms of two arbitrary (state-dependent) functions $\cL(x^+)$ and $\bar{\cL}(x^-)$ and the metric can be written in light-cone coordinates as \cite{Banados:1998gg}
\begin{equation}
\begin{split}\label{aads_metric}
ds^2 =  \ell^2 \bigg\{d\rho^2 - \cL(x^+)(dx^+)^2 - \bar{\cL}(x^-) (dx^-)^2 \qquad \qquad \qquad \qquad  \\ 
- \left( e^{2\rho} + e^{-2\rho}\cL(x^+) \bar{\cL}(x^-) \right) dx^- dx^+  
\bigg\}\,.
\end{split}
\end{equation}
This metric is an exact solution to the Einstein equations in 3D.\footnote{The fact that \eqref{aads_metric} is an exact solution implies that the Fefferman-Graham expansion of AdS$_3$ is finite.} 
It reduces to the BTZ black hole of \cite{Banados:1992wn} with mass $M$ and angular momentum $J$ when we take $\cL$ and $\bar{\cL}$ constant. Explicitly, when $\cL$ and $\bar{\cL}$ are
\begin{equation}
\cL = \frac{2G}{\ell} (J - \ell M)\,, \qquad \bar{\cL} = - \frac{2G}{\ell} (J + \ell M)\,,
\end{equation}
then \eqref{aads_metric} can be written as \cite{Banados:1998gg}
\begin{equation}
ds^2 = - N^2 dt^2 + N^{-2} dr^2 + r^2 \left( d\phi - N^{\phi} dt \right)^2\,, \label{BTZ}
\end{equation}
where
\begin{equation}
N^2(r) = -8 M G + \frac{r^2}{\ell^2} + \frac{16 G^2 J^2}{r^2} \,, \qquad 
N^{\phi}(r) = \frac{4 G J}{r^2}\,.
\end{equation}
The coordinates $(x^{\pm},\rho)$ are related to $(t,\phi,r)$ as
\begin{equation}
\begin{split}
& x^{\pm} = \frac{t}{\ell} \pm \phi\,, \\ 
& r^2 = r_+^2 \cosh^2[ \rho - \rho_0] - r_-^2 \sinh^2[\rho - \rho_0] \,.
\end{split}
\end{equation}
Here $e^{2\rho_0} = (r_+^2 - r_-^2)/ 4 \ell^2 $ and $r_\pm$ are the radii of the BTZ black hole; they are the two solutions to $N^2(r) = 0$.

When $J=0$ and $8 M G = -1$ (or equivalently, when $\cL =\bar{\cL} = 1/4$), the BTZ solution \eqref{BTZ} reduces to anti-de Sitter space. Since the BTZ metric solves the vacuum Einstein equations with negative cosmological constant, it describes a space of constant negative curvature. Locally, the BTZ solution is isometric to anti-de Sitter space. More precisely, as was shown in \cite{Banados:1992gq}, the BTZ black hole can be obtained from identifications of points of AdS$_3$ by a discrete subgroup of $SO(2,2)$.
 
In general, the field equations do not demand that $\cL $ and $\bar{\cL}$ are constant, but rather that $\partial_- \cL = 0$ and $\partial_+ \bar{\cL} = 0$. Solutions with different $\cL$ and $\bar{\cL}$ are related by a symmetry transformation that would have been a gauge transformation, if it was not for the boundary terms in the generators \eqref{Limproved}. Our job now is to find the algebra of symmetry transformations which preserve the metric \eqref{aads_metric}, changing only the values of $\cL$ and $\bar{\cL}$. To this end it convenient to express \eqref{aads_metric} in terms of the connections $A^{\pm\,a} = \omega^a \pm \frac{1}{\ell} e^a$
\begin{equation}
\begin{split}\label{BHbc1}
& A^{+\,0} = ( e^\rho +\cL(x^+) e^{-\rho} ) d x^+\,, \\
& A^{+\,1} = ( e^\rho -\cL(x^+) e^{-\rho} ) d x^+\,, \\
& A^{+\,2} = d \rho\,,
\end{split}
\end{equation}
and
\begin{equation}
\begin{split}\label{BHbc2}
& A^{-\,0} = -( e^\rho +\bar{\cL}(x^-) e^{-\rho} ) d x^-\,, \\
& A^{-\,1} = ( e^\rho - \bar{\cL}(x^-) e^{-\rho} ) d x^-\,, \\
& A^{-\,2} = - d \rho\,.
\end{split}
\end{equation}
It is not so hard to verify that the metric constructed from $e^a = \frac{\ell}{2}( A^{+\,a} - A^{-\,a})$ is \eqref{aads_metric}. 

The set of transformations which preserves the asymptotic structure of boundary conditions \eqref{BHbc1} and \eqref{BHbc2} is given in terms of the gauge parameters $\xi_f^a$ and $\bar{\xi}_{\bar{f}}^a$ and depends on two arbitrary functions $f(x^+)$ and $\bar{f}(x^-)$ as
\begin{equation}\label{asymptdiffs}
\begin{split}
& \xi_f^a = \frac{\ell}{2} \left( f e^{\rho} + e^{-\rho} \left( f \cL + \frac12 f'' \right),  f e^{\rho} - e^{-\rho} \left( f \cL + \frac12 f'' \right) , - f' \right)\,, \\
& \bar{\xi}_{\bar{f}}^a = \frac{\ell}{2} \left( \bar{f} e^{\rho} + e^{-\rho} \left( \bar{f} \bar{\cL} + \frac12 \bar{f}'' \right),  - \bar{f} e^{\rho} + e^{-\rho} \left( \bar{f} \bar{\cL} + \frac12 \bar{f}'' \right) , - \bar{f}' \right)\,.
\end{split}
\end{equation}
Here a prime denotes partial derivation with respect to the functions argument. It is straightforward to check that a symmetry transformation $e \to e + \delta_{+} e + \delta_{-} e$ with 
\begin{equation} \label{diff}
\begin{split}
\delta_+ e_{i} = \{ L_{+}[\xi_f],e_i\} = \left( \partial_i \xi_f +   \left(\omega_{i} + \frac{1}{\ell} e_{i} \right) \times \xi_f \right) \,, \\
\delta_- e_{i} = \{ L_{-}[\bar{\xi}_{\bar f}],e_i\} = \left( \partial_i \bar{\xi}_{\bar f} +  \left(\omega_{i} - \frac{1}{\ell} e_{i} \right) \times \bar{\xi}_{\bar f} \right) \,,
\end{split}
\end{equation}
again gives a metric of the form \eqref{aads_metric}, but with $\cL' = \cL + \delta \cL$ and $\bar{\cL}' = \bar{\cL} + \delta \bar{\cL}$, where
\begin{equation}
\begin{split}
& \delta \cL(x^+) = f(x^+) \cL'(x^+) + 2 f'(x^+)\cL(x^+) + \frac12 f'''(x^+)\,, \\
& \delta \bar{\cL}(x^-) = \bar{f}(x^-) \bar{\cL}'(x^-) + 2 \bar{f}'(x^-)\bar{\cL}(x^-) + \frac12 \bar{f}'''(x^-) \,.
\end{split}
\end{equation}
These are the transformation laws for the holomorphic and anti-holomorphic part of a stress-energy tensor in a conformal field theory.

As an aside, let us briefly remark that the transformations given in \eqref{asymptdiffs} are related to the Killing vectors $\zeta$ which generate the most general diffeomorphisms preserving the asymptotic form of the AdS$_3$ metric \eqref{aads_metric}. From the last section, we know that in the first-order formalism diffeomorphisms are related to local translations with parameters proportional to the dreibein. By invertibility of the dreibein we find the asymptotic Killing vectors $\zeta = (\zeta^{\mu} + \bar{\zeta}^{\mu})\partial_{\mu}$ using $\zeta^{\mu} = e^{\mu} \cdot \xi_f$ and $\bar{\zeta}^{\mu} = e^{\mu} \cdot \bar{\xi}_{\bar{f}}$ to be
\begin{equation}
\begin{split}
\zeta = & \left( f(x^+) + \frac{e^{-2\rho}}{2}\partial_-^2 \bar{f}(x^-) + \cO (e^{-4\rho}) \right) \partial_+  \\
& + \left( \bar{f}(x^-) + \frac{e^{-2\rho}}{2}\partial_+^2 f(x^+) + \cO (e^{-4\rho}) \right) \partial_+ \\
& - \frac12 \left( \partial_+ f(x^+) + \partial_- f(x^-) \right) \partial_{\rho} \,.
\end{split}
\end{equation}
This is consistent with the original Brown-Henneaux results \cite{Brown:1986nw}, which shows that although this approach is different, we are really doing the same thing.

After plugging \eqref{asymptdiffs} into the bulk part of the Poisson brackets, given in \eqref{PBdiffs}, they become
\begin{equation}
\begin{split}\label{Lbrack}
\{ L_{+}[\xi_f], L_{+} [\xi_g]\} & =  - L_{+}[\xi_{\{f,g\}}]  \,, \\
\{ L_{-}[\bar{\xi}_{\bar{f}}], L_{-} [\bar{\xi}_{\bar{g}}]\} & = - L_{-}[\bar{\xi}_{\{\bar{f},\bar{g}\}}] \,,
\end{split}
\end{equation}
where  $ \{f,g\}  = f \partial_{\phi} g - g \partial_{\phi} f$. We now have all the ingredients to integrate the boundary charges \eqref{delQ} to find
\begin{equation}
Q_{+} =  \ell \int d\phi \; \cL(x^+) f(x^+) \,, \qquad
Q_{-} =  - \ell \int d\phi \;  \bar{\cL}(x^-) \bar{f}(x^-) \,.
\end{equation}
Finally, after plugging the restricted gauge transformations \eqref{asymptdiffs} and the Brown-Henneaux boundary conditions \eqref{BHbc1}-\eqref{BHbc2} in the boundary term of the Poisson brackets of the improved generators \eqref{ECcentralext}, we find that
\begin{equation}
\begin{split}
\{ \tilde{L}_{+}[\xi_f], \tilde{L}_{+}[\xi_g] \} = \ldots & - \ell \int d\phi  f \left(2 \cL \partial_+ g + g \partial_+ \cL \right)  + \frac{\ell}{2} \int d\phi\, \partial_{+} f \partial_{+}^2 g \,, \\
\{ \tilde{L}_{-}[\bar{\xi}_{\bar{f}}], \tilde{L}_{-}[\bar{\xi}_{\bar{g}}] \} = \ldots & + \ell \int d\phi  \bar{f}\left(2 \bar{\cL} \partial_- \bar{g} + \bar{g}\partial_- \bar{\cL}\right)  - \frac{\ell}{2} \int d\phi\, \partial_- \bar{f} \partial_{-}^2 \bar{g} \,.
\end{split}
\end{equation}
The dots denote the bulk part \eqref{Lbrack} and the first term on the r.h.s. may be recognized as exactly the boundary charge needed to improve the r.h.s. of \eqref{Lbrack}. The remaining term is a central extension. The bulk part of this algebra vanishes on the constraint surface, hence, after replacing the Poisson brackets by Dirac brackets, only the boundary parts survive. 
After a Fourier expansion of the charges $L_m = -\tilde{L}_{+}[f = e^{im x^+}]$ and $\bar{L}_m = -\tilde{L}_{-}[\bar{f} = e^{imx^-}]$ and reinstating the factor of  $1/ (8 \pi G)$ the boundary part of the algebra reduces to two copies of the Virasoro algebra with a classical central charge\footnote{The more conventional form of the central extension term: $\tfrac{c}{12}m^2(m-1) \delta_{m+n,0}$ can be obtained by a shift in the $L_0$ mode: $L_0 \to L_0 - c/24 $}
\begin{equation}
\begin{split}
i \{L_m,L_n\} = (m-n) L_{m+n} + \frac{c_L}{12} m^3 \delta_{m+n,0}\,, \\
i \{\bar{L}_m, \bar{L}_n\} = (m-n) \bar{L}_{m+n} + \frac{c_R}{12} m^3 \delta_{m+n,0}\,.
\end{split}
\end{equation}
The central charges are given by
\begin{equation}
c_{L/R} = \frac{3 \ell}{2G}\,.
\end{equation}
This is the celebrated result obtained by Brown and Henneaux \cite{Brown:1986nw}. 

This concludes the review of the asymptotic symmetry group and the boundary charges of three dimensional gravity with asymptotically Anti-de Sitter boundary conditions. As was mentioned before, the results reviewed here will become useful in the coming chapters. Most of the theories of massive gravitons discussed in the rest of this thesis are also invariant under local Lorentz transformations and diffeomorphisms and allow for Anti-de Sitter background solutions. Hence, with the same boundary conditions, we will find the same asymptotic symmetry group, only with a modified expression for the central charge. The trick then becomes to identify the first class constraints that generate the asymptotic symmetries of the theory in question and to calculate their boundary contribution to the central charge. That will be the main focus of chapter \ref{chapter:asymptotic_symm}. However, before we get too far ahead of ourselves, let us first investigate how to add degrees of freedom to three dimensional GR without adding too many, or how to construct ghost-free theories of massive gravitons in three dimensions.

%% file: chapter_3/chapter_3.tex
\pagestyle{empty}
\setcounter{chapter}{2}

\chapter[Introducing Degrees of Freedom in 3D Gravity]{Introducing Degrees of Freedom in Three Dimensional Gravity}
\label{chapter:two_company}

\begin{quote}
\em{After looking at pure gravity in three dimensions, we now turn to the problem of adding local degrees of freedom in the bulk in such a way to desribe the two degrees of freedom of a massive spin-2. We look at several different approaches, most of which can be written in a Chern-Simons--like form. After defining what is meant precisely by Chern-Simons--like, we review the free massive spin-2 particle in a curved, maximally symmetric background, as described by Fierz-Pauli theory. We then proceed to discuss non-linear extensions of the free theory, based on parts of the review {\sc [v]}. We then consider different known higher-derivative theories of massive gravity and their CS--like formulation, including Topologically Massive Gravity, New Massive Gravity and a parity violating combination of the two, called General Massive Gravity. The last part of this chapter is devoted to a treatment of Zwei-Dreibein Gravity and contains work which has previously been published in \textsc{[vi,vii]}.}
\end{quote}

\newpage
\pagestyle{headings}

\section{Introduction}\label{sec:introductionc3}
In the last chapter we discussed three dimensional gravity and its relation to Chern-Simons gauge theory. The theory does not propagate any local degrees of freedom and the global characteristics of the theory can be formulated as a two dimensional conformal field theory on the boundary of AdS$_3$. In this chapter we would like to investigate how to add local degrees of freedom for spin-2 excitations in the bulk, i.e. how to make the three dimensional (pure gauge) graviton massive. 

There are two main approaches which can be followed. We can either add an explicit mass term for the graviton, or generate the mass term by adding higher derivative corrections to GR. The first approach introduces the bulk degrees of freedom by breaking the gauge symmetries of the theory. The latter approach can be achieved in a way which preserves the gauge symmetries. The additional degrees of freedom are introduced by the need to specify more initial data to solve the higher derivative field equations. Both approaches have their own advantages and disadvantages, as will become clear throughout the thesis. 

We saw that the first-order formalism, or the noncoordinate basis, led to a very simple Hamiltonian analysis of the theory and a clear relation with gauge theory. These are features which we would like to preserve in the analysis of three dimensional massive gravity theories, however, since the massive three dimensional spin-2 particles do contain local degrees of freedom, we have to depart from the pure Chern-Simons formulation. The desirable features can be preserved when we consider theories which are ``Chern-Simons--like''. These models are defined by a Lagrangian three-form which is written in terms of a collection of $N$ Lorentz vector valued one-form fields $\{ a_{\mu}^{r\,a}dx^{\mu}\}$ as
\begin{equation}\label{LCSlike}
L = \frac12 g_{rs} a^{r} \cdot da^{s} + \frac16 f_{rst} a^{r} \cdot a^{s} \times a^{t}\,.
\end{equation}
Here the indices $r,s,t,\ldots$ label the different one-forms, which define a ``flavor space'' of fields. The symmetric $g_{rs}$ is a metric on the flavor space of fields and we demand that it is invertible. The totally symmetric collection of coupling constants $f_{rst}$ can be seen as a tensor on this flavor space and its indices can be lowered and raised by the flavor space metric $g_{rs}$ and its inverse. Like before, wedge products are implicit and Lorentz indices are suppressed by denoting contractions with $\eta_{ab}$ and $\ve_{abc}$ as dots and crosses respectively.

For a particular choice of the flavor space metric and tensor when $N=2$ and $a^{1\,a} = \omega^{a}$, $a^{2\,a} = e^a$, the theory reduces to a Chern-Simons gauge theory describing pure gravity in three dimensions.\footnote{Note that we do allow the spin connection $\omega^a$ to be one of the Lorentz vector valued one-forms, even though technically the connection is not a Lorentz vector. The treatment of the general theory only depends on the fact that the fields carry only one Lorentz index and not on its transformation properties under local Lorentz transformations.} In fact, whenever the combinations
\begin{equation}
f^{r}{}_{st} \ve^{a}{}_{bc}\,, \qquad  \eta_{ab}g_{rs}\,, 
\end{equation}
are, respectively, the structure constants of a Lie algebra and a group invariant symmetric tensor on this Lie algebra, then the three-form \eqref{LCSlike} is a pure Chern-Simons three-form. However, the class of Chern-Simons models defined by the action \eqref{SCS} is larger than the CS-models obtained in this way, since we only consider Lorentz vector valued one-forms in the definition of the Chern-Simons--like models. The definition of CS--like theories could be extended to include Lorentz scalar or Lorentz tensor valued one-forms, however, the definition \eqref{LCSlike} is sufficient for our purposes.

For $N>2$, we will continue to suppose that two of the one-forms are a dreibein and a spin connection. There are then many ways to depart from pure three dimensional gravity and this chapter will review some of the known theories of massive gravity which fit the general model defined by \eqref{LCSlike}. But before we try to tackle the problem of giving a mass to the graviton in a non-linear theory, we discuss a free massive spin-2 field in a fixed gravitational background.

\section{Fierz-Pauli Theory}
The action and field equations describing a free massive spin-2 particle were given by Fierz and Pauli in 1939 \cite{Fierz:1939ix}. They followed a field theoretical approach, demanding Lorentz invariance and positivity of energy. In flat backgrounds, these requirements can be rephrased in a group theoretical approach by demanding that the particle states form unitary representations of the Poincar\'e group \cite{Wigner:1939cj,Bargmann:1948ck}. For massive fields of integer spin, these requirements are formulated by the Fierz-Pauli equations. A totally symmetric rank-$s$ tensor $\psi_{\mu_1 \ldots \mu_s}$ describes a massive spin-$s$ field when
\begin{subequations}\label{FPeqns}
\begin{align}
 \left( \Box - \cM^2 \right) \psi_{\mu_1 \ldots \mu_s} & = 0\,, \\
 \partial^{\mu_1} \psi_{\mu_1 \ldots \mu_s} & = 0 \,, \\
 \eta^{\mu_1\mu_2} \psi_{\mu_1\mu_2 \ldots \mu_s} & = 0 \,.
\end{align}
\end{subequations}
A Lagrangian formulation for massive bosonic fields of arbitrary spin-$s$ was presented by Singh and Hagen \cite{Singh:1974qz}, for a review and more references see \cite{Bouatta:2004kk}. Here we will focus on the $s=2$ case. As was mentioned in the introduction, the free massless spin-2 field equations are those of linearized general relativity. Hence, to arrive at the massive spin-2 field equations, we start by considering the linearized Einstein equations and add a mass term for the fluctuations around a maximally symmetric background. Since most of this thesis makes use of the first-order CS--like formulation, we will show in detail how the Fierz-Pauli equations for a massive spin-2 can be derived from a first-order Lagrangian.

\subsection{Linear 3D GR}
To study a free massive spin-2 particle we expand the Einstein-Cartan action \eqref{SEC} around a  maximally symmetric background to second order in fluctuations and add a mass term for those perturbations. The background dreibein and spin connection are parameterized by $\be^a$ and $\bo^a$.  
\begin{equation}\label{perturb}
e^a = \be^a + \kappa k^a\,, \quad \omega^a = \bo^a + \kappa v^a\,,
\end{equation}
where $\kappa$ is a small expansion parameter and $k^a$ and $v^a$ are Lorentz vector valued one-form fields parameterizing fluctuations around the background. The linear terms vanish, as the background dreibein and spin connection satisfy the field equations
\begin{equation}
\begin{split}\label{bkgdeom}
&\bar{R}^a - \frac12 \Lambda  (\be \times \be)^a =  d \bo^a + \frac12 (\bo \times \bo)^a - \frac12 \Lambda  (\be \times \be)^a = 0 \,, \\
& \bar{T}^a = \bar{\cD} e^a = d \be^a +  (\bo \times \be)^a = 0\,.
\end{split}
\end{equation}
Here $\bar{\cD}$ is the covariant exterior derivative with respect to the background spin connection $\bo^a$ and bars denote background quantities. Using these relations it can be shown that any Lorentz-vector one-form $f^a$ satisfies
\begin{equation}\label{linBianchi}
\bar{\cD} \bar{\cD} f^a =  (\bar{R} \times f)^a = \frac12 \Lambda \left((\be \times \be) \times f \right)^a= \frac12 \Lambda \ve^{abc}\ve_{bde} \be^d \be^e f_c\,.
\end{equation}
The action quadratic in the fluctuations \eqref{perturb} is
\begin{equation}\label{ECquad}
S^{(2)} = - \int  \left\{ k \cdot \bar{\cD} v + \frac12  \be \cdot \left( v \times v - \Lambda  k \times k \right) \right\}\,,
\end{equation}
which leads to the linearized field equations
\begin{align} \label{lindv}
& \bar{\cD} v^a - \Lambda  (\be \times k)^a = 0\,, \\
& \bar{\cD} k^a + (\be \times v)^a = 0\,. \label{lindh}
\end{align}
These are two coupled first-order differential equations which we can combine into a single second-order differential equation for the field $k^a$. Solving \eqref{lindh} for $v_{\mu}{}^a$ gives
\begin{equation}\label{vsol}
v_{\mu}{}^{a} = - \det(\be)^{-1} \ve^{\mu\nu\rho}\left( \be_{\nu}{}^a \be_{\mu\,b} - \frac12 \be_{\mu}{}^a \be_{\nu\,b} \right) \bar{\cD}_{\rho} k_{\sigma}{}^b\,.
\end{equation}
This we substitute into \eqref{lindv} and after converting the free indices to curved space indices, we find
\begin{equation}\label{linECeom}
\ve_{\mu}{}^{\rho\sigma} \be_{\nu\,a} \left( \bar{\cD_{\rho}} v_{\sigma}{}^a - \Lambda (\be_{\rho} \times k_{\sigma})^a \right) = 2 \det(\be) \cG_{\mu\nu}(k) = 0\,,
\end{equation}
where $\cG_{\mu\nu}(k)$ is the linearized Einstein tensor. It is second order in derivatives on $k_{\mu\nu}$ and defined as
\begin{equation}\label{linEinstein}
\begin{split}
\cG_{\mu\nu}(k) =  & \; \bnabla^{\rho} \bnabla_{(\mu}k_{\nu)\rho} - \frac12 \bBox k_{\mu\nu} - \frac12 \bnabla_{\mu}\bnabla_{\nu} k + \frac12 \bg_{\mu\nu}\left(\bBox k - \bnabla^{\rho} \bnabla^{\sigma} k_{\rho\sigma} \right) \\
& \; + \Lambda \bg_{\mu\nu} k - 2 \Lambda k_{\mu\nu}\,.
\end{split}
\end{equation}
It is transverse ($\bnabla^{\mu} \cG_{\mu\nu}(k) = 0$) and invariant under linearized diffeomorphisms ($\delta k_{\mu\nu} = \bnabla_{(\mu}\xi_{\nu)}$) by construction. The  field $k_{\mu\nu}$ is defined as
\begin{equation}
k_{\mu\nu} \equiv k_{\mu}{}^a \be_{\nu}{}^b \eta_{ab}\,.
\end{equation} 
In principle it contains a symmetric and an anti-symmetric part. However, when substituting \eqref{vsol} into \eqref{lindv}, we see that only the symmetric part of $k_{\mu\nu}$ remains in the final expression, and so the $k$ written in \eqref{linECeom} is a symmetric two-tensor. Furthermore, the anti-symmetric part of $k_{\mu\nu}$ can always be set to zero by a LLT.

\subsection{Adding a mass term}
We can now add a mass term for the fluctuations $k^a$ into the quadratic action \eqref{ECquad}. In this formulation, there is a unique term quadratic in $k$. The resulting action is the Fierz-Pauli action in a first-order form.
\begin{equation}\label{SFP}
S_{\rm FP} = - \int  \left\{ k \cdot \bar{\cD} v + \frac12  \be \cdot \left( v \times v - (\Lambda - \cM^2) k \times k \right) \right\}\,,
\end{equation}
where $\cM$ is the Fierz-Pauli mass. The field equations derived by varying \eqref{SFP} with respect to $k^a$ and $v^a$ are now
\begin{align} \label{FPdv}
& \bar{\cD} v^a - \left( \Lambda - \cM^2\right)( \be \times k)^a  = 0\,, \\
& \bar{\cD} k^a + (\be \times v)^a = 0\,. \label{FPdh}
\end{align}
The presence of the mass term breaks the gauge symmetries of the theory, however, the fields $k_{\mu\nu} \equiv k_{\mu}{}^a \be_{\nu}{}^b \eta_{ab} $ and $v_{\mu\nu} \equiv v_{\mu}{}^a \be_{\nu}{}^b \eta_{ab}$ are still symmetric, which now follows from the equations of motion \eqref{FPdv} and \eqref{FPdh}. By acting on the field equations with a covariant derivative $\bar{\cD}$ and using \eqref{linBianchi}, we can derive the constraints
\begin{equation}
\be^a \be \cdot k = 0\,, \qquad \be^a \be \cdot v = 0\,,
\end{equation}
where it is assumed that $\cM^2 \neq 0$. By invertibility of the background dreibein, these equations imply that $k_{\mu\nu}$ and $v_{\mu\nu}$ are symmetric
\begin{equation}
k_{[\mu\nu]} = 0\,, \qquad v_{[\mu\nu]} = 0\,.
\end{equation}
The two first-order differential equations can again be written as a single second-order differential equation. The result is now:
\begin{equation}\label{FPeom2}
\cG_{\mu\nu}(k) = - \frac12 \cM^2 ( k_{\mu\nu} -\bg_{\mu\nu}k  )\,,
\end{equation}
where $k = \bg^{\mu\nu}k_{\mu\nu}$. Acting on this equation with a covariant derivative $\bnabla^{\mu}$ gives
\begin{equation}\label{deDonder}
\bnabla_{\nu} k = \bnabla^{\mu}k_{\mu\nu}\,.
\end{equation}
After taking the trace of \eqref{FPeom2} we find
\begin{equation}\label{FPtreom}
\frac12 \left( \bBox k - \bnabla^{\mu}\bnabla^{\nu} k_{\mu\nu} + 2 (\Lambda - \cM^2) k \right) = 0\,.
\end{equation}
The condition \eqref{deDonder} reduces \eqref{FPtreom} to
\begin{equation}\label{FPtreom2}
(\Lambda - \cM^2) k= 0\,.
\end{equation}
This implies that, as long as $\Lambda \neq \cM^2$, the trace of $k$ is zero and \eqref{deDonder} gives $\bnabla^{\mu}k_{\mu\nu} =0$. Furthermore, for a transverse traceless $k_{\mu\nu}$, we have that 
\begin{equation}\label{calG}
\cG_{\mu\nu}(k) = - \frac12 (\bBox - 2\Lambda ) k_{\mu\nu}\,.
\end{equation}
Now eqn.~\eqref{FPeom2} reduces to the Fierz-Pauli equation for a free massive spin-2 particle in a maximally symmetric background \eqref{FPeqns}
\begin{equation}\label{FPeom}
\left( \bBox - 2\Lambda - \cM^2 \right)k_{\mu\nu} = 0\,,
\end{equation}
together with the subsidiary conditions
\begin{equation}\label{FPsc}
\bnabla^{\mu} k_{\mu\nu} = 0 \,, \qquad k = 0\,.
\end{equation}
A simple counting argument shows that there are two local degrees of freedom; a symmetric two-tensor in three dimensions has 6 components, and the subsidiary conditions restrict 4 of them. 

\subsection{Parity Violating Massive Fierz-Pauli}
The two degrees of freedom described by the Fierz-Pauli equations correspond to a helicity $+2$ and a helicity $-2$ state. Both helicity states have an equal mass and the theory is even under a parity transformation. We can rewrite the Fierz-Pauli equations to explicitly show both of the helicity states, by making use of the identity
\begin{equation}
\mathcal{G}_{\mu\nu}(k) = - \frac12 \varepsilon_{(\mu}{}^{\alpha\rho} \varepsilon_{\nu)}{}^{\beta\sigma} \bnabla_{\alpha} \bnabla_{\beta} k_{\rho\sigma} - \frac12 \Lambda (k_{(\mu\nu)} - \bar{g}_{\mu\nu} k)\,.
\end{equation}
The Fierz-Pauli equation \eqref{FPeom2} can then be written in terms of linear derivative operators $\cD^M$ as
\begin{equation}\label{linFP}
(\tilde{\cD}^{M} \cD^{M} k )_{\mu\nu} = 0\,.
\end{equation}
Here 
\begin{equation}\label{cDdef}
(\tilde{\cD}^{M} )_{\mu}^{\rho} = \delta_{\mu}^{\rho} - \frac{1}{\det(\be) M} \ve_{\mu}{}^{\alpha\rho} \bnabla_{\alpha} \,,
\qquad 
(\cD^{M} )_{\mu}^{\rho} = \delta_{\mu}^{\rho} + \frac{1}{\det(\be) M} \ve_{\mu}{}^{\alpha\rho} \bnabla_{\alpha} \,,
\end{equation}
and the mass parameter $M$ is related to the Fierz-Pauli mass as
\begin{equation}\label{Mparam}
M = \sqrt{\cM^2 - \Lambda}\,.
\end{equation}
From eqn.~\eqref{linFP} it becomes obvious that two helicity modes are propagated, with the operator $\tilde{\cD}^M$ corresponding to a helicity $+2$ and $\cD^M$ to a helicity $-2$ mode. In the case of Fierz-Pauli theory, both helicities have the same mass and the theory is invariant under $M \to -M$, but we could easily imagine that this need not be the case. Consider for instance
\begin{equation}
(\tilde{\cD}^{M_1} \cD^{M_2} k )_{\mu\nu} = 0\,,
\end{equation}
with $M_1 \neq M_2$. These equations violate parity and the two massive spin-2 modes will have different masses. If we take one of the massive parameters to infinity, the term involving the derivative in \eqref{cDdef} goes to zero and effectively the theory describes a single massive helicity state.

\subsection{The partially massless point}\label{sec:PMpoint}
From the trace of the Fierz-Pauli equation \eqref{FPtreom2} it is obvious that the point $\cM^2 = \Lambda$ is special. At this point the Fierz-Pauli subsidiary condition $k=0$ does not follow from the equation of motion \eqref{FPeom2}. However, this does not imply that the trace of the spin-2 field will propagate. This is the partially massless point and a new gauge symmetry emerges with a scalar gauge parameter $\lambda(x)$ \cite{Deser:1983mm,Deser:2001pe}
\begin{equation}
\delta_{\lambda}k_{\mu\nu} = \bnabla_{\mu} \bnabla_{\nu} \lambda(x) + \cM^2 \bg_{\mu\nu} \lambda(x)\,.
\end{equation}
The field equation for the partially massless mode may be written as
\begin{equation}
( \cD^0 k)_{\mu\nu} = 0\,,
\end{equation}
where $\cD^0$ is defined as
\begin{equation}
\label{cD0def}
(\cD^0)_{\mu}^{\rho} = \frac{1}{\det(\be)} \ve_{\mu}{}^{\alpha\rho} \bnabla_{\alpha}\,.
\end{equation}
The partially massless symmetry is a linearized gauge symmetry, which in most theories remains an artifact of the linear approximation and vanishes at the non-linear level \cite{Blagojevic:2011qc}. This is not the case if the theory possesses conformal invariance, as we will see in section \ref{sec:CSG}.

\subsection{Perturbative Unitarity: Absence of Tachyons and Ghosts}
\label{sec:pertunit}
Thus far we have shown that Fierz-Pauli theory describes two massive helicity $\pm2$ modes in a given background, however, we did not yet specify any conditions on these modes. From a field theoretical point of view, these modes should have positive energy (i.e. they should not be ghosts) and propagate within the light-cone (they should not be tachyonic). From a group theoretical perspective, the modes should be unitary irreducible representations of the isometry group of the background spacetime. Let us review here briefly the conditions imposed by unitarity on the Fierz-Pauli mass.

The Fierz-Pauli equation in AdS$_3$ \eqref{FPeom} can be solved  with the group theoretical approach of \cite{Li:2008dq}. In global coordinates, the AdS metric can be written as
\begin{equation}
ds^2 = \frac{\ell^2}{4}  \left( - du^2- 2 \cosh(2\rho) dudv - dv^2 + 4 d\rho^2 \right)\,,
\end{equation}
where $u$ and $v$ are light-cone coordinates. The solutions of \eqref{FPeom}  form  representations of the $\mathrm{SL}(2,\mathbb{R}) \times \mathrm{SL}(2,\mathbb{R})$
isometry group of $\mathrm{AdS}_3$. These representations can be built up by acting with raising operators of the isometry algebra on a primary state. A primary state was found in \cite{Li:2008dq} and is given by
\begin{equation}\label{psisol}
\psi_{\mu\nu} = e^{-ihu-i\bar{h}v} (\cosh(\rho))^{-(h+\bar{h})}\sinh^2(\rho) F_{\mu\nu}(\rho)\,,
\end{equation}
with $F_{\mu\nu}(\rho)$ given by
\begin{equation}
F_{\mu\nu}(\rho) =
\left(
\begin{array}{ccc}
 \frac{ h - \bar{h}}{4}+\frac{1}{2} & 0 &  \frac{i \left((h-\bar{h})+2\right)}{4 \cosh \rho \sinh \rho} \\
 0 & \frac{1}{2} -\frac{ h - \bar{h}}{4} &\frac{ i \left(2 -(h - \bar{h})\right)}{4 \cosh \rho \sinh \rho}  \\
 \frac{i \left((h-\bar{h})+2\right)}{4 \cosh \rho \sinh \rho} &
\frac{ i \left(2 -(h - \bar{h})\right)}{4 \cosh \rho \sinh \rho}&
 \frac{ -1}{\cosh^2 \rho \sinh^2 \rho} 
\end{array}
\right) \,.
\end{equation}
The constant weights $h$, $\bar{h}$ obey $h - \bar{h} = \pm 2$, as well as the equation
\begin{equation}\label{weights}
 \left(2 h(h-1) + 2\bar{h}(\bar{h}-1) -4 -\ell^2 \cM^2 \right) = 0 \,.
\end{equation}
When the Fierz-Pauli mass $\cM =0$, the modes are massless and obey $( h(h-1) + \bar{h}(\bar{h}-1) -2 )=0$. The weights which satisfy this equation and lead to normalizable modes are $(h,\bar{h}) =(2,0)$ and $(0,2)$. They are solutions of linearized Einstein gravity in AdS$_3$ and correspond to left- and right-moving massless gravitons.

When the Fierz-Pauli mass $\cM \neq 0$, the weights obey \eqref{weights}.
For those primaries that do not blow up at the boundary $\rho \rightarrow \infty$, we obtain the following weights:
\begin{align}
\textrm{left-moving}:\; h & = \frac32 +  \frac12\sqrt{1+\ell^2 \cM^2} \,,
 \qquad\, \bar{h} = -\frac12 + \frac12 \sqrt{1+\ell^2 \cM^2} \,, \label{hleft}\\ 
\textrm{right-moving}: \; h & = -\frac12 + \frac12 \sqrt{1+\ell^2 \cM^2}\,,
\qquad\, \bar{h} = \frac32 +  \frac12\sqrt{1+\ell^2 \cM^2}\,. \label{hright}
\end{align}
These weights correspond to left- and right-moving massive gravitons, with mass $\cM$. The condition that these modes are normalizable implies that the weights of the massive modes are restricted as $h+ \bar h \geq 2$ and hence the masses of the modes must be real:
\begin{equation}\label{notachyonFPM}
\cM^2 \geq 0\,.
\end{equation}
The requirement for unitarity of the irrep in AdS$_3$ thus translates to the positivity of the Fierz-Pauli mass squared. We may therefore interpret this as the no-tachyon condition. 

Note that for scalar fields in AdS$_3$ the absence of tachyons allows for a small negative mass squared, known as the Breitenlohner-Freedman bound \cite{Breitenlohner:1982jf}. For symmetric rank-2 tensors, the bound is \eqref{notachyonFPM} and a small negative mass squared is not allowed \cite{Bergshoeff:2010iy}.

The bound \eqref{notachyonFPM} translates to the massive parameter $M$ defined in \eqref{Mparam} as
\begin{equation}\label{notachyon}
|M\ell| \geq 1\,.
\end{equation}
Hence we can interpret this as the no-tachyon bound for a single massive helicity 2 state in AdS$_3$. 

Next to the no-tachyon condition, we still need to determine that the spin-2 mode is not a ghost. For this the field equations are not sufficient and we need to consider the Fierz-Pauli action \eqref{SFP}. If we solve the first-order field equation for $v^a$ \eqref{FPdh} and back-substitute then we find the following Lagrangian density for the symmetric tensor $k_{\mu\nu} = k_{(\mu}{}^a \bar e_{\nu)}{}^b \eta_{ab}$:
\begin{equation}
\mathcal{L}_{\rm FP} = - A \left\{ k^{\mu\nu} \mathcal{G}_{\mu\nu}(k) + \frac12 \cM^2 \left( k_{\mu\nu} k^{\mu\nu} - k^2\right) \right\} \,,  
\qquad \left(k= \bg^{\mu\nu}k_{\mu\nu}\right)\, , 
\end{equation}
where $A$ is an arbitrary overall factor and $\mathcal{G}_{\mu\nu}(k)$ is the linearized Einstein tensor \eqref{linEinstein}. This is the  Fierz-Pauli  Lagrangian density for a spin-2 field of mass ${\cal M}$ in an AdS$_3$ background; see e.g. \cite{Bergshoeff:2009aq}, where the conventions used are the same as those used here.  From this result we  learn that the no-ghost condition is $A>0$.  Next, we diagonalise the Fierz-Pauli Lagrangian 3-form (\ref{SFP}) by writing it in terms of the new Lorentz-vector one-form fields $k_\pm^a$ defined by 
\begin{equation}
k^a = k_+^a + k_-^a \,, \qquad v = M k_+^a - M k_-^a\, .
\end{equation}
Here $M = \sqrt{\cM^2 - \Lambda}$. We find that 
\begin{eqnarray}\label{FPdecomp}
L_{\rm FP} &=& - A M \left( k_{+} \cdot \bar{\cD} k_+ + M \bar{e} \cdot  k_+ \times k_+ \right) \nonumber \\
&& + \ A M \left( k_{-} \cdot \bar{\cD} k_- - M \bar{e} \cdot k_- \times k_- \right)\,.
\end{eqnarray}
The $k_\pm$ field propagates a single spin-2 mode of helicity $\pm 2$, and for both to have positive energy we require $A>0$. However, the two modes are exchanged by parity, which is a symmetry of the Fierz-Pauli action,  so if the helicity $\pm2$ mode has positive energy then so does the $\mp2$ mode. This means that $A>0$ is the condition for either helicity mode alone to have positive energy, the action for a single mode being obtained by setting either $k_+\equiv0$ or $k_-\equiv0$. Hence the no-ghost condition for either helicity mode is 
\begin{equation}\label{noghost}
A > 0\,.
\end{equation}
These no-ghost and no-tachyon conditions derived here will come to use later when we discuss the perturbative unitarity of various massive gravity theories on AdS$_3$ backgrounds in chapter \ref{chapter:asymptotic_symm}.

\section{Non-linear Fierz-Pauli and Scalar Ghosts}
In the last section we discussed the free massive spin-2 action found by Fierz and Pauli. We started explicitly from the first-order form, where there is a unique mass-term which could be added to the linear Einstein-Hilbert action. In the second-order formulation, the Fierz-Pauli action is
\begin{equation}\label{SFP2}
S_{\rm FP} = - \frac12 \int d^3x \sqrt{-g} \left\{ k^{\mu\nu} \cG_{\mu\nu}(k) + \frac12 \cM^2 \left( k^{\mu\nu} k_{\mu\nu} - k^2 \right) \right\}\,.
\end{equation}
If the action is written in this form, it is not so obvious to see that the mass term is unique. In fact, it seems that the combination appearing in the mass term is fine-tuned to have a coefficient of $-1$ in front of the $k^2$ term. However, this is the unique mass term for a free massive spin-2 as a closer look will verify. Moving away from the Fierz-Pauli tuning will introduce a third degree of freedom, which will propagate with a wrong sign for the kinetic term and hence it is a scalar ghost.

To see the scalar ghost appearing, let us suppose we do not restrict ourselves to the Fierz-Pauli tuning and consider the action \eqref{SFP2} with an arbitrary coefficient $\alpha$ in front of the $k^2$ term. The field equation derived from that action is
\begin{equation}\label{alphaFP1}
\cG_{\mu\nu}(k) = - \frac12 \cM^2 ( k_{\mu\nu} + \alpha \bg_{\mu\nu}k  )\,.
\end{equation}
Acting on this equation with a covariant derivative now gives
\begin{equation}
\bnabla^{\mu} k_{\mu\nu} = -\alpha \bnabla_{\nu}k \,.
\end{equation}
Using this relations, the trace of \eqref{alphaFP1} reads
\begin{equation}\label{alphaFPtr}
\frac12 \left( (1+\alpha) \bBox + 2 (\Lambda - \cM^2) \right) k = 0\,.
\end{equation}
The Fierz-Pauli subsidiary condition $k=0$ only follows from this equation if $\alpha = -1$ (and, like before $\Lambda \neq \cM^2$). If $\alpha\neq -1$, equation \eqref{alphaFPtr} is a Klein-Gordon equation for the scalar quantity $k$. To see that its kinetic term carries the wrong sign, we decompose $k_{\mu\nu}$ into a transverse traceless two-tensor $\tilde{k}_{\mu\nu}$, a vector field $A_{\mu}$ and a scalar field $\phi$, representing the trace of $k_{\mu\nu}$
\begin{equation}\label{stuckelberg}
k_{\mu\nu} = \tilde{k}_{\mu\nu} + \bnabla_{(\mu} A_{\nu)} + 2 \bnabla_{\mu}\bnabla_{\nu} \phi +  \bg_{\mu\nu} \phi \,.
\end{equation}
Substituting this into the Fierz-Pauli action \eqref{SFP2}, we see that the vector part decouples and the kinetic term for the scalar field $\phi$ carries a negative sign with respect to the kinetic term of the transverse traceless field $\tilde{k}_{\mu\nu}$. Hence, if the overall sign of the action is such that the massive spin-2 modes have positive energy, then the scalar mode will be a ghost.

\subsection{The Boulware-Deser ghost}
Gravitation is an inherently non-linear theory and hence to understand the effects of adding a small mass to the graviton, we should investigate non-linear extensions of Fierz-Pauli theory. In the last section we saw that in a second-order formulation, a tuning of parameters was required in order to correctly describe a free massive spin-2 mode. When we consider non-linear extensions of the theory, we may expect this scalar ghost mode to turn up again, only now at higher orders in perturbation theory. In fact, it was shown in \cite{Boulware:1973my} that simply replacing the kinetic term in the Fierz-Pauli action by the non-linear kinetic term of general relativity ($\sqrt{-g} R$) indeed leads to a theory with a scalar ghost in its spectrum. This specific scalar ghost now carries the name of the authors of \cite{Boulware:1973my}: the Boulware-Deser ghost.

To see the origin of the Boulware-Deser ghost, we consider the theory defined by the action
\begin{equation}\label{Snonlin}
S = \frac{1}{\kappa^2} \int d^dx \; \sqrt{-g} \left\{ R[g] - \frac14 m^2 g_{(0)}^{\mu\nu} g_{(0)}^{\rho\sigma} \left( k_{\mu\rho}k_{\nu\sigma} - k_{\mu\nu} k_{\rho\sigma} \right) \right\} \,.
\end{equation}
Here $R$ is the Ricci tensor for the metric $g_{\mu\nu}$. A reference metric $g_{(0)\,\mu\nu}$ is introduced in order to raise and lower indices on the metric perturbations $k_{\mu\nu}$. Usually they are related as $g_{\mu\nu} = g_{(0)\mu\nu} + k_{\mu\nu}$. We consider the theory in $d$ dimensions for the moment, and move back to three dimensions at the end of this section.

We are now dealing with a new theory, and should investigate whether it propagates the right amount of degrees of freedom to describe a massive graviton. To this end one may decompose the metric $g_{\mu\nu}$ in ADM variables; the spatial metric $g_{ij}$, the lapse $N$ and the shift $N_i$ and investigate the phase space of the theory through the Hamiltonian. In general relativity one will find primary constraints that ensure that the canonical momenta of the lapse and shift vanish. These variables then appear in the Hamiltonian as Lagrangian multipliers for a set of secondary constraints. In $d$ dimensions the counting of degrees of freedom proceeds as follows. The spatial component of the metric is a symmetric $d-1$ matrix, so it has $\frac12 d (d-1)$ components. From this we subtract the $d$ constraints enforced by the lapse and shift, leading to a total of $\frac12 d(d-3)$ degrees of freedom. 

The addition of a mass-term in the action \eqref{Snonlin} will introduce terms quadratic in the lapse and shift functions, so that they do not play the role of Lagrange multipliers any more, but instead become auxiliary fields. As a result, the degrees of freedom propagated by the theory are given by the $\frac12 d(d-1)$ components of the spatial metric. In 3 dimensions, these are 3 degrees of freedom, 2 for the massive graviton and in addition there is an extra ghost-like scalar degree of freedom. This is the Boulware-Deser (BD) ghost \cite{Boulware:1973my}.

\subsection{Removing the ghost: dRGT Massive Gravity}
\label{sec:dRGT}


The origin of the ghost mode in \eqref{Snonlin} can be seen in the St\"uckelberg formulation of the theory, where the massive graviton is decomposed into helicity $\pm 2$, helicity $\pm 1$ and helicity $0$ modes, like in \eqref{stuckelberg}. In four spacetime dimensions the leading order interactions for the helicity 0 mode are suppressed by an energy scale $\Lambda_5 = (M_{\rm P} m^4)^{1/5}$. The dynamics of this mode can then be studied in the decoupling limit of \cite{ArkaniHamed:2002sp}: $M_P \to \infty, m \to 0$ and $\Lambda_5$ fixed. This limit sends all operators suppressed by an energy scale higher than $\Lambda_5$ to infinity and thus decouples them from the theory. The kinetic term of the helicity 0 mode now receives four-derivative contributions which signals that the scalar sector propagates two degrees of freedom. One is a massless scalar field and the other the Boulware-Deser ghost, with a mass of order of the energy scale $\Lambda_5$. One may wonder whether this theory will still make sense as an effective theory of massive gravity at energies lower than this cut-off scale, so before the mass of the ghost mode becomes of order one. However, the distance scale set by the BD ghost coincides with the scale at which quantum corrections become important \cite{ArkaniHamed:2002sp}. This scale is parametrically larger than the Vainshtein radius\footnote{The Vainshtein radius is the length scale at which non-linear effects in Fierz-Pauli theory become important. It is given by $r_{\rm V} = \left( \frac{G_N M}{m^4} \right)^{1/5}$ in four spacetime dimensions.}, implying that the quantum theory takes over before the classical non-linearities can restore the proper limit with general relativity. So in order to have a suitable effective theory of massive gravity, we can try to raise the cut-off scale by adding higher-order interactions to the theory.

These ideas were investigated by de Rham, Gabadaze and Tolley (dRGT) in \cite{deRham:2010ik,deRham:2010kj}. They considered an action with an Einstein-Hilbert kinetic term and an interaction potential $U_n(g^{(0)}, k)$, defined as
\begin{equation}
S_{\rm dRGT} \propto \int d^d x \left\{ \sqrt{-g} R - \frac14 \sqrt{-g^{(0)}} m^2 \sum_{n=2}^{\infty} U_n (g^{(0)} , k) \right\}\,,
\end{equation}
with
\begin{align}\label{dRGTinteractions}
U_2 & = [k^2] - [k]^2 \,, \nonumber \\
U_3 & = a_1 [k^3] + a_2 [k][k^2] + a_3 [k]^3\,,  \\
U_4 & = b_1 [k^4] + b_2 [k][k^3] + b_3 [k^2][k^2] + b_4 [k^2][k]^2 + b_5 [k]^4\,, \nonumber \\
	& \qquad \vdots \nonumber
\end{align}
The square brackets denote taking the trace with the reference metric $[k] = g^{(0)\,\mu\nu} k_{\mu\nu}$. Notice that the first term, $U_2$, is just the Fierz-Pauli mass term. After introducing the St\"uckelberg decomposition in this interaction potential, one can see that at every order in $n$ higher derivative kinetic terms for the scalar field will appear. The trick is now to tune the coefficients ($a_i$ for $U_3$, $b_i$ for $U_4$, etc.) in such a way that the higher derivative scalar terms become a total derivative. This is possible at every order in $n$ and fixes all but two coefficients in four dimensions. The tuning of these coefficients removes all scalar interaction terms suppressed by any scale $\Lambda < \Lambda_3 = (m^2 M_P)^{1/3}$. Then the new decoupling limit may be taken with $\Lambda_3$ fixed and the consequent effective Lagrangian has at most second-order derivatives on the scalar field, signalling the absence of the ghost mode in the decoupling limit. 

In order to study the ghost problem beyond the decoupling limit, the authors of \cite{deRham:2010kj} resummed the interaction terms \eqref{dRGTinteractions} into a fully non-linear theory of massive gravity involving elementary symmetric polynomials of the square root matrix $\sqrt{g^{-1} f}$, where $f_{\mu\nu}$ is the reference metric (see also \cite{Hassan:2011vm}). This fully non-linear theory was shown to be ghost free in \cite{Hassan:2011hr} (see also \cite{deRham:2011rn,deRham:2011qq,Hassan:2011ea,Mirbabayi:2011aa,Golovnev:2011aa,Hassan:2012qv,Kluson:2012wf}). This is due to the fact that after an ADM decomposition of the metric, the shift function $N$ remains a Lagrange multiplier and enforces a secondary constraint. The number of degrees of freedom is then $\frac12 d(d-1)-1$ which correctly describes the degrees of freedom of a massive graviton in $d$ dimensions.

In $d$ dimensions the ghost-free dRGT massive gravity theory can be written as \cite{Hassan:2011vm}:
\begin{equation}\label{dRGT}
S_{\rm dRGT} \propto \int d^d x\; \sqrt{-g}\left[ R[g] - \frac{m^2}{4} \sum_{n=0}^{d} \tilde{\beta}_n S_n \left( \sqrt{ g^{-1} f} \right) \right]\,.
\end{equation}
Here $S_n(\mathbb{X})$ is the $n$-th elementary symmetric polynomial for the matrix $\mathbb{X}$. It is given by
\begin{equation}
S_{n}(\mathbb{X}) = \frac{1}{n!(d-n)!}\ve_{\mu_1\mu_2 \dots \mu_d}\ve^{\nu_1\nu_2\dots\nu_d} \mathbb{X}_{\nu_1}^{\mu_1} \cdots \mathbb{X}_{\mu_n}^{\nu_n} \delta_{\nu_{n+1}}^{\mu_{n+1}} \cdots \delta_{\nu_d}^{\mu_d}\,.
\end{equation}
It vanishes identically for $n>d$, hence in principle, there are $d+1$ free parameters $\tilde{\beta}_n$ in \eqref{dRGT}. However,  the $d$-th symmetric polynomial is just $\sqrt{-\det(f)}$ and does not contribute to the equations of motion, reducing the parameter space by one. Furthermore, another parameter is fixed by demanding that flat space is a solution to the field equations, leaving $d-1$ free parameters.

In this formulation, dRGT massive gravity seems very removed from the CS--like models we introduced in chapter \ref{sec:introductionc3}. However, in ref.~\cite{Hinterbichler:2012cn} a first-order form of the action \eqref{dRGT} was presented in which the interaction terms greatly simplify. Under certain assumptions, the dRGT model may be written in terms of a vielbein $e^a$ and a ``reference vielbein'' $\bar{e}^a$ as
\begin{equation}
\begin{split}
S_{\rm dRGT} \propto \int d^dx\; & \det(e) R[e] \\
 & - \frac{m^2}{4} \int \sum_{n=0}^{d} \beta_{n} \ve_{a_1a_2 \ldots a_d}  e^{a_1} \wedge \cdots \wedge e^{a_n} \wedge \bar{e}^{a_{n+1}} \wedge \cdots \wedge \bar{e}^{a_d} \,.
\end{split}
\end{equation}
The parameters $\beta_n$ are related to the $\tilde{\beta}_n$ as
\begin{equation}
\beta_n = \frac{\tilde{\beta}_n}{n!(d-n)!}\,.
\end{equation}
The interaction terms are such that, assuming that the product $\bar{e}_{\mu}{}^a e_{\nu\,a}$ is symmetric\footnote{This condition, which is also important for the Hamiltonian analysis of the theory, does not generically follow from the equations of motion of the theory as shown in \cite{Deffayet:2012nr}. In three dimensions, it follows from the field equations only if the linear combination $\beta_1e^a + \beta_2 \bar{e}^a$ is invertible.} \cite{Deffayet:2012zc}, the interaction terms reproduce the dRGT interaction terms in \eqref{dRGT} when $\bar{e}^a$ is associated with the reference metric and $e^a$ with the dynamical metric. In three dimensions, the dRGT massive gravity model can then be written (in our usual conventions with implicit wedge products and Lorentz indices) as 
\begin{equation}\label{3DdRGT}
L_{\rm dRGT} = - e \cdot R(\omega) - \alpha_1 \frac{m^2}{6} e \cdot e \times e + \frac{m^2}{2} \left(\beta_1' e \cdot e \times \bar{e} + \beta_2' e \cdot \bar{e} \times \bar{e} \right)\,. 
\end{equation}
This formulation is a lot closer to the Chern-Simons--like form of \eqref{LCSlike}. In fact, it would fall into this class of model, if it was not for the fact that the flavor space metric $g_{rs}$ is not invertible. This is due to the fact that the Lagrangian three-form \eqref{3DdRGT} depends on a fixed reference metric, which does not have any kinetic terms. This feature is responsible for the broken diffeomorphism invariance in dRGT massive gravity, which is the cause for the introduction of local degrees of freedom.

\section{Auxiliary fields and Higher Derivatives}
\label{section:auxfields}

We shall now return to study theories with the general Chern-Simons--like form of \eqref{LCSlike}. First we will review some known three-dimensional gravity theories which fit this form. In particular, this section will focus on higher-derivative theories of massive gravity in 3D. The idea here is somewhat different from the dRGT approach of the last section; now theories are studied which describe a massive spin-2 in a dynamical gravitational setting. Hence, the gauge symmetries of 3D GR are not broken and the degrees of freedom of a massive spin-2 are introduced by the presence of higher-derivative terms in a metric formulation. These higher-derivative terms are introduced by auxiliary fields in the Chern-Simons--like formulation of three dimensional gravity.

Let us suppose that we start with a general Chern-Simons--like theory involving a dreibein and a spin connection. In order to preserve the local Lorentz symmetry, the spin connection may only come in specific combinations. These are, as the covariant derivative of some field
\begin{equation}
\cD X^a = d X^a + (\omega \times X)^a \,,
\end{equation}
in the Riemann tensor
\begin{equation}
R^a = d \omega^a + \frac12 (\omega \times \omega)^a\,,
\end{equation}
or as a Lorentz-Chern-Simons term in the action
\begin{equation}\label{LCS}
L_{\rm LCS} = \frac12 \omega \cdot \left( d \omega + \frac13 \omega \times \omega \right)\,.
\end{equation}
In the last chapter, we already analyzed three dimensional gravity, which contains only the dreibein and a spin connection. Here we will explore theories which extend these models by using extra, auxiliary fields. 

\subsection{Conformal Gravity} \label{sec:CSG}
For a general model with $N=3$ fields there is a combination which is also a Chern-Simons gauge theory; one of the conformal group in 3D, called conformal gravity (CG) \cite{Horne:1988jf}. It is defined as
\begin{equation}\label{LCSG}
L_{\rm CG} = \frac{1}{\mu} L_{\rm LCS} + \frac{1}{\mu} f^a \cD e_a\,,
\end{equation}
where $f^a$ is an auxiliary field. It serves as a Lagrange multiplier for the torsion constraint. 

The above action is a Chern-Simons theory \eqref{SCS} for the Lie algebra valued connection one-form \cite{Horne:1988jf}
\begin{equation}\label{confcon}
A = e^a P_a + \omega^a J_a + f^aK_a + b D\,.
\end{equation}
Here $P_a, J_a, K_a$ and $D$ span the conformal $SO(3,2)$ group and are the generators of translations, local Lorentz transformations, special conformal transformations and dilatations, respectively. Their commutators are given by
\begin{equation}\label{SO32}
\begin{split}
[P_a,J_b] = \ve_{abc}P^c\,,& \qquad [K_a,D]= -K_a\,, \qquad  [P_a,D] = P_a\,, \\
 [J_a, J_b] = \ve_{abc} J^c\,,& \qquad [P_a,K_b] = - \ve_{abc}J^c + \eta_{ab} D\,, \\
[K_a,J_b] = \ve_{abc} K^c\,,&  \qquad [P_a ,P_b] = [J_a,D]=[K_a,K_b] = 0\,.
\end{split}
\end{equation}
The corresponding non-degenerate bilinear form is
\begin{equation}
\tr(J_a,J_b) = \eta_{ab}\,, \qquad \tr(P_a,K_b) =- \eta_{ab}\,, \qquad \tr(D,D) = 1\,.
\end{equation}
Using these relations in \eqref{SCS}, the Chern-Simons action with the connections \eqref{confcon} can be written as
\begin{equation}
S = \frac{k}{4\pi} \int \left\{ L_{\rm LCS} - f \cdot \cD e + b d b + 2 e \cdot f b \right\} \,.
\end{equation}
For an invertible dreibein, it is possible to partially fix the gauge symmetries to set $b = 0$ \cite{Horne:1988jf}. For this gauge choice, the action is equivalent to \eqref{LCSG} and the field equations for $f^a$, $\omega^a$ and $e^a$ are respectively
\begin{equation}\label{CSGeom}
\begin{split}
& \cD e^a = 0\,, \\
& R^a + (e \times f)^a = 0\,, \\
& \cD f^a = 0\,.
\end{split}
\end{equation}
The first of these equations is the torsion constraint and allows one to solve the spin connection in terms of derivatives of the dreibein. The second equation can be solved for $f^a$ in terms of derivatives acting on the dreibein. Using the identity \eqref{EinsteinID}, we obtain
\begin{equation}\label{Sdef}
f_{\mu\nu} \equiv f_{\mu}{}^a e_{\nu}{}^b \eta_{ab} = - \left( R_{\mu\nu} -\frac14 g_{\mu\nu} R \right) \equiv - S_{\mu\nu}\,. 
\end{equation}
Here $S_{\mu\nu}$ is called the Schouten tensor. Using the solution of $f^a$, we may express the final equation of motion as:
\begin{equation}\label{Cdef}
\det(e)^{-1} \ve_{\mu}{}^{\alpha \rho} \nabla_{\alpha} S_{\rho \nu} \equiv C_{\mu\nu} = 0\,,
\end{equation}
where $C_{\mu\nu}$ is the Cotton tensor.

We will now show that the linear theory contains two massless modes and a partially massless mode. To this end, we perturb the dreibein and spin connection around a maximally symmetric background as in \eqref{perturb}. The auxiliary field is proportional to the Schouten tensor, whose background value is $\bar{S}_{\mu\nu} = \frac{\Lambda}{2} \bg_{\mu\nu}$. Hence we expand the auxiliary field as:
\begin{equation}\label{fperturb}
f^a = - \frac{\Lambda}{2}\left( \be^a + \kappa k^a \right) + \kappa p^a\,.
\end{equation}
The quadratic Lagrangian of CG then becomes
\begin{equation}
L_{\rm CG}^{(2)} = \,\mu^{-1} \left\{   \left( p - \frac{\Lambda}{2} k \right) \cdot \left(\bar{\cD} k + \bar e \times v \right)+\frac12 v \cdot \left(\bar{\cD} v - \Lambda \bar e \times k \right) \right\}  \,. \label{LinCSG}
\end{equation}
The $p$ equation of motion enforces the equation \eqref{lindh} and hence we can substitute the solution to this equation, given by \eqref{vsol}, into the action. The result is the Lagrangian density
\begin{equation}
\cL_{\rm CG} = \mu^{-1}\epsilon^{\mu\alpha}{}_{\rho} \bar{\nabla}_{\alpha} k^{\rho\nu} \cG_{\mu\nu}(k) \,.
\end{equation}
The linear field equations then become
\begin{equation}
(\cD^0 \cG(k))_{\mu\nu} = 0\,.
\end{equation}
In AdS, where $\Lambda = - 1 /\ell^2$, we may write
\begin{equation}
\cG_{\mu\nu} (k) =  \frac{1}{2 \ell^2} (\cD^{L} \cD^{R} k)_{\mu\nu}\,,
\end{equation}
where $\cD^{L/R}$ is given by \eqref{cDdef} with $M = \pm 1/\ell$
\begin{equation}\label{cDLRdef}
(\cD^{L/R})^{\rho}_{\mu} = \delta_{\mu}^{\rho} \pm \frac{\ell}{\det(\be)} \ve_{\mu}{}^{\alpha \rho} \bnabla_{\alpha}\,.
\end{equation}
The linear field equations can then be written as 
\begin{equation}
(\cD^0 \cD^L \cD^R k)_{\mu\nu} = 0\,.
\end{equation}
In this form it is very clear that the linear theory describes a partially massless mode and two massless gravitons, all of which are pure gauge modes and do not propagate any local degrees of freedom. In that respect, the conformal invariance of the CG model can be seen as a non-linear generalization of the partially massless gauge symmetry introduced in the last section.

The metric form of the CG action is related to a gravitational Chern-Simons action for the Christoffel connection
\begin{equation}
S_{\rm gCS} = \frac{k}{4\pi} \int_{\cM} d^3x\; \ve^{\mu\nu\rho} \Gamma_{\mu\sigma}^{\lambda} \left( \partial_{\nu} \Gamma_{\rho \lambda}^{\sigma} + \frac{2}{3} \Gamma_{\nu \tau}^{\sigma} \Gamma_{\rho \lambda}^{\tau} \right) \,.
\end{equation}
This formulation differs from the first-order form by a total derivative bulk term and a boundary term \cite{Kraus:2005zm}
\begin{equation}
S_{\rm CG} - S_{\rm gCS} =  \frac{k}{12\pi} \int_{\cM} \tr(e^{-1} d e)^3 - \frac{k}{4\pi} \int_{\partial \cM} \tr(\omega d e e^{-1} )\,.
\end{equation}
For more details on conformal gravity and its relation to holography, we refer to \cite{Afshar:2011yh,Afshar:2011qw,Afshar:2014rwa}.

\subsection{Topologically Massive Gravity}\label{sec:TMG}
The simplest model which is not a pure CS gauge theory, but propagates a single degree of freedom is called Topologically Massive Gravity (TMG) \cite{Deser:1981wh,Deser:1982vy}. The theory is defined as the sum of conformal gravity and Einstein-Cartan theory. Its first-order CS--like formulation is \cite{Carlip:2008qh,Blagojevic:2008bn} 
\begin{equation}\label{LTMG_cslike}
L_{\rm TMG} = - \sigma e \cdot R + \frac{\Lambda}{6} e \cdot e \times e +  \frac{1}{\mu} f \cdot \cD e +  \frac{1}{2\mu}\omega \cdot \left( d \omega + \frac13  \omega \times \omega \right)\,.
\end{equation}
Here $\sigma = \pm1$ is a sign parameter, which is introduced for later convenience. The inclusion of Einstein-Cartan theory into CG introduces a mass scale in the theory. Hence the conformal symmetry of the theory is broken. In the linear theory this gives a mass to the partially massless mode and the linear spectrum consists out of a single massive mode of helicity $+ 2$ (or $-2$, depending on the sign of $\mu$).

Again the field equations for $f^a$, $\omega^a$ and $e^a$ may be solved to obtain a single, higher-derivative field equation for the dreibein. In this case, the field equations are 
\begin{equation}\label{TMGeom}
\begin{split}
& \cD e^a = 0\,, \\
& R^a + (e \times f)^a = 0\,, \\
& \frac{1}{\mu} \cD f^a + \sigma (e \times f)^a + \frac{\Lambda}{2} (e \times  e)^a = 0\,.
\end{split}
\end{equation}
The first two equations remain identical to the CG case, and hence the solutions for $\omega(e)$ and $f(e)$ are the same. The final equation of motion becomes
\begin{equation}
\sigma G_{\mu\nu} + \Lambda g_{\mu\nu} + \frac{1}{\mu} C_{\mu\nu} = 0\,.
\end{equation}
This field equation can also be derived from an equivalent, metric form of the TMG action. This is the sum of the Einstein-Hilbert action and a gravitational Chern-Simons term. Its Lagrangian density is
\begin{equation}
\cL_{\rm TMG}  = \sqrt{-g} \left( \sigma R - 2 \Lambda + \frac{1}{\mu}  \ve^{\mu\nu\rho} \Gamma_{\mu\sigma}^{\lambda} \left( \partial_{\nu} \Gamma_{\rho \lambda}^{\sigma} + \frac{2}{3} \Gamma_{\nu \tau}^{\sigma} \Gamma_{\rho \lambda}^{\tau} \right) \right)\,.
\end{equation}
The quadratic action can be found by expanding the fields around a maximally symmetric background. The dreibein and spin connection are expanded as in \eqref{perturb} and the auxiliary field $f^a$ as in \eqref{fperturb}. The quadratic Lagrangian then becomes
\begin{align} \label{TMGquadaction}
L_{\rm TMG}^{(2)} = & - \bigg\{ \sigma \left( k \cdot \bar{\cD} v + \frac12  \be \cdot \left( v \times v - \Lambda  k \times k \right) \right) - \frac{1}{2\mu} v \cdot \left( \bar{\cD} v - \Lambda  \be \times k \right) \nonumber \\
& - \left(p - \frac{\Lambda}{2 \mu}k \right) \cdot \left( \bar{\cD} k +  \be \times v \right)
\bigg\} \,.
\end{align}
The equation of motion of $p^a$ is equivalent to \eqref{lindh} and hence the solution for $v^a$ in terms of $k^a$ is unchanged with respect to \eqref{vsol}. Substituting this into the action, we find a third-order derivative action for $k_{\mu\nu}$,
\begin{equation}
L_{\rm TMG}^{(2)} = \sigma k^{\mu\nu} \cG_{\mu\nu}(k) + \frac{1}{\mu} \ve^{\mu\alpha}{}_{\rho} \bnabla_{\alpha} k^{\rho\nu} \cG_{\mu\nu}(k)\,.
\end{equation}¥
In AdS, where $\Lambda = - \frac{1}{\ell^2}$, the linear field equations can be written as
\begin{equation}
(\cD^{M} \cD^L \cD^R k)_{\mu\nu} = 0\,,
\end{equation}
where $M = \sigma \mu$. It is now clear that the theory propagates a single massive state with helicity $+2$ or $-2$, depending on the sign of $M$.

\subsection{New Massive Gravity}\label{sec:NMG}
The examples of conformal gravity and Topologically Massive Gravity nicely illustrate how we can use auxiliary vector valued one-forms in a Chern-Simons--like theory to obtain higher-derivative terms in a metric formulation of gravity. The field equations \eqref{CSGeom} and \eqref{TMGeom} in this form have an iterative quality about them. The first equation ($De^a=0$) allows one to solve the spin connection in terms of the derivatives on the dreibein. The second equation gives the auxiliary field in terms of derivatives on the spin connection (or in terms of the Riemann curvature two-form). We could continue this process and introduce a fourth field $h^a$ and demand its field equation fixes it in terms of derivatives on $f^a$. A minimal way would be to write:
\begin{equation}\label{twoaux}
\begin{split}
& \cD e^a = 0\,, \\
& R^a + (e \times f)^a = 0\,, \\
& \cD f^a - m^2 (e \times h)^a = 0\,, \\
\end{split}
\end{equation}
where the convention for the new mass parameter $m$ and its sign are chosen for later convenience. 
The auxiliary fields may be solved for in terms of derivatives of the dreibein. The result is
\begin{equation}
f_{\mu\nu} = - S_{\mu\nu}\,, \qquad h_{\mu\nu} = - \frac{1}{m^2} C_{\mu\nu}\,.
\end{equation}
Hence the field $f$ contains two derivatives of the metric, while $h$ has three. The equations \eqref{twoaux} can be integrated to a parity even action with no more than four derivatives of the metric. Since $h$ is odd under parity, we use it to multiply the torsion term in the action, which results in a parity even term. The auxiliary field $f$, even under parity, must then multiply the second equation in \eqref{twoaux}. The third equation will follow from varying $\omega^a$. The result, plus the Einstein-Cartan Lagrangian, is
\begin{equation}
L_{\rm NMG} = - \sigma e \cdot R + \frac{\Lambda_0}{6} e \cdot e \times e + h \cdot \cD e - \frac{1}{m^2} f \cdot \left( R + \frac12  e \times f \right) \,.
\end{equation}
Here $\sigma= \pm1$ is a sign parameter. This is the first-order form of the New Massive Gravity action \cite{Blagojevic:2010ir,Hohm:2012vh}. Its variation with respect to $h^a, f^a$ and $\omega^a$ gives the equations \eqref{twoaux}, while variation with respect to $e^a$ gives
\begin{equation}
\cD h^a - \sigma R^a + \frac{\Lambda_0}{2} (e \times e)^a - \frac{1}{2m^2} (f \times f)^a = 0\,.
\end{equation}
Substituting the solutions for $h$ and $f$ into this equation gives a fourth-order differential equation, which can equivalently be derived from varying the Lagrangian density
\begin{equation}
\cL_{\rm NMG} = \frac{1}{2} \left\{\sigma R - 2 \Lambda_0 + \frac{1}{m^2}\left( R^{\mu\nu}R_{\mu\nu} - \frac38 R^2 \right) \right\} \,.
\end{equation}
This is the original NMG Lagrangian as it was introduced in \cite{Bergshoeff:2009hq,Bergshoeff:2009aq}. 

To show that the linear theory contains a massive spin-2 mode, we expand the fields $e^a$ and $\omega^a$ as in \eqref{perturb} and, since the auxiliary field $f^a$ is still proportional to the Schouten tensor, we expand it as \eqref{fperturb}. The second auxiliary field, $h^a$ is proportional to the Cotton tensor, which vanishes on maximally symmetric backgrounds, so we take
\begin{equation}
h^a = \kappa q^a\,.
\end{equation}
The background field equations \eqref{bkgdeom} are satisfied when $\Lambda$ solves the quadratic equation
\begin{equation}
\Lambda_0 = \Lambda\left(\sigma  + \frac{\Lambda}{4m^2}\right)\,.
\end{equation}
The quadratic Lagrangian three-form reads,
\begin{equation}
\begin{split}
L_{\rm NMG}^{(2)} =  & - \left(\sigma - \frac{\Lambda}{2m^2} \right) \left(k \cdot \bar{\cD} v + \frac12 \be \cdot \left( v \times v  - \Lambda  k \times k \right) \right) \\
&- \frac{1}{m^2}  \bigg\{ p \cdot \left( \bar{\cD} v - \Lambda \bar{e} \times k + \tfrac12  \bar{e} \times p \right) 
 +  q \cdot \left(  \bar{\cD} k + \bar{e} \times v \right) \bigg\} \,.
 \end{split} 
\end{equation}
After eliminating $v^a$ and $q^a$ by using their equations of motion, the Lagrangian reduces to the density,
\begin{equation}\label{LNMG}
\cL_2^{(2)} =  -\bar{\sigma} \, k^{\mu\nu} \cG_{\mu\nu}(k)- \frac{2}{m^2}  p^{\mu\nu} \cG_{\mu\nu}(k) - \frac{1}{2m^2}( p_{\mu\nu} p^{\mu\nu} - p^2)\,,
\end{equation}
where
\begin{equation}
\bar{\sigma} =\sigma - \frac{\Lambda}{2m^2} \,.
\end{equation}
The Lagrangian may be diagonalized by a shift in the $k$ field
\begin{equation}
k_{\mu\nu} \rightarrow k_{\mu\nu}+ \frac{1}{m^2 \bar{\sigma}}\, p_{\mu\nu}\,.
\end{equation}
It reduces to the sum of a massless mode, plus a Fierz-Pauli Lagrangian:
\begin{equation}\label{NMGqa}
\cL_{\rm NMG}^{(2)} = - \bar{\sigma} k^{\mu\nu} \cG_{\mu\nu}(k) + \frac{1}{m^4\bar{\sigma}} \left( p^{\mu\nu} \cG_{\mu\nu}(p) + \frac12 \cM^2 \left( p_{\mu\nu}p^{\mu\nu} - p^2 \right) \right)\,,
\end{equation}
with Fierz-Pauli mass
\begin{equation}\label{FPNMG}
\cM^2 = - \sigma m^2 + \frac{\Lambda}{2} = - \bar{\sigma} m^2\,.
\end{equation}
The NMG action is a non-linear extension of Fierz-Pauli theory in three dimensions. However, there is a problem. The two kinetic terms in \eqref{NMGqa} come with opposite signs. This means that, depending on the sign of $\bar{\sigma}$, either the massive or the massless mode will have negative energy. In flat space, where $\Lambda = 0$, this problem can be resolved by taking $\sigma = -1$. Then the kinetic term for the Fierz-Pauli part carries the correct sign. The pure gauge mode $k_{\mu\nu}$ has negative energy, but since it does not propagate, this is no serious problem. In AdS, this does lead to a more serious problem, since the massless modes correspond to the boundary gravitons and a wrong sign kinetic term will lead to a negative mass for the BTZ black hole and a negative central charge, as we will see in more detail in chapter \ref{chapter:asymptotic_symm}.

\subsection{General Massive Gravity}\label{sec:GMG}
New Massive Gravity, as obtained in the last section, is the unique way to obtain a parity even CS--like action with two auxiliary fields resulting in a fourth-order derivative extension of GR. But if we relax the assumption of preserving parity, we have more options. We can, for instance, supplement NMG with the parity violating Lorentz-Chern-Simons term \eqref{LCS}, resulting in General Massive Gravity (GMG). Adding the Lorentz Chern-Simons term to NMG does not affect the number of local degrees of freedom and the linear theory now describes two helicity $\pm 2$ states with different masses. 

The GMG Lagrangian three-form is given in a CS--like form as \cite{Hohm:2012vh}:
\begin{equation}\label{LGMGsum}
L_{\rm GMG} = L_{\rm NMG} + \frac{1}{\mu} L_{\rm LCS}\,.
\end{equation}
In the limit where $m \to \infty$, this reduces to the TMG Lagrangian and a limit where $\mu \to \infty$ gives NMG. The field equations are now
\begin{equation}\label{GMGeom}
\begin{split}
& \cD e^a = 0\,, \\
& R^a +  (e \times f)^a = 0\,, \\
& \cD f^a - m^2 (e \times h)^a - \frac{m^2}{\mu} R^a = 0\,, \\
&\cD h^a  - \sigma R^a + \frac{\Lambda_0}{2} (e \times e)^a - \frac{1}{2m^2} (f \times f)^a  = 0\,,
\end{split}
\end{equation}
which gives for the auxiliary fields
\begin{equation}
f_{\mu\nu} = -S_{\mu\nu}\,, \qquad h_{\mu\nu} = - \frac{1}{\mu} S_{\mu\nu} - \frac{1}{m^2} C_{\mu\nu}\,.
\end{equation}
The linearized equations of motion now describe the two helicity states of a massive mode with masses $m_+$ and $m_-$:
\begin{equation}
(\cD^{m_+} \cD^{m_-} \cD^L \cD^R k)_{\mu\nu} = 0\,.
\end{equation}
The masses $m_{\pm}$ are related to the parameters of the theory as
\begin{equation}
m_{\pm} = \frac{m^2}{2 \mu} \pm \sqrt{\frac{m^4}{4\mu^2} - \frac{\Lambda}{2} - m^2 \sigma }\,.
\end{equation}
In the limit to NMG, where $\mu \to \infty$ this gives
\begin{equation}
m_{\pm}^2 = - \left( \sigma m^2 + \frac12 \Lambda \right) = \cM^2 - \Lambda\,,
\end{equation}
with $\cM^2$ the Fierz-Pauli mass in NMG, given by \eqref{FPNMG}. In the TMG limit, $m^2 \to \infty$ (and $\mu >0$), we have that $m_+ \to \infty$. One of the massive modes thus becomes infinitely heavy and the theory describes a single massive helicity state with mass $m_- = \sigma \mu$.

\section{Zwei-Dreibein Gravity}
\label{section:ZDG}
As we saw in the preceding sections, the non-linear extensions of Fierz-Pauli discussed thus far are in some way or another dissatisfactory. In the dGRT model, we found a scalar ghost-free model which is fully non-linear, but at the expense of introducing a reference field. Furthermore, we broke the gauge symmetries of general relativity, which we know how to handle so well in the Chern-Simons formulation of gravity. This complicates a further classification of the field theory dual to dRGT in AdS spacetimes (for attempts see however \cite{Vegh:2013sk,Davison:2013jba,Blake:2013bqa,Blake:2013owa}).

NMG is also a ghost-free construction of three dimensional massive gravity in flat space.\footnote{The absence of the Boulware-Deser ghost in NMG, along with its Hamiltonian analysis will be treated in the next chapter.} The problem here arises when we consider the theory in AdS space, where positive energy for the massive mode cannot be reconciled with a positive boundary central charge (see for more details chapter \ref{chapter:asymptotic_symm}). Hence, it is worth the effort to look for different, or more general non-linear theories of massive spin-2 modes, while retaining the computational advantages of the Chern-Simons--like model.  

One such example is Zwei-Dreibein Gravity (ZDG) \cite{Bergshoeff:2013xma}. It extends both dRGT and NMG to a more general class of theories in the following sense. In ZDG, the reference dreibein of dRGT is promoted to a fully dynamical dreibein, with corresponding spin connection. The theory can thus be thought of as a first-order formulation of a bimetric gravity \cite{Isham:1971gm,Hassan:2011zd,Hinterbichler:2012cn}. 
Simultaneously, ZDG extends NMG in the Chern-Simons--like form by loosening the assumption that the fields $f^a$ and $h^a$ are auxiliary. By auxiliary here we mean, that they can be solved for in terms of a finite number of derivatives acting on the dreibein, and that they can be eliminated from the action in favor of higher-derivative terms. 

In this section we will study the Zwei-Dreibein Gravity model in full detail, following the discussion presented in \cite{Bergshoeff:2013xma} and \cite{Bergshoeff:2014eca}. We will present the model and give its linear theory. Then, we will show how the ZDG model is related to dRGT massive gravity and NMG. Finally, we show the relation between ZDG and a higher derivative theory. We show that the four coupled first-order field equations of ZDG can be written in terms of a single equation for a metric with an infinite amount of higher-derivate terms. However, this reformulation of the ZDG field equations cannot be derived from varying an action containing an infinite amount of higher-derivative contributions with respect to the metric; the ZDG action can only be written in terms of the two dreibeine.

\subsection{The ZDG model}
In three dimensions, Zwei-Dreibein Gravity can be described as a family of actions with Lagrangian three-form \cite{Bergshoeff:2013xma}:
\begin{equation}\label{Lbimetric}
\begin{split}
L_{ZDG} =  - M_P \bigg\{ & \sigma e_{1} \cdot R_1 +  e_{2} \cdot R_2 +  \frac{\alpha_1}{6} m^2 e_1 \cdot e_1 \times  e_1
  +  \frac{\alpha_2}{6} m^2 e_2 \cdot e_2 \times e_2 \\ & - \frac12 m^2\beta_1  e_1 \cdot e_1 \times e_2 - \frac12 m^2\beta_2 e_1 \cdot e_2 
  \times e_2 \bigg\} \,.  
\end{split}
\end{equation}
The basic variables of this model are two Lorentz vector valued one-forms, the drei\-/beine $e_{I}{}^a$ with $I= 1,2$, and a pair of Lorentz vector valued connection one-forms $\omega_{I}{}^a$, whose curvature two-forms $R_I{}^a$ are given by:
\begin{equation}\label{Rdefinition}
R_I{}^a = d \omega_I{}^a + \frac12 (
\omega_{I} \times \omega_{I})^a\,.
\end{equation}
The independent parameters of \eqref{Lbimetric} are two cosmological parameters $\alpha_I$, two coupling constants $\beta_I$ and the Planck mass $M_P$. Besides these, we have introduced a convenient, but redundant parameter $m^2$ with dimension mass-squared and a sign parameter $\sigma = \pm1 $.

The equations of motion for $e_{1}{}^a, e_{2}{}^a$ and $\omega_{I}{}^a$, derived from the Lagrangian density \eqref{Lbimetric} are given by:
\begin{align}
0  = & \;  \sigma R_1^a+ \frac12 m^2 \left[
 \alpha_1 (e_{1} \times e_{1})^a - 2 \beta_1 (e_{1} \times e_{2})^a - \beta_2 (e_{2} \times  e_{2})^a \right]\,,
\label{R1eom} \\[.3truecm]
0  = &  \;  R_2^a + \frac12 m^2  \left[
\alpha_2 (e_{2} \times e_{2})^a - \beta_1  (e_{1} \times e_{1})^a - 2 \beta_2 (e_{1} \times e_{2})^a \right]\,,
\label{R2eom} \\[.3truecm]
0  = & \; T_{I}^a\,. \label{Teom}
\end{align}
Here, the $T_I^a$ are the torsion 2-forms, given by:
\begin{equation}\label{Tdefinition}
T_I^a = \cD_I e_I^a
\equiv d e_I^a + (\omega_{I} \times  e_{I})^a\,,
\end{equation}
where $\cD_I$ is the covariant derivative with respect to $\omega_I{}^a$ for $I= 1,2$. 
Note that the curvature and torsion 2-forms satisfy the Bianchi identities:
\begin{equation}\label{Bianchi}
\cD_I R_I^a = 0\,, \qquad 
\cD_I T_I^a = (R_{I} \times  e_{I})^a\,.
\end{equation}
Each of the kinetic terms of the dreibeine is invariant under their own diffeomor\-/phisms and local Lorentz transformations. Due to the presence of the interaction term, these symmetries are broken to their diagonal subgroups, defined by identifying the two sets of gauge parameters.

ZDG allows for maximally symmetric vacua, given by:
\begin{align} \label{AdSsol}
e_1{}^a = \bar{e}^a\,, \qquad e_2{}^a = \gamma \bar{e}^a\,, \qquad \omega_I{}^a = \bar{\omega}^a \,,
\end{align}
where $\gamma$ is a scaling parameter. The $\bar{e}^a$ and $\bar{\omega}^a$ are a dreibein and spin connection for a maximally symmetric spacetime with cosmological constant $\Lambda$ and as such obey:
\begin{align}
& d\bar{\omega}^a + \frac12 (\bar{\omega} \times \bar{\omega})^a - \frac12 \Lambda  (\bar{e} \times\bar{e})^a = 0\,,\\
& \bar{\cD} \bar{e}^a \equiv d\bar{e}^a + ( \bar{\omega} \times \bar{e})^a= 0\,.
\end{align} 
Indeed, it can then be seen that \eqref{AdSsol} is a solution of the ZDG equations of motion, provided that the scaling parameter $\gamma$ and cosmological constant $\Lambda$ obey:
\begin{align}\label{cc12}
  \alpha_1 =  2 \gamma \beta_1 + \gamma^2 \beta_2 -\sigma \frac{ \Lambda}{m^2}  \,, && 
 \gamma^2 \alpha_2 =   \beta_1 + 2 \gamma \beta_2 -  \frac{\Lambda}{m^2} \,. 
\end{align}
For given values of the parameters $\alpha_I$, $\beta_I$, these can be seen as two equations that can be solved to express $\gamma$ and $\Lambda$ in terms of the ZDG parameters.

\subsection{Linearized theory}\label{sec:linZDG}

We will now linearize ZDG around an AdS$_3$ spacetime with cosmological constant $\Lambda$, characterized by the dreibein $\bar{e}^{a}$ and spin connection $\bar{\omega}^{a}$ as described above. We thus expand the two dreibeine and spin connections, taking the scaling parameter $\gamma$ of \eqref{AdSsol} into account, as follows:
\begin{align}
e_{1}{}^a & =  \bar{e}^a + \kappa k_{1}{}^a\,, &
\omega_{I}{}^a & = \bar{\omega}^{a} + \kappa v_{I}{}^a\,, \\
e_{2}{}^a & = \gamma \left( \bar{e}^a + \kappa k_{2}{}^a \right)\,, \nonumber
\end{align}
where $\kappa$ is a small expansion parameter.  The linear terms in the expansion of the Lagrangian density \eqref{Lbimetric} cancel when eqs. \eqref{cc12} hold.
The quadratic Lagrangian for the fluctuations $k_{I\,\mu}{}^{a}$ and $v_{I\,\mu}{}^a$ is given by:
\begin{align} \label{lin_L}
L^{(2)} =  - \sigma M_P & \left[k_{1} \cdot \bar{\cD}v_{1} + \frac12  \bar{e} \cdot \left(v_{1} \times v_{1} - \Lambda k_{1} \times k_{1}\right) \right] \nonumber \\
 - \gamma M_P & \left[k_{2} \cdot \bar{\cD}v_{2}+ \frac12 \bar{e} \cdot \left(v_{2} \times v_{2} - \Lambda k_{2} \times k_{2}\right) \right] \\
\nonumber   - \frac12 m^2 \gamma & (\beta_1+\gamma \beta_2) M_P \bar{e} \cdot  \left( k_{1} -  k_{2} \right) \times  \left(  k_{1} -  k_{2}\right) \,.
\end{align}
Provided $\sigma + \gamma \neq 0$, this Lagrangian can be diagonalized by performing the linear field redefinitions
\begin{align} \nonumber
(\sigma + \gamma)  k_{+}{}^a &  = \sigma  k_{1}{}^a +  \gamma  k_{2}{}^a\,, &
k_{-}{}^a  & =  k_{1}{}^a -  k_{2}{}^a\,, \\
(\sigma + \gamma) v_{+}{}^a  & = \sigma v_{1}{}^a +  \gamma  v_{2}{}^a \,, &
v_{-}{}^a & =  v_{1}{}^a -  v_{2}{}^a \,.
  \label{lin_redef}
\end{align}
In terms of these fields the linearized Lagrangian becomes:
\begin{align} \label{linear_bi_lagrangian}
L^{(2)}   =  - (\sigma + \gamma) M_P & \left[ k_{+} \cdot \bar{\cD} v_{+} + \frac12 \bar{e} \cdot \left( v_{+} \times v_{+} - \Lambda k_{+} \times k_{+} \right) \right]  \nonumber \\[.2truecm]
  - \frac{\sigma \gamma }{(\sigma + \gamma)} M_P & \bigg[ k_{-} \cdot \bar{\cD} v_{-} + \frac12 \bar{e} \cdot \left( v_{-} \times v_{-} - \Lambda k_{-} \times  k_{-} \right)  \\
  & \;\; + \frac12 \cM^2 \bar{e} \cdot k_- \times k_{-} \bigg]\,,
\nonumber\end{align}
where the mass parameter $\cM$ is given in terms of the ZDG parameters as:
\begin{equation}\label{MFP}
\cM^2 =  m^2 (\beta_1 + \gamma \beta_2)\frac{\sigma + \gamma}{\sigma} \,.
\end{equation}
The first line in \eqref{linear_bi_lagrangian} is just the linearized Einstein-Cartan Lagrangian \eqref{ECquad} for the fields $k_+{}^a$ and $v_+{}^a$. The remainder is the Fierz-Pauli Lagrangian \eqref{SFP} for a massive spin-2 with Fierz-Pauli mass \eqref{MFP}.

When $\sigma + \gamma = 0$, the massive mode becomes massless and the linearized Lagrangian cannot be diagonalized. At this point, a new solution can be found with logarithmic fall-off behavior towards the AdS$_3$ boundary and it can be argued that the dual theory is a logarithmic CFT. We will return to this special case in chapter \ref{chapter:AdS_LCFT}.

\subsection{Scaling Limits of ZDG}
Since ZDG can be seen as both an extension of dRGT massive gravity and of New Massive Gravity, it is not so surprising that both theories are contained within ZDG as a scaling limit. Here we show this relation explicitly. 

\subsubsection{dRGT limit}
To obtain the dRGT model discussed in chapter \ref{sec:dRGT}, we need to fix one of the two dreibeine of ZDG to be come a fixed reference dreibein. We will choose this to be $e_2{}^{a}$, but a similar limit exists where $e_1{}^a$ is held fixed. Consider expanding $e_2{}^a$ and $\omega_2{}^a$ in powers of $\lambda$ as
\begin{equation}
\begin{split}
& e_2{}^a = \frac{1}{\lambda^2} \bar{e}^a + \frac{1}{\lambda} k_2{}^a \,, \\
& \omega_2{}^a = \bar{\omega}^a + \lambda v_2{}^a \,.
\end{split}
\end{equation}
Here $\bar{e}^a$ is the fixed reference dreibein of dRGT and $\bo^a$ its corresponding (torsion free) spin connection. We parametrize the ZDG coupling constants as
\begin{equation}
\beta_1 = \lambda^2 \beta_1'\,, \qquad \beta_2 = \lambda^4 \beta_2'\,, \qquad \alpha_2 = \lambda^4 \frac{\bar{\Lambda}}{m^2}\,.
\end{equation}
Taking the $\lambda \to 0$ limit reduces the ZDG Lagrangian 3-form to
\begin{equation}
L =  M_P L^{(2)}(k_2,v_2) + \sigma M_p L_{\rm dRGT}(e_1,\omega_1)\,.
\end{equation}
The first term is the Lagrangian three-form of the quadratic approximation to the 3D EC action in the reference background. The action constructed from this Lagrangian is given in \eqref{ECquad}. The second term is the dRGT Lagrangian three-form \eqref{3DdRGT} for the dreibein $e_1$ and spin connection $\omega_1$. It also depends on the reference dreibein $\bar{e}^a$, which is a solution to Einstein's equations with cosmological constant $\bar{\Lambda}$.

\subsubsection{NMG Limit}
\label{sec_NMGLim}
ZDG is similar to NMG in the sense that it is also a four flavor Chern-Simons--like theory and it shares its linear spectrum. Both theories contain two parity even fields ($e_1$ and $e_2$ in ZDG, $e$ and $f$ in NMG) and two parity odd fields ($\omega_1$ and $\omega_2$ in ZDG and $\omega$ and $h$ in NMG). The crucial difference is that NMG is defined such that the fields $f$ and $h$ are auxiliary; they can be solved for in terms of derivatives on the dreibein and eliminated from the action. Although in ZDG it is possible to solve for $e_2$ and $\omega_2$ in terms of derivatives on the dreibein, they cannot be eliminated from the action and the resulting field equation for $e_1$ contains an infinite amount of higher-derivative contributions, as will become clear in the next subsection. 

Still, ZDG contains NMG as a scaling limit \cite{Paulos:2012xe}, which we will now show. In our convention for the parameters of the Lagrangian of ZDG \eqref{Lbimetric} the limit to NMG is slightly different from the one given in \cite{Bergshoeff:2013xma}. To flow to NMG, we should take $\sigma = -1$ and make the field redefinition
\begin{equation}\label{NMGlim}
e_{2}{}^a = \Gamma e_{1}{}^a + \frac{\lambda}{m^2} f^a\,, \qquad \omega_{2}{}^a = \omega_{1}{}^a - \lambda h^a \,.
\end{equation}
NMG can then be obtained from the flow:
\begin{align}\label{NMGlim2}
M_P(\lambda) & =  \frac{1}{\lambda} M' \,, &&  \Gamma (\lambda) = 1 + \sigma' \lambda \,,  \nonumber \\
\alpha_1(\lambda) & = \left( 6 - \frac{\Lambda_0}{m^2}\right) \lambda + \frac{2(1+\sigma'\lambda)}{\lambda} \,,
&& \beta_1(\lambda) = \frac{1}{\lambda} + \lambda\,, \\
 \alpha_2 (\lambda) & = \frac{1}{\lambda} - 2 \sigma' \,, && \beta_2 = 0\,, \nonumber
\end{align}
and sending $\lambda \to 0$. Here $\sigma' = \pm 1$ is a new sign parameter and $\Lambda_0$ is a new cosmological parameter. The Lagrangian 3-form \eqref{Lbimetric} then becomes:
\begin{equation}\label{CSNMG}
\begin{split}
\cL_{\rm NMG} = M'  \bigg\{ & - \sigma' e_{1} \cdot R_{1} + \frac{\Lambda_0}{6} e_{1} \cdot e_{1} \times e_{1}+ h \cdot T_1  \\
& - \frac{1}{m^2}\left( f \cdot R_{1} + \frac12 e_1 \cdot f \times f \right)   \bigg\}\,.
\end{split}
\end{equation}
This action is the Chern-Simons--like formulation of New Massive Gravity as considered in \cite{Hohm:2012vh} and reviewed in section \ref{sec:NMG}, with a Planck mass $M' = (8 \pi G)^{-1}$.

\subsection{ZDG as a higher derivative theory}
\label{section:ZDGHD}

In \cite{Hassan:2013pca}, it was observed that bimetric theories can be thought of as higher-derivative theories. In this section we make this connection explicit in the ZDG case with $\beta_2 =0$.\footnote{This parameter choice is motivated by the removal of the Boulware-Deser ghost in ZDG theories, as we will discuss at length in chapter \ref{chapter:three_crowd}. However, it is also possible to find a higher derivative formulation of ZDG without this parameter restriction.} First we observe that we can solve the equations of motion (\ref{R1eom}) to obtain an expression for $e_2{}^{a}$ in terms of $e_1{}^{a}$. Using the property $\varepsilon^{\rho \sigma \tau} R_{1\, \sigma \tau}{}^a =  \det({e_1}) e_1{}^{\sigma\, a} G_1{}^{\rho}{}_{\sigma}$, we obtain the following expression for $e_2{}^{a}$:
\begin{align}
	e_{2\, \mu}{}^{a} &= \dfrac{\alpha_1}{2\beta_1} e_{1\, \mu}{}^{a} + \dfrac{\sigma}{m^2 \beta_1} S_{1\,\mu}{}^a \,,
	\label{e2}
\end{align}
where $ S_{1\,\mu}{}^a \equiv S_{1\, \mu\nu} e_{1}^{\nu\, a}$ and $S_{1\, \mu\nu} = R_{1\,\mu\nu} - \frac14 R_1 g_{1\,\mu\nu}$ is the Schouten tensor of $g_{1\,\mu\nu} \equiv e_{1\,\mu}{}^a e_{1\,\nu}{}^b \eta_{ab}$. In this approach to ZDG, we identify $g_{1\,\mu\nu}$ with the physical metric, as we have to assume $e_1{}^a$ to be invertible for the absence of ghosts (see chapter \ref{chapter:three_crowd}) and $e_2{}^a$ can be expressed as a function of $e_1{}^a$ and its derivatives\footnote{Since $e_2$ can be expressed in terms of $e_1$ and its derivatives on-shell, it is possible to find the inverse of $e_2$ as an infinite expansion in $1/m^2$.}. The field $e_2$ represents the higher-derivative content of the theory. It is similar to the auxiliary field in the two-derivative formulation of NMG in the sense that it can be solved for algebraically upon using the equations of motion. It is different in the sense that after solving $e_2$ for $e_1$, we cannot eliminate $e_2$ from the action, as this would now change the $e_1$ field equation which we used to express $e_2$ in terms of $e_1$. 

Using the relation \eqref{e2}, we can solve the torsion equation $T_2{}^a =0$ for $\omega_2{}^a(e_1)$ as a power series in $1/ m^2$. By expressing $\omega_2{}^a$ as:
\begin{equation} \label{seriesomega2}
\omega_{2\,\mu}{}^a = \sum_{n=0}^{\infty} \frac{1}{m^{2n}} \Omega^{(2n)}_{\mu}{}^a\,,
\end{equation}
and solving $T_2{}^a = d e_2{}^a + \epsilon^a{}_{bc} \omega_2{}^b e_2{}^c = 0$ order by order in $1/m^2$ we find:
\begin{equation}
\begin{split}
\Omega^{(0)}_\mu{}^a = & \; \omega_{1\,\mu}{}^a\,, \qquad \Omega^{(2)}_{\mu}{}^a = - \frac{2 \sigma}{\alpha_1}  C_{1\,\mu}{}^a \,, \\
\Omega^{(2k)}_{\mu}{}^a = & \; - \frac{2 \sigma}{\alpha_1} \det(e_1)^{-1} \varepsilon^{\nu \rho \sigma} \epsilon_{bcd} \left(e_{1\,\nu}{}^a e_{1\,\mu}{}^b -\frac12 e_{1\,\nu}{}^b e_{1\,\mu}{}^a \right) \Omega^{(2k-2)}_{\rho}{}^c S_{1\,\sigma}{}^d \,, 
\end{split} \label{Omega}
\end{equation}
for $k >1$. Here  $C_{1\,\mu}{}^a \equiv C_{1\,\mu\nu} e_{1}^{\nu\, a} $ and $C_{1\, \mu\nu} = \det(e_1)^{-1} \varepsilon_{\mu}{}^{\alpha \beta} \cD_{\alpha} S_{1\, \beta \nu}$ is the Cotton tensor associated with $g_{1\,\mu\nu}$. This result enables us to write $R_2{}^a$ as a series in $1/m^2$:
\begin{equation}\label{R2exp1}
R_2{}^a = \sum_{n=0}^{\infty} \frac{1}{m^{2n}} R_2^{(2n)\,a}\,,
\end{equation}
where the coefficients in this expansion are given by:
\begin{equation}\label{R2exp2}
\begin{split}
& R_2^{(0)\,a} = R_1{}^a\,, \qquad R_2^{(2)\, a} = -\frac{2\sigma}{\alpha_1}  \cD C_1{}^a \,, \\
& R_2^{(2k)\,a} = \cD \Omega^{(2k)\,a} + \frac12 \sum_{i=1}^{k-1} \left( \Omega^{(2i)} \times  \, \Omega^{(2k-2i)} \right)^a\,.
\end{split}
\end{equation}
The covariant derivative $\cD$ is defined with respect to $\omega_1{}^a$. Replacing these expressions into the equation of motion \eqref{R2eom} for $e_2{}^a$ we obtain a higher-order differential equation for $e_1{}^a$ as a power series in $1/m^2$.  The resulting equation of motion, up to order $1/m^4$ is given by:
\begin{align} \label{HDeom}
0 = \sqrt{-g} M_P & \left\{ \left(1+ \frac{\alpha_1\alpha_2\sigma}{2\beta_1^2}\right) G_{\mu\nu} - \left( \frac{\alpha_1^2\alpha_2}{4\beta_1^2}-\beta_1\right) m^2 g_{\mu\nu} + \frac{1}{m^2} E_{\mu\nu} + \frac{1}{m^4} F_{\mu\nu} \right.  \nonumber \\
&\quad  \left. + \cO\left(\frac{1}{m^6} \right)  \right\}\,,
\end{align}
where here and in the following we have omitted the label 1, used to denote the dreibein and we have rewritten it in a second-order form using the metric $g_{\mu\nu} = e_{1\,\mu}{}^a e_{1\,\nu}{}^b \eta_{ab}$ to raise and lower indices.
The symmetric tensors $E_{\mu\nu}$ and $F_{\mu\nu}$ carry terms with four and six derivatives respectively. They are:
\begin{align}
E_{\mu\nu} =  & - \frac{2\sigma}{\alpha_1} \bigg[ \Box R_{\mu\nu} - \frac14 (g_{\mu\nu} \Box R + \nabla_{\mu} \nabla_{\nu} R ) - 3 R_{\mu\rho}R^{\rho}{}_{\nu} + g_{\mu\nu} R_{\rho\sigma}R^{\rho\sigma} - \frac12 g_{\mu\nu} R^2 \nonumber \\ & + \dfrac{3}{2} R R_{\mu\nu}\bigg]
+\frac{\alpha_2}{2 \beta_1^2} \left[ g_{\mu\nu} R_{\rho\sigma}R^{\rho\sigma} - \frac58 g_{\mu\nu} R^2 + \frac32 RR_{\mu\nu} - 2 R_{\mu\rho}R^{\rho}{}_{\nu} \right] \,, \label{Eterms} \\
F_{\mu\nu} = & \; \frac{4}{\alpha_1^2} \left\{ \nabla^{\rho} \left[ S_{(\mu}{}^{\sigma} \nabla_{\nu)} S_{\rho \sigma} -  S_{(\mu|}{}^{\sigma}\nabla_{\rho} S_{|\nu) \sigma} - S_{\rho}{}^{\sigma} \nabla_{(\mu}S_{\nu) \sigma} + 2 S^{\sigma}{}_{\mu} \nabla_{[\rho} S_{\sigma] \nu}  \right. \right. \nonumber \\
& \qquad \qquad  \left. \left. + S^{\sigma}{}_{\rho} \nabla_{\sigma} S_{\mu\nu} \right] + \nabla^{\rho} S^{\lambda \sigma} \nabla_{[\lambda} S_{\rho] \sigma} g_{\mu\nu} - 2 \nabla^{\rho} S^{\sigma}{}_{\nu} \nabla_{[\sigma} S_{\rho] \mu} \right\} \label{Fterms}.
\end{align} 
The equation of motion for $e_1{}^a$ thus obtained contains an infinite number of terms and  features infinitely many derivatives. Note that higher-derivative actions with more than four derivatives can, unlike ZDG, propagate two or more massive gravitons (see \cite{Bergshoeff:2012ev} for an example). Typically however, in cases where this happens, there are terms with more than four derivatives acting on the metric field. This is not the case here as $\Omega^{(2n)}$ contains contractions of the Cotton tensor with $(2n-2)$ Schouten tensors and hence $R_2$ is an infinite series of terms that are products of terms that have at most four derivatives acting on the metric tensor. The resulting equations of motion thus do contain an infinite amount of derivatives, but the maximum amount of derivatives acting on the metric is four. Also note that in the case of an infinite number of derivatives, the initial value problem and ensuing counting of degrees of freedom is subtle (see e.g. \cite{Barnaby:2007ve} for a discussion on these issues). The above higher-derivative formulation is thus not at odds with the fact that ZDG propagates a single massive graviton.

As mentioned before, since the field equation for $e_1$ was used to obtain the solution $e_2(e_1)$ it is not allowed to substitute \eqref{e2} back into the action. The higher-derivative equation of motion \eqref{HDeom} can in general not be integrated to an action. However, at order $1/m^2$, there is a parameter relation for which this is possible:
\begin{equation} \label{actionrelation}
	-\frac{\sigma}{\alpha_1} = \dfrac{\alpha_2}{2\beta_1^2} .
\end{equation}
If the ZDG parameters are restricted in this way, the $1/m^2$ contributions to \eqref{HDeom}, given explicity in \eqref{Eterms}, can be integrated to an action proportional to $R_{\mu\nu}R^{\mu\nu} - \frac38 R^2$. This combination of $R^2$ terms corresponds to the higher-curvature part of the NMG action. 

This is not a coincidence as can be seen by explicitly performing the NMG limit. Indeed, after substituting \eqref{NMGlim2} into the coefficients in eqn.~\eqref{HDeom}, we see that the terms at order $1/m^4$ scale as $\lambda$ and hence vanish in the $\lambda \to 0$ limit, while the remaining coefficients become:
\begin{align}
& M_P \left(1 + \frac{\alpha_1\alpha_2 \sigma}{2\beta_1^2} \right)  = \sigma' + \cO(\lambda)\,, \nonumber \\
& - M_P\left( \frac{\alpha_1^2\alpha_2}{4\beta_1^2}-\beta_1\right) m^2 = \Lambda_0 + \cO(\lambda)\,, \\
& - M_P\frac{2 \sigma}{\alpha_1}  = 1 + \cO(\lambda)\,, \qquad M_P \frac{\alpha_2}{2\beta_1^2} = \frac12 + \cO(\lambda)\,. \nonumber
\end{align}
In particular, the last two equations show that the $\lambda \rightarrow 0$ limit enforces the parameter relation \eqref{actionrelation} 
and as a consequence the NMG equations of motion  \cite{Bergshoeff:2009hq,Bergshoeff:2009aq} that result from eqn.~\eqref{HDeom} in the $\lambda \rightarrow 0$ limit
\begin{equation}
\begin{split}
0 = \sigma' G_{\mu\nu} + \Lambda_0 g_{\mu\nu} + \frac{1}{m^2} & \left[ \Box R_{\mu\nu} - \frac14 (g_{\mu\nu} \Box R + \nabla_{\mu}\nabla_{\nu}R ) - 4 R_{\mu\rho}R^{\rho}{}_{\nu} \right. \\
& \left. + \frac94 RR_{\mu\nu} + \frac32 g_{\mu\nu} R_{\rho\sigma}R^{\rho\sigma} - \frac{13}{16} g_{\mu\nu}R^2 \right] \,,
\end{split}
\end{equation}
can be integrated to an action, even if for the generic ZDG equations of motion \eqref{HDeom} this is not possible order by order in $m^2$. 

Note that the utility of the higher-derivative formulation of ZDG will very much depend on the specific application one has in mind. For instance, the higher\-/derivative formulation can be used to find solutions to the non-linear equations of motion, as we will show in chapter \ref{chapter:AdS_LCFT}. For other applications, e.g. in AdS/CFT, an action for which the variational principle is well-defined is required and one will have to resort to the two-derivative Zwei-Dreibein formulation. Note that even if the higher-derivative terms could be integrated to an action, a formulation without higher derivatives is still more useful in order to set up a well-defined variational principle, as is discussed in the NMG case in \cite{Hohm:2010jc}.

\section{Conclusions}
In this chapter we have investigated how to describe a free massive spin-2 mode in three dimensions starting from an action which is first order in derivatives. We then discussed various non-linear extensions of the free Fierz-Pauli theory, most of which have a very natural formulation in terms of a Chern-Simons--like model, defined by \eqref{LCSlike}. In the case where some of the Lorentz vector-valued one-form fields are auxiliary and in the ZDG model, we have shown how these theories are related to higher-derivative models of gravity in a metric formulation. We have also discussed how the CS--like models reduce to Fierz-Pauli theory upon linearisation around a maximally symmetric background.

What we did not discuss here is whether there are extra, ghost-like degrees of freedom which are hidden from the linear analysis. We did discuss the Boulware-Deser mode in a metric formulation of dRGT massive gravity, but what about the Chern-Simons--like models? Would they have an additional, third degree of freedom when taking non-linearities into account?  To answer this question, we will investigate the degrees of freedom of the non-linear theory, independently of the chosen background, by performing a Hamiltonian analysis of the theory. This is where the advantage of the Chern-Simons--like formulation really becomes clear, as it is possible to treat the general formalism at great length before specifying the specific model. This will be the subject of the next chapter.

%% file: chapter_4/chapter_4.tex
\pagestyle{empty}
\setcounter{chapter}{3}
\chapter[Hamiltonian Formulation of 3D CS--like Models]{The Hamiltonian Formulation of Three Dimensional Chern-Simons--like Gravity Models}
\label{chapter:three_crowd}

\pagestyle{headings}

\begin{quote} \em
The fact that many of the higher-derivative extensions of general relativity in three dimensions may be brought in a Chern-Simons--like form greatly simplifies the Hamiltonian analysis of these theories. In this chapter, we first consider the Hamiltonian form of the general Chern-Simons--like model. We show that the presence of secondary constraints is directly related to the presence of invertible fields in the theory. After discussing the general procedure, we apply it to some of the examples discussed in the previous chapter, being General Massive Gravity (GMG), Zwei-Dreibein Gravity (ZDG) and a further parity violating extension of the latter, called General Zwei-Dreibein Gravity (GZDG). We show that for GMG, the presence of secondary constraints is guaranteed by invertibility of the dreibein. For ZDG this is not the case and the general theory contains three degrees of freedom. However, by making use of the general procedure, we are able to remove the third degree of freedom by a simple assumption; we need to require that a linear combination of the two dreibeine is invertible.  The content of this chapter is based on \textsc{[viii]}.
\end{quote}

\newpage
\section{Introduction: CS--like gravity theories}
\label{sec:1}

As discussed in chapter \ref{chapter:GRin3D}, General Relativity (GR) in three space-time dimensions can be interpreted as a Chern-Simons (CS) gauge theory of the 3D Poincar\'e,  de Sitter (dS) or anti-de Sitter (AdS) group, depending on the value of the cosmological constant \cite{Achucarro:1987vz,Witten:1988hc}. The action is the integral of a Lagrangian three-form $L$ constructed from the wedge products of Lorentz-vector valued one-form fields:  the dreibein $e^a$ and the dualised spin-connection $\omega^a$. Using a notation in which the wedge product is implicit, and a ``mostly plus'' metric signature convention, we have 
\begin{equation}\label{CSGR}
L = -e_a  R^a + \frac{\Lambda}{6} \varepsilon^{abc} e_a e_b e_c \, ,
\end{equation}
where $R^a$ is the dualised Riemann two-form:
\begin{equation}
R^a= d\omega^a + \frac{1}{2} \varepsilon^{abc} \omega_b \omega_c\, . 
\end{equation}
This action is manifestly local Lorentz invariant,  in addition to its manifest  invariance under diffeomorphisms, which are on-shell equivalent to local translations. The field equations are zero field strength conditions for the Poincar\'e or (A)dS group.

Strictly speaking, the CS gauge theory is equivalent to 3D GR only if one assumes  invertibility of the dreibein; this is what allows the Einstein field equations to be written as zero field-strength conditions, and it is one way to see that 3D GR has no local degrees of freedom, and hence no gravitons. However, there are variants of 3D GR that do propagate gravitons.  The simplest of these are 3D ``massive gravity'' theories found by including certain higher-derivative terms in the action\footnote{It is possible, at least in some cases, to take a massless limit but since ``spin'' is not defined for massless 3D particles, one cannot get a theory of  ``massless gravitons''  this way, if by ``graviton'' we mean a particle of spin-2.}.
The best known examples were reviewed in chapter \ref{chapter:two_company}. They are: Topologically Massive Gravity (TMG),  which leads to third-order field equations that propagate a single spin-2 mode \cite{Deser:1981wh} and New Massive Gravity (NMG) which leads to parity-preserving fourth-order equations that propagate a parity-doublet of massive spin-2 modes. Combining TMG and NMG we get a parity-violating fourth-order General Massive Gravity (GMG) theory that propagates two massive gravitons, but with different masses \cite{Bergshoeff:2009hq}.

Although the field equation of GMG is fourth order in derivatives, it is possible to introduce auxiliary tensor fields to get a set of equivalent first-order equations \cite{Hohm:2012vh}; in this formulation the fields can all be taken to be Lorentz vector-valued one-forms, and the action takes a form that is ``CS--like''  in the sense that it  is the integral of a Lagrangian three-form defined without an explicit space-time metric (which appears only on the assumption of an invertible dreibein). The general model of this type can be constructed as follows \cite{Hohm:2012vh}. We start from a collection of $N$ Lorentz-vector valued one-forms $a^{r \, a} = a_{\mu}^{r \, a} dx^{\mu}$, where $r,s,t\dots$ are ``flavor'' indices; the generic Lagrangian three-form constructible from these
one-form fields is
\begin{equation}\label{Lgeneral}
L = \frac12 g_{rs}  a^r \cdot da^s + \frac16 f_{rst} a^r \cdot (a^s \times a^t)\,,
\end{equation}
where $g_{rs}$ is a symmetric constant metric on the flavor space which we will require to be invertible\footnote{If $g_{rs}$ is not invertible, then not all primary constraints are independent and a separate analysis is required. One example is the dRGT model discussed in chapter \ref{sec:dRGT} which does not contain kinetic terms for the reference dreibein.}, so it can be used  to raise and lower flavor indices, and the coupling constants $f_{rst}$ define a  totally symmetric ``flavor tensor''.  We now use  a 3D-vector algebra notation for Lorentz vectors in which contractions with $\eta_{ab}$ and $\ve_{abc}$ are represented by  dots and crosses respectively. The  three-form  (\ref{Lgeneral}) is a CS three-form when the constants
\begin{equation}
f^{ar} {}_{bs \ ct} \equiv \ve^a{}_{bc}f^r{}_{st}\,, \quad \& \quad  g_{ar\ bs} \equiv \eta_{ab} g_{rs}\,,
\end{equation}
are, respectively,  the structure constants of a Lie algebra, and a group invariant symmetric tensor on this Lie algebra\footnote{There are CS gauge theories for which the Lagrangian three-form is {\it not} of the form  (\ref{Lgeneral}) because not all of the generators of the Lie algebra of the gauge group are Lorentz vectors. If  we wish the class of  CS gravity theories to be a subclass of the class of   CS--like gravity theories, we should define the latter by a larger  class of three-form Lagrangians, as in \cite{Hohm:2012vh}, but  (\ref{Lgeneral})  will be sufficient for our purposes.}.  For example, with $N=2$ we may choose $a^{e\,a} =e^a$ and $a^{\omega\, a} =\omega^a$, and then a choice of the flavor metric and coupling constants that ensures local Lorentz invariance will yield a CS three-form equivalent, up to field redefinitions, to  (\ref{CSGR}). For $N>2$, we will continue to suppose that $a^{e\,a}=e^a$ and $a^{\omega\,a}=\omega^a$, and that the flavor metric and coupling constants are such that the action is local Lorentz invariant, but even with this restriction the generic $N>2$ model will be only CS--like. In particular, TMG has a CS--like formulation with $N=3$ and both NMG and GMG have CS--like formulations with $N=4$ (see chapter \ref{section:auxfields}). Since these models have local degrees of freedom they are strictly CS--like, and not CS models.

The generic $N=4$ CS--like gravity model also includes the recently analyzed Zwei-Dreibein Gravity (ZDG) \cite{Bergshoeff:2013xma}. This is a parity preserving massive gravity model with the same local degrees of freedom as NMG (two propagating spin-2 modes of equal mass in a maximally-symmetric vacuum background) but has advantages in the context of the AdS/CFT correspondence since, in contrast to NMG, it leads to a positive central charge for a possible dual CFT at the AdS boundary, as we will discuss in detail in chapter \ref{chapter:asymptotic_symm}.  We shall show here that there is a parity-violating extension of ZDG, which we call ``General Zwei-Dreibein Gravity'' (GZDG).

We focus here on the Hamiltonian formulation of CS--like gravity models for a number of reasons. One is that the CS--like formulation allows us to discuss various 3D massive gravity models as special cases of a generic model, and this formulation is well-adapted to a Hamiltonian analysis. Another is that there are some unusual features of the Hamiltonian formulation of massive gravity models that are clarified by the CS--like formalism.  One great advantage of the Hamiltonian approach is that it allows a determination of the number of local degrees of freedom independently of a linearized approximation (which can give misleading results). In particular,  massive gravity models typically have an additional local degree of freedom, the Boulware-Deser ghost \cite{Boulware:1973my}.  It is known that GMG has no Boulware-Deser ghost, and this is confirmed by its Hamiltonian analysis,  but ZDG does have a Boulware-Deser ghost for generic parameters \cite{Banados:2013fda} (see also \cite{Deffayet:2012nr}), even though it is ghost-free in a linearized approximation. Fortunately, this problem can be avoided by a imposing an additional assumption on the theory as we will present here in detail.  We also present a parity-violating CS--like extension of ZDG, and we show that it has the same number of local degrees of freedom as  ZDG. 

\section{Hamiltonian Analysis}

It is straightforward to put the CS--like model defined by (\ref{Lgeneral}) into Hamiltonian form. We perform the space-time split
\begin{equation}\label{spacetimesplit}
a^{r\,a} = a_0^{r\,a} dt  +  a_i^{r\,a} dx^i \,,
\end{equation}
which leads to the Lagrangian density
\begin{equation}\label{gentimedecomp}
\cL = - \frac12  \varepsilon^{ij} g_{rs} a_{i}^r \cdot \dot{a}_{j}^s + a_{0}^r  \cdot \phi_r\, , 
\end{equation}
where $\varepsilon^{ij} = \varepsilon^{0ij}$. The time components of the fields, $a_0^{r\,a}$, become Lagrange multipliers for the primary constraints $\phi_r^a$:
\begin{equation}\label{constraints}
\phi_r^a = \varepsilon^{ij} \left(g_{rs} \partial_i a_j^{s\, a} + \frac12 f_{rst} \left( a_{i}{}^s \times a_{j}{}^t \right)^a \right)\,.
\end{equation}
The Hamiltonian density is just the sum of the primary constraints, each with a Lagrange multiplier $a_0^{r\,a}$,
\begin{equation}\label{canH}
\mathcal{H} = - \frac12\varepsilon^{ij} g_{rs} a_i^r \cdot \partial_0 a_j^s - \cL = - a_0{}^r \cdot \phi_r\,.
\end{equation}
We must now work out the Poisson brackets of the primary constraints. Then, following Dirac's procedure \cite{dirac-lecture}, we must consider any secondary constraints. 
We consider these two steps in turn.

\subsection{Poisson brackets and the primary constraints}

The Lagrangian is first order in time derivatives, so the Poisson brackets of the canonical variables can be determined by inverting the first term of \eqref{gentimedecomp}; this gives
\begin{equation}\label{poissonbr}
\left\{ a_{i\, a}^r (x) , a_{j\, b}^s (y) \right\}_{\rm P.B.} = \varepsilon_{ij}g^{rs} \eta_{ab} \delta^{(2)} (x-y)\,.
\end{equation}
Using this result we may calculate the Poisson brackets of the primary constraint functions. It will be convenient to first define the ``smeared'' functions $\phi[\xi]$ associated to the constraints \eqref{constraints} by integrating them against a test function $\xi_a^r(x)$ as follows
\begin{equation}\label{phi}
\phi[\xi] = \int_{\Sigma} d^2 x \; \xi_a^r(x) \phi_r^a (x) \,,
\end{equation}
where $\Sigma$ is space-like hypersurface.
In general, the functionals $\phi[\xi]$ will not be differentiable, but we can make them so by adding boundary terms. Varying \eqref{phi} with respect to the fields $a_{i}{}^s$ gives
\begin{equation}\label{gen_varphi}
\delta \phi [\xi] = \int_{\Sigma} d^2 x \; \xi^{r}_{a} \frac{\delta \phi_r^a}{\delta a_{i}^{s\, b}} \delta a_{i}{}^{s\,b} + \int_{\partial \Sigma} dx \; B[\xi, a,\delta a]\,.
\end{equation}
A non-zero $B[\xi,a,\delta a]$ could lead to delta-function singularities in the brackets of the constraint functions. To remove these, we can choose boundary conditions which make $B$ a total variation
\begin{equation}
\int_{\partial \Sigma} dx \; B[\xi,a, \delta a] = - \delta Q[\xi, a]\,.
\end{equation}
We then work with the `improved' quantities
\begin{equation}\label{varphi}
\varphi[\xi] = \phi[\xi] + Q[\xi, a]\,,
\end{equation}
which have well-defined variations, with no boundary terms. In our case, after  varying $\phi[\xi]$ with respect to the fields $a_{i}{}^{s}$, we find
\begin{equation}\label{gen_varbc}
\delta Q = - \int_{\partial \Sigma} dx^i  g_{rs} \xi^r \cdot \delta a_{i}{}^{s}\,.
\end{equation}

The Poisson brackets of the constraint functions can now be computed by using equation \eqref{poissonbr}. They are given by
\begin{align} \label{gen_poissonbr}
\left\{ \varphi[\xi] , \varphi[\eta]  \right\}_{\rm P.B.} = & \; \phi[[\xi, \eta]] + \int_{\Sigma} d^2x \; \xi^r_a \eta^s_b \, \cP_{rs}^{ab}
\nonumber \\
& - \int_{\partial \Sigma} dx^i \; \xi^r \cdot \left[g_{rs}  \partial_i \eta^s + f_{rst} (a_{i}{}^s \times \eta^t)   \right]\, ,
\end{align}
where
\begin{equation}
[\xi ,\eta]^t_c  = f_{rs}{}^{t} \ve^{ab}{}_{c} \xi^r_a \eta^s_b\, ,
\end{equation}
and
\begin{align}
\cP_{rs}^{ab} & = f^t{}_{q[r} f_{s] pt} \eta^{ab} \Delta^{pq}  +  2f^t{}_{r[s} f_{q]pt} (V^{ab})^{pq}\,, \label{Pmat_def} \\[.2truecm]
V_{ab}^{pq} & =  \varepsilon^{ij} a^p_{i\, a} a^q_{j\, b}\,, \hskip 1truecm \Delta^{pq} = \varepsilon^{ij} a_i^p \cdot a_j^q\,. \hskip 1truecm
\end{align}
In general, adopting non-trivial boundary conditions may lead to a (centrally extended) asymptotic symmetry algebra spanned by the first-class constraint functions if the corresponding test functions $\xi_a^r(x)$ are the gauge parameters of boundary condition preserving gauge transformations. The boundary term in \eqref{gen_poissonbr} will then split into a part proportional to the boundary term required to improve $\phi[[\xi,\eta]]$ and a possible central extension \cite{Banados:1994tn}.  In this chapter we will focus on the bulk theory and assume that the test functions $\xi_a^r(x)$ do not give rise to boundary terms in \eqref{gen_varphi} and \eqref{gen_poissonbr}. We will return to consider the boundary terms in the next chapter.

The consistency conditions guaranteeing time-independence of the primary constraints are
\begin{equation}
\frac{d}{dt} \phi^b_s = \{\mathcal{H}, \phi^b_s \}_{\rm P.B.} \approx - a_{0\,a}^r \cP_{rs}^{ab} \approx 0\,.
\end{equation}
This expression is equivalent to a set of ``integrability conditions'' which can be derived from the equations of motion. The field equations of \eqref{Lgeneral} are
\begin{equation}\label{covEOM}
g_{rs}d a^{s\,a} + \frac12 f_{rst} (a^s \times a^t)^a = 0\,.
\end{equation}
Taking the exterior derivative of this equation and using $d^2 = 0$, we get the conditions
\begin{equation}\label{Intcon}
f^t{}_{q[r}f_{s]pt}a^{r\,a} a^p \cdot a^q = 0\,.
\end{equation}
Using the space-time decomposition \eqref{spacetimesplit} we have
\begin{equation}\label{Intconspacetime}
0= f^t{}_{q[r}f_{s]pt}a^{r\,b} a^p \cdot a^q = a_0^{r}{}_a\cP_{rs}^{ab}\,,
\end{equation}
the right hand side being exactly what is required to vanish for time-independence of the primary constraints. These conditions are three-form equations in which each three-form necessarily contains a Lagrange multiplier one-form factor, so they could imply that some linear combinations of the Lagrange multipliers is zero.

If the matrix $\cP_{rs}^{ab}$ vanishes identically then  all primary constraints are first-class and there is no constraint  on any Lagrange multiplier.  In this case the model is actually a Chern-Simons theory, that of the Lie algebra with structure constants $\ve^a{}_{bc}f^{r}{}_{st}$.  In general, however, $\cP_{rs}^{ab}$ will not vanish and  $\rm{rank}(\cP)$ will be non-zero. We can pick a basis of constraint functions such that  $3N - \rm{rank}(\cP)$ have zero Poisson bracket with all constraints, while the remaining $\rm{rank}(\cP)$ constraint functions have non-zero Poisson brackets amongst themselves.  At this point, it might appear that  the Lagrange multipliers for the latter set of constraints will be set to zero by the conditions
(\ref{Intconspacetime}). However, when one of the fields is a dreibein, this may involve setting $e_0{}^a = 0$. This is not acceptable for a theory of gravity, as the dreibein must be invertible! When specifying a model, we must therefore be clear whether we are assuming invertibility of any fields as it affects the Hamiltonian analysis. In general, if we require invertibility of any one-form field  then we may need to impose further, secondary,  constraints.

In other words, the consistency of the primary constraints is equivalent to satisfying the integrability conditions (\ref{Intconspacetime}). If some one-form is invertible, then some integrability condition may reduce to a two-form constraint on the canonical variables, which we must add as a secondary constraint in our theory. We now turn to an investigation of these secondary constraints.

\subsection{Secondary constraints}
\label{sec:seccon}

To be precise, consider for fixed $s$ the expression $f^{t}{}_{q[r}f_{s]pt}a^{r\,a}$. If the sum over $r$ is non-zero for only one value of $r$, say $a^{e\, a} = e^a$, and $e^a$ is invertible, then the integrability conditions \eqref{Intcon} imply that
\begin{equation}
f^{t}{}_{q[e}f_{s]pt} a^p \cdot a^q = 0\,.
\end{equation}
In particular, taking the space-space part of this two-form, we find
\begin{equation}
\varepsilon^{ij} f^{t}{}_{q[e}f_{s]pt} a_i^p \cdot a_j^q = 0\,,
\end{equation}
which depends only on the canonical variables and is therefore a new, secondary, constraint. One invertible field may lead to several constraints if the above equation  holds for multiple values of $s$. The secondary constraints arising in this way\footnote{Here we should issue a warning: a linear combination of invertible one-forms is not in general invertible, so if $f^{t}{}_{q[r}f_{s]pt}a^{r\,a}$ sums over multiple values of $r$ with each corresponding one-form invertible, this does not in general imply a new constraint. } are therefore the inequivalent components of the flavor space vector $\psi_{s} = f^t{}_{q[e}f_{s]pt} \Delta^{pq}$. Let $M$ be the number of these secondary constraints, and let us write them as
\begin{equation}\label{seccon}
\psi_I = f_{I,pq}\Delta^{pq}\, , \quad I=1,\dots,M\, .
\end{equation}
We now have a total of $3N+M$ constraints.

According to Dirac, after finding the secondary constraints we should add them to the Hamiltonian with new Lagrange multipliers \cite{dirac-lecture}. However, in general this can change the field
equations. To see why let us suppose that we have a phase-space action $I[z]$ for some phase space coordinates $z$, and that  the equations of motion imply the constraint
$\phi(z)=0$. If we add this constraint to the action with a Lagrange multiplier $\lambda$ then we get a new action for which the equations of motion are
\begin{equation}
\frac{\delta I}{\delta z} = \lambda\frac{\partial \phi}{\partial z}\, , \qquad \phi(z)=0\, .
\end{equation}
Any solution of the original equations of motions, together with $\lambda=0$, solves these equations, but there may be more solutions for which $\lambda\ne0$. This is precisely what happens for NMG and GMG (although not for TMG) \cite{Hohm:2012vh}; the field equations of these models lead to a (field-dependent) cubic equation for one of the secondary constraint Lagrange multipliers, leading to two possible non-zero solutions for this Lagrange multiplier\footnote{This problem appears to be distinct from the problem of whether the ``Dirac conjecture'' is satisfied, since that concerns the values of Lagrange multipliers of first-class constraints. It may be related to the recently discussed ``sectors'' issue \cite{Dominici:2013lba}.}. In this case, Dirac's procedure  would appear to lead  us to a Hamiltonian formulation of  a theory that is more general than the one we started with (in that its solution space is larger). Perhaps more seriously, adding the secondary constraints to the Hamiltonian will generally lead to a violation of symmetries of the original model.

Because of this problem, we will omit the secondary constraints from the total Hamiltonian. This omission could lead to difficulties. The  first-class constraints are found by consideration of  the matrix of Poisson brackets of {\it all} constraints, so it could happen that some are linear combinations of primary with secondary constraints. We would then have a situation in which not all  first-class constraints are imposed by Lagrange multipliers in the (now restricted) total Hamiltonian, and this would appear to lead to inconsistencies.  Fortunately, this problem does not actually arise for any of the CS--like gravity models that we shall consider, as they satisfy conditions that we now spell out.

The  Poisson brackets of the primary with the secondary constraint functions are
\begin{align}
\label{PBsecondary}
\left\{ \phi[\xi], \psi_I \right\}_{\rm P.B.} =
&\, \varepsilon^{ij} \left[  f_{I,rp} \partial_i (\xi^r) \cdot a^p_{j} +  f_{rs}{}^t f_{I, pt} \xi^r \cdot \left(a_{i}^{s} \times a_{j}^{p} \right) \right]\,,
\end{align}
and the Poisson brackets of the secondary constraint functions amongst themselves are
\begin{equation}
\{\psi_I, \psi_J\}_{\rm P.B.} = 4 f_{I,pq}f_{J,rs} \Delta^{pr} g^{qs}\,.
\end{equation}
We now make the following two assumptions, {\it which hold for all our examples}:

\begin{itemize}

\item We assume that  all  Poisson brackets  of secondary constraints with other secondary constraints
vanish on the full constraint surface. It then follows that the total matrix of Poisson brackets of all $3N+M$ constraint functions takes the form
\begin{equation}\label{totalP}
\mathbb{P} = \left( \begin{array}{cc} \cP' & -\left\{ \phi, \psi \right\}^T   \\ \left\{ \phi, \psi \right\} & 0 \end{array}\right) \,,
\end{equation}
where $\cP^\prime$ is  the  matrix of Poisson brackets of the $3N$ primary constraints evaluated on the new constraint surface defined by all $3N+M$ constraints.

\item We assume that inclusion of the secondary constraints in the set of all constraints does not  lead to new  first-class constraints. This means that the secondary constraints must all be second-class, and any linear combination of secondary
constraints and the $\rm{rank}(\cP')$ primary constraints with non-vanishing Poisson brackets on the full constraint surface must be second-class.

\end{itemize}

The rank of $\mathbb{P}$, as given in (\ref{totalP}),  is the number of its linearly independent columns. By the second assumption, this  is $M$ plus the number of linearly independent columns of
\begin{equation}
\left( \begin{array}{c} \cP'  \\ \left\{ \phi, \psi \right\} \end{array}\right) \,.
\end{equation}
The number of linearly independent columns of this matrix, as for any other matrix,  is the same as the number of linearly independent rows, which by the second assumption is $\rm{rank}(\cP') + M$. The rank of $\mathbb{P}$, and therefore the number of second-class constraints, is then $\rm{rank}(\cP') + 2M$.

In principle one should now check for tertiary constraints. However, in this procedure the invertibility of certain fields will be guaranteed by the secondary constraints. The consistency of the primary constraints under time evolution can be guaranteed by fixing rank($\mathcal{P}'$) of the Lagrange multipliers. The consistency of the secondary constraints under time evolution, $a^r_{0a}\{\phi^a_r,\psi_I\} \approx 0$ can be guaranteed, under the second assumption, by fixing a further $M$ of the Lagrange multipliers. The fact that these $M$ multipliers are distinct from the rank($\mathcal{P}'$) multiplier fixed before follows from the second assumption. The remaining consistency condition, $\{\psi,\psi\} \approx 0$, is guaranteed by the first assumption.

We therefore have $3N$ - rank($\mathcal{P}'$) - $M$ undetermined Lagrange multipliers, corresponding to the $3N$ - rank($\mathcal{P}'$) - $M$ first-class constraints. The remaining rank($\mathcal{P}'$) + 2$M$ constraints are second-class. The dimension
of the physical phase space per space point is the number of
canonical variables $a_i^{ra}$, minus twice the number of
first-class constraints, minus the number of second-class
constraints, or
\begin{equation}
\label{dimcounting}
D = 6N - 2 \times \left( 3N - \rm{rank}(\cP') - M \right) - 1 \times \left( \rm{rank}(\cP') + 2M \right) = \rm{rank}(\cP') \,.
\end{equation}

We will now apply this procedure to determine the number of local degrees of freedom of various 3D gravity models with a CS--like formulation.

\section{Specific Examples}
\label{sec:2}

We will now derive the Hamiltonian form of a number of three-dimensional CS--like gravity models of increasing complexity following the above general procedure.

\subsection{Einstein-Cartan Gravity}

To illustrate our formalism we will start by using it to analyze 3D General Relativity with a cosmological constant $\Lambda$, in its first-order Einstein-Cartan form. 
There are two flavors of one-forms: the dreibein, $a^{e\,a} = e^a$, and the dualised spin-connection $a^{\omega \,a}= \omega^a = \frac12\varepsilon^{abc}\omega_{bc}$. The Lagrangian three-form is that of  (\ref{CSGR}). This takes the general form of \eqref{Lgeneral}, with the flavor index $r,s,t, \ldots = \omega, e$. The first step is to read off $g_{rs}$ and $f_{rst}$, and for later convenience we also determine the components of the inverse metric $g^{rs}$ and the structure constants with one index raised, $f^{r}{}_{st}$. The non-zero components of these quantities are:
\begin{align}
g_{\omega e} = -1\,, && f_{eee}=  \Lambda\,, && f_{e\omega \omega} = -1\,, \\ \nonumber
g^{\omega e} = -1\,, && f^{\omega}{}_{ee} = - \Lambda\,, && f^{\omega}{}_{\omega \omega} = 1\,, && f^{e}{}_{e \omega} = 1\,.
\end{align}
These constants define a Chern-Simons three-form, as mentioned in the introduction; the structure constants are $\varepsilon^a{}_{bc}f^{r}{}_{st}$.  This algebra is spanned by the Hamiltonian constraint functions, which are all first-class.  In three-dimensions, General Relativity, like any Chern-Simons theory, has no local degrees of freedom.

To see how the details  work in our general formalism, we can work out the matrix \eqref{Pmat_def} and find that it vanishes. Then, by equation \eqref{dimcounting} the dimension of the physical phase space is
\begin{equation}
D = 12 - 2\times 6 = 0\,,
\end{equation}
as expected. Using \eqref{gen_poissonbr} we can also verify that
\begin{equation}\nonumber
\{ \phi^a_\omega, \phi^b_{\omega} \}_{\rm P.B.} = \ve^{ab}{}_c \, \phi^c_\omega \,, \quad
\{ \phi^a_e, \phi^b_{\omega} \}_{\rm P.B.} = \ve^{ab}{}_c \, \phi^c_e \,, \quad
\{ \phi^a_e, \phi^b_e \}_{\rm P.B.} = - \Lambda \ve^{ab}{}_c \, \phi^c_{\omega}\,,
\end{equation}
which is the $SO(2,2)$ algebra for $\Lambda < 0$, $SO(3,1)$ for $\Lambda > 0$ and $ISO(2,1)$ for $\Lambda = 0$, as expected.

\subsection{General Massive Gravity}\label{sec:GMG_Hamil}

General Relativity was a very simple application of our general formalism; as a Chern-Simons theory the Poisson brackets of the constraint functions formed a closed algebra, so it did not require our full analysis. We will now turn to a more complicated example, General Massive Gravity (GMG). This theory does have local degrees of freedom; it propagates two massive spin-2 modes at the linear level. It contains two well known theories of 3D massive gravity as limits: Topologically Massive Gravity (TMG) \cite{Deser:1981wh} and New Massive Gravity (NMG) \cite{Bergshoeff:2009hq}. 

We can write the Lagrangian three-form of GMG in the general form \eqref{Lgeneral}. There are four flavors of one-form, $a^{r\,a} = (\omega^a, h^a, e^a, f^a)$, the dualised spin-connection and dreibein and two extra fields $h^a$ and $f^a$. The Lagrangian three-form was given in equation \eqref{LGMGsum}. It is, in full form: 
\begin{equation}
\begin{split} \label{LGMG}
L_{\rm GMG} = & - \sigma e \cdot R + \frac{\Lambda_0}{6} e \cdot e \times e + h \cdot T + \frac{1}{2\mu} \left[ \omega \cdot d\omega + \frac13  \omega \cdot \omega \times \omega \right]  \\
& - \frac{1}{m^2} \left[ f \cdot R + \frac12 e \cdot f \times f  \right]\,,
\end{split}
\end{equation}
where we recall that $R^a$ is the dualised Riemann two-form. 
The flavor-space metric $g_{rs}$ and the structure constants $f_{rst}$ can again be read off:
\begin{align}
g_{\omega e} = -\sigma\,, && g_{eh} = 1\,, && g_{f\omega} = - \frac{1}{m^2}\,, && g_{\omega\omega} = \frac{1}{\mu}\,, \nonumber \\
f_{e\omega \omega} = - \sigma\,, && f_{eh\omega} = 1\,, && f_{eff} = - \frac{1}{m^2}\,, && f_{\omega\omega\omega} = \frac{1}{\mu}\,, \\ \nonumber
f_{eee}= \Lambda_0\,, &&  && f_{\omega\omega f} = - \frac{1}{m^2}\,.
\end{align}
The next step is to work out the integrability conditions \eqref{Intcon}. We find three inequivalent three-form relations,
\begin{equation}
 e^a e \cdot f = 0\,, \quad
 f^a \left(\frac{1}{\mu } e \cdot f + h \cdot e \right)-h^a e \cdot f = 0\,, \quad
  e^a \left( \frac{1}{\mu } e \cdot f + h \cdot e \right) = 0\,. \\
\end{equation}
We will demand that the dreibein, $e^a$, is invertible. Following our general analysis, we find the two secondary constraints
\begin{equation}\label{secconGMG}
\psi_1 = \Delta^{eh} = 0\,, \qquad \psi_2 = \Delta^{ef} = 0\,.
\end{equation}
Next, we compute the matrix $\cP^{ab}_{rs}$ as defined in \eqref{Pmat_def}. All the $\Delta^{pq}$ terms drop out because of the secondary constraints, and in the basis $(\omega, h, e, f)$ we get
\begin{equation}\label{PmatGMG}
(\cP'_{ab})_{rs} = \left(
\begin{array}{cccc}
 0 & 0 & 0 & 0 \\
 0 & 0 & V_{ab}^{ef} & -V_{ab}^{ee} \\
 0 & V_{ab}^{fe} & -2 V_{[ab]}^{hf}  +\frac{1}{\mu }V_{ab}^{ff} & V_{ab}^{he}-\frac{1}{\mu }V_{ab}^{fe} \\
 0 & -V_{ab}^{ee} & V_{ab}^{eh}-\frac{1}{\mu }V_{ab}^{ef} & \frac{1}{\mu }V_{ab}^{ee} \\
\end{array} 
\right) \,.
\end{equation}
We must now determine the rank of this matrix at an arbitrary point in space-time. A Mathematica calculation shows that the rank of $\cP'$ is 4. According to \eqref{dimcounting} this is also the dimension of the physical phase-space, however, to complete the analysis, we need to verify if the assumptions stated in the last section hold. To this end we consider the Poisson brackets of the secondary 
constraint functions $\psi_I$ ($I=1,2$) with themselves and with the 
primary constraint functions. The Poisson bracket $\{\psi_1,\psi_2\}$ is zero on the 
constraint surface, which verifies the first assumption. The Poisson brackets of $\psi_I$ with the primary constraint functions are
\begin{equation} \label{GMGpsi1}
\begin{split}
\{ \phi[\xi],\psi_1 \}_{\rm P.B.}& = \ve^{ij} \bigg[ \partial_i \xi^h \cdot e_j - \xi^h \cdot (\omega_i \times e_j) -\partial_i \xi^e \cdot h_j + \xi^e \cdot (\omega_i \times h_j)  \\
& + \left(\sigma \xi^e + \frac{1}{m^2} \xi^f \right) \cdot (e_i \times f_j) + \left(\sigma \xi^f + \Lambda_0 \xi^e\right) (e_i \times e_j)
\bigg]\,,
\end{split}
\end{equation}
\begin{equation}\label{GMGpsi2}
\begin{split}
\{ \phi[\xi],\psi_2 \}_{\rm P.B.}& =  \ve^{ij} \bigg[ \partial_i \xi^f \cdot e_j - \xi^f \cdot (\omega_i \times e_j) -\partial_i \xi^e \cdot f_j + \xi^e \cdot (\omega_i \times f_j) \\
&  + \left( m^2 \xi^h -\frac{m^2}{\mu} \xi^f \right) (e_i \times e_j)  + m^2 \xi^e \cdot \left(e_i \times \left( h_j - \frac{1}{\mu} f_j \right) \right)
\bigg]\,.
\end{split}
\end{equation}
The full matrix of Poisson brackets is a $14\times 14$ matrix $\mathbb{P}$ given by
\begin{equation}\label{Pbrackets}
\mathbb{P} =
\left( \begin{array}{ccc}
\cP' & v_1 & v_2 \\
-v_1^T & 0 & 0 \\
-v_2^T & 0 & 0
\end{array} \right)\,,
\end{equation}
where the $v_I$, $(I = 1,2)$, are column vectors with components
\begin{equation} \label{column}
v_I =  \left( \begin{array}{c}
\{\phi^a_{\omega} , \psi_I \}_{\rm P.B.}\\
\{\phi^a_{h} , \psi_I \}_{\rm P.B.}\\
\{\phi^a_{e} , \psi_I \}_{\rm P.B.}\\
\{\phi^a_{f} , \psi_I \}_{\rm P.B.}
\end{array} \right)\,.
\end{equation}
These brackets can be read off from equations \eqref{GMGpsi1} and \eqref{GMGpsi2}. The vectors \eqref{column} are linearly independent from each other and from the columns of $\cP'$. This verifies the second assumption in the last section and the rank of $\mathbb{P}$ is increased by $4$. The full $(14 \times 14)$ matrix therefore has rank 8, so eight constraints are second-class and the remaining six are first-class. By eqn.~\eqref{dimcounting}, the dimension of the physical phase space per space point is
\begin{equation}
D = 24 - 8 - 2 \times 6 = 4 \,,
\end{equation}
consistent with the rank of $\cP'$. This means there are two local degrees of freedom, and we conclude that the non-linear theory has the same degrees of freedom as the linearized theory, two massive states of helicity $\pm 2$.

\subsection{Zwei-Dreibein Gravity}\label{sec:ZDG_Hamil}

We now turn our attention to another theory of massive 3D gravity, the recently proposed Zwei-Dreibein Gravity (ZDG) \cite{Bergshoeff:2013xma} which was reviewed in chapter \ref{section:ZDG}. It is a theory of two interacting dreibeine, $e_1^a$ and $e_2^a$, each with a corresponding spin-connection, $\omega_1{}^a$ and $\omega_2{}^a$. It also has a Lagrangian three-form of our general CS--like form  \eqref{Lgeneral}. Like NMG, ZDG preserves parity and has two massive spin-2 degrees of freedom when linearized about a maximally-symmetric vacuum background, but this does not exclude the possibility of additional local degrees of freedom appearing in other backgrounds. In fact, it was shown by  \cite{Banados:2013fda} that the generic ZDG model  does have an additional local degree of freedom, the Boulware-Deser ghost. We will see why this is so, and also how it can be removed by including a defining assumption to the theory. 

The Lagrangian three-form is
\begin{equation}\label{LZDG}
\begin{split}
L_{\rm ZDG} =  - M_P \bigg\{ &  \sigma e_{1} \cdot R_1 + e_{2} \cdot R_2  + \frac{m^2}{6}  \left( \alpha_1 e_{1} \cdot  e_{1} \times e_{1\, c}  +  \alpha_2 e_{2} \cdot e_{2} \times e_{2}\right) \\
&  - \cL_{12}(e_1,e_2) \bigg\} \,,
\end{split}
\end{equation}
where $R_1{}^a$ and $R_2{}^a$ are the dualised Riemann two-forms constructed from $\omega_1{}^a$ and $\omega_2{}^a$ respectively, and the interaction Lagrangian three-form $L_{12}$ is given by
\begin{equation}\label{Lint}
 L_{12}(e_1,e_2) =   \frac12 m^2 \left( \beta_1 e_{1} \cdot e_{1} \times e_{2} + \beta_2 e_{1} \cdot e_{2} \times e_{2} \right) \,.
\end{equation}
Here $\sigma = \pm 1$ is a sign parameter, $\alpha_1$ and $\alpha_2$ are two dimensionless cosmological parameters and $\beta_1$ and $\beta_2$ are two dimensionless coupling constants. The parameter $m^2$ is a redundant, but convenient, dimensionful parameter. 

From \eqref{LZDG} we can read off the components of $g_{rs}$ and $f_{rst}$. We will ignore the overall factor $M_P$ to simplify the analysis; after this step they become
\begin{eqnarray}\nonumber \label{ZDGfieldmetric}
g_{e_1\omega_1} = g_{\omega_1 e_1} = - \sigma  \,, & & g_{e_2\omega_2} = g_{\omega_2 e_2} =  - 1 \,, \\
f_{e_1 \omega_1 \omega_1} =  - \sigma  \,, & & f_{e_2 \omega_2 \omega_2} = - 1  \,, \\
f_{e_1e_1e_2} =  m^2 \beta_1 \,, & & f_{e_1e_2e_2} =  m^2 \beta_2 \,,\nonumber
\\
f_{e_1e_1e_1} = - m^2 \alpha_1 \,, & & f_{e_2e_2e_2} = -  m^2  \alpha_2 \,. \nonumber
\end{eqnarray}
We also work out the inverse metric $g^{rs}$ and the structure constants $f^r{}_{st}$,
\begin{eqnarray}\nonumber
g^{e_1\omega_1} = g^{\omega_1 e_1} = - \frac{1}{\sigma } \,, & & g^{e_2\omega_2} = g^{\omega_2 e_2} =  - 1  \,, \\
f^{\omega_1}{}_{\omega_1 \omega_1} = f^{e_1}{}_{\omega_1 e_1} = 1 \,, & & f^{\omega_2}{}_{\omega_2 \omega_2} = f^{e_2}{}_{\omega_2 e_2} = 1\,, \\
f^{\omega_1}{}_{e_1e_2} = f^{\omega_1}{}_{e_2e_1} = -\frac{ m^2 \beta_1 }{\sigma}  \,, & & f^{\omega_1}{}_{e_2e_2} =  - \frac{ m^2 \beta_2}{\sigma} \,,\nonumber
\\
f^{\omega_1}{}_{e_1e_1} = \frac{m^2 }{\sigma }\alpha_1 \,, & & f^{\omega_2}{}_{e_2e_2} = m^2 \alpha_2 \,, \nonumber \\
f^{\omega_2}{}_{e_1e_2} = f^{\omega_2}{}_{e_2e_1} = - m^2 \beta_2 \,, & & f^{\omega_2}{}_{e_1e_1} = - m^2 \beta_1 \,.\nonumber
\end{eqnarray}
Equipped with these expressions, we can evaluate the $12\times 12$ matrix of Poisson brackets \eqref{gen_poissonbr}, in the flavor basis $(\omega_1,\omega_2,e_1,e_2)$
\begin{align}\label{Pmat}
\hspace{-1truecm} \nonumber (\cP_{ab})_{rs} =  & \;\; m^2 \eta_{ab} \left(
\begin{array}{cccc}
 0 & 0 & - \beta _1 \Delta^{e_1e_2} & - \beta _2 \Delta^{e_1e_2} \\[.1truecm]
 0 & 0 &  \beta _1 \Delta^{e_1e_2} &  \beta _2 \Delta^{e_1e_2} \\[.1truecm]
  \beta _1 \Delta^{e_1e_2} & - \beta _1 \Delta^{e_1e_2} & 0 &  - \beta_1 \Delta^{\omega_-e_1}- \beta _2 \Delta^{\omega_-e_2} \\[.1truecm]
  \beta _2 \Delta^{e_1e_2} & - \beta _2 \Delta^{e_1e_2} &  \beta_1 \Delta^{\omega_-e_1} +\beta _2 \Delta^{\omega_-e_2} & 0
\end{array}
\right) \nonumber \\[.2truecm]
& + m^2 \beta_1
\left(
\begin{array}{cccc}
 0 & 0 &  V_{ab}^{e_1e_2} & - V_{ab}^{e_1e_1} \\[.1truecm]
 0 & 0 & -V_{ab}^{e_1e_2} & V_{ab}^{e_1e_1}  \\[.1truecm]
  V_{ab}^{e_2e_1} & - V_{ab}^{e_2e_1} & - (V_{[ab]}^{\omega_1e_2}-V_{[ab]}^{\omega_2e_2}) &
   (V_{ab}^{\omega_1e_1} - V_{ab}^{\omega_2e_1}) \\[.1truecm]
 - V_{ab}^{e_1e_1} &  V_{ab}^{e_1e_1} & (V_{ab}^{e_1\omega_1}-V_{ab}^{e_1\omega_2}) & 0 
\end{array}
\right) \\[.2truecm]
& + m^2 \beta_2
\left(
\begin{array}{cccc}
 0 & 0 & V_{ab}^{e_2e_2} & - V_{ab}^{e_2e_1} \\[.1truecm]
 0 & 0 & - V_{ab}^{e_2e_2} &  V_{ab}^{e_2e_1} \\[.1truecm]
 V_{ab}^{e_2e_2} & - V_{ab}^{e_2e_2} & 0 &
  - (V_{ab}^{e_2\omega_1}-V_{ab}^{e_2\omega_2}) \\[.1truecm]
- V_{ab}^{e_1e_2} &  V_{ab}^{e_1e_2} & -( V_{ab}^{\omega_1e_2} -  V_{ab}^{\omega_2e_2}) &
 (V_{[ab]}^{\omega_1e_1}-V_{[ab]}^{\omega_2e_1}) 
\end{array}
\right) \nonumber\,.
\end{align}
Where $\omega_- \equiv \omega_1 - \omega_2$. We determine the rank of this matrix as before using Mathematica, and find it to be 6. This means that there are $12 - 6 = 6$ gauge symmetries in the theory.

To find the secondary constraints we must study the integrability conditions \eqref{Intcon}. There are three independent equations
\begin{align}
(\beta_1 e_1{}^a + \beta_2 e_2{}^a ) e_1 \cdot e_2 = 0\,, \label{e1e2} \\
e_2{}^a \omega_- \cdot (\beta_1 e_1 + \beta_2 e_2) - \beta_1 \omega_-{}^a e_1 \cdot e_2 = 0\,, \label{omega1}  \\
e_1{}^a \omega_- \cdot (\beta_1 e_1 + \beta_2 e_2) + \beta_2 \omega_-{}^a e_1 \cdot e_2 = 0 \,. \label{omega2}
\end{align}
Assuming invertibility of both dreibeine, $e_1^a$ and $e_2^a$, is not sufficient to generate a secondary constraint; from \eqref{e1e2} we need that $(\beta_1 e_1{}^a + \beta_2 e_2{}^a)$ is invertible. This does not follow from the invertibility of the two separate dreibeine. Without any secondary constraints, the dimension of the physical phase space, using eqn.~\eqref{dimcounting}, is 6. This corresponds to 3 local degrees of freedom, one massive graviton and the other presumably a scalar ghost.

We are interested in theories of massive gravity without ghosts, so we must restrict our general model to ensure secondary constraints. The analysis above suggests that if instead of the two separate dreibeine, the linear combination $\beta_1 e_1{}^a + \beta_2 e_2{}^a$ is invertible, we can derive the needed secondary constraints. A special case of this assumption is when $\beta_1\beta_2 = 0$, but one of them non-zero. In that case, the invertibility of one of the original dreibeine is sufficient. We will investigate both possibilities.

\subsubsection{The case $\beta_1\beta_2 = 0$}

In the case that we set to zero one of the two parameters $\beta_i$ we may choose, without loss of generality, to set
\begin{equation}
\beta_2=0\, .
\end{equation}
In this case the invertibility of $e_1{}^a$ alone implies the two secondary constraints.
\begin{equation}\label{ZDGSeccon}
\psi_1 = \Delta^{e_1e_2}  = 0\,, \qquad  \psi_2 = \Delta^{\omega_- e_1} = 0 \,.
\end{equation}
These constraints and parameter choices cause the first and last matrices in eqn.~\eqref{Pmat} to vanish. The remaining matrix $\cP'{}_{rs}^{ab}$ has rank 4, which is equal to the dimensionality of the physical phase space, provided the assumptions of the section \ref{sec:seccon} hold. We will now verify that this is the case. 

The secondary constraints \eqref{ZDGSeccon} are in involution with each other which confirms the first assumption. Their brackets with the primary constraint functions are given by
\begin{equation} \label{ZDGpsi1}
\begin{split}
\{ \phi[\xi], \psi_1 \}_{\rm P.B.} = \varepsilon^{ij} \bigg[ & \partial_i \xi^{e_1} \cdot e_{2\,j} - \xi^{e_1} \cdot \omega_{1\,i} \times e_{2\, j} - \partial_i \xi^{e_2} \cdot e_{1\,j} + \xi^{e_2} \cdot \omega_{2\,i} \times e_{1\,j} \\
& - \left(\xi^{\omega_1} - \xi^{\omega_2} \right) \cdot e_{1\,i} \times e_{2\,j}
\bigg]\,,
\end{split}
\end{equation}
and
\begin{align} \label{ZDGpsi2}
\{ \phi[\xi], \psi_2 \}_{\rm P.B.} = & \varepsilon^{ij} \bigg[  ( \partial_i \xi^{\omega_1} - \partial_i \xi^{\omega_2}) \cdot e_{1\,j} - (\xi^{\omega_1} - \xi^{\omega_2}) \cdot  (\omega_{2\,i} \times e_{1\, j})  - \partial_i \xi^{e_1} \cdot \omega_{-\,j} \nonumber \\
& + \xi^{e_1} \cdot (\omega_{1\,i} \times \omega_{-\,j}) + m^2 \left( \sigma \beta_1  \xi^{e_1} + \alpha_2 \xi^{e_2}  \right) \cdot (e_{1\,i} \times e_{2\,j})  \\
& - m^2 \left( (\sigma \alpha_1 + \beta_1) \xi^{e_1} - \sigma \beta_1 \xi^{e_2}  \right) \cdot (e_{1\,i} \times e_{1\,j})
\bigg]\,. \nonumber
\end{align}
The full matrix of Poisson brackets is again a $14\times 14$ matrix $\mathbb{P}$ given by \eqref{Pbrackets}, where the $v_I$ with $I = 1,2$ are now
\begin{equation} \label{columnZDG}
v_I =  \left( \begin{array}{c}
\{\phi^a_{\omega_1} , \psi_I \}_{\rm P.B.}\\
\{\phi^a_{\omega_2} , \psi_I \}_{\rm P.B.}\\
\{\phi^a_{e_1} , \psi_I \}_{\rm P.B.}\\
\{\phi^a_{e_2} , \psi_I \}_{\rm P.B.}
\end{array} \right)\,.
\end{equation}
These brackets can be read off from equations \eqref{ZDGpsi1} and \eqref{ZDGpsi2}. The vectors \eqref{columnZDG} are linearly independent from each other and with the columns of $\mathbb{P}$, so this confirms that the second assumption also holds and increases the rank of $\mathbb{P}$ by $4$. The total number of second-class constraints is 8, leaving 6 first-class constraints. Using \eqref{dimcounting} we find that for general values of the parameters $\alpha_1$, $\alpha_2$ and $\beta_1$ the dimension of the physical phase space per space point is 4. This corresponds to the 2 local degrees of freedom of a massive graviton.

\subsubsection{The case of invertible $\beta_1 e_1{}^a + \beta_2 e_2{}^a$}

The second option is to assume invertibility of the linear combination of the two dreibeine, $\beta_1 e_1{}^a + \beta_2 e_2{}^a$. In this case, to keep track of the invertible field, we make a field redefinition in the original Lagrangian \eqref{LZDG}. We define, for $\beta_1 + \sigma \beta_2 \neq 0$,
\begin{equation}\label{ZDGredef1}
e^a = \frac{2}{\beta_1 + \sigma \beta_2} \left( \beta_1 e_1{}^a + \beta_2 e_2{}^a \right)\,, \qquad
f^a = \sigma e_1{}^a - e_2{}^a \,.
\end{equation}
For convenience we will work with the sum and difference of the spin connections\footnote{Note that the sum of the two connections also transforms as a connection, while the difference transforms as a tensor under the diagonal gauge symmetries}
\begin{equation}\label{ZDGredef2}
\omega^a = \frac12 \left(\omega_1{}^a + \omega_2{}^a \right)\,, \qquad
h^a = \frac12 \left(\omega_1{}^a - \omega_2{}^a \right)\,.
\end{equation}
In terms of these new fields, the ZDG Lagrangian three-form is
\begin{align}
L = &  - M_P \bigg\{  \sigma e \cdot R(\omega) + c f \cdot R(\omega) + f \cdot \mathcal{D} h + \frac12  (\sigma e + c f) \cdot h \times h \, \nonumber \\
& + m^2 \left( \frac{a_1}{6} e \cdot e \times e^c - \frac{b_1}{2} e \cdot e \times f - \frac{b_2}{2} e \cdot f \times f \right. \\
& \qquad \qquad \quad \left. + \frac{(c^2 -1)b_1 - 2\, c\, \sigma b_2}{6} f \cdot f \times f \right) \bigg\}\,, \nonumber
\end{align}
where $\mathcal{D}$ is the covariant derivative with respect to $\omega$. The new dimensionless constants $(a_1, b_1, b_2, c)$ are given in terms of the old $(\alpha_I, \beta_I)$ parameters as follows
\begin{align}
a_1 = & \; \frac{1}{8} \left( \alpha_1 - 3 \sigma \beta_1 - 3 \beta_2 + \sigma \alpha_2 \right)\,, & b_1 = & \; \frac{\alpha_2 \beta_1 + \beta_2^2 - \beta_1^2 - \alpha_1 \beta_2}{4(\beta_1 + \sigma \beta_2)}\,, \\
b_2 = & - \frac{\alpha_1\beta_2^2 + \sigma \beta_1\beta_2^2 + \beta_1^2 \beta_2 + \sigma \alpha_2 \beta_1^2}{2(\beta_1 + \sigma \beta_2)^2}\,, & c = & \;  \frac{\sigma \beta_2 - \beta_1}{\sigma \beta_2 + \beta_1}\,. \nonumber
\end{align}
By construction, this theory has two secondary constraints for invertible $e^a$. Indeed, when we calculate the integrability conditions \eqref{Intcon} for this theory we find the three equations
\begin{equation}\label{IntconexZDG}
\frac12(\beta_1 + \sigma \beta_2) e^a  f \cdot e =  0\,, \qquad   \frac12 (\beta_1 + \sigma \beta_2)e^a h \cdot e = 0\,, 
\end{equation}
and
\begin{equation} 
\frac12(\beta_1 + \sigma \beta_2) \left( h^a f \cdot e + f^a h \cdot e\right) =  0 \,.
\end{equation}
From \eqref{IntconexZDG} we can derive two secondary constraints, since we assumed that $e^a$ was invertible and that $\beta_1 + \sigma \beta_2 \neq 0$. The secondary constraints are
\begin{equation}
\psi_1 = \Delta^{fe} = 0 \,, \qquad \psi_2 = \Delta^{he} = 0\,.
\end{equation}
After imposing these constraints, the matrix of Poisson brackets in the basis $(\omega, h, f, e)$ reduces to
\begin{equation}
 (\cP'_{ab})_{rs} = \frac12 m^2 (\beta_1 + \sigma \beta_2) \left(
\begin{array}{cc}
0   &   0   \\
0   &   Q
\end{array} \right)\,,
\end{equation}
where
\begin{equation}\label{PmatexZDG}
Q =
\left(
\begin{array}{ccc}
 0 &  V_{ab}^{ee} &  - V_{ab}^{ef} \\
 V_{ab}^{ee} & 0  &  - V_{ab}^{eh} \\
 - V_{ab}^{fe} & - V_{ab}^{he} & V_{[ab]}^{hf} \\
\end{array}
\right) \,. 
\end{equation}
Using the same techniques as previously, we find that this matrix has rank 4.

The secondary constraints are again in involution with themselves, and their brackets with the primary constraint functions are given by
\begin{align} \label{psi1PB}
\{ \phi[\xi], \psi_1 \}_{\rm P.B.}  = & \; \varepsilon^{ij} \bigg[ \partial_i \xi^{f} \cdot e_{j} - \xi^{f} \cdot \omega_{i} \times e_{j} - \partial_i \xi^{e} \cdot f_{j} + \xi^{e} \cdot \omega_{i} \times f_{j}  \nonumber \\
& - \left(\sigma \xi^{e} + c\, \xi^{f} \right) \cdot e_{i} \times h_{j}   - \left(c\, \xi^{e} + \sigma (c^2 -1) \xi^{f} \right) \cdot f_{i} \times h_{j} \\
&- \xi^h \cdot \left( \sigma e_i \times e_j + 2 c\, e_i \times f_j + \sigma (c^2 -1 ) f_i \times f_j \right)
\bigg]\,, \nonumber
\end{align}
and
\begin{align} \label{psi2PB}
\{ \phi[\xi], \psi_2 \}_{\rm P.B.} = \; & \varepsilon^{ij} \bigg[  \partial_i \xi^{h} \cdot e_{j} - \xi^{h} \cdot \omega_{i} \times e_{j}  - \partial_i \xi^{e} \cdot h_{j}  + \xi^{e} \cdot \omega_{i} \times h_{j} \nonumber \\
&  + m^2 \left( (c\, \sigma a_1 +  b_1) \xi^{e} - (c\, \sigma b_1 -  b_2) \xi^{f}  \right) \cdot e_{i} \times e_{j}  \\ \nonumber
&  -  m^2 \left( (c\, \sigma b_1 - b_2)  \xi^{e}  +  ( (c^2-1) b_1 - c\, \sigma b_2) \xi^{f}  \right) \cdot e_{i} \times f_{j} \\
 - (c\, \xi^e + \sigma & (c^2 -1) \xi^f) \cdot h_i \times h_j   - \xi^h \cdot \left(c\, e_i \times h_j + \sigma (c^2-1) f_i \times h_j \right)
\bigg]\,. \nonumber
\end{align}
For generic values of the parameters these constraints increase the rank of the total matrix of Poisson brackets, $\mathbb{P}$, by 4, leading to a $14 \times 14$ matrix of rank 8. This implies that there are eight second-class constraints and six first-class constraints, leading to two degrees of freedom, those of two massive spin-2 modes in 3 dimensions.

To summarize, demanding the presence of secondary constraints in ZDG to remove unwanted degrees of freedom forces us to make an additional assumption about the theory. We must assume invertibility of a linear combination of the two dreibeine. This assumption reduces to the invertibility of a single dreibein if one of the two coupling constants is set to zero. In each scenario, only one dreibein need be assumed invertible. This suggests that we identify its square as the ``physical'' metric with which distances are measured. This suggestion is supported by the fact that the second dreibein may be solved for in terms of the invertible dreibein and its derivatives, leading to an equation of motion for a single dreibein containing an infinite sum of higher derivative contributions, as was shown in \cite{Bergshoeff:2014eca} and reviewed in chapter \ref{section:ZDGHD}.

\subsection{General Zwei-Dreibein Gravity}
\label{sec:GZDG}
It is natural to look for a parity violating generalization of ZDG, just as GMG is a parity violating version of NMG. To this end we add to the ghost-free, $\beta_2 = 0$ ZDG Lagrangian three-form \eqref{LZDG} a Lorentz-Chern-Simons (LCS) term for the spin-connection $\omega_{1}{}^a$.\footnote{It is also possible to include a LCS term for $\omega_2{}^a$, in this case the expressions presented in this subsection are only slightly modified and lead to the same conclusion.}
\begin{equation}
L_{\rm GZDG} = L_{\rm ZDG}(\beta_2 = 0) + \frac{M_P}{2\mu} \omega_{1} \cdot \left( d \omega_1 + \frac13  \omega_{1} \times \omega_{1} \right) \,.
\end{equation}
The introduction of the LCS term for $\omega_1{}^a$ introduces non-zero torsion for $e_1{}^a$. One might consider adding a torsion constraint for $e_1{}^a$, enforced by a Lagrange multiplier field $h^a$, but this introduces new degrees of freedom \cite{Bergshoeff:2013xma}. In any case, the equations of motion for General ZDG are such that the torsion constraint is not needed in order to solve for the spin-connections, and there exists a scaling limit similar to the NMG-limit presented in \cite{Bergshoeff:2013xma} where the General ZDG Lagrangian reduces to the GMG Lagrangian \eqref{LGMG}.

From the point of view of our general formalism, the addition of the LCS term adds the following non-zero components to $g_{rs}$ and $f_{rst}$
\begin{equation}
g_{\omega_1 \omega_1} = \frac{1}{\mu} \,, \qquad f_{\omega_1 \omega_1 \omega_1} = \frac{1}{\mu}\,.
\end{equation}
The integrability conditions now read
\begin{align}
e_1{}^a  e_1 \cdot e_2 = 0\,, \label{e1e2GZDG} \\
e_1{}^a \left( \omega_- \cdot e_1 + \frac{\beta_1 m^2}{\mu} e_1 \cdot e_2 \right) = 0\,, \label{omega1GZDG} \\
e_2{}^a \omega_- \cdot e_1  + \left( \frac{\beta_1 m^2}{\mu} e_2{}^a  -  \omega_-{}^a\right) e_1 \cdot e_2 = 0\,. \label{omega2GZDG}
\end{align}
Invertibility of $e_1{}^a$ implies the same secondary constraints as in eqn.~\eqref{ZDGSeccon}, and the counting of degrees of freedom proceeds analogously. After a linear redefinition of the constraints to $\phi_{\omega'} = \phi_{\omega_1} + \phi_{\omega_2}$, the matrix of Poisson brackets reduces to
\begin{equation}
(\cP'_{ab})_{rs} =  m^2 \beta_1 \left(
\begin{array}{cc}
0   &   0   \\
0   &   Q
\end{array} \right)\,,
\end{equation}
where
\begin{equation}\label{PmatGZDG}
Q =
\left(
\begin{array}{ccc}
 0 & - V_{ab}^{e_1e_2} &  V_{ab}^{e_1e_1} \\
 - V_{ab}^{e_2e_1} &  -(V_{[ab]}^{\omega_1e_2}-V_{[ab]}^{\omega_2e_2}) +  \frac{\beta_1 m^2}{\mu} V_{ab}^{e_2e_2} &
   (V_{ab}^{\omega_1e_1}-V_{ab}^{\omega_2e_1}) -  \frac{\beta_1 m^2}{\mu} V_{ab}^{e_2e_1} \\
  V_{ab}^{e_1e_1} & ( V_{ab}^{e_1\omega_1} -  V_{ab}^{e_1\omega_2}) -  \frac{\beta_1 m^2}{\mu} V_{ab}^{e_1e_2} &
 \frac{\beta_1 m^2}{\mu} V_{ab}^{e_1e_1}
 \\
\end{array}
\right). \nonumber
\end{equation}
We find that this matrix has rank 4. The Poisson brackets of the secondary constraints with the primary ones are now:
\begin{align} \label{GZDGpsi1}
\{ \phi[\xi], \psi_1 \}_{\rm P.B.} = \varepsilon^{ij} \bigg[ & \partial_i \xi^{e_1} \cdot e_{2\,j} - \xi^{e_1} \cdot \omega_{1\,i} \times e_{2\, j} - \partial_i \xi^{e_2} \cdot e_{1\,j} + \xi^{e_2} \cdot \omega_{2\,i} \times e_{1\,j}  \nonumber \\
& - \left(\xi^{\omega_1} - \xi^{\omega_2} + \frac{\alpha_1 m^2}{\mu} \xi^{e_1} - \frac{m^2 \beta_1}{\mu} \xi^{e_2} \right) \cdot e_{1\,i} \times e_{2\,j} \\
& \nonumber  + \frac{\beta_1 m^2  }{\mu } \xi^{e_1} \cdot e_{2\,i} \times e_{2\,j}
\bigg]\,,
\end{align}
and
\begin{align} \label{GZDGpsi2}
\{ \phi[\xi], \psi_2 \}_{\rm P.B.} = & \varepsilon^{ij} \bigg[  ( \partial_i \xi^{\omega_1} - \partial_i \xi^{\omega_2}) \cdot e_{1\,j} - (\xi^{\omega_1} - \xi^{\omega_2}) \cdot  (\omega_{2\,i} \times e_{1\, j})  - \partial_i \xi^{e_1} \cdot \omega_{-\,j} \nonumber \\
& + \xi^{e_1} \cdot (\omega_{1\,i} \times \omega_{-\,j}) + m^2 \left( \sigma \beta_1  \xi^{e_1} + \alpha_2 \xi^{e_2}  \right) \cdot (e_{1\,i} \times e_{2\,j})  \\
& - m^2 \left( (\sigma \alpha_1 + \beta_1)\xi^{e_1} - \sigma \beta_1 \xi^{e_2}  \right) \cdot (e_{1\,i} \times e_{1\,j}) \nonumber \\
& + m^2 \left( \frac{\alpha_1}{\mu} \xi^{e_1} - \frac{\beta_1}{\mu} \xi^{e_2}\right) \cdot (e_{1\,i} \times \omega_{-\,j}) - m^2 \frac{\beta_1}{\mu} \xi^{e_1} \cdot (e_{2\,i} \times \omega_{-\,j})
\bigg]\,. \nonumber
\end{align}
Again, the secondary constraints are in involution, and the new columns are linearly independent from each other and the original columns. The usual analysis shows that there are 8 second-class constraints and 6 first-class constraints. The total dimension of the physical phase space remains 4, and so the model has the same number of local degrees of freedom as GMG.

\section{Conclusions}

It is a remarkable fact that many of the 3D ``massive gravity'' models that have been found and analyzed in recent years have a CS--like formulation in which the action is an integral over a Lagrangian three-form constructed from wedge products of one-forms  that include an invertible dreibein. One example not considered here is Topologically Massive Supergravity \cite{Routh:2013uc}. 

Many of these CS--like models have an alternative formulation as a higher\-/derivative extension of 3D General Relativity, and it is certainly not the case that all such higher-derivative extensions can be recast as CS--like models. It appears that the unitary (ghost-free) 3D massive models are also special in this respect. Whatever the reason may be for this, it is fortunate because the CS--like formalism is well-adapted to a
Hamiltonian analysis, which we have reviewed, and refined, extending the results of \cite{Hohm:2012vh} for General Massive Gravity (GMG) to include the recently proposed  Zwei-Dreibein Gravity (ZDG) \cite{Bergshoeff:2013xma}.  

This Hamiltonian analysis leads to a simple determination of the number of local degrees of freedom, independent of any linearization about a particular background. This allows one to establish that a class of 3D massive gravity models, including ZDG, is free of the Boulware-Deser ghost that typically afflicts massive gravity models \cite{Boulware:1973my}. Conversely, the CS--like formulation of these models can be used as a starting point to find higher-derivative extensions of New Massive Gravity which are guaranteed to be free of scalar ghosts \cite{Afshar:2014}, a subject which we will return to in chapter \ref{chapter:ENMG_VDG}.

The techniques developed in this chapter can only partly be used to analyze the 3D dRGT theory reviewed in chapter \ref{sec:dRGT}, since in that case the flavor space metric is not invertible. However, it is still possible to derive a set of integrability conditions similar to \eqref{Intcon} for the dRGT model and to analyze the presence of secondary constraints in that theory. A situation similar to the ZDG case follows. The theory only contains secondary constraints if the linear combination $\beta'_1 e^a + \beta'_2 \bar{e}^a$ is invertible. In that case, the number of degrees of freedom is compatible with a single massive spin-2 mode. 

We have also discussed a parity-violating extension of ZDG; it has some similarities to GMG (and has a limit to GMG for a certain range of its parameters) so it  could be called ``General Zwei-Dreibein Gravity'' (GZDG). We have shown that it has exactly the same number of local degrees of freedom as GMG. We know that ZDG propagates two spin-2 modes of equal mass in a maximally symmetric vacuum, so it seems that GZDG will propagate two spin-2 modes of different masses, like GMG. Furthermore, there exists a limit of the parameters of GZDG that sends one mass to infinity keeping the other fixed. This results in a model similar to TMG  but with better behavior in relation to the AdS/CFT correspondence. This model, called Minimal Massive Gravity \cite{Bergshoeff:2014pca}, will be treated in full detail in chapter \ref{chapter:MMG}. But in order to analyze the boundary theories, we should first set up the proper machinery to do so. This requires to take into account the boundary terms which were omitted in the analysis of this chapter. This will be the subject on the next chapter.

%% file: chapter_5/chapter_5.tex
\pagestyle{empty}
\setcounter{chapter}{4}

\chapter[Asymptotic Symmetries]{Asymptotic Symmetries and Bulk and Boundary Unitarity}
\label{chapter:asymptotic_symm}

\pagestyle{headings}

\begin{quote}\em
The first-order Chern-Simons--like formulation of gravity has an additional advantage; it is relatively easy to find the asymptotic symmetry group when imposing Brown-Henneaux boundary conditions. In this chapter, we continue the canonical analysis initiated in the last chapter, paying special attention to the boundary terms which were omitted there. The boundary terms for the set of first-class constraints constitute the conserved global charges of the theory, whose Poisson bracket algebra are two copies of the Virasoro algebra with a central extension, just like in the case of pure gravity, discussed in chapter \ref{chapter:GRin3D}. The only difference in this case for Chern-Simons--like theories of gravity is the explicit expression for the central charge, which we compute for the case of GMG and ZDG in this chapter. Next we discuss the requirement of positivity of the central charge in these models of gravity and the compatibility with positive energy and mass for the massive modes in the bulk AdS$_3$ space-time. 
\end{quote}
\newpage

\section{Introduction}

As was shown in chapter \ref{chapter:GRin3D}, pure gravity in three dimensions contains no bulk degrees of freedom. However, it was shown in \cite{Brown:1986nw} that in AdS$_3$ the theory contains non-trivial boundary excitations. These boundary gravitons fall into representations of the asymptotic symmetry algebra which is two copies of the Virasoro algebra with a classical central extension. The asymptotic symmetry algebra is generated by diffeomorphisms which preserve the AdS background asymptotically. In the last two chapters of this thesis, we have discussed theories which depart from general relativity in three dimensions and are no longer pure CS theories, but CS--like. However, the theories are constructed in such a way that the number of gauge symmetries is preserved. This was verified by a Hamiltonian analysis in chapter \ref{chapter:three_crowd}. Furthermore, all theories admit an AdS$_3$ vacuum, and hence imposing Brown-Henneaux boundary conditions should be consistent.  

In this chapter we will compute the algebra of conserved global charges when adopting Brown-Henneaux boundary conditions, making use of our general formalism where we can. For theories which are no longer pure Chern-Simons, i.e. contain bulk degrees of freedom, the calculation of the asymptotic symmetries group depends on the structure of the specific theory as we need to identify the first class constraints explicitly (or at least, in the neighborhood of $\partial \Sigma$). This prevents us from a general treatment of the global charges, which is possible for pure CS gauge theories (see for instance \cite{Banados:1994tn}). The analysis here is a generalization of Carlip's analysis of TMG in \cite{Carlip:2008qh}, modified to fit the specific scenario of the examples in the last chapter.

\section{Generators of Symmetry Transformations}\label{sec:generators}
In contrast to the pure Chern-Simons gauge theories, not all constraint functions are first-class in the Chern-Simons--like models. In order to properly analyze the asymptotic symmetries, we should look at the algebra of first-class constraint functions which generate gauge symmetries. Hence the first step to take is to identify which (combination of) constraint functions generate the gauge symmetries of the theory. Fortunately, with the general formalism discussed in detail in the last chapter, we are well equipped to do just so.

The general CS--like theory defined by \eqref{Lgeneral} is manifestly diffeomorphism invariant. For what follows, we shall also assume that the specific CS--like theory of our interest is also manifestly invariant under local Lorentz transformations. The most general model \eqref{Lgeneral} certainly is not, so this assumption implies the following. The CS--like model of our interest contains a (dualised) spin connection $\omega^a$, which we will take to be the gauge field for local Lorentz transformations. This means that it may only appear in the combination $R^a$ (the field strength for $\omega^a$) or in a Lorentz-Chern-Simons term in the action. All derivatives of the other one-form fields $a^{r\,a}$ with $r \neq \omega$ should appear as covariant derivatives $\cD a^{r\,a}$. Translated to components of the field flavor space metric $g_{rs}$ and flavor tensor $f_{rst}$ this assumption is equivalent to the following statement
\begin{equation}\label{omega_assumpt}
\text{For every element of } g_{rs}\,, \text{ there is a non-zero } f_{rs \omega}\, \text{ such that: } f_{rs \omega} = g_{rs}\,.
\end{equation}
Equipped with this assumption we should expect the CS--like models defined by \eqref{Lgeneral} to have at least six gauge symmetries, corresponding to three diffeomorphism and three local Lorentz transformations. 

To identify the constraint functions which generate these symmetries, it is instructive to look at the Poisson brackets of the primary constraints with the dynamical components of the theory. In full generality (but omitting boundary terms for the moment), they can be computed using the general formulas \eqref{constraints}, \eqref{poissonbr} and \eqref{phi} as
\begin{equation}\label{transgen}
\{ \phi[\xi], a_i^{r\,a}\} = \partial_i \xi^{r\,a} +  f^{r}{}_{st} ( a_{i}^{s} \times \xi^{t})^a\,.
\end{equation}
From this result, we can deduce that a local Lorentz transformation with gauge parameter $\chi^a$ is generated by the constraint function
\begin{equation}\label{phiLL}
\phi_{\rm LL}[\chi] = \phi[\xi] \,, \text{ with: } \xi^{\omega \,a} = \chi^a \,, \text{ and } \xi^{r\,a} = 0\, \text{ for } r \neq \omega\,.
\end{equation}
In this case, we recover the usual transformation for the fields under local Lorentz transformations from \eqref{transgen}:
\begin{equation}
\begin{split}
& \{\phi_{\rm LL}[\chi] , \omega_{i}{}^{a} \} = \partial_i \chi^a +  (\omega_i \times \chi)^a\,, \\
& \{\phi_{\rm LL}[\chi] , a_i^{r\,a} \} =  (a_i{}^{r} \times \chi)^a\,. \\
\end{split}
\end{equation}
In the last line $r\neq\omega$ and we have used the fact that by the assumption \eqref{omega_assumpt} we may write $f^r{}_{s \omega} \equiv g^{rp}f_{ps \omega} = g^{rp}g_{ps} = \delta^r_s$.

In chapter \ref{chapter:GRin3D} we saw that on-shell, diffeomorphisms are generated by an appropriate combination of constraint functions with parameters proportional to the fields. In the general CS--like theory this is still true. Let us define
\begin{equation}\label{phidiff}
\phi_{\rm diff}[\zeta] = \phi[a_{\mu}^r \zeta^{\mu}]\,.
\end{equation}
Then, by equation \eqref{transgen} we find that
\begin{equation}
\begin{split}
\{\phi_{\rm diff}[\zeta], a_{i}^{r\,a} \} = \cL_{\zeta} a_{i}^{r\,a} + & \, \zeta^j \left( \partial_i a_j^{r\,a} - \partial_{j} a_i^{r\,a} +  f^r{}_{st} (a_i^{s} \times a_{j}^{t})^a \right) \\
+ & \, \zeta^0 \left( \partial_i a_0^{r\,a} - \partial_{0} a_i^{r\,a} +  f^r{}_{st} (a_i^{s} \times a_{0}^{t})^a \right)\,.
\end{split}
\end{equation}
Here $\cL_{\zeta}$ is the Lie derivative with respect to the vector field $\zeta^{\mu}$. The expressions in the parentheses are equivalent to the equations of motion of the general space-time decomposed model \eqref{gentimedecomp}. Hence on-shell, we have that
\begin{equation}
\{\phi_{\rm diff}[\zeta], a_{i}^{r\,a} \} = \cL_{\zeta} a_{i}^{r\,a}\,. 
\end{equation}
We have now identified the constraint functions which give the correct transformation rules on the dynamical variables of the theory. Of course, we should check whether these combinations of constraint functions are truly first-class, i.e. if they have weakly vanishing Poisson brackets with all other constraint functions. This can be verified on a case by case basis.

Once the first-class constraint functions have been identified, we would like to write them in a basis of mutually commuting $SL(2,\mathbb{R})$ generators. This is hardly possible in general, however it does work on the AdS background as the generators of gauge symmetries should reflect the isometries of the AdS vacuum solution. At the boundary of an asymptotically AdS space-time, the AdS background identities hold and hence the new basis of first-class constraint functions is also valid there. In general, we can define the new basis as
\begin{equation}\label{Lpm}
L_{\pm} [\zeta] = \phi_{\rm diff}'[\zeta] + a_{\pm} \phi_{\rm LL}[e_{\mu}\zeta^{\mu}]\,,
\end{equation}
where we have defined $\phi_{\rm diff}'[\zeta] \equiv \phi_{\rm diff}[\zeta] - \phi_{\rm LL} [\omega_{\mu}\zeta^{\mu}]$, since we would like to add the generator of LLTs independently to the $L_{\pm}$ with the gauge parameter $\chi = e_{\mu}\zeta^{\mu}=\xi^e$. The constants $a_{\pm}$ are defined such that
\begin{equation}
\{L_+[\xi],L_{-}[\eta]\} = 0\,,
\end{equation}
on the AdS background. 

At this point one should reinstate the boundary terms and investigate the Poisson brackets algebra of the improved generators
\begin{equation}\label{Lpmimproved}
\tilde{L}_{\pm}[\xi] = L_{\pm}[\xi] + Q_{\pm}[\xi]\,,
\end{equation}
under asymptotically AdS (or Brown-Henneaux) boundary conditions. The variation of the boundary terms $\delta Q_{\pm}$ can be obtained from the general formula \eqref{gen_varbc} and it is given by
\begin{equation}
\delta Q_{\pm} = - \int_{\partial \Sigma} dx^i \; \left( g_{rs} \xi^r + a_{\pm} g_{\omega s} \xi^e \right) \cdot \delta a^s_i\,,
\end{equation}
where $r$ sums over all values except $\omega$. In all the examples we treat, after imposing Brown-Henneaux boundary conditions \eqref{BHbc1}-\eqref{BHbc2} and restricting the gauge parameters $\xi^e$ to be the boundary condition preserving gauge transformations \eqref{asymptdiffs}, the computation will become identical to the pure gravity case, except for a modified expression of the central charge. It is exactly this piece of information we are looking for.

\section{Specific Examples}
Here we will treat two examples specified in earlier chapters. First we will look at General Massive Gravity, since the theory contains both Topological Massive Gravity and New Massive Gravity as limits. After that we consider the asymptotic symmetries of the more recent Zwei-Dreibein Gravity with Brown-Henneaux boundary conditions. 

\subsection{General Massive Gravity}
\label{sec:AS_GMG}
General Massive Gravity was defined in section \eqref{sec:GMG}. It was given in terms of the dreibein $e^a$ and corresponding spin-connection $\omega^a$. The fields $f^a$ and $h^a$ are auxiliary and can be solved in terms of $e^a$ and derivatives. The GMG equations of motion imply:
\begin{equation}
f_{\mu}{}^a = - e^{\nu\,a}  S_{\mu\nu}(e) \,, \qquad h_{\mu}{}^a = - \frac{1}{\mu} e^{\nu\,a}S_{\mu\nu}(e) - \frac{1}{m^2}e^{\nu\,a} C_{\mu\nu} (e)  \,.
\end{equation}
Here $S_{\mu\nu}(e)$ and $C_{\mu\nu}(e)$ are the Schouten and Cotton tensors defined in \eqref{Sdef}-\eqref{Cdef}. The theory has an AdS vacuum with a cosmological constant $\Lambda = - 1/ \ell^2$ where $\Lambda$ satisfies the quadratic equation:
\begin{equation}
\label{GMGcc}
\sigma \Lambda - \Lambda_0 + \frac{\Lambda^2}{4m^2} = 0\,.
\end{equation}
On this background the Schouten tensor is proportional to the AdS metric and the Cotton tensor vanishes. The background values of the fields $f^a$ and $h^a$ are then given by:
\begin{equation}\label{GMGbkgd}
\bar{f}^a = - \frac{\Lambda}{2} \bar{e}^a \,, \qquad \bar{h}^a = - \frac{\Lambda}{2\mu} \bar{e}^a \,.
\end{equation}

\subsubsection{Algebra of first-class constraints}
Here we analyze the algebra of first-class constraints in GMG. First let us verify that the constraint functions defined in \eqref{phiLL} and \eqref{phidiff} are truly first-class by considering their Poisson brackets with all other constraint functions. 

It is straightforward to verify that \eqref{phiLL} has weakly vanishing Poisson brackets by inspection of the matrix \eqref{PmatGMG}, which does not contain any non-zero entries in the column and row corresponding to $r = \omega$. Furthermore, the Poisson brackets of the primary constraint functions with the secondary constraints \eqref{GMGpsi1}-\eqref{GMGpsi2} do not depend on $\xi^{\omega\,a} = \chi^a$ and hence they vanish weakly.

To verify that the Poisson brackets of \eqref{phidiff} vanish weakly a little more work is required. As was outlined above, diffeomorphisms are generated when the gauge parameters are proportional to the fields. 
\begin{equation}\label{diffparam}
\xi^r_a = a_{\mu\,a}^r \zeta^{\mu}\,.
\end{equation}
In the case of GMG, we could also derive from the integrability conditions that
\begin{equation}
e_{[\mu} \cdot f_{\nu]} = 0\,, \qquad \text{and} \qquad e_{[\mu} \cdot h_{\nu]} = 0\,.
\end{equation}
Using these identities, one can show that for gauge parameters chosen as in \eqref{diffparam}, the following identities hold:
\begin{equation}\label{GMGgaugeparam}
e_{\mu}{}^a \xi^f_a = f_{\mu}{}^a \xi^e_a\,, \quad \text{and:} \quad e_{\mu}{}^a \xi^h_a = h_{\mu}{}^a \xi^e_a\,.
\end{equation}
Using these identities in the general formula \eqref{gen_poissonbr} with the GMG specifics of \eqref{PmatGMG} it is possible to show that the Poisson brackets of $\phi_{\rm diff}$ with all other primary constraint functions vanish weakly. The Poisson brackets of $\phi_{\rm diff}[\xi^r_a]$ with the secondary constraints \eqref{GMGpsi1} and \eqref{GMGpsi2} can alternatively be written as
\begin{align}
\{ \phi_{\rm diff}[\xi], \psi_1 \} = \ldots  + \epsilon^{ij} & \left( \partial_i \left( \xi^h \cdot e_{j} - \xi^e{} \cdot h_{j} \right) +   \xi^e{} \cdot \left( \frac{1}{2m^2} f_i \times f_j - 2 \sigma e_i \times f_j \right. \right.  \nonumber \\ 
& \left. \left.  - \frac32 \Lambda_0 e_j \times e_j \right)  + \xi^f \cdot \left( \frac{1}{m^2} e_i \times f_j - \sigma  e_j \times e_j \right)
\right)\,, \\
\{\phi_{\rm diff}[\xi], \psi_2 \} = \ldots  + \epsilon^{ij} & \left( \partial_i \left( \xi^f \cdot e_{j} - \xi^e \cdot f_{j} \right) + 2 \xi^e \cdot \left( m^2 e_i \times h_j - \frac{m^2}{\mu} e_i \times f_j \right) \right. \nonumber \\ 
& \left.+ \left(\xi^h - \frac{1}{\mu} \xi^f \right) \cdot e_j \times e_j 
\right) \,.
\end{align}
Now the dots denote terms proportional to primary constraints which vanish weakly. In the neighborhood of the AdS boundary, the relations \eqref{GMGbkgd} hold and eqn.~\eqref{GMGgaugeparam} implies that:
\begin{equation}\label{GMGparambkgd}
\xi^{f\,a} = - \frac{\Lambda}{2} \xi^{e\,a}\,, \qquad \xi^{h\,a} = - \frac{\Lambda}{2\mu} \xi^{e\,a}\,.
\end{equation}
Using these relations and the quadratic equation for $\Lambda$ \eqref{GMGcc}, we see that the Poisson brackets of $\phi_{\rm diff}[\zeta]$ with the secondary constraints vanish weakly near the AdS boundary. 

Now that we have a basis of first-class constraint functions, we can investigate their Poisson bracket algebra. After defining $\phi_{\rm diff}'[\zeta] = \phi_{\rm diff}[\zeta] - \phi_{\rm LL}[\omega_{\mu}\zeta^{\mu}]$ we find that the first-class constraints form the $SO(2,2)$ algebra in the neighborhood of the AdS boundary. 
\begin{align}
\{\phi_{\rm diff}'[\xi], \phi_{\rm diff}'[\eta]\} = &\, -\Lambda \phi_{\rm LL}[[\xi , \eta]]\,, \qquad 
\{\phi_{\rm diff}'[\xi], \phi_{\rm LL}[\eta]\} = \phi_{\rm diff}'[[\xi , \eta]]\,, \\ \nonumber
\{\phi_{\rm LL}[\xi], \phi_{\rm LL}[\eta]\} = & \, \phi_{\rm LL}[[\xi , \eta]]\,.
\end{align}
Here we have used the AdS background relations to write all the gauge parameters in terms of $\xi^{e\,a} = e_{\mu}{}^a \zeta^{\mu} = \xi^a$ (and similarly for $\eta^a$). Like in Einstein-Cartan gravity, this enables us to split the constraint functions into two mutually commuting sectors. If we define
\begin{equation}
L_{\pm}[\xi] = \phi_{\rm diff}'[\xi] \pm \frac{1}{\ell} \phi_{\rm LL}[\xi]\,,
\end{equation}
then the Poisson bracket algebra becomes
\begin{align}
\{ L_{\pm}[\xi], L_{\pm} [\eta]\} & = \pm \frac{2}{\ell} L_{\pm}[[\xi , \eta]]\,, \\
\{ L_{+}[\xi], L_{-} [\eta]\} & = 0\,. \nonumber
\end{align}
Now the analysis becomes equivalent to the Einstein-Cartan example reviewed in section \ref{sec:ECASG}. The only difference is that the generators of gauge transformations are a bit more involved and hence when we reinstate the boundary terms using \eqref{gen_varbc} and \eqref{gen_poissonbr} we will find a different overall factor for the global charges and hence a different central charge. In this case, the variation of the boundary charge in \eqref{Lpmimproved}, which is needed to improve the generators $L_{\pm}[\xi]$ is
\begin{equation}
\begin{split}
\delta Q_{\pm} [\xi] = & - \int_{\partial \Sigma} dx^i \; \left( \xi^e g_{es} + \xi^f g_{fs} + \xi^h g_{hs} \pm \frac{1}{\ell} \xi^e  g_{\omega s} \right)\cdot  \delta a_{i}^{s}\,, \\
= & \left( \sigma + \frac{1}{2\ell^2m^2} \mp \frac{1}{\mu\ell} \right) \int_{\partial \Sigma} d\phi\; \xi \cdot \left(\delta \omega_{\phi} \pm \frac{1}{\ell} \delta e_{\phi} \right) \,.
\end{split}
\end{equation}
In the last line, we have used the background AdS identities \eqref{GMGbkgd} and \eqref{GMGparambkgd} and we denote $ \xi^e_a = \xi_a$. Reinstating the boundary terms into the Poisson brackets of the first-class constraints, using the general formula \eqref{gen_poissonbr} and the background identities, we find that at the AdS boundary:
\begin{equation}\nonumber
\{ \tilde{L}_{\pm}[\xi], \tilde{L}_{\pm}[\eta] \} = \ldots \pm \frac{2}{\ell} \left(\sigma + \frac{1}{2 \ell^2 m^2} \mp \frac{1}{\mu \ell} \right) \int d\phi\, \xi \cdot \left[ \partial_{\phi} \eta +  \left( \omega_{\phi} \pm \frac{1}{\ell} e_{\phi}\right) \times \eta \right]\,.
\end{equation}
Both the boundary term in the Poisson brackets and the variation of the global charges are proportional to what we found for pure 3D gravity in chapter \ref{sec:ECASG}. Hence, we may adopt the same (Brown-Henneaux) boundary conditions \eqref{BHbc1} and \eqref{BHbc2}, together with the same boundary condition preserving gauge transformations \eqref{asymptdiffs}. This leads, of course, to the same asymptotic symmetry algebra, only with a modified expression for the central charge. Reintroducing a factor of $(8 \pi G)^{-1}$ we find for GMG
\begin{equation}\label{ccGMG}
c^{\text{\tiny GMG}}_{\rm L/R} = \frac{3 \ell}{2 G} \left(\sigma + \frac{1}{2 \ell^2 m^2} \mp \frac{1}{\mu \ell} \right)\,.
\end{equation}
This expression agrees with the one found in \cite{Bergshoeff:2009aq} using different methods. As we noted in section \ref{sec:GMG}, there is a limit from GMG to NMG ($\mu \to \infty$) and to TMG ($m^2 \to \infty$). These limits can also be taken in the above central charges to give for NMG
\begin{equation}
\label{ccNMG}
c^{\text{\tiny NMG}}_{\rm L/R} = \frac{3 \ell}{2 G} \left(\sigma + \frac{1}{2 \ell^2 m^2} \right)\,.
\end{equation}
and for TMG
\begin{equation}\label{ccTMG}
c^{\text{\tiny TMG}}_{\rm L/R} = \frac{3 \ell}{2 G} \left(\sigma \mp \frac{1}{\mu \ell} \right)\,.
\end{equation}
We will discuss the implications of this result on the unitarity of the theories in section \ref{sec:bbunitarity} below.

\subsection{Zwei-Dreibein Gravity}\label{sec:AS_ZDG}

We will now repeat the computation of the boundary terms and central charges for ZDG. In the last chapter we saw that the removal of the extra degree of freedom in ZDG required the additional assumption of an invertible linear combination of the two dreibeine. Here we will analyze the asymptotic symmetries for ZDG with this assumption. For all practical purposes this amounts to analyzing the theory after the redefinitions \eqref{ZDGredef1}-\eqref{ZDGredef2} with an invertible $e^a$.

\subsubsection{AdS background in ZDG}

In section \ref{section:ZDG} we saw how there are anti-de Sitter vacua in ZDG, which are parametrized by $\gamma$, the proportionality factor of the second dreibein. The background values of the four fields in ZDG are
\begin{equation}
e_1{}^a = \bar{e}^a \,, \qquad e_2{}^a = \gamma \bar{e}^a \,, \qquad \omega_1{}^a = \omega_2{}^a = \bar{\omega}^a\,.
\end{equation}
Here $\bar{e}^a$ denotes the AdS dreibein and $\bar{\omega}^a$ is the AdS spin connection. Plugging this in the field redefinitions \eqref{ZDGredef1}-\eqref{ZDGredef2} gives the AdS background values for our new fields $e^a, f^a, \omega^a $ and $h^a$:
\begin{equation}
e^a = \frac{2(\beta_1+\gamma \beta_2)}{(\beta_1 + \sigma \beta_2)} \bar{e}^a \equiv A \bar{e}^a\,, \qquad f^a = (\sigma - \gamma)\bar{e}^a \,, \qquad \omega^a = \bar{\omega}^a\,, \qquad h^a = 0\,.
\end{equation}
This background solves the ZDG equations of motion when $\Lambda = -1/\ell^2$ and $\gamma$ are solutions of
\begin{equation}\label{bkgdZDG}
\begin{split}
& \sigma \frac{\Lambda}{m^2} +  a_1 A^2 - 2 b_1 A(\sigma -\gamma) - b_2(\sigma -\gamma)^2 = 0\,, \\
& c \frac{\Lambda}{m^2} - b_1 A^2 - 2 b_2 A(\sigma-\gamma) + \left((c^2-1)b_1-2c \sigma b_2\right) (\sigma-\gamma)^2 = 0\,.
\end{split}
\end{equation}
These two equations are equivalent to the ZDG background relations \eqref{cc12} in the original formulation of the theory. We will refer to this solution as the ${\rm AdS} + \gamma {\rm AdS}$ background.

\subsubsection{Algebra of first class constraints}

Like before, due to the integrability conditions \eqref{IntconexZDG}, we have that on the background the gauge parameters for diffeomorphisms satisfy
\begin{equation}
\xi^e_a = A \xi^{\bar{e}}_a\,, \qquad \xi^f_a = (\sigma - \gamma)\xi^{\bar{e}}_a\,, \qquad \xi^h_a = 0\,.
\end{equation}
Due to this relation and the background relations \eqref{bkgdZDG} the algebra of first-class constraints \eqref{phiLL} and \eqref{phidiff} derived from \eqref{gen_poissonbr} on the ${\rm AdS} + \gamma {\rm AdS}$ background becomes the $SO(2,2)$ algebra:
\begin{align}
\{\phi_{\rm diff}'[\xi], \phi_{\rm diff}'[\eta]\} = &\, -\Lambda \phi_{\rm LL}[[\xi , \eta]]\,, \qquad 
\{\phi_{\rm diff}'[\xi], \phi_{\rm LL}[\eta]\} = \phi_{\rm diff}'[[\xi , \eta]]\,, \\ \nonumber
\{\phi_{\rm LL}[\xi], \phi_{\rm LL}[\eta]\} = & \, \phi_{\rm LL}[[\xi , \eta]]\,,
\end{align}
where now all the gauge parameters are expressed in terms of $\xi^{\bar{e}\,a} = \bar{e}_{\mu}{}^a \zeta^{\mu} = \xi^a$ (and similarly for $\eta^a$). 

We can once again define the $SL(2, \mathbb{R})$ generators
\begin{equation}
L_{\pm}[\xi] = \phi_{\rm diff}'[\xi] \pm \frac{1}{\ell} \phi_{\rm LL}[\xi]\,,
\end{equation}
and the Poisson bracket algebra becomes
\begin{align}\label{pbAdS}
\{ L_{\pm}[\xi], L_{\pm} [\eta]\} & = \pm \frac{2}{\ell} L_{\pm}[[\xi , \eta]]\,, \\
\{ L_{+}[\xi], L_{-} [\eta]\} & = 0\,. \nonumber
\end{align}
At this point we may readily proceed in analogy to the GMG example and the pure gravity analysis of section \ref{sec:ECASG}. In the case of ZDG, using the relation
\begin{equation}
\sigma A + c (\sigma - \gamma) = (\sigma + \gamma)\,,
\end{equation} 
we find the variation of the boundary charges in \eqref{Lpmimproved} to be
\begin{equation}
\delta Q_{\pm} [\xi] =  \left( \sigma + \gamma \right) \int_{\partial \Sigma} d\phi\; \xi \cdot\left(\delta \bar{\omega}_{\phi} \pm \frac{1}{\ell} \delta \bar{e}_{\phi} \right) \,.
\end{equation}
Once again the result is proportional to the Einstein-Cartan case, now with proportionality factor $(\sigma + \gamma)$. The boundary term in the Poisson brackets of the first-class constraints \eqref{pbAdS} becomes
\begin{equation}
\{ \tilde L_{\pm}[\xi], \tilde L_{\pm}[\eta] \} = \ldots \pm \frac{2}{\ell} \left(\sigma + \gamma \right) \int d\phi\, \xi \cdot \left[ \partial_{\phi} \eta +  \left( \bar{\omega}_{\phi} \pm \frac{1}{\ell} \bar{e}_{\phi} \right) \times \eta \right]\,,
\end{equation}
After adopting Brown-Henneaux boundary conditions \eqref{BHbc1} and \eqref{BHbc2} for $\bar{e}^a$ and $\bar{\omega}^a$, together with the boundary condition preserving gauge transformations \eqref{asymptdiffs}, we find that the asymptotic symmetry algebra is again two copies of the Virasoro algebra, now with a semi-classical central charge given by
\begin{equation}\label{ccZDG}
c^{\text{\tiny ZDG}}_{\rm L/R} = \frac{3 \ell}{2 G} \left(\sigma + \gamma \right)\,.
\end{equation}
The conserved global charges for ZDG are also proportional to the Einstein-Cartan ones
\begin{equation}
\begin{split}
& Q^{\rm ZDG}_{+} =  \ell(\sigma+\gamma) \int d\phi \; \cL(x^+) f(x^+) \,, \\
& Q^{\rm ZDG}_{-} =  - \ell(\sigma + \gamma) \int d\phi \;  \bar{\cL}(x^-) \bar{f}(x^-) \,.
\end{split}
\end{equation}
We thus find that the central charge in ZDG is independent of the coupling constants $\beta_1$ and $\beta_2$.

\section{Bulk and Boundary Unitarity}
\label{sec:bbunitarity}

Now that we have derived the central charges for GMG and ZDG, we can investigate whether there are regions in the parameter space of the theories where perturbative unitarity in the bulk is compatible with a positive central charge on the boundary. The latter is a requirement for unitarity of the dual CFT.\footnote{Note that these requirements are necessary but not sufficient for the unitarity of the theory.} In this section we will discuss the parameter restriction imposed by the absence of tachyons and ghosts for the massive mode in the bulk, and positivity of the boundary central charges for TMG, NMG and ZDG. 

\subsection{Topologically Massive Gravity}
The conditions for a single helicity 2 mode to be `physical' were derived in section \ref{sec:pertunit}. In the case of TMG, the absence of tachyons \eqref{notachyon} translates to
\begin{equation} \label{TMGnotachyon}
\mu \ell \geq 1\,,
\end{equation}
For the absence of ghosts we need the coefficient in front of the massive mode in the quadratic action. To this end we diagonalize the quadratic action \eqref{TMGquadaction} by a linear field redefinition
\begin{equation}
\begin{split}\label{TMGredef}
& k = \ell f_+ + \ell f_- + \frac{\ell^2 \mu}{\ell^2 \mu^2 - 1} p\,, \\
& v = f_+ - f_- + \frac{\sigma \ell^2 \mu^2}{\ell^2 \mu^2 -1} p\,.
\end{split}
\end{equation}
The quadratic TMG action then becomes
\begin{align}\label{TMGqadiag}
L_{\rm TMG}^{(2)} = & \; \left(-\sigma \ell + \frac{1}{\mu} \right) \left[ f_{+} \cdot \bar{\cD} f_+ + \frac{1}{\ell}  \be \cdot f_+ \times f_+ \right]  \nonumber \\
& +  \left(\sigma \ell + \frac{1}{\mu} \right) \left[ f_{-} \cdot \bar{\cD} f_- - \frac{1}{\ell}  \be \cdot f_- \times f_- \right] \\ \nonumber
& + \frac{\mu \ell^2}{2(\mu^2\ell^2 -1)} \left[ p \cdot \bar{\cD} p + \sigma \mu \be \cdot p \times p \right] \,.
\end{align}
The first two lines represent the massless modes which do not propagate any degrees of freedom. The last line is the action for a single massive spin-2 state. By comparison with the diagonalized Fierz-Pauli action \eqref{FPdecomp} we see that the energy of the massive mode is positive whenever
\begin{equation}\label{ATMG}
A = - \frac{\sigma \ell^2}{2(\mu^2 \ell^2 -1)} > 0\,.
\end{equation}
For $\sigma = +1$ this is in direct contradiction with the no-tachyon condition \eqref{TMGnotachyon}. Hence for the two to be compatible, we need to take $\sigma = -1$, or in other words, we need to take the `wrong sign' in front of the Ricci scalar in the action. 

This conclusion poses a problem for the positivity of the central charges \eqref{ccTMG}. They are never positive for $\sigma = -1$ and $\mu \ell >1$! In fact, positive central charges and $\mu\ell >1$ requires use to take $\sigma = +1$. In that case, the best we can do is to saturate the bound \eqref{TMGnotachyon} \cite{Li:2008dq}. At this point the massive mode becomes massless and $p^a$ degenerates with $f_+^a$. Another consequence is that the field redefinitions \eqref{TMGredef} are not well-defined and the quadratic action becomes non-diagonalizable. The field equations, however, remain third order and a new logarithmic solution appears \cite{Grumiller:2008qz}. At this point the Brown-Henneaux boundary conditions can be modified, to allow for this logarithmic behavior towards the AdS boundary \cite{Grumiller:2008es}. This has led to the conjecture that TMG at this critical point is dual to a logarithmic conformal field theory (LCFT) \cite{Maloney:2009ck,Skenderis:2009nt,Grumiller:2009mw}. 

Note that the coefficients in front of the massless modes in \eqref{TMGqadiag} are proportional to the central charges. By comparing them with the diagonalized Fierz-Pauli action \eqref{FPdecomp} with mass parameter $M = 1/\ell$, we see that requiring positive energy for the massless modes is equivalent to the condition of positive central charges. This is compatible with the results of \cite{Li:2008dq}, obtained by analyzing the quadratic action in a Hamiltonian form.

\subsection{New Massive Gravity}
In NMG, the no-tachyon and no-ghost conditions \eqref{notachyon} and \eqref{noghost} can be read off directly from the diagonalized quadratic action \eqref{NMGqa}. They are, respectively:
\begin{equation}
- m^2 \left(\sigma + \frac{1}{2\ell^2m^2} \right) = - m^2 \bar{\sigma} > 0\,,
\end{equation}
and
\begin{equation}
- m^4 \bar{\sigma} > 0\,.
\end{equation}
In principle the massive parameter $m^2$ could be chosen with a relative minus sign, but we see from the combined no-tachyon and no-ghost conditions that we should take $m^2>0$. We are then forced to conclude that requiring the massive spin-2 to be physical restricts the parameters of the theory as $\bar{\sigma} < 0$. This is in direct conflict with positivity of the central charges \eqref{ccNMG}. Hence the theory is not unitary in AdS$_3$. 

Once again the situation is more subtle at a special point in the parameter space, where $\bar{\sigma} = 0$. The two massive modes degenerate with the two massless modes and new logarithmic modes appear. We will discuss the appearance of these logarithmic modes in more detail in chapter \ref{chapter:AdS_LCFT}.

Since both TMG and NMG do not have a region in their parameter space where perturbative bulk unitarity is compatible with positive central charges, it should not be surprising that their combination into GMG suffers from the same problem. This can indeed be verified with the methods presented in this thesis, and we will leave this as an exercise for the interested reader.

Also in NMG  the coefficient in front of the massless modes in the quadratic action is proportional to the central charge. In a sense this should not be a big surprise, since the massless, pure gauge modes correspond to the boundary gravitons. As we saw in chapter \ref{sec:ECASG} the global charges which generate the asymptotic conformal algebra are the boundary terms for the generators of gauge symmetries. When the massless sector gets a specific pre-factor in its action, then this will reflect in the global charges and finally in the central charge.

\subsection{Zwei-Dreibein Gravity}
We will now investigate the same parameter constraints for the Zwei-Dreibein model. The fact that the ZDG action can not be written as a higher-derivative action for a single metric (this only works at the level of equations of motion) indicates that the situation may be improved in this theory. Higher-derivative actions typically suffer from the Ostrogradsky instability, which states that systems whose Lagrangians depend on second and higher-order derivatives of time are necessarily unstable. In ZDG, there are two dreibeine, which both come with second-order time derivatives in the Lagrangian and hence there is no direct indication for an Ostrogradsky instability. However, we did find a set of higher-derivative field equations for ZDG in chapter \ref{section:ZDGHD} and we know that NMG is contained within ZDG as a scaling limit.

The conditions for perturbative unitarity of the massive modes in ZDG are positive Fierz-Pauli mass \eqref{MFP}
\begin{equation}\label{ZDGnotachyon}
\cM^2 = m^2(\beta_1 + \gamma \beta_2)\frac{\sigma + \gamma}{\sigma} > 0\,,
\end{equation}
and a positive coefficient in front of the Fierz-Pauli Lagrangian \eqref{linear_bi_lagrangian}
\begin{equation}\label{ZDGnoghost}
\frac{\sigma \gamma}{\sigma + \gamma} >0\,.
\end{equation}
Positivity of the central charge \eqref{ccZDG} requires
\begin{equation}\label{ZDGpcc}
\sigma + \gamma > 0\,,
\end{equation}
where we remind you that the AdS background parameter $\gamma$ should solve the background equations \eqref{cc12}. 

The Hamiltonian analysis of the last chapter revealed that in addition to these inequalities, the absence of the Boulware-Deser ghost required us to assume the invertibility of the linear combination of dreibeine $\beta_1e_1^a + \beta_2 e_2^a$. On the AdS background, where $e_1^a = \bar{e}^a$ and $e_2^a = \gamma \bar{e}^a$, this implies that we should restrict $\beta_1 + \gamma \beta_2 \neq 0$. Consequently, any small perturbation around this background cannot render the linear combination noninvertible. 

From these inequalities combined, we see that $\sigma = -1$ is not allowed, since \eqref{ZDGnoghost} would then imply that $\gamma <0$ and this is in clear contradiction with \eqref{ZDGpcc}. Hence we should restrict to the positive sign $\sigma = +1$ and $\gamma >0$.\footnote{This is consistent with the NMG results, since the NMG limit to ZDG required us to take $\sigma = -1$} In that case, all inequalities can be satisfied simultaneously, as there are two of them ($\gamma >0$ and \eqref{ZDGnotachyon}) for a four dimensional parameter space $(\alpha_1, \alpha_2, \beta_1, \beta_2)$. An explicit example satisfying all constraints is 
\begin{equation}
\beta_1 = \beta_2 = 1\,, \qquad \alpha_1 = \alpha_2 = 3 + \zeta\,,
\end{equation}
where $\zeta$ is a positive constant. For this choice of parameters there is an AdS vacuum satisfying \eqref{cc12} with $\gamma =1$ and $(\ell m)^{-2} = \zeta$. 
Furthermore, $\gamma \approx 1$ for any `nearby' ZDG model, with slightly different parameters, which are themselves constrained only by inequalities that have been satisfied but not saturated. It follows that the above explicit model is one of an open set of models in the ZDG parameter space with similar "good" properties; these properties are not the result of any fine-tuning of parameters that could be destabilized by perturbative quantum corrections. 

\section{Discussion}
The appearance of a ``bulk-boundary clash'' (the incompatibility of positive boundary central charge with positive energy and mass of the bulk spin-2 modes) is a general feature of higher-derivative gravity theories. We have shown here that there are alternatives, within the class of CS--like models; ZDG does not show this clash for some region in its parameter space. A related issue concerns the mass of the BTZ black holes in these theories. As was shown in chapter \ref{sec:ECASG}, the mass and angular momentum of the BTZ black holes are related to the zero modes of the Virasoro generators. As we have seen here, in the CS--like theories, the conserved global charges pick up a factor proportional to the central charge and hence, the mass of the BTZ black hole will receive the same overall factor. This implies that whenever the central charge is negative, the BTZ black hole will have negative mass. Due to this argument, the term ``bulk-boundary clash'' is technically not very fitting, as negative mass black holes are a clear indication of a bulk problem. However, to be consistent with some of the literature, we will continue to use this phrase.

An interesting question is whether there exists a CS--like modification of TMG which could avoid the bulk-boundary clash, much like ZDG extends and generalizes the NMG action in a CS--like formulation. An early indication that this could be possible follows from a counting argument along the lines of chapter \ref{chapter:three_crowd}. The generic CS--like model with three one-form fields will always have 9 primary constraints and 18 dynamical variables. If we assume, as always, a locally Lorentz symmetric theory and diffeomorphism invariance, then at least 6 constraints are first-class, leaving at most 3 second-class constraints. However, the number of second class constraints cannot be odd, hence either one (or all) of them are actually first-class, or there exists at least one secondary second-class constraint. In both of these cases, a counting of the dimensionality of the physical phase space following \eqref{dimcounting} shows that there can be at most 1 local degree of freedom. Of course, when there are no degrees of freedom, the theory is actually a CS gauge theory. Hence not only TMG, but {\it all} $N=3$ CS--like models have one local degree of freedom. This is a clear indication that in the CS--like framework, TMG is not the unique theory for a single massive helicity 2 state and that there may be alternatives, perhaps with better behavior in light of the AdS/CFT-correspondence. This turns out to be the case as we will discuss at length in the next chapter.
\enlargethispage{\baselineskip}

%% file: chapter_6/chapter_6.tex
\pagestyle{empty}
\setcounter{chapter}{5}

\chapter[Minimal Massive 3D Gravity]{Minimal Massive 3D Gravity}
\label{chapter:MMG}

\pagestyle{headings}

\begin{quote}\em
Although the higher-derivative models of massive gravity generally give rise to a non-unitary theory in AdS$_3$, this is not true for some of the Chern-Simons--like theories of gravity. In the last chapter we saw how ZDG has regions in its parameter space where perturbative bulk unitarity is consistent with positive boundary central charge. In this chapter we investigate a Chern-Simons--like model with three flavor fields and with the same "minimal" bulk properties as Topologically Massive Gravity; a single local degree of freedom, realized as a massive graviton in linearization about an Anti-de Sitter vacuum. This "minimal massive gravity" has both a positive energy for the graviton and positive central charges for the asymptotic symmetry group.  The content of this chapter is based on \textsc{[ix]}.
\end{quote}

\newpage

\section{Introduction}

Topologically Massive Gravity (TMG), discussed in chapter \ref{sec:TMG}, is a parity-violating  extension of three\-/dimensional (3D) General Relativity (GR) that propagates, on linearization about a maximally symmetric vacuum, a single  massive spin-2 mode \cite{Deser:1981wh}. Its action augments the Einstein-Hilbert action (plus ``cosmological'' term with cosmological parameter $\Lambda_0$)  by a  Chern-Simons action for  the Levi-Civita affine connection one-form $\Gamma$. Omitting a positive  factor proportional to the inverse of the 3D Newton constant, which has dimensions of inverse mass, the TMG action is\footnote{We use a ``mostly plus'' metric signature convention.}
\begin{equation}\label{TMGaction}
S_{\rm TMG} \ = \  \int \!d^3x\, \sqrt{-\det g}\,  (\sigma R -2\Lambda_0)+ \frac{1}{2\mu}\int \! \tr \left\{ \Gamma d\Gamma + \frac{1}{3} \Gamma^3\right\}  \,, 
\end{equation}
where $\mu$ is a mass parameter,  and $\sigma$ is a sign (plus for GR  but minus for TMG if we insist on positive energy for the spin-2 mode, as we saw in the last chapter).  The  TMG field equation derived from this action is
\begin{equation}\label{TMGfield}
\frac{1}{\mu} C_{\mu\nu} + \sigma  G_{\mu\nu}  + \Lambda_0 g_{\mu\nu} =0 \, ,
\end{equation}
where $G_{\mu\nu}$ is the Einstein tensor,  and $C_{\mu\nu}$  the (symmetric traceless) Cotton tensor, defined as
\begin{equation}
C_{\mu\nu} \equiv \frac{1}{\sqrt{-\det g}} \, \varepsilon_\mu{}^{\tau\rho} \nabla_\tau S_{\rho\nu}\, , \qquad S_{\mu\nu} \equiv R_{\mu\nu} - \frac{1}{4}g_{\mu\nu} R\, .
\end{equation}
Here, $\nabla$ is the covariant derivative defined with $\Gamma$, and $S_{\mu\nu}$  the 3D Schouten tensor.  As  $G_{\mu\nu}= -\Lambda g_{\mu\nu}$
for maximally symmetric vacua with cosmological constant $\Lambda$, and since $C_{\mu\nu}$ is zero in such vacua, the relation between $\Lambda_0$ and $\Lambda$ for TMG is $\Lambda= \sigma\Lambda_0$. 

TMG is a minimal theory of massive 3D gravity in the sense that a field propagating a single spin-2 mode in a Minkowski vacuum defines a unitary irrep of the 3D Poincar\'e group. 
An obvious question is whether TMG could be the semi-classical approximation to some  3D quantum gravity theory; in particular, it is natural to wonder whether there might be a holographically dual conformal field theory (CFT) on the boundary  of an AdS$_3$ vacuum of TMG with cosmological constant  $\Lambda=-1/\ell^2$ for AdS$_3$ radius $\ell$. 
The trouble with this idea was discussed in chapter \ref{sec:bbunitarity}; the central charge of such a CFT, computed in a semi-classical approximation, is negative whenever the  bulk spin-$2$ mode has positive energy, implying a non-unitary CFT.  A closely related problem is that the  Ba\~nados-Teitelboim-Zanelli (BTZ)  black hole solutions  (which exist for any 3D gravity theory with an AdS$_3$ vacuum)  have negative mass whenever the energy of bulk graviton modes is positive.  It was suggested in \cite{Li:2008dq} that this problem might be circumvented by first choosing $\sigma=1$, to ensure positive mass BTZ black holes, and then tuning the dimensionless parameter $\mu\ell$  to a critical point  at which the  bulk mode  is absent, and the boundary CFT is chiral, for sufficiently strong boundary conditions. However, another bulk mode appears at the critical point \cite{Carlip:2008jk}, and it was soon realized that this is  a chirality-violating ``logarithmic''  mode that  is compatible with consistent AdS$_3$ boundary conditions \cite{Grumiller:2008qz}, implying a non-unitary ``logarithmic'' boundary  CFT. 

Until recently, a similar state of affairs held for the parity-preserving New Massive Gravity (NMG) \cite{Bergshoeff:2009hq,Bergshoeff:2009aq},  which is a 4th-order extension of 3D
GR that propagates a parity doublet of spin-2 modes (and is therefore minimal with respect to the product of the Poincar\'e group with parity). This too suffers from the defect 
that the central charge of the boundary CFT dual is negative whenever the bulk spin-2 modes have positive energy, and tuning to critical points again leads only to non-unitary
``logarithmic''  boundary CFTs.  Extensions of TMG or NMG with  higher powers  of curvature have been discussed in the literature \cite{Sinha:2010ai,Paulos:2010ke} but the 
``bulk-boundary clash'' persists. Other attempts to evade this conflict (e.g. \cite{Banados:2009it}) typically introduce, as a by-product,    the Boulware-Deser ghost 
\cite{Boulware:1973my} (which is  invisible in a linearized approximation). However,  the recently constructed ``Zwei Dreibein Gravity'' (ZDG) shows that there is a viable alternative to NMG \cite{Bergshoeff:2013xma,Bergshoeff:2014bia}. 

At present  there is no known alternative to TMG that resolves the  ``bulk-boundary clash''  while preserving  the minimal bulk properties. The main purpose of this chapter is to present such an alternative.  Since it has the same minimal local structure as TMG (and also for another reason to be explained later)  we shall call it ``Minimal Massive Gravity'' (MMG);
the essential difference is that the field equation of MMG includes the additional, curvature-squared, symmetric tensor
\begin{eqnarray}\label{Jtensor}
J_{\mu\nu} &=&  \frac{1}{2\det g}\ \varepsilon_\mu{}^{\rho\sigma} \varepsilon_\nu{}^{\tau\eta} S_{\rho\tau}S_{\sigma\eta}  \nonumber \\
 &=&   R_\mu{}^\rho R_{\rho\nu} - \frac{3}{4} RR_{\mu\nu} - \frac{1}{2} g_{\mu\nu} \left(R^{\rho\sigma}R_{\rho\sigma} - \frac{5}{8} R^2\right)  \, .
\end{eqnarray}
In other words, the MMG field equation is
\begin{equation}\label{modTMG}
\frac{1}{\mu} C_{\mu\nu} + \bar\sigma  G_{\mu\nu}  + \bar\Lambda_0 g_{\mu\nu} =- \frac{\gamma}{\mu^2} J_{\mu\nu} \, ,
\end{equation}
where $\gamma$ is some non-zero dimensionless constant\footnote{We shall see later that it must be positive.}; we have replaced $\sigma$ by $\bar\sigma$ since it is no longer obvious why it should be just a sign, and we have replaced $\Lambda_0$ by $\bar\Lambda_0$ since we  should not expect it to equal the cosmological parameter when $\bar\sigma=1$. 

Using the Bianchi identities satisfied by the Einstein and Cotton tensors, we see that consistency requires
$\nabla_\mu J^{\mu\nu} =0$, but a direct computation shows that
\begin{equation}\label{Jid}
\sqrt{-\det g}\, \nabla_\mu J^{\mu\nu} =    \varepsilon^{\nu\rho\sigma} S_\rho{}^\tau C_{\sigma\tau}\, .
\end{equation}
The right hand side is not identically zero, but it is only required to be zero as a consequence of the field equation (\ref{modTMG}); using this
to eliminate the Cotton tensor, we find that
\begin{equation}\label{onshellid}
\sqrt{-\det g}\, \nabla_\mu J^{\mu\nu}=  \frac{\gamma}{\mu} \varepsilon^{\nu\rho\sigma} S_\rho{}^\tau J_{\tau\sigma} \equiv 0\, .
\end{equation}
The identity holds because the tensor $J$ can be written as 
\begin{equation}
J_{\mu\nu} =  S_\mu{}^\rho S_{\rho\nu} -S S_{\mu\nu} - \frac{1}{2}g_{\mu\nu}\left(S^{\rho\sigma}S_{\rho\sigma}-S^2\right)\, , 
\end{equation}
where $S\equiv g^{\mu\nu}S_{\mu\nu}$. This tells us that ``$\varepsilon SJ$'' is a linear combination  of terms of the form $\varepsilon S^n$ for $n=1,2,3$, which are all zero because $S^n$ is symmetric.  Thus, remarkably, the modified field equation (\ref{modTMG}) is consistent. 

The manner in which the MMG field equation (\ref{modTMG}) evades inconsistency is novel; we are not aware of any other example in which
consistency is achieved in this way.  We elaborate on the implications of this novelty  in the final section of this chapter. One implication  is that  the MMG field equation (for the metric alone) cannot be obtained from any conventional higher-curvature  modification of the TMG action (\ref{TMGaction}).  Nevertheless,  there is a very  simple action with auxiliary 
fields that yields  precisely the equation (\ref{modTMG}), and the required auxiliary fields are those  already present in the dreibein formulation of TMG \cite{Grumiller:2008pr,Carlip:2008qh}.  These auxiliary fields can be eliminated from the field equations derived from the MMG action, leaving only the MMG equation (\ref{modTMG}). However, in contrast to the 
usual situation for auxiliary fields, back-substitution into the action is not legitimate  (for $\gamma\ne0$)  so the MMG action with auxiliary fields does not imply the existence of one without them.

These results would be no more than curiosities if the proposed modification of TMG were to change the local structure in an unacceptable way. A first indication
that this will not happen is that the $J$ tensor does not contribute to linearization about a Minkowski vacuum, but this is a rather weak test. It  would seem to require
a miracle for this property to be maintained for other vacua, leaving aside the Boulware-Deser ghost.  Nevertheless, the miracle occurs, as we explain in detail using the Hamiltonian methods developed in chapter \ref{chapter:three_crowd}. Furthermore,  not  only is the local structure of MMG exactly that of TMG, but the freedom allowed in MMG permits a resolution
of the ``bulk-boundary clash'', as we show by a computation of the algebra of the asymptotic conformal group in an asymptotically AdS$_3$
spacetime.

\section{Chern-Simons--like formulation}
\label{sec:CSLM}

We were led to the MMG model of massive gravity by considering possible modifications of TMG in the context of its formulation as a Chern-Simons--like model of gravity \eqref{LCSlike}.
For present purposes, we focus on the case with $N=3$ and we take $(e,\omega,h)$ as the three Lorentz vector-valued one-form fields. The $h$ field has the same parity (odd) and dimension as $\omega$ and it appears in the TMG action as a Lagrange multiplier for the zero torsion constraint. The Lagrangian three-form for TMG was given in \eqref{LTMG_cslike}. After redefining the auxiliary field $f = \mu h$ it is given by:
\begin{equation}
L_{TMG} = -\sigma e \cdot R + \frac{\Lambda_0}{6}e\cdot e\times e + h \cdot T(\omega) + \frac{1}{2\mu}\left(\omega \cdot d \omega + \frac{1}{3}\omega\cdot \omega \times \omega\right) \,,
\end{equation}
where  $\Lambda_0$ is a  cosmological parameter with dimensions of mass-squared, and $\sigma$ a sign. The last term, with a factor of $1/\mu$, is the ``Lorentz Chern-Simons'' (LCS) term, but here for the independent dual spin connection $\omega$. The mass parameter  $\mu$ could have either sign since a parity transformation would effectively flip the sign of $\mu$. In the limit that $|\mu|\to\infty$ the TMG action reduces (for $\sigma=1$) to an action for 3D GR.  

Because  the 3D Newton constant has  dimensions of inverse mass, the Lagrangian three-form should have dimension of mass-squared, as it does if we assign zero dimension to $e$ and dimensions of mass to  both $\omega$ and $h$. With these assignments of parity and dimension,  and given the requirement of  local Lorentz invariance, the TMG Lagrangian three-form is  almost unique, up to field redefinitions, if we suppose that parity is broken only by the LCS term. However, there is one further parity-even term that could be included; this is an ``$e\cdot h \times h$'' term. This leads us to consider the following
one-parameter family of ``Minimal Massive Gravity'' (MMG) Lagrangian three-forms
\begin{equation}\label{Letmg}
L_{MMG} = L_{TMG} + \frac{\alpha}{2} e \cdot h \times h\, ,
\end{equation}
where $\alpha$ is a dimensionless parameter.  In the absence of the parity-violating LCS term, the new ``ehh'' term is innocuous; it leads only to an alternative action for 3D GR. 
However, when combined with the LCS term it leads, as we shall now show,  to a modification of TMG that is equivalent to the one described  in tensor form in the introduction (with the constant $\gamma$ being a function of $\alpha$). 

The field equations derived from the Lagrangian three-form \eqref{Letmg} are
\begin{equation}
\begin{split} \label{etmgeom}
0 &= T(\omega) + \alpha e \times h \,, \\
0 &= R(\omega) + \mu e\times h  - \sigma \mu T(\omega) \,, \\
0 &= - \sigma R(\omega)+ \frac{\Lambda_0}{2} e\times e + \cD(\omega) h + \frac{\alpha}{2} h\times h \,.
\end{split}
\end{equation}
An equivalent set of equations is
\begin{eqnarray}\label{inOm}
0 &=& T(\Omega)\, , \nonumber\\
0 &=& R(\Omega) + \frac{\alpha\Lambda_0}{2} e\times e + \mu\left(1+ \sigma\alpha\right)^2 e\times h\, , \\
0 &=& \cD(\Omega)h - \frac{\alpha}{2} h\times h + \sigma\mu \left(1+\sigma\alpha\right) e\times h  + \frac{\Lambda_0}{2}\, e\times e\, , \nonumber
\end{eqnarray}
where the new dual spin connection one-form is
\begin{equation}
 \Omega = \omega + \alpha h\, .
 \end{equation}
In what follows we shall assume that the dreibein is invertible.
Using the identities
\begin{equation}
\cD(\Omega)T(\Omega) \equiv R(\Omega)\times e\, , \qquad \cD(\Omega)R(\Omega) \equiv 0\,,
\end{equation}
and assuming that 
\begin{equation}\label{alphacon}
1+ \sigma\alpha\ne 0\, ,
\end{equation}
one finds that the field equations imply
\begin{equation}\label{second}
0= e\cdot h \equiv e^a h^b \eta_{ab} \, .
\end{equation}

The first of  equations (\ref{inOm}) implies that $\Omega= \Omega(e)$,  the usual torsion free spin connection, which can be traded for the Levi-Civita affine connection.  
The second of equations (\ref{inOm}) can be solved for $h$:
\begin{equation}\label{hsol}
h_{\mu\nu} \equiv h_\mu{}^a e_\nu{}^b \eta_{ab} =  - \frac{1}{\mu\left(1+\alpha\sigma\right)^2} \left[S_{\mu\nu} + \frac{\alpha\Lambda_0}{2} g_{\mu\nu}\right]\, ,
\end{equation}
where $S_{\mu\nu}$ is the 3D Schouten tensor. Since this is a symmetric tensor, we learn that $h_{\mu\nu}$ is symmetric; this is precisely the content of (\ref{second}).
At this point, we have used the first two of equations (\ref{inOm}) to solve for $\Omega$ and $h$ in terms of $e$. Because of local Lorentz invariance, back-substitution 
into the action will produce an action for the metric alone. However, this back-substitution is not legitimate when $\alpha\ne0$ because the equations used are {\it not} equivalent, jointly, to the two equations found by varying the action with respect to $\Omega$ and $h$; to get them one needs (if $\alpha\ne0$) to use the  $e$-equation in addition to the $\Omega$ and $h$ equations. 

Although it is not legitimate to back-substitute into the action, it is legitimate to substitute the expressions for $\Omega$ and $h$ into the third equation of  (\ref{inOm})  to arrive at an equation for the metric\footnote{The situation is similar for ZDG as was emphasized in chapter \ref{section:ZDGHD}; the equations allow the elimination of one dreibein in terms of the other one but back-substitution into the action is not legitimate. One can still substitute into the ZDG equations to get a field equation for only one dreibein but  it  involves an infinite series that must be constructed order by order  \cite{Bergshoeff:2014eca}.}.  Doing so, we  find that this equation is equivalent to the MMG tensor equation (\ref{modTMG}) with
coefficients
\begin{equation}
\bar\sigma = \sigma +\alpha\left[1 + \frac{\alpha\Lambda_0/\mu^2}{2\left(1+\sigma\alpha\right)^2}\right]\, , \qquad \gamma= -\frac{\alpha}{\left(1+ \sigma\alpha\right)^2} \, ,
\end{equation}
and\footnote{Recall  that $\Lambda_0$ is the cosmological parameter occurring in the Lagrangian,  $\bar\Lambda_0$ the constant coefficient of $g_{\mu\nu}$ in the field equation
and $\Lambda$  the cosmological constant of a vacuum solution of these equations. For TMG  $\Lambda= \sigma\bar\Lambda_0 =\sigma\Lambda_0$  but for  other 3D gravity models, including MMG, the relation between these parameters is more complicated.}
\begin{equation}
\bar\Lambda_0 = \Lambda_0 \left[1+ \sigma\alpha - \frac{\alpha^3 \Lambda_0/\mu^2}{4\left(1+\sigma\alpha\right)^2}\right]\, .
\end{equation}
We have now found an action that yields the MMG equation (\ref{modTMG}). In its CS--like form it is a very simple and natural modification of the TMG action. However, {\it it cannot be reduced to an action for the metric alone by elimination of the auxiliary fields}. This result  accords with our  earlier observation that the MMG equation cannot be obtained from an action for the metric alone. 

We shall  later give a detailed proof that MMG has the same number of local degrees of freedom as TMG,  but the essence of the analysis is  as follows.
Starting from the CS--like action,  a time/space split leads directly to a constrained Hamiltonian system with the time components  acting  as Lagrange multipliers for $9$ primary constraints. The  $18$ space components  are the canonical variables. By construction,  $6$ of the primary constraints are first-class, generating diffeomorphisms and local Lorentz transformations, which leaves $3$ primary constraints. If any of these are first-class there will be additional gauge invariances, so any such model will be exceptional.  For example, if  all $3$ remaining primary constraints are first-class then the CS--like theory is actually a CS theory with no local degrees of freedom. This possibility is realized when  $\alpha\sigma=-1$ because $L_{MMG}$ is then the sum of a CS three-form for $(e,\Omega)$ (with gauge group depending on $\Lambda_0$) and an $SL(2,\bR)$ CS three-form for $\omega$; we exclude this case by imposing the restriction (\ref{alphacon}).

One can go systematically through the other possibilities for the general $3$-flavor model (allowing for the possibility of secondary constraints) to show that the dimension of the  physical phase space, per space point, must be either $0$ or $2$. As we shall see, only the latter possibility is consistent with a linearized analysis of MMG (assuming $1+\sigma\alpha\ne0$)  so the generic case  is the one of relevance here.  This is the case for which the three remaining primary constraints form a second-class set together  with one secondary constraint, the space component of   (\ref{second}). This yields a total of $6$ first-class and $4$ second-class constraints, implying a physical phase space with dimension per space point of $(18 - 2\times 6 - 1\times 4) = 2$, corresponding to one local degree of freedom, exactly as for TMG.

\section{Linearized analysis}

We now look for maximally symmetric vacuum solutions of the MMG equations in the form (\ref{inOm}). These are solutions for which
\begin{equation}
R(\Omega) = \frac{1}{2}\Lambda\,  e \times e\, , \qquad \left( \ \Rightarrow\quad S_{\mu\nu}= \frac{1}{2}\Lambda g_{\mu\nu}\right)\, ,
\end{equation}
where $\Lambda$ is the cosmological constant.  Substitution into the equation of motion \eqref{modTMG} yields
\begin{equation}\label{ha}
h = C\mu \, e \, , \qquad \sigma\Lambda/\mu^2= \Lambda_0/\mu^2 - \alpha\left(1+\sigma\alpha\right)C^2 \, ,
\end{equation}
for the dimensionless constant
\begin{equation}
C= - \frac{\left(\Lambda +\alpha\Lambda_0\right)/\mu^2}{2\left(1+\sigma\alpha\right)^2} = - \frac{\Lambda}{2\mu^2} + \mathcal{O}(\alpha)\, .
\end{equation}

When $\alpha=0$ we have $\sigma\Lambda=\Lambda_0$; otherwise we have a quadratic equation for
$\Lambda$, for which the solution is
\begin{equation}\label{lambdasol}
\Lambda = -\alpha\Lambda_0 - \frac{2\mu^2\left(1+ \sigma\alpha\right)^3}{\sigma\alpha}
 \left[1 \mp \sqrt{1+ \frac{\alpha\Lambda_0/\mu^2}{\left(1+\sigma\alpha\right)^2}}\right]\, .
\end{equation}
To recover the $\alpha=0$ case in the $\alpha\to0$ limit, one must choose the upper sign. From now on, we will reserve the upper sign for the TMG branch, i.e.~the branch that contains the TMG model in the $\alpha \rightarrow 0$ limit. The lower sign denotes the non-TMG branch.

For real $\Lambda$ we must restrict $\Lambda_0$ such that
\begin{equation}\label{restrict}
\alpha\Lambda_0/\mu^2 + \left(1+\sigma\alpha\right)^2\ge 0\, .
\end{equation}

\subsection{Linear equations}

We now linearize about an AdS background, for which
\begin{equation}
\Lambda = - 1/\ell^2\, ,
\end{equation}
where $\ell $ is the AdS radius.  Let  $\bar{e}$ be the background dreibein, and $\bar{\omega}\equiv \Omega(\bar e)$ the background spin connection.  We expand about this background by writing
\begin{equation}
e^a = \bar{e}^a + k^a\,, \qquad \Omega^a = \bar{\omega}^a +  v^a\,, \qquad
h^a = C\mu \left( \bar{e}^a +  k^a \right) + p^a\,,
\end{equation}
where $(k^a,v^a,p^a)$ are perturbations; $k^a$ has even parity whereas $v^a$ and $p^a$ have odd parity. The expansion of $h^a$ breaks parity due to the presence of a term linear in $\mu$.
To first order in these perturbations, the field equations are\,\footnote{These equations break parity but only because of terms involving odd powers of the mass $\mu$.}
\begin{equation}\label{lin}
\begin{split}
& \bar{\cD} k + \bar{e}\times v = 0\,, \\
& \bar{\cD} v - \Lambda \, \bar{e}\times k = - \mu (1 + \alpha \sigma)^2 \bar{e}\times p \,, \\
& \bar{\cD} p + M \bar{e}\times p = 0\,,
\end{split}
\end{equation}
where $\bar \cD$ is the covariant exterior derivative for spin connection $\bar\omega$, and 
\begin{equation}\label{mass}
M = \left[\sigma(1+ \sigma\alpha) - \alpha C\right]\mu =  \pm\, \sigma\mu \left(1+\sigma\alpha\right)\sqrt{1 + \frac{\alpha\Lambda_0}{\mu^2\left(1+\sigma\alpha\right)^2}} \, .
\end{equation}
The sign here is the same as the one appearing in (\ref{lambdasol});  in other words, the top sign allows an $\alpha\to0$ limit, whereas the bottom sign does not.  

Notice that   the condition (\ref{restrict}), required for reality of $\Lambda$, is equivalent to $M^2\ge0$.  Let us also record here, for future use, the identity
\begin{equation}\label{CM}
1-2C = \frac{(\ell M)^2-1}{(1+\sigma\alpha)^2 (\ell\mu)^2}\, .
\end{equation}

The integrability conditions of the equations (\ref{lin}) may be found by using  the fact that for any Lorentz-vector valued one-form $a$, 
\begin{equation}
\bar \cD^2 a=  \frac{1}{2} \Lambda (\bar e\times \bar e) \times a =\Lambda \, \bar e \left(\bar e \cdot a\right)\, . 
\end{equation}
This leads to the conclusion that the equations (\ref{lin}) imply 
\begin{equation}
\bar e\cdot p =0\, . 
\end{equation}
Equivalently, these equations imply that the tensor field $p_{\mu\nu}\equiv p_\mu{}^a \bar e_\nu{}^b\eta_{ab}$ on AdS$_3$ is symmetric.

Provided that   $|\ell M|\ne1$ (equivalently, $2C \ne1$)  the  set of three first-order equations  (\ref{lin}) may be diagonalized. This is achieved by introducing the new variables 
$f_\pm^a$ defined by
\begin{equation}
\label{cov}
k^a = \ell(f^a_{+} + f^a_{-}) + \frac{1}{\mu\left(1 - 2C\right)}\,  p^a\,, \qquad v^a = f^a_{+} - f^a_{-} + \frac{M}{\mu\left(1 -2C\right)}\,  p^a\, .
\end{equation}
This leads to the three equations
\begin{equation}\label{three-eq}
\begin{split}
&\bar{\cD} f_+ +  \ell^{-1}  \bar{e}\times  f_+ = 0\,, \\
&\bar{\cD} f_- -  \ell^{-1}  \bar{e}\times  f_- = 0\, , \\
&\bar{\cD} p + M  \, \bar{e}\times p \ = 0\,.
\end{split}
\end{equation}
Parity now exchanges $f_+$ with $f_-$, so the equations for these fields are exchanged by parity. Taken together, these two equations preserve parity.
The equation for $p$ breaks parity, as expected because $M\propto\mu$; this is the AdS$_3$ version of the ``self-dual'' equation for a single massive spin-2 mode \cite{Aragone:1986hm}.

For any Lorentz vector-valued one-form field $a$ the first-order equation 
\begin{equation}\label{first-order}
\bar \cD a + m\,  \bar e\times a =0\,,
\end{equation} 
is equivalent, given that $\bar e\cdot a=0$ and hence that $a_{\mu\nu}$ is symmetric\footnote{This condition is a consequence of (\ref{first-order}) when $|\ell m|\ne1$, and when $\ell m=\pm 1$ it may be imposed as a gauge condition.},  to the equation
\begin{equation}
( \cD^m  a)_{\mu\nu} =0 \, , \qquad 
( \cD^m)_\mu{}^\nu \equiv  \delta_\mu{}^\nu + \frac{1}{m \det(\be)}\,  \ve_\mu{}^{\tau\nu}{} {\bar \nabla}_\tau \, , 
\end{equation}
combined with the condition that the symmetric tensor $a$ is traceless.  Using this result, we may rewrite the equations (\ref{three-eq}) in tensor form as 
\begin{equation}\label{another-three}
\cD^L f_+=0\, , \qquad \cD^R f_- =0\, , \qquad  \cD^M p=0\,,
\end{equation}
for symmetric traceless tensors $(f_\pm,p)$. Here $\cD^{L/R}$ are defined in \eqref{cDLRdef} as $\cD^m$ with $m = \pm 1/\ell$. More generally, without assuming that $|\ell M|\ne1$, we may use the equations (\ref{lin})  to eliminate $p$ and $v$ and thus obtain the following third-order equation 
\begin{equation}\label{third-order}
(\cD^M \cD^L\cD^R k)_{\mu\nu}= 0\, , 
\end{equation}
where the tensor $k$ is both symmetric and traceless. Evidently, this third-order equation is equivalent to the three first-order equations (\ref{another-three}) when $|\ell M|\ne1$. The $|\ell M|=1$ case yields the linearized equations of a ``critical''  MMG model with a ``logarithmic'' bulk mode; we shall not study this critical case here.

\subsection{Absence of tachyons}

The solutions of the first-order equation $\cD^m a=0$, for symmetric traceless second-order tensor $a$,  define an irrep of the AdS$_3$ isometry group, which  is unitary
provided that $|m \ell|\ge1$, with $m \ell =\pm1$ corresponding to the singleton irreps that have no bulk support (see chapter \ref{sec:pertunit} and \cite{Bergshoeff:2010iy} for a review).
It follows that of the three equations  (\ref{another-three})  only the one with $\cD^M$ propagates a bulk mode, which has spin-$2$ because $p$ is a 
symmetric traceless second-order tensor. The condition for unitarity of the irrep defined by this equation is 
\begin{equation} \label{notach}
|\ell M| >1 \,  \qquad \left(\Leftrightarrow \quad 1-2C >0\right)\, . 
\end{equation}
An immediate consequence is that $M^2>0$, so the condition (\ref{restrict}) required for reality of $\Lambda$ will be satisfied. The more stringent condition (\ref{notach}) is equivalent to positivity of the graviton mass-squared. This is because 
\begin{equation}
\cD^M k=0 \quad \Rightarrow \quad \cD^M \tilde \cD^M k=0\, , 
\end{equation}
where $\tilde \cD^M$ is defined in \eqref{cDdef} as $\cD^M$ with $M \to -M$. This is the Fierz-Pauli spin-2 field equation in AdS$_3$ for a spin-2 field $k$ with mass ${\cal M}$ given by
\begin{equation}
\ell^2{\cal M}^2 = \ell^2 M^2 -1 \, . 
\end{equation}
We may therefore interpret the condition (\ref{notach}) as a ``no-tachyon'' condition.

\subsection{Absence of ghosts}

We have still to determine the condition that the spin-2 bulk mode is not a ghost, but to do this we need to consider the quadratic action for the perturbations about the AdS$_3$ vacuum, not just the linearized field equations.  When the action (\ref{Letmg}) is expanded about the AdS$_3$ vacuum in terms of the one-form field fluctuations $(f_+,f_-,p)$ one finds, to quadratic order, an action that is the integral of the Lagrangian $3$-form 
\begin{eqnarray}\label{quad}
L^{(2)} &=& \frac{\lambda_+}{\mu} \left[f_+ \cdot \bar \cD f_+  + \ell^{-1} \bar e\cdot f_+\times f_+ \right] +
\frac{\lambda_-}{\mu} \left[f_- \cdot \bar \cD f_-  - \ell^{-1} \bar e\cdot f_-\times f_- \right] \nonumber \\
&& + \frac{1}{2\mu\left(1-2C\right)}\ \left[p\cdot \bar \cD p + M \, \bar e\cdot p\times p\right]\, ,
\end{eqnarray}
where
\begin{equation}
\lambda_\pm = 1\mp \left(\sigma + \alpha C\right)\mu\ell\, .
\end{equation}
We shall see later that these coefficients  are directly related to the boundary central charges; in fact
\begin{equation}
c_\mp \ \propto  \ \mp \frac{\lambda_\pm}{\mu\ell}  =  \sigma \mp \frac{1}{\mu\ell} + \alpha C\, ,
\end{equation}
which agrees with the TMG result for $\alpha=0$.

Notice that
\begin{equation}
-\lambda_+\lambda_- =  \ell^2\mu^2\left(1-2C\right) >0\, ,
\end{equation}
where the final inequality is a consequence of the no-tachyon condition (\ref{notach}). This inequality implies that  $\lambda_+$ and $\lambda_-$ must have opposite sign, which  means that we can rescale the  $f_\pm$ fields to bring their contribution to $L^{(2)}$ into the form of the difference of two linearized $SL(2,\bR)$ CS  three-forms.  Up to an overall sign,  this is the linearization of 3D GR with negative cosmological constant in its  CS-formulation, so the $f_\pm$ fields have no  local degrees of freedom, in agreement 
with the analysis of the previous subsection. 

Let us now focus on the term in $L^{(2)}$ that is quadratic in $p$; this is 
\begin{equation}\label{quadp}
L_p^{(2)} =  -AM \left(p\cdot \bar \cD p + M \, \bar e\cdot p\times p\right)\, , \qquad A= -\frac{1}{2M\mu\left(1-2C\right)}\, . 
\end{equation}
By the analysis of chapter \ref{sec:pertunit} we conclude that the no-ghost condition for this Lagrangian three-form is \eqref{noghost}: $A>0$ with $A$ given in (\ref{quadp}); i.e. 
\begin{equation}
M\mu(1-2C)<0\, . 
\end{equation}

\subsection{Combined no-tachyon/no-ghost conditions}\label{subsec:combined}

The  no-tachyon and no-ghost conditions combined are equivalent to the two conditions
\begin{equation}
1-2C>0 \quad \& \quad  M/\mu <0\, . 
\end{equation}
These are equivalent to the two conditions
\begin{equation}
{\cal M}^2 >0 \quad \& \quad \pm \sigma(1+ \sigma\alpha) <0\, , 
\end{equation}
where the upper sign must be chosen if the AdS$_3$ vacuum is the one allowing an $\alpha\to0$ limit. The latter equation leads to the following possibilities:

\begin{enumerate}

\item {\it Top sign}:  $\sigma=-1$ and $0<\alpha<1$.  

\item {\it Top sign}:  $\sigma=-1$ and $\alpha=0$. This is the TMG case, for which the no-tachyon condition is $|\ell\mu|>1$. The fact that 
$\sigma=-1$ is the origin of the negative BTZ black hole mass and negative boundary central charge for TMG. 

\item {\it Top sign}:  $\sigma=-1$ and $\alpha<0$.

\item {\it Top sign}:  $\sigma=1$ and $\alpha<-1$.

\item {\it Bottom sign}: $\sigma=-1$ and $\alpha>1$.

\item {\it Bottom sign}: $\sigma=1$ and $-1<\alpha<0$.

\item {\it Bottom sign}: $\sigma=1$ and $\alpha >0$.

\end{enumerate}
We shall see later that only three of these seven possibilities survive when we add the condition of positive boundary central charges.

\section{Hamiltonian analysis}

We will  now analyze the constraint structure to show that, quite generally, the Lagrangian three-form (\ref{Letmg}) defines a model describing a single bulk degree of freedom.
The analysis for the general CS--like model was treated in detail in chapter \ref{chapter:three_crowd} and \ref{chapter:asymptotic_symm}. In this analysis it is not necessary to consider any particular background but we will be interested in spacetimes that are asymptotic to an AdS vacuum, so we will pay attention to boundary terms. This will allow us to find the central charges in the sum of Virasoro algebras spanned by the conserved charges at the AdS boundary \cite{Brown:1986nw}.

\subsection{Local degrees of freedom of MMG}
Here we will determine the local degrees of freedom of the MMG model, following the general procedure of chapter \ref{chapter:three_crowd}.
Specializing to the Lagrangian three-form \eqref{Letmg}, we find, as a consequence of the assumed invertibility of the dreibein, that the integrability conditions \eqref{Intcon} give that
\begin{equation}
\mu(1 +\sigma\alpha)^2\Delta^{eh} = 0\,.
\end{equation}
As we also assume $(1 + \sigma\alpha) \neq 0$,  this equation gives the additional (secondary) constraint $0= \Delta^{eh} \equiv \psi$. Taking this into account in the matrix of Poisson brackets \eqref{gen_poissonbr},
we can omit the $\Delta^{eh}$ term from the right hand side of \eqref{Pmat_def}, and there is then no  $\Delta^{pq}$ term. In the basis ($\omega,e,h $), the remaining term gives the $9\times 9$ matrix
\begin{equation}\label{matPB}
\cP = \mu(1 + \sigma\alpha)^2\left( \begin{array}{ccc}
0 & 0 & 0 \\
0 & - V^{hh}_{ab} & V^{he}_{ab} \\
0 & V^{eh}_{ab} & V^{ee}_{ab}
\end{array} \right) \, .
\end{equation}
We will also need the  Poisson brackets of the primary constraint functionals $\phi[\xi]$ with the one secondary constraint function $\psi$; this is
\begin{align}\label{PBSecCon}
\{ \phi[\xi], \psi\}_{\rm PB} = \varepsilon^{ij} \big[ & \cD_{i} \xi^e\cdot  h_j - \cD_i \xi^h \cdot  e_j - \alpha\,  \xi^e\cdot h_i\times  h_j + \mu \sigma (1 + \alpha\sigma) \xi^e \cdot e_i \times h_j \nonumber \\
&  +  \left( \Lambda_0\, \xi^e+ \mu \sigma(1+\alpha\sigma) \xi^h \right) \cdot e_i \times e_j \big] \,,
\end{align}
where
\begin{equation}
\cD_i\xi^r = \partial_i \xi^r + \omega_i \times \xi^r\, .
\end{equation}
The $(9\times 9)$ matrix ${\cal P}$ has rank 2. When we combine this with the Poisson brackets of the secondary constraints, this increases the dimension of the matrix by one and the rank by two, since the brackets \eqref{PBSecCon} are independent of the column space defined by \eqref{matPB}. The final $(10\times10)$ matrix has rank 4, meaning that four constraints are second-class and the remaining six are first-class. The dimension of the physical phase space, per space point, is then $3\times 6 -2\times 6 -4 =2$, implying a single bulk degree of freedom. This is consistent with the linear analysis of the last section, but we now know that this  is  a background independent property of the fully non-linear theory.

\section{Boundary central charge}

To extract the boundary central charge from the Poisson bracket algebra  \eqref{gen_poissonbr} it is sufficient to consider the AdS$_3$ boundary terms for the two sets of mutually commuting first-class constraints  \cite{Carlip:2008qh}. We may identify the first-class constraints as in chapter \ref{sec:generators}, where \eqref{phiLL} generate local Lorentz transformations and \eqref{phidiff} generate diffeomorphisms. 
%
We now aim to find a new basis \eqref{Lpm} such that the Poisson-bracket algebra of first-class constraints becomes a direct sum of isomorphic 
algebras close to the AdS boundary.  This new basis can be constructed by considering the linear combinations
\begin{eqnarray}
L_{\pm} [\xi] &=& \phi_{\rm diff}'[\zeta] + a_{\pm} \phi_{\rm LL}[\zeta^\mu e_\mu] \,,
\end{eqnarray}
for constants $a_\pm$. 
By making use of the general result \eqref{gen_poissonbr} and the fact that $h=\mu C e$ in the AdS vacuum, and hence close to the boundary of any asymptotically-AdS 
spacetime,  we find that
\begin{align}\label{Lpmbrack}
\{L_+[\xi], L_-[\eta]\} = &  \left( 2 \alpha \mu C + a_+ + a_-\right) (\phi_e[[\xi,\eta]]+ \mu C\, \phi_h[[\xi,\eta]])  \nonumber \\
& + \left( 2\mu^2 (1+\alpha \sigma) C + a_+a_-\right) \phi_{\omega}[[\xi,\eta]] + \ldots\,, 
\end{align}
where the dots denote boundary terms which will vanish after choosing suitable boundary conditions. Here we have also used that by virtue of the symmetry of $h_{\mu\nu} = e_{\mu} \cdot h_{\nu}$ we have that:
\begin{equation}
e_{\mu}{}^a\xi^h_a = e_{\mu} \cdot h_{\nu} \zeta^{\nu} = h_{\mu} \cdot e_{\nu}\zeta^{\nu} = h_{\mu}{}^a \xi^e_a\,,
\end{equation}
and hence, on the AdS background, we may write
\begin{equation}
\xi^h_a = \mu C \xi^e_a\,.
\end{equation}
The remainder of the right hand side of \eqref{Lpmbrack} vanishes when
\begin{equation}
a_{\pm} = - \alpha \mu C \pm \frac{1}{\ell}\,.
\end{equation}
Using this parametrization for $a_{\pm}$ and the identities
\begin{equation}
2\alpha\mu C + 2 a_{\pm} = \pm \frac{2}{\ell}\,, \qquad a_{\pm}^2 + 2\mu^2(1+\sigma\alpha)C = \pm \frac{2a_{\pm}}{\ell}\,,
\end{equation}
we find the Poisson brackets for the improved generators $\tilde L_{\pm}[\xi]$ \eqref{Lpmimproved} 
\begin{equation}\begin{split}\label{PBLs}
\{\tilde L_{\pm}[\xi], \tilde L_{\pm}[\eta] \}  = &  \pm \frac{2}{\ell} L_{\pm}[[\xi,\eta]]  \\
& \pm \frac{2}{\ell}\left(\sigma  \pm \frac{1}{\mu \ell} + \alpha C\right) \int_{\partial \Sigma}\! d\phi\, \xi \cdot \left[ \partial_{\phi} \eta +  \left( \bar{\omega}_{\phi}{} \pm \frac{1}{\ell} \bar{e}_{\phi}{}\right)\times \eta \right]\,,
\end{split}
\end{equation}
where $\phi$ is the angular coordinate parametrizing  $\partial\Sigma$, which is the intersection of the boundary of AdS$_3$ with the equal-time slice $\Sigma$. After choosing suitable (Brown-Henneaux) boundary conditions \eqref{BHbc1}-\eqref{BHbc2} and restricting the gauge transformations to those which preserve these boundary conditions \eqref{asymptdiffs}, we can see that the boundary term in \eqref{PBLs} is responsible for a central extension in the asymptotic symmetry algebra of global charges. The expression obtained here is equivalent to the pure three dimensional gravity case \eqref{ECcentralext}, with a modified expression for the central charge. After including the proper normalizations, we find that the asymptotic symmetry algebra consists of two copies of the Virasoro algebra with a central charge
\begin{equation}\label{app:centralcharge}
c_{\pm}  =  \frac{3 \ell}{2 G_3}\left(\sigma  \pm \frac{1}{\mu \ell} + \alpha C\right)\, ,
\end{equation}
where $G_3$ is the 3D Newton constant. 
Note that in the TMG limit $\alpha \to 0$ the central charges reduces to the TMG expressions. Unitarity of the dual CFT requires
these central charges to be positive.

\subsection{Positivity of the central charges}

We saw in subsection \ref{subsec:combined} that there are seven distinct choices of (i) AdS$_3$ vacuum branch (ii) sign of $\sigma$ and (iii) range of $\alpha$ for which the propagating spin-2 mode is neither a ghost nor a tachyon, and for each of these we require ${\cal M}^2>0$.  We shall now investigate the compatibility of these constraints with the requirement of positive central charges $c_\pm$; we shall see that only three of the seven cases survive.  For clarity, we summarize here the conditions that we wish to fulfil simultaneously:
\begin{align}\label{inequalities}
\text{No-tachyon\ \& No-ghost:} &&& 1-2C > 0\,, \quad \& \quad 
 \pm \sigma (1+\alpha\sigma) < 0 \, . \nonumber\\
\text{Positive\ central\ charges}: &&& \sigma  - \frac{1}{|\mu \ell|} + \alpha C >0\,. 
\end{align}
Here the $\pm$ in the no-ghost condition depends on the sign in \eqref{lambdasol} and hence differentiates the two branches of AdS vacua, which we shall now consider in turn. 
The  parameters are $\Lambda_0/\mu^2$ and $\alpha$, both of which may, a priori, be any real numbers, and $\mu\ell$, which we may assume to be positive without loss of generality. We find that the conditions (\ref{inequalities}) are satisfied simultaneously in the following three cases:
\begin{itemize}

\item {\it Top sign}: $\sigma=-1$,  $\alpha<0$ and 
\begin{equation} \label{paramTMG1}
\Lambda_0/\mu^2 = \frac{1}{\alpha ( \mu\ell)^2} + \frac{2(1-\alpha)^3}{\alpha^3} \left( 1+ \sqrt{1+\frac{\alpha^2/(\mu\ell)^2}{(1-\alpha)^2} } \right)\, . 
\end{equation}
As expected, this is singular at $\alpha=0$.  This is case 3 of subsection \ref{subsec:combined}. In this case the dimensionless parameter $\gamma$ in the MMG field equation (\ref{modTMG}) 
is restricted to the range
\begin{equation}
0< \gamma \le  \frac{1}{4}\, . 
\end{equation}

\item {\it Top sign}: $\sigma=1$,  $\alpha<-1$ and 
\begin{equation} \label{paramTMG2}
 \Lambda_0/\mu^2 = \frac{1}{\alpha ( \mu\ell)^2} + \frac{2(1+\alpha)^3}{\alpha^3} \left( 1 - \sqrt{1+\frac{\alpha^2/(\mu\ell)^2}{(1+\alpha)^2} } \right) \, .
\end{equation}
This is case 4 of subsection \ref{subsec:combined}. In this case the dimensionless parameter $\gamma$ is restricted to the range
\begin{equation}
\gamma>0\, . 
\end{equation}

\item  {\it Bottom sign}: $\sigma = 1$, $-1< \alpha < 0$ and 
\begin{equation}
\Lambda_0/\mu^2 = \frac{1}{\alpha ( \mu\ell)^2} + \frac{2(1+\alpha)^3}{\alpha^3} \left( 1+ \sqrt{1+\frac{\alpha^2/(\mu\ell)^2}{(1+\alpha)^2} } \right) \,. 
\end{equation}
This is case 6 of subsection \ref{subsec:combined}.  In this case the dimensionless parameter $\gamma$ is again restricted to the range
\begin{equation}
\gamma>0\, . 
\end{equation}

\end{itemize}
Notice that $\alpha<0$, necessarily, and there are no ``bottom-sign'' cases with  $\sigma = -1$. Notice too that $\gamma>0$ in all cases. 

We remark that inversion of the formula (\ref{lambdasol}) giving $\Lambda$ as a function of $\Lambda_0$ yields
\begin{equation}\label{alLamzero}
\alpha\Lambda_0 = -\Lambda + \frac{2\mu^2\left(1+\sigma\alpha\right)^3}{\alpha^2} \left[1 \mp \sqrt{1- \frac{\alpha^2\Lambda/\mu^2}{\left(1+\sigma\alpha\right)^2}}\right]\, , 
\end{equation}
but there is no simple correlation of the sign with the vacuum branch sign of (\ref{lambdasol}). Given a choice of this branch, TMG (top sign)  or non-TMG (bottom sign), 
there is a definite value of $\Lambda$ for given $\Lambda_0$, and the sign in (\ref{alLamzero}) must then be chosen such that upon substitution for $\Lambda$ one recovers the given $\Lambda_0$. The signs in the above expressions for $\Lambda_0/\mu^2$ are such that this is the case, as may be checked by considering the first terms 
in an expansion in powers of $1/(\mu\ell)^2$ for $\mu\ell \gg 1$.

\section{Discussion}

We have presented a new, multi-parameter,  massive 3D gravity theory  that we have called ``Minimal Massive Gravity'' (MMG).  It is ``minimal'' in essentially two
different ways. 

One is that it shares with the well-known ``Topologically Massive Gravity'' (TMG) the property that it describes, when linearized about a flat or AdS$_3$ vacuum, 
a single massive graviton mode. Like TMG, this mode is physical for some parameter range,  in the sense that it is neither a tachyon nor a ghost, and  there
are no other local degrees of freedom. TMG has a problem, however, when considered as a possible semi-classical limit of some quantum theory defined 
holographically via a dual CFT on an AdS$_3$ boundary. The asymptotic symmetry algebra  is a direct sum of two  Virasoro algebras \cite{Brown:1986nw} which must have positive central charges for unitarity of the CFT, and the parameters of TMG do not allow this condition to be satisfied while maintaining physical properties of the bulk mode.  In contrast, the one additional parameter of MMG allows a resolution of this ``bulk-boundary'' clash; in fact we found three disjoint 
regions of parameter space  for which this is possible.  

It might appear from this summary that our resolution of the ``bulk-boundary'' clash of TMG  has been achieved in a  rather obvious way (by the inclusion of extra terms in the action, leading to extra parameters and hence more freedom) and that MMG is really just a variant of TMG with more parameters. However, this is very far from being the case. 
As observed in the introduction, the MMG equation for the metric alone (after elimination of other, auxiliary, fields) cannot be found by variation of an action for the metric alone, 
so it does not correspond to any conventional extension of the TMG equation. In fact, MMG is {\it qualitatively} different from TMG, in various ways.  One of them leads to the conclusion that MMG is indeed ``minimal''  in another sense. 

Let us write the MMG equation (\ref{modTMG}) in the form $E_{\mu\nu}=0$, where
\begin{equation}
E_{\mu\nu} = \bar\Lambda_0 g_{\mu\nu}  + \bar\sigma  G_{\mu\nu}  + \frac{1}{\mu} C_{\mu\nu}  + \frac{\gamma}{\mu^2} J_{\mu\nu} \, . 
\end{equation}
Using the identity (\ref{Jid}), one finds that 
\begin{equation}
\sqrt{-\det g}\, \nabla_\mu E^{\mu\nu}  = \frac{\gamma}{\mu} \varepsilon^{\nu\rho\sigma} S_\rho{}^\tau E_{\tau\sigma}\, .
\end{equation}
Consistency of the MMG field equation requires the left hand side to be zero as a consequence of the MMG field equation, and this consistency condition is satisfied. 
However, let us now attempt to couple MMG to ``matter'' fields, which we suppose to have a (symmetric) stress tensor $T$ with the usual property that 
\begin{equation}\label{conserved}
\nabla_\mu T^{\mu\nu}=0\, ,
\end{equation}
as a consequence of the matter field equations.  Let us now consider (in some convenient units for  the 3D Newton constant) the equation 
\begin{equation}\label{E=T}
E_{\mu\nu} = T_{\mu\nu}\, . 
\end{equation}
In the case that $\gamma=0$, this is just the TMG field equation in the presence of matter with stress tensor $T$. Consistency of this equation requires that
\begin{equation}
\nabla_\mu\left[E^{\mu\nu}-T^{\mu\nu}\right] =0\, , 
\end{equation}
and this is an identity for TMG, given (\ref{conserved}). However, for MMG ($\gamma\ne0$) we find, using (\ref{E=T}), that 
\begin{equation}
\sqrt{-\det g}\, \nabla_\mu\left[E^{\mu\nu}-T^{\mu\nu}\right] = \frac{\gamma}{\mu} \varepsilon^{\nu\rho\sigma}S_\rho{}^\tau T_{\tau\sigma}\, . 
\end{equation}
This is not zero unless the spacetime is an Einstein space, which is not required by the field equations\footnote{Alternatively, we could require the stress tensor $T$ to be a linear combination of the metric and Einstein tensors, but this would just change the coefficients in the source-free equation.}.  We conclude that the standard coupling of gravity to matter
is not possible for MMG. It appears that MMG is also ``minimal'' in the sense that the graviton can couple consistently only to itself.

The difficulty here is that we are effectively assuming the standard minimal coupling of the metric to matter, such that variation of the matter action with respect to the metric yields, 
in a vacuum spacetime, the usual symmetric (Belinfante) stress tensor,  but there is no action for the metric alone to which this matter action could be added. It may be that some 
consistent matter couplings can be found by an extension of the CS--like action for MMG itself; we suspect that a coupling to spin-3/2 fields can be achieved in this way, leading to a locally supersymmetric extension of  MMG that is consistent in a similarly novel way. However, all that is clear at present  is  that the standard minimal coupling of gravity to matter is not possible for MMG; this will presumably have significant implications for the boundary CFT.\footnote{It has recently been shown that the MMG equation can be extended
to include matter via a particular source tensor that is quadratic in the matter stress tensor \cite{Arvanitakis:2014yja}}

We conclude with one further comment on MMG, which was actually our point of departure. As shown in chapter \ref{sec:GZDG} \cite{Bergshoeff:2014bia}, the Zwei-Dreibein Gravity model of \cite{Bergshoeff:2013xma}, which resolves the ``bulk-boundary clash'' for NMG,  has a parity-violating extension to a model that propagates two spin-2 modes of opposite 3D-helicity with different masses. By sending the mass of one of the two modes to infinity,  one arrives at an alternative to TMG. This limit is rather subtle, but it leads to the MMG model described in this chapter. This embedding of MMG into a model with additional degrees of freedom may be a useful source of further insight into its novel features.

%% file: chapter_7/chapter_7.tex
\pagestyle{empty}
\setcounter{chapter}{6}

\chapter[Extensions of Massive 3D Gravity]{Extensions of Massive 3D Gravity}
\label{chapter:ENMG_VDG}

\pagestyle{headings}

\begin{quote}\em
We have now established that the Chern-Simons--like description for 3D gravity models is well suited to describe theories of massive spin-2 modes. All of the known scalar ghost-free higher-derivative extensions of 3D general relativity fall into the class of CS--like theories with $N \leq 4$ fields. We have also discussed other, novel, theories of gravity within this class, which show improved behavior in the context of the AdS/CFT correspondence. In this chapter we build on these ideas to extend the discussion to CS--like theories with $N > 4$. First we will construct a set of scalar ghost-free theories of higher-derivative gravity which propagate multiple massive spin-2 modes. However, the higher-derivative nature of these theories introduces pathologies, as not all of the massive spin-2 modes are physical. We then turn to resolve this problem, much like ZDG resolves the bulk-boundary clash in NMG, by considering a more general class of CS--like theories with $N>4$. These theories have a structure similar to ZDG, only with more interacting dreibeine, hence we name them Viel-Dreibein Gravities. This chapter is based on two papers which are in preparation at the time of writing this thesis \textsc{[x,xi]}.
\end{quote}

\newpage

\section{Introduction}
So far in this thesis, we have concentrated on Chern-Simons--like theories with $N\leq 4$ Lorentz vector valued one-forms. This was sufficient to describe a wide class of massive gravity models whose different characteristics were discussed at length. One prevalent assumption was that the number of gauge symmetries remained the same as in the Einstein-Cartan theory. The CS--like models are manifestly diffeomorphism invariant and we have always assumed the presence of a (dual) spin connection $\omega^a$, resulting in local Lorentz symmetric field equations. 

Besides a dreibein $e^a$ and a spin connection $\omega^a$, the models we have discussed contain additional fields which are auxiliary. What we mean precisely by auxiliary is that upon using the field equations, they can be expressed in terms of other fields in the theory, usually the dreibein and its derivatives. However, the fact that these fields are auxiliary does not imply that they are redundant and can be eliminated from the action. This is only true in specific cases, and this feature can be seen as a characteristic of the specific model which divides the general CS--like theories into two categories: those with an action in terms of an invertible dreibein (and hence a metric) alone, and those who do not have such an action.

All of the higher-derivative theories of gravity studied in chapter \ref{section:auxfields} belong to the first category; there is an equivalent formulation of the theory in terms of a higher-derivative extension of general relativity. This is because the field equations which determine the auxiliary fields are obtained by varying the action with respect to those auxiliary fields themselves. Furthermore, if we use those equations in the action, the auxiliary fields are completely eliminated from the action, at the expense of introducing higher-derivatives of the dreibein. At the end of the day, we can vary the action with respect to the dreibein and obtain a higher-derivative equation of motion for the dreibein which is equivalent to the original field equation for the dreibein, after plugging in the solutions for the auxiliary fields. 

To the second category belong the ``novel'' theories of massive 3D gravity: Zwei-Dreibein gravity (ZDG), introduced in chapter \ref{section:ZDG}, and the Minimal Massive Gravity (MMG) model of chapter \ref{chapter:MMG}. For both of these models, it is possible to solve the additional fields ($e_2$ and $\omega_2$ in the case of ZDG, $h$ in the case of MMG) in terms of the dreibein and its derivatives, but to do so, one needs to use the field equation obtained by varying the dreibein. Back-substitution of this equation into the action is illegal, since this will change the equation of motion for the dreibein, which was used to solve for the auxiliary fields in the first place! The resulting action would have different field equations and will not describe the same theory. However, it is possible to substitute the solution of the auxiliary fields into the other field equations to obtain a single, higher-derivative equation for the metric alone. The novel feature of these theories is that they do not follow from an action for the metric alone and hence posses other, new characteristics. Most notably, they resolve the bulk-boundary clash present in the higher-derivative extensions of GR. On the other hand, as discussed in the last chapter, these theories cannot be coupled to matter in the conventional, minimal way. 

In this chapter, we will consider extending both categories of CS--like models to theories with more than $N =4$ Lorentz vector-valued one-form fields. In the first category this will result in higher-derivative extensions of GR with more than four derivatives on the metric. The extensions are constructed in a manner similar to New Massive Gravity, and give exactly the higher-order derivative combinations which do not lead to additional scalar degrees of freedom in the theory. The linear spectrum now contains multiple massive spin-2 modes, the number of modes depend on the number of derivatives in the final theory. We will show that the construction of these theories is compatible with a holographic $c$-theorem when minimally coupled to matter. However, the higher-derivative nature of the theory does lead to pathologies similar to those in NMG on AdS$_3$ backgrounds; the bulk-boundary clash persists and is in a sense even worse, since not all massive spin-2 modes are physical (i.e. tachyon- and ghost-free), irregardless of the value of the boundary central charge. 

This novel clash, which is in fact solely a bulk clash, is once again resolved by considering theories with $N >4$ fields belonging to the second category. We consider a class of theories with $N=6$ modeled after ZDG, but now with three pairs of dreibeine and spin connections, called Drei-Dreibein Gravity. After constraining the interaction terms of the theory to ensure the presence of secondary constraints, we see that these models describe two massive spin-2 modes, just like the sixth-order derivative extension of GR. Unlike these higher-derivative extensions, they do allow for regions in their parameter space where all bulk modes are physical and the boundary central charges are positive. However, a minimal coupling to matter for these theories is problematic, as there is no action principle for one metric alone. We conclude the chapter with some general remarks on going beyond three dreibeine and define the scalar-ghost free interaction terms which make out a Viel-Dreibein  theory of gravity.

\section{Extended Massive Gravity Models}\label{Extended}
In this section, we propose a procedure to derive higher-order derivative extensions of 3D GR which propagate multiple massive spin-2 particles. The extensions are obtained from an auxiliary field formalism which guarantees the freedom of scalar ghosts. However, the higher-derivative nature of the theory does lead to the presence of massive spin-2 ghosts. We discuss the pathologies in section \ref{linearth}. We work out the extensions up to eight order in derivatives explicitly.

\subsection{General Procedure}
There are two parity preserving gravity models in three dimensions which are both purely topological and do not propagate any local degrees of freedom. They are Einstein-Cartan (EC) gravity, of even parity (see chapter \ref{chapter:GRin3D}), and conformal gravity (CG) (see chapter \ref{sec:CSG}), which is odd under parity:
\begin{align}
&\text{Parity Even},\qquad S_{\text{\tiny{EC}}}\equiv S_0 = - \frac{1}{\kappa^2} \int  e\cdot \left( R - \tfrac{\Lambda_0}{6}  e \times e\right)\,, \label{EHG}\\
&\text{Parity Odd},\qquad S_{\text{\tiny{CG}}}\equiv S_1=\frac{1}{2\kappa^{2}\mu} \int \left\{ \omega \cdot \left(d \omega + \tfrac{1}{3}\omega \times \omega\right)+ 2 f \cdot  \cD e \right\}\,, \label{CSG}
\end{align}
where $\kappa^{-2} = 8 \pi G$ is the three dimensional Planck mass and $\Lambda_0$  the cosmological constant with the dimension of (mass)$^2$ while $\mu$ has the dimension of mass. This amounts to a dimensionless coupling for CG which is a conformally invariant theory. 

Due to lack of any local degrees of freedom, these models can be written purely as Chern-Simons gauge theories for SO(2,2) and SO(2,3) respectively \cite{Achucarro:1987vz,Horne:1988jf,Witten:1988hc}, where $e$, $\omega$ and $f$ correspond to the gauge fields for translation, rotation and special conformal transformation in three dimensions, see \cite{Afshar:2013bla} for a more recent treatment.

The dualised curvature two-form is $R=d\omega+\frac{1}{2}  \omega \times \omega$ and the field equations will ensure vanishing of the torsion two-form, 
\begin{equation}\label{torsion-const}
T=\mathcal{D}e =de+ \omega \times e  =0\,.
\end{equation}
Assuming the invertibility of the dreibein it is possible to solve this equation for the spin connection in terms of the dreibein, $\omega^a = \omega^a (e)$.
Varying the action \eqref{EHG} w.r.t. $e^a$ gives the equation $R = \frac{1}{2}\Lambda_0 e \times e$, which can be written in the metric form
as, 
$ G_{\mu\nu} + \Lambda_0\,g_{\mu\nu} = 0$, where $G_{\mu\nu}=R_{\mu\nu}-\tfrac{1}{2}Rg_{\mu\nu}$ is the Einstein tensor.

Varying the action \eqref{CSG} w.r.t. $f^a$, $\omega^a$ and $e^a$ gives the field equations,
\begin{align}
& \cD e = 0 \,, \nonumber \\
& R + f \times e =0\,,\label{Schouten}\\
& \cD f=0 \nonumber \,.
\end{align}
Assuming the invertibility of dreibein, the auxiliary field $f^a$ can be solved in terms of the curvature two-form as
\begin{equation}\label{Schoutenten}
f_{\mu\nu} \equiv f_{\mu}{}^a e_{\nu\,a} = -\left(R_{\mu\nu}-\tfrac{1}{4}R\,g_{\mu\nu}\right)\equiv - S_{\mu\nu}(e)\,,
\end{equation}
and the last equation in \eqref{Schouten} then gives a third-order differential equation for dreibein; $C_{\mu\nu}(e)\equiv e^{-1} \ve_{(\mu|}{}^{\alpha\beta} \nabla_{\alpha} S_{\beta |\nu)}=0$. 
Here $S_{\mu\nu}$ and $C_{\mu\nu}$ are the symmetric Schouten and Cotton tensors respectively, constructed from dreibein $e^a$. 

This simple example shows how we can construct a higher-derivative action -- CG in this case with three derivatives -- in a first-order CS-like form by introducing an auxiliary field $f^a$ \cite{Afshar:2011yh,Afshar:2011qw}.
As we discussed in chapter \ref{sec:NMG}, this approach can be extended to fourth order by introducing two extra one-form fields ($f^a$, $h^a$) to obtain New Massive Gravity (NMG), denoted by $S_2$
\begin{equation}\label{NMG}
S_2 = S_0 - \frac{1}{\kappa^2m^2} \int \left\{ f \cdot \left(R + e\times f \right)+h\cdot \cD e   \right\}\,.
\end{equation}
The field $h$ guarantees the torsion constraint \eqref{torsion-const}, and the following field equations arise
\begin{equation}
\begin{split} \label{NMGeom}
& \cD e = 0 \,, \\
& R +  e \times f = 0 \,, \\
& \cD f +  e \times h = 0\,, \\
& \cD h +\tfrac{1}{2} \left(f \times f  - 2  m^2 e \times f - m^2\Lambda_0 e \times e \right) = 0\,.
\end{split}
\end{equation}
The first equation in \eqref{NMGeom} is solved as in \eqref{Schoutenten} and the second equation gives
\begin{equation}\label{scho-cot}
h{}_{\mu\nu} \equiv h{}_{\mu}{}^a e_{\nu\,a} = e^{-1} \ve_{(\mu|}{}^{\alpha\beta} \nabla_{\alpha} S_{\beta \,|\nu)}= C_{\mu\nu}(e)\,.
\end{equation}
Looking at equations \eqref{Schouten} and \eqref{NMGeom} suggests that we can continue this logic to obtain arbitrarily higher-derivative extensions of these theories. 
Consider extending the equations of motion to have the schematic form
\begin{equation}
\begin{split} \label{ENMGeom}
1 \qquad \qquad & \cD e = 0\,, \\
2 \qquad \qquad & R + e \times f_{1} = 0 \,, \\
3 \qquad \qquad & \cD f_1+ e \times h_{1} = 0\,, \\
4 \qquad \qquad & \cD h_1 +  e \times f_{2} + \ldots = 0\,, \\
& \qquad \quad \vdots \\
2N+1 \qquad & \cD f_M +  e \times h_{M}+ \ldots = 0\,,\\
2N+2 \qquad & \cD h_M + \ldots = 0\,.
\end{split}
\end{equation}
The structure of these equations is such that they may be solved one after the other in terms of derivatives acting on the dreibein. The number appearing before the equation denotes the maximum number of derivatives of the dreibein which may appear in the equation after all fields have been solved. The dots denote terms which may be lower or equal in derivatives on $e^a$.  

The first equation in \eqref{ENMGeom} solves the spin connection in terms of the dreibein. The next two equations are already solved as in \eqref{Schoutenten} and \eqref{scho-cot}. The other auxiliary fields ($f^a_I$, $h^a_I$) can be obtained in terms of  $e^a$ and derivatives on it, such that the final equation is a higher derivative field equation for the dreibein. This set of equations may terminate on an equation for $\cD h_M{}^a$ or $\cD f_{M+1}{}^a$ and the final equation then becomes, respectively, an even or an odd order partial differential equation for the dreibein, corresponding to a parity even or odd theory. 

We can diagramatise the even and odd cases as follows
\begin{align}\label{Even_diagram}
\text{Even:}\quad&\mathlarger{\underbrace{\overbrace{\overset{e}{\bullet}\mathlarger\longrightarrow\overset{\omega}{\circ}}^\text{EC}\mathlarger\longrightarrow\overbrace{\overset{f_1}{\circ}\mathlarger\longrightarrow\overset{h_1}{\circ}}^\text{2 dof}}_\text{NMG}\mathlarger\longrightarrow\cdots\mathlarger\longrightarrow\overset{f_{M}}{\circ}\mathlarger\longrightarrow\overset{h_{M}}{\circ}}  
\end{align}
\begin{align}
\text{Odd:}\quad&\mathlarger{\overbrace{\overset{e}{\bullet}\mathlarger\longrightarrow\overset{\omega}{\circ}\mathlarger\longrightarrow\overset{f_1}{\circ}}^\text{CG}\mathlarger\longrightarrow\overbrace{\overset{h_1}{\circ}\mathlarger\longrightarrow\overset{f_2}{\circ}}^\text{2 dof}\cdots\mathlarger\longrightarrow\overset{h_{M}}{\circ}\mathlarger\longrightarrow\overset{f_{M+1}}{\circ}} \label{Odd_diagram} 
\end{align}
The sequential form of the diagram shows which field are solved in terms of which. The filled circle denotes the assumption of invertibility of the dreibein, while all other fields (open circles) need not to be invertible. Every set of auxiliary fields $(f, h)$ adds two local degrees of freedom to the theory. The parity even theory contains $2M$ degrees of freedom, where $M>0$ denotes the number of $(f,h)$  pairs. 

In the parity odd sector, the starting point $S_1$ already contains one auxiliary field, which is responsible for the partially massless mode in CG (see chapter \ref{sec:CSG}). The presence of additional auxiliary fields will break the conformal symmetry of CG and hence the partially massless mode will no longer be pure gauge, but propagate a local degree of freedom. So adding a pair of auxiliary fields $(h, f)$ to CG will initially increase the degrees of freedom by three and any subsequent pair will raise this number by two more. Hence, CG plus $M$ pairs of auxiliary fields will have $2M+1$ local degrees of freedom.

\subsection{Action Principle}

In both even and odd cases the set of equations \eqref{ENMGeom} can be integrated to an action by the same general procedure. The field with the highest number of derivatives on the dreibein ($h_M{}^a$ for even parity, $f_{M+1}{}^a$ for odd parity) can be used as a multiplier for the torsion constraint. The field with one derivative less will be used to multiply the second equation, and so on, until half of the field equations have been used. The rest of the field equations then follow from the action by varying the fields with a lower number of derivatives on the dreibein. This procedure guarantees that the highest number of derivatives appearing in the action after solving for all the auxiliary fields is $2M+2$ for the parity even models and $2M+3$ for the parity odd models.

\subsubsection*{Parity Even Models}\label{sec:2even}

The parity preserving extensions of EC gravity \eqref{EHG} in this CS--like form can be obtained from the following recursive action,

\begin{align}\label{SNeven}
S_{2M}=S_{2M-2}+\frac{\kappa^{-2}}{(m^2)^{M}}\int\bigg[\sum_{I+J=M}& f_I\cdot\mathcal Dh_{J} + \sum_{I+J+K=M+1}\alpha_{IJK}  f_I \cdot f_J \times f_K \nonumber\\
+&\sum_{\underset{J,K\neq0}{I+J+K=M}}\beta_{IJK} f_I \cdot h_J \times h_{K}\bigg]\,,
\end{align}
where $I,J,K=0,1,2,\cdots, M$ with $f_0\equiv e$, $h_0\equiv\omega$ and $R=\mathcal{D}\omega$ is the curvature two-form.  
The starting value in 
this recursive relation is given by \eqref{EHG}. As an example $S_2$ is already constructed in \eqref{NMG}.     
\subsubsection*{Parity Odd Models}\label{sec:2odd} 

The parity preserving extension of CG \eqref{CSG} in this CS-like form can be obtained from the following recursive action,

\begin{align}\label{SNodd}
S_{2M+1}=S_{2M-1}+&\frac{\kappa^{-2}}{\mu(\mu^2)^{M}}\int\bigg[\sum_{I+J=M} h_I\cdot \cD h_{J} + \sum_{\underset{I,J,K\neq0}{I+J+K=M}} \alpha_{IJK}  h_I\cdot  h_J \times h_K  \nonumber\\
+&\sum_{I+J=M+1}  f_I \cdot \cD f_{J} + \sum_{\underset{K\neq0}{I+J+K=M+1}}\beta_{IJK} f_I \cdot f_J\times h_{K}\bigg]\,,
\end{align}
The indices of $f_I$ in this odd sector run from zero to $M+1$ and of $h_I$ from zero to $M$ with $f_0\equiv e$ and $h_0\equiv\omega$.  The starting value in 
this recursive relation is given by \eqref{CSG}. The next example, denoted by $S_3$, is
\begin{equation}\label{S3}
S_3 = S_1 + \frac{1}{\kappa^2\mu^3} \int \left\{ e \cdot \cD f_2+h_1\cdot \left(R + e \times f_1 \right)+ \tfrac{\alpha}{2}  f_1 \cdot \cD f_1   \right\}\,.
\end{equation}
onumberot all couplings $\alpha_{IJK}$ and $\beta_{IJK}$ in \eqref{SNeven} and \eqref{SNodd} are physical. For a given $M$ we have $2M$ auxiliary fields in the even sector and $2M+1$ in the odd case
which can be rescaled such that the same number of coefficients may be set to unity. 
In \eqref{SNeven}  and \eqref{SNodd} we have already exhausted $M+1$ and $M+2$  of the rescalings respectively, by canonically normalizing the $M+1$ and $M+2$ kinetic terms.

Similarly, we also retain the freedom to redefine the auxiliary fields $f_{M}{}^a = f'_{M}{}^a + a m^2 f_{M-1}{}^a + \ldots$ for some arbitrary constant $a$ (and likewise for $h_{M}{}^a$). Such a field redefinition can always be used to simplify or cancel terms appearing in $S_{2M}$ or $S_{2M+1}$. In the concrete examples coming later, we will use such shifts to cancel the kinetic terms of the lower-order action. 

In this work we will only analyze extensions which preserve parity, however, it is straightforward to extend the analysis to parity-violating models by taking the sum of an even and odd parity theory. In fact, Topologically Massive Gravity can be defined as the sum of $S_0$ and $S_1$, while General Massive Gravity is (up to a field redefinition) the sum of $S_0$, $S_1$ and $S_2$. We will explicitly construct the even and odd parity models up to the eight-derivative extension of general relativity in metric formalism in the next section. In section \ref{linearth} we will linearize the sixth-order theory around AdS$_3$ and confirm that adding a set of auxiliary fields will add 2 massive spin-2 degrees of freedom. However, before doing so we will comment on the absence of scalar ghosts and the growth of local degrees of freedom by adding a $(f,h)$ pair of auxiliary fields, independently of any linearized approximation.

\subsection{Absence of Scalar Ghosts}\label{counting}
The advantage of the first-order formulation over the metric form is that it is relatively easy to count the number of local degrees of freedom (dof) and see the second class constraints which remove the Boulware-Deser scalar ghost. They arise from the symmetry of the auxiliary fields $h_{I\,\mu\nu} \equiv h_{I\,\mu} \cdot e_{\nu}$ and $f_{I\,\mu\nu} \equiv f_{I\,\mu}\cdot e_{\nu}$,
\begin{equation}\label{fhantisym}
f_{I\,[\mu\nu]} = 0 \,, \qquad\qquad h_{I\,[\mu\nu]} = 0\,.
\end{equation}
These constraints can be derived directly from the equations of motion \eqref{ENMGeom} by acting on them with an exterior derivative and using $d^2 =0$. This is in essence equivalent to deriving the integrability conditions \eqref{Intcon}. By invertibility of the dreibein, the first equation in \eqref{ENMGeom} simplifies to $f_{1\,a} e^a = 0$ and the second gives $h_{1\,a} e^a = 0$, whose spatial projections are secondary constraints in a Hamiltonian formulation of the theory \cite{Hohm:2012vh,Bergshoeff:2014bia}. A similar procedure ensures the symmetry of $h_{I \, \mu\nu}$ and $f_{I \mu\nu}$ for $I >1$.
The counting of degrees of freedom was shown in \cite{Hohm:2012vh} for NMG but it can be generalized to all CS--like theories constructed as outlined in the preceding section, 
provided that the secondary constraints \eqref{fhantisym} are second-class and do not lead to further tertiary constraints\footnote{
In this counting we assume that adding these new auxiliary fields does not change gauge symmetries; this actually happens for the auxiliary field in CG through the presence of additional symmetries which cancel the degrees of freedom introduced by $f^a$, see \cite{Afshar:2011qw}.} (see also chapter \ref{chapter:three_crowd}). 

After a space-time decomposition of the fields, the time components $f^a_0$ and $h^a_0$, become Lagrange multipliers for a set of six primary constraints 
and the spatial components of the fields, $f^a_i$ and $h^a_i$, add to the canonical variables of the theory. Along with the additional secondary constraints $f^a_{[ij]}=0$ and $h^a_{[ij]}=0$,
each pair of auxiliary fields will add
\begin{equation}\label{ENMGcounting}
\frac{1}{2}\left( 12 - 6 - 2\right)  = 2\,,
\end{equation}
degrees of freedom to the theory. These two degrees of freedom correspond to the two helicity states of a massive spin-2 mode in three dimensions. This counting works for all vector valued one-form pairs ($f_I$, $h_J$). 
Hence, any action which gives the equations of motion with the general structure of \eqref{ENMGeom} is guaranteed to produce a scalar ghost-free, higher-derivative extension of gravity in three dimensions.

\section{Extended New Massive Gravity}\label{ExtendedNMG}
In this section we construct the extensions in the parity even sector up to eight derivatives. A similar analysis for the parity odd sector is done in appendix \ref{ExtendedCSG}.

The fourth-order action $S_2$ is NMG \eqref{NMG}. The next step for the parity even models is $S_4$, which is sixth-order in derivatives and its Lagrangian three-form can be derived using \eqref{SNeven}
\begin{equation}\label{LENMG}
\begin{split}
 L_4 & = - e \cdot \left(\sigma  R - \frac{\Lambda_0}{6} e \times e \right) + \frac{1}{2m^2}  e \cdot f_1{} \times f_1{}  - \frac{1}{m^4}\Big[e \cdot \cD h_2    \\
 & + \frac{a}{6}  f_1 \cdot f_1{} \times f_1 + f_{2} \cdot \left( R + e \times f_{1} \right) + b\,  h_{1} \cdot \left(\cD f_1 + \tfrac{1}{2}e \times h_{1} \right)\Big]  \,.
\end{split}
\end{equation}
Here we have introduced a sign parameter $\sigma = \pm 1$ and two arbitrary dimensionless parameters $a,b $. The dimensionful parameters $\Lambda_0$  and $m^2$ were introduced before in \eqref{EHG} and \eqref{NMG}. Up to field redefinitions and rescaling of the auxiliary fields, this is the most general parity-even Lagrangian which produces field equations of the form \eqref{ENMGeom} with 2 pairs of auxiliary fields.
Explicitly, the equations of motions for this Lagrangian, obtained by varying with respect to $h_2$, $f_2$, $h_1$, $f_1$, $\omega$ and $e$ respectively, are,
\begin{equation}\label{ENMGeom2}
\begin{split}
& \cD e = 0\,,\\
& R + e \times f_{1} = 0\,,\\
& \cD f_1 +  e \times h_{1} = 0\,, \\
& b\, \cD h_1 + \tfrac{1}{2} \left( a \,f_{1} \times f_{1} + 2 e \times f_{2} -  2m^2 e \times f_{1}  \right) = 0\,,  \\
& \cD f_2 +  \left( b\,f_{1} \times h_{1} + e \times h_{2} \right) =0\,,  \\
& \cD h_2 + \tfrac{1}{2}  \left( 2f_{2} \times f_{1} + b\, h_{1} \times h_{1} - m^2 f_{1} \times f_{1} - \Lambda_0 m^4 e \times e - 2m^4 \sigma \, e \times f_{1} \right) = 0 \,.
\end{split}
\end{equation}
The first equation imposes zero torsion and allows the spin-connection to be solved in terms of the dreibein. Moving down the line, we find for the fields $f_{1\, \mu\nu}$ and $h_{1\,\mu\nu}$
\begin{equation}
f_{1\,\mu\nu}= -S_{\mu\nu}(e) \,, \qquad \text{and} \qquad 
h_{1\, \mu\nu} =  C_{\mu\nu}(e)\,. \label{f1h1}
\end{equation}
which in turn fixes $f_{2\,\mu\nu}$ and $h_{2\,\mu\nu}$,
\begin{align}
f_{2\,\mu\nu} &=  - b \,D_{\mu\nu} + a \left( P_{\mu\nu}-\tfrac{1}{4}P  g_{\mu\nu}  \right)  - m^2 S_{\mu\nu}\,,\\
h_{2\, \mu\nu} &=  - E_{\mu\nu} - 2 b\left(   Q_{\mu\nu} - \tfrac{1}{4}  Q g_{\mu\nu}\right) + b\,SC_{\mu\nu}  \,. 
\end{align}
where
\begin{align}\label{DP}
D_{\mu\nu}&\equiv e^{-1} \ve_{(\mu|}{}^{\alpha\beta} \nabla_{\alpha} C_{\beta\, |\nu)}\,,\;\qquad
P_{\mu\nu}\equiv G_{\mu}{}^{\rho} S_{\nu \rho}\,,\\
E_{\mu\nu}&\equiv e^{-1} \ve_{(\mu|}{}^{\alpha\beta} \nabla_{\alpha} f_{2\,\beta\, |\nu)}
\,,\qquad Q_{\mu\nu}\equiv C_{(\mu}{}^{\rho} S_{\nu) \rho}\,.
\end{align}
Back substitution in the action leads to the following, Extended New Massive Gravity Lagrangian (ENMG) density
\begin{equation} \label{NMG2c7}
 \mathcal{L}_{\text{\tiny{ENMG}}}\equiv\mathcal{L}_4= \frac{1}{2}\left\{\sigma R-2\Lambda_0 - \frac{1}{m^2} P + \frac{1}{m^4}\left( 2 a \det(S) - b \,C^{\mu\nu}C_{\mu\nu}\right)\right\}\,.
\end{equation}
At order $1/m^2$ we have the NMG combination of $R^2$ terms,
\begin{equation}
P =R_{\mu\nu}R^{\mu\nu}-\frac{3}{8}R^2,
\end{equation}
and at order $1/m^4$ we find two combinations of sixth-order terms;
\begin{equation}
\begin{split}\label{Cubic}
- 6 \det(S)& = 2R_{\mu}^{\nu}R_{\nu}^{\rho}R_{\rho}^{\mu}-\frac{9}{4}RR_{\mu\nu}R^{\mu\nu}+\frac{17}{32}R^3\,, \\
C^{\mu\nu}C_{\mu\nu}& = R_{\mu\nu}\square R^{\mu\nu} - \frac{3}{8}R\square R - 3R_{\mu}^{\nu}R_{\nu}^{\rho}R_{\rho}^{\mu} + \frac{5}{2}RR_{\mu\nu}R^{\mu\nu} - \frac{1}{2}R^3\,,
\end{split}
\end{equation}
where  the last identity is up to total derivatives. This theory is free of scalar ghosts and has four local degrees of freedom by construction, as was verified in section \ref{counting}. 

We can systematically continue this program and find the scalar ghost-free eight-order derivative theory. It is
\begin{align}\label{EENMG}
 \mathcal{L}_6=
\mathcal{L}_{\text{\tiny{ENMG}}}+\frac{e}{m^6}\bigg\{&\kappa_1\left(P_{\mu\nu}P^{\mu\nu}-\tfrac{3}{8}P^2\right)+\kappa_2\left(S^{\rho}{}_{\mu}C_{\nu\rho} C^{\mu\nu}-\tfrac{1}{2}SC_{\mu\nu}C^{\mu\nu}\right)\nonumber\\
&+\kappa_3\left(C_{\mu\nu}\square C^{\mu\nu}+3S^{\rho}{}_{\mu}C_{\nu\rho} C^{\mu\nu} +SC_{\mu\nu}C^{\mu\nu}\right)\bigg\}\,.
\end{align}
One can reduce the $\kappa_1$-term using the Schouten identity, $\slashed{ S}_{\mu\nu}^4=\frac{1}{2}(\slashed{S}_{\mu\nu}^2)^2$ where
$\slashed{S}$ is the traceless Schouten tensor, $\slashed{S}_{\mu\nu}=S_{\mu\nu}-\frac{1}{3}Sg_{\mu\nu}$;
\begin{equation}
-\frac{\kappa_1}{12}\left(16\,SS_\mu^\rho S_\rho^\sigma S_\sigma^\mu -3(S_{\mu\nu}S^{\mu\nu})^2-18S_{\mu\nu}S^{\mu\nu}S^2+5S^4\right)\,.
\end{equation}
Interestingly, the $a$- and $\kappa_1$-term above are precisely the combination of $R^3$ and $R^4$ terms found by Sinha in \cite{Sinha:2010ai} by demanding the presence of a holographic $c$-theorem in higher-derivative extensions for New Massive Gravity, see also \cite{Paulos:2010ke}. The $b$-term,  $\kappa_2$ and $\kappa_3$ terms were not considered in their considerations regarding holographic $c$-theorem; we will comment on this in section \ref{sec:ctheorem}.

\subsection{Linearization}\label{linearth}
In this section we study the higher-derivative theory constructed above by linearizing it around a maximally symmetric vacuum parametrized by the background dreibein $\bar{e}$, the spin connection $\bar{\omega}$ and the cosmological constant $\Lambda$ 
satisfying,
\begin{equation}\label{vacuum}
\bar{R} = d \bar \omega + \frac12 \bo \times \bo = \frac{\Lambda}{2} \bar{e} \times \bar{e}\,,\qquad \qquad \bar{\cD}\bar{e} = 0\,.
\end{equation}
All barred quantities refer to the background. The background values for the auxiliary fields can be determined by their background equations of motion. Since the parity even and odd models have the same field equations, 
these fields have the same background values in both models. The $f$ fields all have background values proportional to the background dreibein and the $h$ fields, which are constructed from the Cotton tensor, vanish on this background. 
We expand the one-form fields around the vacuum as,
\begin{align}\label{fluctuations}
    e &= \bar{e} + \kappa \,k\,, & \omega &= \bar{\omega} + \kappa\, v\,, \nonumber \\
    f_1{} &= -\tfrac{\Lambda}{2}\,\bar{e}  + \kappa \,k_1{} \,, &      h_1{} &= \kappa\, v_1{} \,, \\ \nonumber
    f_2{} &=-\tfrac{\Lambda}{2} \left( m^2 + \tfrac{a \Lambda}{4} \right)  \bar{e}  + \kappa \,k_2{} \,, &      h_2 &= \kappa\, v_2{} \,,
\end{align}
where we used $\kappa$ as a small expansion parameter. In the parity odd case $m$ and $a$ are replaced by $\mu$ and $\alpha$.

The quadratic Lagrangians $L_0^{(2)}$, $L_1^{(2)}$ and $L_2^{(2)}$ corresponding to Einstein-Cartan gravity, conformal gravity and New Massive Gravity respectively, were reviewed in chapter \ref{chapter:two_company}. The linear theory of the parity odd model defined by $S_3$ is treated in appendix \ref{ExtendedCSG}. Here we consider the quadratic Lagrangian $L_4^{(2)}$ by inserting the fluctuations \eqref{fluctuations} into the Lagrangian \eqref{LENMG}. The terms linear in $\kappa$ vanish when the cosmological parameter $\Lambda_0 $ is related to the physical cosmological constant $\Lambda$ as
\begin{equation}\label{ENMGbkgd}
 \Lambda_0 = \Lambda\left(\sigma  + \frac{\Lambda}{4 m^2} + \frac{a \Lambda^2}{8 m^4}\right)\,.
\end{equation}
After a shift in the fields,
\begin{align}
k_2\rightarrow k_2-\tfrac{\Lambda}{2}(m^2+\tfrac{a\Lambda}{4})k\,,\qquad
k_1\rightarrow k_1-\tfrac{\Lambda}{2}k\,,\qquad
v_2\rightarrow v_2+\tfrac{b\Lambda}{2}v_1\,,
\end{align}
the quadratic Lagrangian three-form of $S_4$ reduces to,
\begin{equation}
\begin{split}
L_4^{(2)} = & \; - \left(\sigma - \frac{\Lambda}{2m^2} - \frac{a\Lambda^2}{8m^4} \right) \left( k \cdot \bar{\cD} v + \frac12  \bar{e} \cdot ( v \times v^c - \Lambda k \times k) \right)  \\ 
 + &\frac{1}{2m^2}\left(1 + \frac{a \Lambda}{2m^2} \right) \bar{e} \cdot  k_1 \times k_1 
-  \frac{1}{m^4}\bigg\{  k_{2} \cdot \left( \bar{\cD} v - \Lambda  \bar{e} \times k +  \bar{e} \times k_1 \right) 
  \\ 
   + & v_{2} \cdot \left(  \bar{\cD} k +  \bar{e} \times v \right) 
- b \,v_{1} \cdot \left(  \bar{\cD} k_1 +  \tfrac12 \bar{e}\times  v_1{} \right)   \bigg\} \,.
 \end{split} 
\end{equation}
Upon eliminating the auxiliary fields $v{}^a, v_1{}^a$ and $v_2{}^a$ by using their equations of motion, the quadratic Lagrangian density may be written as
\begin{align} \label{linShk}
\cL_4^{(2)} = &\, -\sigma_4 \, k^{\mu\nu} \cG_{\mu\nu}(k)  - \frac{2}{m^4} k^{\mu\nu}_2 \cG_{\mu\nu}(k) -  \frac{b}{m^4} k_1^{\mu\nu} \cG_{\mu\nu}(k_1) \nonumber \\ 
& + \frac{1}{2m^2}\,\Theta \left(k_1^{\mu\nu} k_{1\mu\nu} - k_1^2\right) - \frac{1}{m^4} \left( k_1^{\mu\nu}k_{2\mu\nu} - k_1 k_2 \right)  \,.
\end{align}
where
\begin{equation}
\sigma_4=\sigma - \frac{\Lambda}{2m^2}- \frac{a\Lambda^2}{8m^4}\,,\qquad\text{and}\qquad\Theta=1  + \frac{a \Lambda}{2m^2} - \frac{b\Lambda}{ m^2} \,.
\end{equation}
For general values of the parameters this quadratic Lagrangian leads to a sixth-order differential equations for $k_{\mu\nu}$. The matrix for the kinetic terms and the mass terms in the basis of $|k\rangle$, $m^2|k_1\rangle$ and $m^4|k_2\rangle$ can be written as,
\begin{equation}
\mathcal K=-\left( \begin{array}{ccc}
\sigma_4  & 0 & 1 \\
0 & b & 0 \\
1 & 0 & 0 \end{array} \right)\,,\qquad\text{and}\qquad
\mathcal M^2=m^2\left( \begin{array}{ccc}
0  & 0 & 0 \\
0 & -\Theta & 1 \\
0 & 1 & 0 \end{array}\right)\,.
\end{equation}
When $\sigma_4 \neq 0$ and $b \neq 0$, the quadratic Lagrangian can be diagonalized by redefining the fields
\begin{align}
k & = k^0 - \frac{1}{\sigma_4} \left(k^+ - k^-\right)\,, \\
k_1 & = - \frac{m^4}{b }\left(\frac{1}{ \cM_-^2} k^+- \frac{1}{\cM_+^2}  k^-\right) \,, \\
k_2 & = m^4 \left( k^+ -  k^- \right)\,,
\end{align}
The quadratic Lagrangian now becomes the sum of the linearized Einstein-Hilbert term, denoted by $\cL_0^{(2)}$ and two Fierz-Pauli terms $\cL_{\rm FP}$,
\begin{align}
\cL_4^{(2)}&= \sigma_4 \cL_{0}^{(2)}(k^0) + \mathcal K_+ \cL_{\text{\tiny{FP}}}(k^+,\mathcal M_+) + \mathcal K_-\,\cL_{\text{\tiny{FP}}}(k^-,\mathcal M_-)\,,
\end{align}
where $\sigma_4$ and $\mathcal K_{\pm}$ satisfy,
\begin{equation}\label{KK}
\sigma_4\,\mathcal K_{+} \mathcal K_{-}= -\frac{\Theta^2-4 b\, \sigma_4}{b \, \sigma_4^2} = \mathcal K_{+}+ \mathcal K_{-}\,,
\end{equation}
and the corresponding Fierz-Pauli masses $\cM_{\pm}^2$ satisfy,
\begin{equation}\label{MM}
\cM_+^2 \cM_-^2=\frac{m^4}{b}\sigma_4  \,\qquad\text{and}\qquad \cM_+^2 - \cM_-^2= \frac{m^2}{b} \sqrt{\Theta^2 - 4 b \sigma_4}\,.
\end{equation}
The numerator in \eqref{KK} should be positive, otherwise the square of the masses becomes imaginary. Hence, it is not possible for both the kinetic terms and the masses to be positive simultaneously. When $\sigma_4 > 0$,  from \eqref{KK} and \eqref{MM} we see that there is either a negative mass squared ($b <0$) or a wrong-sign kinetic term ($b>0$) in the theory. For $\sigma_4 < 0$, one ($b<0$) or both ($b>0$) kinetic terms $\mathcal K_{\pm}$ have the wrong sign. Hence for all values where $\sigma_4 \neq 0$, one of the massive modes is either tachyonic or a ghost.

\subsection{Critical Lines and the Tricritical Point}\label{sec:ENMG_crit}
In above analysis we disregarded a number of special points in the parameter space. Here we present them separately;
\begin{itemize}
\item {$\boldsymbol{b=0}$}: At this point  the rank of the matrix $\mathcal K$ is reduced by one. The action \eqref{LENMG} is now independent of the auxiliary field $h_1{}^a$, and $f_2{}^a$ is algebraically given in terms of $f_1{}^a$. This reduces the  number of local degrees of freedom from four to two, representing a single massive graviton. From eq.~\eqref{NMG2c7} we see that the term involving the Cotton tensor has disappeared and the action reduces to the `cubic extended'  NMG model described in \cite{Sinha:2010ai}. 

\item {$\boldsymbol{b=\Theta=0}$}: At this special point the linearized field equations for $k_{\mu\nu}$ becomes second order in derivatives and the massive mode disappears from the linearized spectrum. The linearized spectrum hence only contains a pure gauge mode, which could lead to a dual CFT with positive central charge if $\sigma_4 >0$ \cite{Sinha:2010ai}. However, it is unclear that this fine-tuning persists at the non-linear level.

\item {$\boldsymbol{\sigma_4=0}$}:
On this critical line, 
one of the FP masses become zero, but the linear field equations remain sixth order in derivatives. Consequently, a new, logarithmic (log)-mode appears and it forms a Jordan cell of rank two with the massless mode.
The Lagrangian \eqref{linShk} is not diagonalizable any more. 

\item {$\boldsymbol{\sigma_4=\Theta=0}$}:
This is a `tricritical' point, 
where both FP masses vanish and  the corresponding massless gravitons form a Jordan cell of rank three. The spectrum now contains one log-mode and a log$^2$-mode (see for instance \cite{Bergshoeff:2012ev}). 

\item {$\boldsymbol{\Theta^2 = 4 b \sigma_4}$}:
This is another critical line where the two non-zero FP masses degenerate and form a Jordan cell of rank two. At this point the spectrum contains one massive mode and a massive log-mode.
\end{itemize}
These critical points are interesting from AdS/LCFT point of view. For a more detailed 
treatment along the lines of \cite{Bergshoeff:2012ev}, we refer to chapter \ref{sec:ENMG_log}.

\section{Anti-de~Sitter Holography}\label{Holography}
All of the extended massive gravity models admit an AdS vacuum and hence it is possible to study the holographic dual theory by imposing suitable asymptotically AdS boundary conditions. In section \ref{sec:generators} we investigated the first-class constraints of the general CS--like models which generate local Lorentz transformations and diffeomorphisms. To proceed with an analysis of the asymptotic symmetry group we would like to write them in a basis of mutually commuting $SL(2,\mathbb{R})$ generators \eqref{Lpm}.
\begin{equation}\label{Lpm2}
L_{\pm} [\zeta] = \phi_{\rm diff}'[\zeta] + a_{\pm} \phi_{\rm LL}[e_{\mu}\zeta^{\mu}]\,,
\end{equation}
where $\phi_{\rm diff}'[\zeta] = \phi_{\rm diff}[\zeta] - \phi_{\rm LL} [\omega_{\mu}\zeta^{\mu}]$. The constants $a_{\pm}$ are defined such that
\begin{equation}\label{LplusLminus}
\{L_+[\xi],L_{-}[\eta]\} = 0\,, 
\end{equation}
on the AdS background. 

One can then reinstate the boundary terms introduced in \eqref{varphi} and investigate the Poisson bracket algebra of the generators \eqref{Lpm2} subject to asymptotically AdS (or Brown-Henneaux) boundary conditions. This computation requires us to consider the specific models on a case by case basis, as we need the explicit flavor space metric $g_{rs}$ and the structure constants $f_{rst}$, as well as the AdS background values of the auxiliary fields. We will analyze the parity-odd $S_3$ theory in full detail in appendix \ref{ExtendedCSG} and here we will treat the parity-even theory defined by $S_4$. 

\subsection{Central Charges}\label{sec:cc}
Here we will construct explicitly the algebra of first-class constraints for the extended new massive gravity model introduced in the last section. To this end it is convenient to first define the constraint functions
\begin{equation}
\phi_t[\xi^t] = \int d^2x \; \xi^t_a \phi^a_t \,,
\end{equation}
where here there is {\it no sum over $t$}. We now set out to find the generators \eqref{Lpm2} such that \eqref{LplusLminus} holds. Let us first comment that, quite generally, by the fact that \eqref{fhantisym} holds, the gauge parameters for diffeomorphisms satisfy
\begin{equation}
\begin{split}
& e_{\mu}{}^a \xi^{f_I}_a = e_{\mu}{}^a f_{I\,\nu\,a} \zeta^{\nu} = f_{I\,\mu}{}^a e_{\nu\,a} \zeta^{\nu} = f_{I\,\mu}{}^a \xi^e_{a}\,, \\
& e_{\mu}{}^a \xi^{h_I}_a = h_{I\,\mu}{}^a \xi^e_a\,.
\end{split}
\end{equation}
Moreover, on the AdS background, the auxiliary fields $f_I{}^a$ are proportional to the AdS dreibein and the auxiliary fields $h_I{}^a$ vanish. Hence we have that on the AdS background
\begin{equation}\label{gpbkgd}
\xi^{f_I}_a = \bar{f}_I \xi^e_a \,, \qquad \xi^{h_I}_a = 0\,.
\end{equation}
Where $\bar{f}_I$ is the constant background value of the auxiliary fields, i.e. $f_I{}^a = \bar{f}_I\, \bar{e}^a$ on the AdS background. The values for $\bar{f}_1$ and $\bar{f}_2$ can be read off from \eqref{fluctuations}. This allows us to express all the gauge parameters in $\phi_{\rm diff}'[\zeta]$ in terms of $\xi^e_a$. Using these relations and the specific values of $g_{rs}$ and $f_{rst}$ for ENMG, which may be read of from \eqref{LENMG}, we can compute the Poisson brackets of $\phi_{\rm diff}[\zeta]$ and $\phi_{\rm LL}[\xi]$. We find
\begin{align}
& \{\phi_{\rm diff}'[\xi] ,\phi_{\rm diff}'[\eta]\} =    \phi_t[ f^t{}_{rs} \xi^r \times \eta^s] \\
& =  \; \bigg(- \Lambda_0 m^4 +  \sigma \Lambda m^4  + \frac{m^2 \Lambda^2}{4} + \frac{a\Lambda^3}{8} \bigg) \phi_{h_2} [\xi^e \times \eta^e] - \Lambda \phi_{\rm LL} [\xi^e \times \eta^e] \nonumber \\
& =  \; - \Lambda \phi_{\rm LL}[\xi^e \times \eta^e]\,.
\end{align}
In the first line here the sum over $t$ is over all flavors, while $r$ and $s$ sum over all flavors except $\omega$. In the second line, we have used \eqref{gpbkgd} to express all gauge parameters in terms of $\xi^e_a$ and $\eta^e_a$ and the last line follows from the ENMG background equation for the cosmological constant \eqref{ENMGbkgd}.

The other Poisson brackets are
\begin{equation}
\{ \phi_{\rm diff}'[\xi] , \phi_{\rm LL} [\eta]\} = \phi_{\rm diff}' [\xi \times \eta ]\,, \qquad \{ \phi_{\rm LL}[\xi] , \phi_{\rm LL} [\eta]\} = \phi_{\rm LL}[\xi\times \eta]\,.
\end{equation}
This shows that the first-class constraint functions $\phi_{\rm diff}' [\zeta]$ and $\phi_{\rm LL}[\xi]$ span the $SO(2,2)$ algebra when $\Lambda = -1/\ell^2<0$. It is now simple to define the $SL(2,\bR)$ generators as in \eqref{Lpm2} with
\begin{equation}
a_{\pm} = \pm \frac{1}{\ell}\,.
\end{equation}
The algebra of Poisson brackets on the AdS background then becomes
\begin{equation}\label{Lpmbrackets}
\{ L_{\pm} [\xi], L_{\pm} [\eta] \} = \pm \frac{2}{\ell} L_{\pm}[\xi \times \eta] \,, \qquad \{L_+[\xi],L_{-}[\eta]\} = 0\,.
\end{equation}
We can now consider reinstating the boundary terms for \eqref{Lpm2} and adopting suitable asymptotically AdS boundary conditions

The boundary charges $Q_{\pm}$ of the generators $L_{\pm}[\xi]$ can be found trough the general formula \eqref{gen_varbc}. Their variation is
\begin{equation}
\begin{split}\label{Qpm}
\delta Q_{\pm} = & - \int_{\partial \Sigma} dx^i \; \left( \xi^r_a g_{rs} + a_{\pm} \xi^e_a g_{\omega s}\right) \delta a_{i}^{s\,a} \\
= & - \int_{\partial \Sigma} dx^i \; \left(  g_{es} + \bar{f}_1 g_{f_1s} + \bar{f}_2 g_{f_2s} + \ldots + a_{\pm} g_{\omega s}\right) \xi^e_a\delta a_{i}^{s\,a}\,,
\end{split}
\end{equation}
where in the first line the sum over $r$ does not include $\omega$ and we have used \eqref{gpbkgd} in the last line. In general, after plugging in the explicit flavor space metric and AdS background values of the fields, the result may be written as
\begin{equation}
\delta Q_{\pm}[\xi^{\pm}] =  \frac{\hat k}{2\pi} \int_{\partial \Sigma} dx^i \; \xi_a^{\pm} \left( \delta \omega_i{}^a \pm \frac{1}{\ell} \delta e_i{}^a \right)\,.
\end{equation}
where $\hat k$ is an effective coupling determined by the elements of $g_{rs}$ and the $\bar{f}_I$'s. We have also distinguished the gauge parameters for the left and right moving sectors explicitly. 

In order to integrate this expression to the boundary charges, we impose Brown-Henneaux boundary conditions \eqref{BHbc1}-\eqref{BHbc2} on the dreibein and the spin connection.
Together with the boundary condition preserving gauge transformations \eqref{asymptdiffs} they lead to the following expression for the conserved charge $Q = Q_{+} + Q_{-}$ at the boundary
\begin{equation}
Q = \frac{\hat k}{2\pi}\int d\varphi \left[f (x^+)\mathcal{L}(x^+)-\bar{f}(x^-)\bar{\mathcal{L}}(x^-)\right]\,.
\end{equation}
We can compute the Poisson brackets \eqref{Lpmbrackets} with the boundary term from  \eqref{gen_poissonbr} after suitably identifying $\xi$ and $\eta$ from \eqref{asymptdiffs} and defining the improved generators as in \eqref{Lpmimproved}. After Fourier expanding the improved generators as
\begin{equation}
 L_n= \tilde L_+[f=e^{inx^+}]\,,\qquad\text{and}\qquad \bar L_n= \tilde L_-[\bar f=e^{inx^-}]\,,
\end{equation}
we find two copies of Virasoro algebra as expected,
\begin{align}
 i\{L_m,L_n\}&=(m-n)L_{m+n}+\frac{c}{12}m(m^2-1)\delta_{m+n,0}\,,\\
 i\{\bar L_m,\bar L_n\}&=(m-n)\bar L_{m+n}+\frac{\bar c}{12}m(m^2-1)\delta_{m+n,0}\,,
\end{align}
where $c=\bar c=6\hat k$ for parity even models and $c=-\bar c=6\hat k$ for parity odd models. After explicitly computing the effective couplings $\hat{k}$ for the parity odd theory $S_3$ (see appendix \ref{ExtendedCSG}) we find the central charges
\begin{equation}
c_{3}  = - \bar{c}_{3} =   \frac{3}{2\mu G} \,.
\end{equation}
This result is independent of the new coupling constant $\alpha$.
A similar computation for the extended New Massive Gravity model $S_4$, defined by \eqref{LENMG} shows that
\begin{equation}\label{ENMGcc}
c_{\text{\tiny ENMG}} = \frac{3\ell}{2G}\sigma_4  = \frac{3\ell}{2G} \left( \sigma + \frac{1}{2\ell^2 m^2} - \frac{a}{8m^4 \ell^4} \right)\,.
\end{equation}
The central charge is proportional to the earlier defined parameter $\sigma_4$ and independent of the coupling constant $b$.

 \subsection{Holographic $c$-theorem}
 \label{sec:ctheorem}
RG flows between fixed points in a matter theory with stress tensor $\mathcal{T}_{\mu\nu}$ coupled to gravity and with AdS vacua can be described by a metric of the form
\begin{equation}\label{anst}
ds^2=e^{2\mathcal{A}(r)}(-dt^2+d\mathbf{x}^2)+dr^2\,.
\end{equation}
 Assuming that the null energy condition holds for the matter sector, i.e.~$\mathcal{T}_{\mu\nu}\xi^\mu\xi^\nu\geq0$ for any null vector  $\xi^{\mu}$, it was shown in \cite{Freedman:1999gp} that a monotonically increasing holographic $c$-function can be found in terms of $\mathcal{A}(r)$, such that it satisfies Zamolodchikov's $c$-theorem \cite{Zamolodchikov:1986gt} with the radial coordinate $r$ as the measure of the energy.
The null energy condition simplifies to
\begin{equation}
-\mathcal{T}^t_t+\mathcal{T}^r_r\geq0\,,\qquad\text{for}\qquad(\xi^t,\xi^r,\xi^x)=\left(e^{-\mathcal{A}},1,0\right)\,.
\end{equation}
A monotonically increasing holographic $c$-function can then be obtained from
\begin{equation}\label{cprime}
c'(r)=-\frac{\mathcal{T}^t_t-\mathcal{T}^r_r}{\kappa^2\mathcal{\mathcal{A}}'^2}\geq0\,.
\end{equation}
If the bulk field equations are given by $\mathcal {E}_{\mu\nu}=\kappa^2\mathcal{T}_{\mu\nu}$, the null energy condition can equivalently be written as $\mathcal {E}_{\mu\nu}\xi^\mu\xi^\nu\geq0$.  In \cite{Sinha:2010ai,Myers:2010tj} it is argued that
one way to make $c'(r)$ fulfill the  inequality \eqref{cprime} is to have $c(r)$ be only a function of $\mathcal{A}'$ which implies that $\mathcal {E}^t_t-\mathcal {E}^r_r$ should only be a function of $\mathcal{A}'$ and $\mathcal{A}''$.  They used this logic to constrain higher-derivative interactions by demanding the presence of such a monotonically increasing function.

Here we show that the construction with field equations \eqref{ENMGeom} is consistent with this assumption. The ansatz \eqref{anst} is conformal to AdS spacetime which is an Einstein metric and all solutions of the Einstein equations in three dimensions are also solutions of $C_{\mu\nu}=0$. This has the following two consequences:
 \begin{enumerate}\label{C=0}
 \item All fields which are constructed from the Cotton tensor and its derivatives are zero on the background \eqref{anst}. In other words, all fields $h_I$ and all $\mathcal Df$ terms vanish on the ansatz \eqref{anst} and all equations \eqref{ENMGeom} reduce to a set of algebraic equations among the $f_I$ fields. These equations can be solved in terms of the Schouten tensor, which only contains up to two derivatives of $\mathcal{A}$. Hence, after solving for all auxiliary fields the bulk field equation $\mathcal E_{\mu\nu}$ involves only $\mathcal{A}'$ and $\mathcal{A}''$ by construction.
 \item We can afford terms in the action constructed from the Cotton tensor as higher-derivative corrections without affecting the $c$-function. This also suggests that the only consistent way to include $\nabla R$ terms in the action is to use the Cotton tensor.
 \end{enumerate}
If we only focus on the bulk actions  \eqref{SNeven} and \eqref{SNodd}, this suggests that terms containing the fields $h_I$ and $\mathcal{D}f$ terms  do not directly contribute to the one-point functions around the AdS vacuum. The variation of the action \eqref{SNeven} around the background \eqref{anst} is only affected by $ (\bar f \cdot \bar  f \times \delta f)$-terms because fluctuations in other terms are always
 proportional to a power of $h$ or a $\mathcal{D}f$ term, which are zero for \eqref{anst}. In the metric formulation this means that the linearized theory around \eqref{anst} is not affected by terms where graviton fluctuations are proportional to a power of Cotton tensor which is zero for this background\footnote{
The value of the central charge is not fully determined by the equations of motion. There is always a total derivative ambiguity which should be fixed by adding suitable boundary terms to the action and imposing suitable boundary conditions.}.
In fact, this is confirmed by direct calculation of the central charge for the first few parity-even models \eqref{NMG2c7} and \eqref{EENMG}, 
\begin{equation}
c_{\text{\tiny even}}=\frac{3\ell}{2G}\left(\sigma+\frac{1}{2m^2\ell^2}-\frac{a}{8m^4\ell^4}+\frac{\kappa_1}{16m^6\ell^6}+ \cdots\right)\,.
\end{equation}
The dots here refer to higher-derivative contributions to the central charge. By the same reasoning, the variation of the action \eqref{SNodd} around the background \eqref{anst} is only affected by $(\bar f \cdot \bar f \times \delta h)$-terms. But this term also vanishes for any maximally symmetric spacetime such as AdS. In the metric formulation this is more transparent from the fact that $\bar{g}^{\mu\nu} \delta C_{\mu\nu}=0$. Hence the interaction terms in the odd sector do not contribute to the central charge. This means that the central charge in the parity-odd models is universal and not affected by any higher-derivative term,
\begin{equation}\label{centodd}
c_{\text{\tiny odd}}=-\bar c_{\text{\tiny odd}}=\frac{3}{2\mu G}\,.
\end{equation}
We conclude that only interaction terms constructed solely from the Schouten tensor can contribute to the central charge. This is consistent with earlier studies of the holographic $c$-theorem in this context \cite{Paulos:2010ke}. Terms involving the Cotton tensor are allowed by the holographic $c$-theorem and do not contribute to the central charge --- these terms however can contribute to the graviton masses and the two point functions.

\section{Drei-Dreibein Gravity}\label{sec:DDG}
In this section we introduce the Drei-Dreibein Gravity (DDG) model. It is a straight-forward generalization of the Zwei-Dreibein model introduced in chapter \ref{section:ZDG} to a theory with three interacting dreibeine and three `spin connections'.\footnote{Since there is only one overall local Lorentz invariance, there is also only one `real' spin connection as the gauge field for this diagonal gauge symmetry.} As we discussed at length in chapter \ref{chapter:three_crowd}, the presence of secondary constraints (and correspondingly, the absence of additional local d.o.f.) in ZDG required us to assume a linear combination of the two dreibeine to be invertible. A special case of this assumption was to restrict the parameters of the model such that the invertibility of a single dreibein was sufficient to derive the necessary secondary constraints. 

In Drei-Dreibein Gravity, similar considerations will play a role and we will deduce the restrictions on the parameters space of the theory which remove the additional scalar degrees of freedom in section \ref{sec:constraints}. We proceed to analyze the linear theory and give the expression for the two Fierz-Pauli masses and the dual central charge. We discuss a scaling limit to the ENMG model presented in chapter \ref{ExtendedNMG} and find further restrictions on the DDG parameter space by demanding the two masses, the energy of the massive modes and the dual central charge to be positive. 

\subsection{The Model}
The most general Lagrangian three-form constructed from three dreibeine $e_I{}^a$ with $ I = 1,2,3$ and three dualised spin connections $\omega_I{}^a$ invariant under diagonal diffeomorphisms and local Lorentz invariance is given by:
\begin{equation}\label{DDG}
L = - M_P \sum_{I=1}^{3}  \left( \sigma_I \, e_{I} \cdot R (\omega_I) + \frac16 m^2 \alpha_I e_I{} \cdot e_I{} \times e_I{} \right) + L_{\rm int}\,,
\end{equation}
where
\begin{equation}
\begin{split}
L_{\rm int} = \,\,  & \frac12 m^2 M_P \bigg\{ \beta_{12} e_1 \cdot e_1 \times e_2  +  \beta_{21} e_2 \cdot e_2 \times e_1  +  \beta_{13} e_1 \cdot e_1 \times e_3 \\ & + \beta_{31} e_3 \cdot e_3 \times e_1  + \beta_{23} e_2 \cdot e_2 \times e_3 + \beta_{32} e_3 \cdot e_3 \times e_2 +  \beta_{123} e_1 \cdot e_2 \times e_3 \bigg\}\,.
\end{split}
\end{equation}
Here the $\beta$'s are dimensionless, free coupling constants and $\alpha_I$, with $I = 1,2,3$ are three cosmological parameters. The mass parameter $m^2$ is inessential, but convenient and the coefficients $\sigma_I$ are in principle arbitrary free ratios of Planck masses, however, we can always rescale one of the dreibeine $e_I{}^a$ such that $\sigma_I = 1$. 

This theory is in essence a three dreibein extension of Zwei-Dreibein Gravity (ZDG) presented in \cite{Bergshoeff:2013xma}. In ZDG the presence of secondary constraints in the Hamiltonian formalism is essential for removing the Boulware-Deser ghost \cite{Bergshoeff:2013xma,Banados:2013fda,Bergshoeff:2014bia}. As was shown in chapter \ref{chapter:three_crowd}, the needed secondary constraints are in turn closely related to the presence of invertible fields in the theory. If the inverse of a dreibein can be used to construct a two-form equation from the Bianchi identities\footnote{The Bianchi identities for this theory are jointly equivalent to the integrability conditions \eqref{Intcon} derived in chapter \ref{chapter:three_crowd}.} ( $\cD_I R_I{}^a =0$, and $\cD_I T_{I}{}^a = \ve^{abc} R_{I\,b} e_{I\,c}$), then the theory possesses secondary constraints \cite{Bergshoeff:2014bia}. Conversely, demanding the presence of secondary constraints to remove unwanted degrees of freedom may impose a restriction on the parameter space of the theory or require us to assume the invertibility of linear combinations of the dreibeine.

The equations of motion derived from the Lagrangian \eqref{DDG} are
\begin{align}
\sigma_1 R(\omega_1) =  \frac12 m^2 \bigg[ & - \alpha_1 e_{1} \times e_{1} + \beta_{123} e_{2} \times e_{3} + \left( 2 \beta_{12} e_{1} \times e_{2} + \beta_{21} e_{2} \times e_{2} \right) \nonumber \\
&  + \left( 2 \beta_{13} e_{1} \times e_{3} + \beta_{31} e_{3} \times e_{3} \right) \bigg]\,, \label{ddgeom1} \\
\sigma_2 R(\omega_2)=  \frac12 m^2 \bigg[ & - \alpha_2 e_{2} \times e_{2} +  \beta_{123} e_{1} \times e_{3} + \left( \beta_{12} e_{1} \times e_{1} + 2\beta_{21} e_{1} \times e_{2} \right) \nonumber \\
&  + \left( 2 \beta_{23} e_{2} \times e_{3} + \beta_{32} e_{3} \times e_{3} \right) \bigg]\,,  \label{ddgeom2} \\
\sigma_3 R (\omega_3) =  \frac12 m^2 \bigg[ & - \alpha_3 e_{3} \times e_{3} +  \beta_{123} e_{1} \times e_{2} +  \left( \beta_{13} e_{1} \times e_{1} + 2\beta_{31} e_{1} \times e_{3} \right) \nonumber \\
&  + \left( \beta_{23} e_{2} \times e_{2} + 2\beta_{32} e_{2} \times e_{3} \right) \bigg]\,,  \label{ddgeom3}
\end{align}
together with three torsion constraints (for fixed $I$)
\begin{equation}
T_I = \cD_I e_{I} \equiv de_I + \omega_{I} \times e_{I} = 0\,.
\end{equation}
The three curvature two-forms satisfy three Bianchi identities $\cD_I R_I =0$, and the three torsions satisfy three Cartan identities $\cD_I T_{I} = R_{I} \times  e_{I}$. These equations are all three-form equations and in a Hamiltonian analysis their space-time decomposition is equivalent to the consistency equations which the primary constraints of this theory need to satisfy \cite{Bergshoeff:2014bia}. If these equations can be turned into two-form equations by the invertibility of (some combination of) the dreibeine, then the theory possesses secondary constraints.

\subsection{Constraint Analysis}
\label{sec:constraints}
Let us first investigate the number of secondary constraints needed to remove any unwanted degrees of freedom in the theory. The above Lagrangian has one diagonal diffeomorphism invariance and one diagonal local Lorentz symmetry. This suggests that there is one overall massless spin-2 mode coupled to two massive spin-2 modes. There should thus be 4 local degrees of freedom, leading to an 8 dimensional physical phase space. 

After a space-time decomposition in the Lagrangian \eqref{DDG}, the time-components of the fields become Lagrange multipliers for 18 primary constraints. The matrix of Poisson brackets for these constraints has rank 12, since 6 constraints are first class, reflecting the six gauge symmetries of the theory. In the absence of secondary constraints, the dimension of the physical phase space would be $(3\times 2 \times 6) = 36$ components in the spatial parts of the 6 fields, minus $(2 \times 6) =12$ first class constraints and minus 12 second class constraints, leading to 12 dynamical components in the physical phase space, i.e. 6 local degrees of freedom. These are 2 degrees of freedom too many, so we need 4 secondary scalar constraints to remove the unwanted degrees of freedom.

From the Cartan identities $\cD_I T_{I}{}^a = \ve^{abc} R_{I\,b} e_{I\,c}$ we can derive 3 three-form equations which are satisfied on-shell. They are
\begin{align}\label{Cartan1}
\big( \beta_{12} e_{1}{}^a + \beta_{21} e_{2}^a + \beta_{123} e_3{}^a \big) e_1 \cdot e_2 
 + \big(\beta_{13} e_{1}{}^a + \beta_{31} e_{3}^a + \beta_{123} e_2{}^a \big) e_1 \cdot e_3 = 0\,, \\
\big( \beta_{12} e_{1}{}^a  + \beta_{21} e_{2}^a +  \beta_{123} e_3{}^a \big) e_1 \cdot e_2  \label{Cartan2} - \big(  \beta_{23} e_{2}{}^a + \beta_{32} e_{3}^a  + \beta_{123} e_1{}^a \big) e_2 \cdot e_3 = 0\,,  \\
\big(  \beta_{13} e_{1}{}^a  + \beta_{31} e_{3}^a   + \beta_{123} e_2{}^a \big) e_1 \cdot e_3 \label{Cartan3} + \big(  \beta_{23} e_{2}{}^a + \beta_{32} e_{3}^a  +  \beta_{123} e_1{}^a \big) e_2 \cdot e_3  = 0\,. 
\end{align}
There are only secondary constraints if these equations can be turned into 2-form identities. This can be achieved by setting to zero some of the coupling constants of the theory and assuming invertibility of (some of) the dreibeine. In order to derive two secondary constraints from this system of equations, we must restrict the parameters of the theory such that exactly one of the following combinations vanish, while the other two should have an inverse.
\begin{equation}\label{combinations}
 \beta_{12} e_{1} + \beta_{21} e_{2}+\beta_{123} e_{3}\,, \qquad \beta_{13} e_{1} + \beta_{31} e_{3} +\beta_{123} e_{2}\,, \qquad 
  \beta_{23} e_{2} + \beta_{32} e_{3} +\beta_{123} e_{1}\,.
\end{equation}
Requiring one of these combinations to vanish implies that the coupling constants appearing in the combination should vanish.\footnote{In principle, one could impose a combination to be zero as an additional constraint enforced by a Lagrange multiplier $\lambda$ in the action. This would change the field equations and only the $\lambda=0$ sub-sector of the theory would be equivalent to our original action. This condition is, however, not enforced by the new field equations and hence the resulting theory would be different. See \cite{Hohm:2012vh} for related issues in higher-derivative theories of 3D massive gravity.} This is a very strong statement since 
it implies that a non-zero $\beta_{123}$ would not lead to secondary constraints, which means interaction terms with more than two dreibeine coupled to each other are not permitted. In addition, we must set to zero (at least) two of the $\beta_{IJ}$ parameters. Here we choose to take
\begin{align}\label{zerocouplings}
  \beta_{23}=\beta_{32}=\beta_{123}=0\,.
\end{align}
Of course, we could have picked any other combination in \eqref{combinations} to be zero, but this would lead to the same theory, as these choices are related to each other by a transformation of the discrete symmetry group of dreibein label permutations. Only once a specific choice has been made the $S_3$ group is broken to the subgroup which leaves this combination zero. Now, the supplementary assumption is to have,
\begin{align}
\label{invertddg}
 \beta_{12} e_{1} + \beta_{21} e_{2} \, \qquad\text{and} \qquad\beta_{13} e_{1} + \beta_{31} e_{3} \,, 
\end{align}
invertible. 
The three equations \eqref{Cartan1}-\eqref{Cartan3}  then reduce to
\begin{align}\label{SecConcov}
&( \beta_{12} e_{1} + \beta_{21} e_{2}) e_1 \cdot e_2 + (\beta_{13} e_{1} + \beta_{31} e_{3}  ) e_1 \cdot e_3 = 0\,, \nonumber \\
&(\beta_{13} e_{1} + \beta_{31} e_{3} ) e_1 \cdot e_3 = 0\,,  \nonumber \\
&(\beta_{12} e_{1} + \beta_{21} e_{2} ) e_1 \cdot e_2 = 0\,. 
\end{align}
From which two secondary constraints on the spatial variables of the theory follow
\begin{equation}\label{seccon1}
\varepsilon^{ij} e_{1\,i} \cdot e_{2\,j} \equiv \Delta^{e_1e_2} = 0\,, \qquad 
\varepsilon^{ij} e_{1\,i} \cdot e_{3\,j} \equiv \Delta^{e_1e_3} = 0\,.
\end{equation}
The assumption in \eqref{zerocouplings} and \eqref{invertddg} (and the ultimate theory) is invariant under the stabilizer of $e_1$ which is a reflection symmetry ($2\leftrightarrow3$). 

\begin{figure}
\centering
\includegraphics[height=0.2\textwidth]{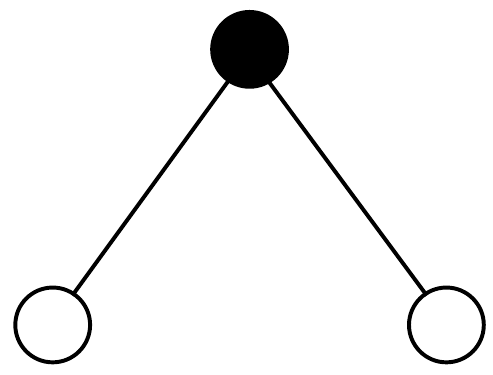}
\caption{The ghost-free interaction terms present in DDG represented in a theory diagram. The absence of interaction terms mixing more than two dreibeine demands the diagram to be a graph. The absence of ghosts further requires the graph to be a tree and the lines between the nodes represent the two invertible linear combinations of dreibeine, as in \eqref{invertddg}. In addition, the dreibein which couples to the other two dreibeine should have an inverse, denoted as a solid circle.} \label{fig:dreibein1}
\end{figure}

Let us now turn our attention to the 3 Bianchi identities ($\cD_I R_I=0$) and their consequence on the field equations. After acting on \eqref{ddgeom1}-\eqref{ddgeom3} with an exterior derivative and doing some algebra, we can derive the following 3 three-form equations:
\begin{equation}\label{ddgbianchi}
\begin{split}
& e_1 \left[\omega_{12}\cdot (\beta_{12} e_1 + \beta_{21} e_2)\right]  = 0\,, \\
& e_1 \left[\omega_{13}\cdot (\beta_{13} e_1 + \beta_{31} e_3)\right]  = 0 \,, \\
& e_2 \left[\omega_{12}\cdot (\beta_{12} e_1 + \beta_{21} e_2)\right]  + e_3 \left[\omega_{13}\cdot (\beta_{13} e_1 + \beta_{31} e_3)\right]  = 0 \,. 
\end{split}
\end{equation}
Here $\omega_{IJ} = \omega_I - \omega_J$ and we have used the parameter restrictions \eqref{zerocouplings} and the identities $e_1 \cdot e_2 = e_1 \cdot e_3 =0$. These equations are the $N=3$ generalizations of the ZDG integrability conditions \eqref{omega1}-\eqref{omega2}. In the ZDG case it was possible to take a linear combination of the two equations such that the effective dreibein $\beta_1 e_1 + \beta_2 e_2$ would carry the free Lorentz index and its inverse can be used to derive another secondary constraint. In this case this is no longer possible and we see that in addition to the invertibility of the linear combinations \eqref{invertddg}, we should also require $e_1$ to have an inverse. The DDG theory can now be represented by the diagram shown in figure \ref{fig:dreibein1}. 
Another two secondary constraints can be derived from \eqref{ddgbianchi} and they read
\begin{equation}\label{seccon2}
\begin{split}
\varepsilon^{ij} \omega_{12\,i}  \cdot (\beta_{12}e_{1\,j} + \beta_{21} e_{2\,j}) \equiv \beta_{12}\Delta^{\omega_{12}e_1} - \beta_{21}\Delta^{\omega_{12} e_2} = 0\,,  \\
\varepsilon^{ij} \omega_{13\,i}  \cdot (\beta_{13}e_{1\,j} + \beta_{31} e_{3\,j}) \equiv \beta_{13}\Delta^{\omega_{13}e_1} - \beta_{31}\Delta^{\omega_{13} e_3} = 0\,. 
\end{split}
\end{equation}
These constraints together with \eqref{seccon1} are necessary and sufficient to remove all the unwanted degrees of freedom in the theory.
After adding the secondary constraints to the primary constraints, the total amount of constraints grows to 22. There are still 6 first class constraints (FCC), reflecting the six gauge symmetries present. The remaining 16 constraints are second class (SCC) and the dimension of the physical phase space, per space point, is
\begin{equation}
36\; \{\text{canonical var.}\} - 2 \times 6\; \{\text{FCC}\} - 16\;  \{\text{SCC}\} = 8\,.
\end{equation}
The total number of degrees of freedom is then four, which corresponds to the four helicity $\pm2$ modes of two massive gravitons. For more details and the explicit computation of the matrix of Poisson brackets and its rank, we refer to appendix \ref{app:Hamil}.

\begin{figure}
\centering
\subfigure[]{\includegraphics[height=0.2\textwidth]{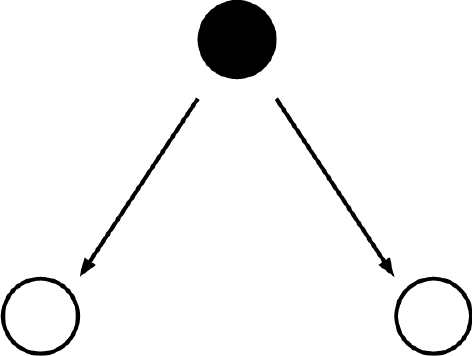}\label{fig:dreibein2a}} \quad
\subfigure[]{\includegraphics[height=0.2\textwidth]{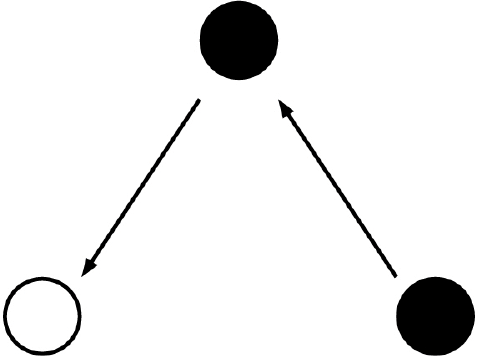}\label{fig:dreibein2b}} \quad
\subfigure[]{\includegraphics[height=0.2\textwidth]{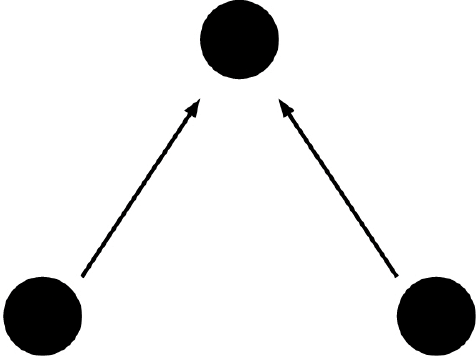}\label{fig:dreibein2c}}
\caption{Three inequivalent ghost-free interaction terms in DDG when assuming one (a), two (b) and three (c) of the original dreibeine to be invertible, while a linear combinations of them may have zero determinant. The theory graphs are oriented trees and each of them require a different number of invertible frame fields, denoted here by solid circles} \label{fig:dreibein2}
\end{figure}

There are several special cases where the invertibility of only the original dreibeine, and not of some linear combination of them, is sufficient. 
For instance, if we assume invertibility of only $e_1$, there is a unique parameter choice which leads to secondary constraints
\begin{equation}\label{paramchoice}
\beta_{12} \neq 0\, \qquad\text{and}\qquad \beta_{13} \neq 0\,,
\end{equation}
while all other $\beta$-parameters should be set to zero. 
A similar choice of parameters exists if we take $e_2$ or $e_3$ to be invertible, as is indicated in figure \ref{fig:dreibein2a}. 

The parameter choice \eqref{paramchoice} is unique if we assume the invertibility of only one dreibein. If instead there are two invertible dreibeine, then there are two other possibilities of choosing two non-zero $\beta$-parameters:
\begin{equation}\label{paramchoice2}
\beta_{12}\,, \beta_{31} \neq 0\,,\qquad
\beta_{13}\,, \beta_{21} \neq 0\,.
\end{equation}
These choices are equivalent theories, as they are related to each other by the residual reflection symmetry which transforms the dreibein labels ($2 \leftrightarrow 3$). The corresponding diagram is depicted in figure \ref{fig:dreibein2b}. Figure \ref{fig:dreibein2c} corresponds to a ghost-free theory with all three dreibeine invertible and
\begin{equation}
\beta_{31} ,\beta_{21} \neq 0\,.
\end{equation}
When assuming all three dreibeine to be invertible, the number of ghost-free DDG theories becomes equal to the number of oriented trees with three unlabeled nodes.\footnote{Note that when not all of the dreibeine are invertible, like in figure \ref{fig:dreibein2a} and \ref{fig:dreibein2b}, the arrows are always pointing from the solid circle to the empty ones. This limits the number of oriented diagrams in these cases, i.e. one for the first diagram and two for the second diagram.}

As was discussed in \cite{deRham:2014naa}, multi-gravity theories can be coupled to matter by making use of an effective metric. This effective metric should be constructed such that its determinant does not reintroduce the interaction terms that have been set to zero. Hence matter can be coupled to any of the original dreibeine which are assumed to be invertible. In addition to this, an effective metric can be constructed out of two of the three dreibeine. An effective metric containing all three dreibeine is not allowed, since its determinant would reintroduce the $\beta_{123}$ term. Coincidently, there are also two linear combinations in DDG which are required to be invertible in order to derive the necessary secondary constraints. They are related by a reflection symmetry (${\mathbb Z}_2$) transformation ($2\leftrightarrow3$) so they can be regarded as the effective dreibein of the theory up to a symmetry transformation, 
\begin{align}\label{e3eff}
e_{\rm eff}=\beta_{1J} e_{1} + \beta_{J1} e_{J} \qquad\text{with}\qquad J=2,3\,.
\end{align}
The effective metric constructed from this dreibein is a good candidate for coupling to matter, since it is invertible by assumption and its determinant does not reintroduce the couplings which were set to zero in \eqref{zerocouplings}. 
Moreover the single internal dreibein (in this case $e_1$) should also be invertible and one may couple matter directly to the metric constructed from it. This means that in contrast to the $N=2$ case, in the $N=3$  case with the most general allowed interaction terms, there are two choices of invertible metrics for matter coupling; either the effective linear combination \eqref{e3eff} or the internal dreibein itself ($e_1$ in this case).

To summarize, the analysis of the presence of secondary constraints in DDG severely restricts the number of allowed interaction terms. In full generality, to derive the secondary constraints needed to remove the extra degrees of freedom, we must:
\begin{itemize}
\item Set to zero $\beta_{123}$, i.e. there is no interaction term mixing three dreibeine. (The theory can be pictorially represented as a graph)
\item Assume invertibility of two out of the three linear combinations in \eqref{combinations}, while the parameters of the theory should be restricted such that the third combination vanishes. (The theory graph is a tree)
\item In addition to the invertible linear combinations, the dreibein coupling to two other dreibeine should be invertible. (Internal nodes should correspond to invertible frame fields)
\end{itemize}
The second rule implies that only one dreibein is allowed to couple to the other two dreibeine, i.e. there can be no loops in the diagrams in figure \ref{fig:dreibein2}. This is compatible with the analysis of ref.~\cite{Nomura:2012xr}, obtained by different means.
In a restricted case where the original dreibeine themselves are assumed to be invertible, and not necessarily their linear combinations, the allowed theory graphs become all possible oriented trees with 3 nodes in figure \ref{fig:dreibein2}. The internal node should always be invertible.

\subsection{Linearized theory}
Before selecting a specific ghost-free theory, we analyze the linear theory around a common maximally symmetric background with cosmological constant $\Lambda =  - 1/ \ell^2$, described by the dreibein $\bar{e}^a$ and spin-connection $\bo^a$. We allow for two arbitrary scale parameters $\gamma_2$ and $\gamma_3$ for the backgrounds of $e_{2}{}^a$ and $e_3{}^a$ respectively.
\begin{align}
e_{1}{}^a & =  \bar{e}^a + \kappa k_{1}{}^a\,, &
e_{2}{}^a & = \gamma_2 \left( \bar{e}^a + \kappa k_{2}{}^a \right)\,, \\
e_{3}{}^a & = \gamma_3 \left( \bar{e}^a + \kappa k_{3}{}^a \right)\,, &
\omega_{I}{}^a & = \bar{\omega}^{a} + \kappa v_{I}{}^a\,.
\end{align}
The linear terms in the $\kappa$ expansion cancel when
\begin{align}\label{gammas}
\sigma_1 \frac{\Lambda}{m^2}  & =   2\beta_{12}\gamma_2 + \beta_{21}\gamma_2^2 + 2\beta_{13}\gamma_3 + \beta_{31}\gamma_3^2 - \alpha_1  \,, \nonumber \\
 \sigma_2 \frac{\Lambda}{m^2}  &=   \beta_{12} + 2\beta_{21}\gamma_2 + 2\beta_{23}\gamma_2\gamma_3 + \beta_{32}\gamma_3^2 - \gamma_2^2 \alpha_2\,, \\
 \sigma_3 \frac{\Lambda}{m^2} &=   \beta_{13} + 2\beta_{31}\gamma_3 + \beta_{23}\gamma_2^2 + 2\beta_{32}\gamma_2\gamma_3 - \alpha_3\gamma_3^2 \,. \nonumber
\end{align}
These three equations fix $\gamma_2, \gamma_3$ and $\Lambda$ in terms of the DDG parameters.

The quadratic Lagrangian for the fluctuations $k_I{}^a$ and $v_{I}{}^a$ is
\begin{align}\label{Lquad}
L^{(2)} =  - M_P \sum_{I=1}^{3} \sigma_I \gamma_I & \left\{ k_{I} \cdot \bar{\cD}v_{I} + \frac12 \bar{e} \cdot \left( v_{I} \times v_{I} - \Lambda k_{I} \times k_I \right) \right\} \nonumber \\
 - \frac12 m^2 M_P & \bar{e} \cdot  \left( \gamma_2 (\beta_{12} + \gamma_2 \beta_{21}) (k_1 - k_2) \times (k_1 - k_2) \right. \\
 & +  \gamma_3(\beta_{13} + \gamma_3\beta_{31})  (k_1 - k_3) \times (k_1 - k_3)  \nonumber \\ 
&  \left. + \gamma_2 \gamma_3(\gamma_2 \beta_{23} + \gamma_3 \beta_{32})  (k_2 - k_3) \times (k_2 - k_3) \right) \,. \nonumber
\end{align}
where $\gamma_1 = 1$. In the last section, we showed that one of the three combinations appearing in \eqref{combinations} should vanish by a restriction of the $\beta$ parameters. This implies that only one of the three dreibeine can couple to the other two simultaneously. After choosing this dreibein to be $e_2{}^a$, we see that we have to take $\beta_{13} = \beta_{31} = 0$ and the $(k_1-k_3)^2$ term is not present in the above quadratic Lagrangian. Since we may always relabel the dreibein indices $I =1,2,3$, this implies that the specific ghost-free parameter choice does not influence the linear theory and that there are only two non-diagonal mass terms in \eqref{Lquad}. For simplicity, we will proceed the analysis of the linear theory by also taking $\beta_{21} =0$ and $\beta_{32} = 0$, since in the linear theory these two coupling constants can be absorbed in $\beta_{12}$ and $\beta_{23}$.

\subsubsection{Diagonalisation}
The above quadratic Lagrangian has diagonal kinetic terms for the fields $k_I$ and $v_I$, but it contains mass-terms for the differences $k_1 - k_2$ and $k_2-k_3$. Here we diagonalize the theory and write it in terms of a massless field and two massive fields. 
The first step is to define two new fields equal to the difference appearing in the mass terms
\begin{align}
f_1 = k_1 - k_2\,, && f_2 = k_2 - k_3\,, \\
w_1 = v_1 - v_2 \,, && w_2 = v_2 - v_3\,.
\end{align}
We also redefine the fields $k_2$ and $v_2$ as
\begin{equation}
\begin{split}
k_2 = k_{(0)} - \frac{\sigma_1}{\gamma_{\rm crit}} f_1 + \frac{\sigma_3\gamma_3 }{\gamma_{\rm crit}} f_2 \,, \\
v_2 = v_{(0)} - \frac{\sigma_1 }{\gamma_{\rm crit}} w_1 + \frac{\sigma_3\gamma_3}{\gamma_{\rm crit}} w_2 \,,
\end{split}
\end{equation}
where
\begin{equation}
\label{gammacrit}
\gamma_{\rm crit} = \sigma_1 + \sigma_2 \gamma_2 + \sigma_3 \gamma_3 \,.
\end{equation} 
Assuming $ \gamma_{\rm crit} \neq 0$, the quadratic Lagrangian becomes:
\begin{align}\label{Lquad2}
L^{(2)} & =  - \gamma_{\rm crit} M_P \left\{ k_{(0)} \cdot \bar{\cD}v_{(0)} + \frac12 \bar{e} \cdot \left( v_{(0)} \times v_{(0)} - \Lambda k_{(0)} \times k_{(0)} \right) \right\} \nonumber \\
& - \frac{\sigma_1 (\gamma_2 \sigma_2 + \gamma_3\sigma_3) M_P}{\gamma_{\rm crit}} \left\{ f_{1} \cdot \bar{\cD}w_{1} + \frac12 \bar{e} \cdot \left( w_{1} \times w_{1} - \Lambda f_{1} \times f_1 \right) \right\} \nonumber\\
& - \frac{\gamma_3\sigma_3 (\sigma_1 + \gamma_2\sigma_2 )M_P }{ \gamma_{\rm crit}} \left\{ f_{2} \cdot \bar{\cD}w_{2} + \frac12  \bar{e} \cdot \left( w_{2} \times w_{2} - \Lambda f_{2} \times f_2 \right) \right\}   \\
& - \frac{\gamma_3 \sigma_1 \sigma_3 M_P}{ \gamma_{\rm crit} } \left\{ f_{1} \cdot \bar{\cD}w_{2} + f_{2} \cdot \bar{\cD}w_{1}  +  \bar{e} \cdot \left( w_{1} \times w_{2} - \Lambda f_{1} \times f_2 \right) \right\} \nonumber \\
& - \frac12 m^2 M_P \gamma_2 \bar{e} \cdot  \left( \beta_{12}  f_1 \times f_1  + \gamma_2 \gamma_3 \beta_{23} f_2 \times f_2 \right) \,. \nonumber
\end{align}
We have now traded an off-diagonal mass-term for off-diagonal kinetic terms, but at least we are able to identify the Lagrangian for the massless spin-2 mode, which corresponds to the first line of \eqref{Lquad2}. 

We now diagonalize the massive part of the linearized Lagrangian, which corresponds to the last four lines of \eqref{Lquad2}. We redefine the fields $f_1, f_2,w_1$ and $w_2$ as a linear combination of two massive spin-2 modes
\begin{equation}
\begin{split}
f_1 =   k_{(\cM_1)} +   k_{(\cM_2)}\,, \qquad &
f_2 = A_+ k_{(\cM_1)} + A_- k_{(\cM_2)} \,, \\
w_1 =   v_{(\cM_1)} +   v_{(\cM_2)} \,, \qquad &
w_2 = A_+ v_{(\cM_1)} + A_- v_{(\cM_2)} \,.
\end{split}
\end{equation}
The massive part of \eqref{Lquad2} is diagonal if the (non-zero) dimensionless coefficients $A_{\pm}$ are given by
\begin{equation}
\begin{split}
A_{\pm} = & \; \frac12 \left(  \frac{\beta_{12}(\sigma_1 + \sigma_2 \gamma_2)}{\beta_{23} \sigma_1\gamma_2\gamma_3} - \frac{\gamma_2\sigma_2}{\gamma_3\sigma_3} - 1 \right) \\
& \pm \sqrt{\frac{\beta_{12}}{\beta_{23}\gamma_2\gamma_3} + \frac14 \left(  \frac{\sigma_1+\gamma_2\sigma_2}{\sigma_1\gamma_2\gamma_3} \frac{\beta_{12}}{\beta_{23}} -   \frac{\gamma_2\sigma_2 + \gamma_3\sigma_3}{\gamma_3 \sigma_3} \right)^2}\,.
\end{split}
\end{equation}
The quadratic Lagrangian factorizes into a part describing a linearized massless spin-2 plus 2 massive Fierz-Pauli Lagrangians. 
\begin{align}\label{Lquad3}
L^{(2)} = & - \gamma_{\rm crit} M_P \left\{ k_{(0)} \cdot \bar{\cD}v_{(0)} + \frac12  \bar{e} \cdot \left( v_{(0)} \times v_{(0)} - \Lambda k_{(0)} \times k_{(0)} \right) \right\}  \\
& + C_1 M_P\, L_{\rm FP}(k_{\cM_1}, v_{\cM_1},\cM_1) + C_2 M_P \, L_{\rm FP}(k_{\cM_2},v_{\cM_2},\cM_2) \,, \nonumber
\end{align}
where
\begin{equation}
L_{\rm FP}(k,v, \cM) = - \left\{ k \cdot \bar{\cD}v + \frac12 \bar{e} \cdot \left( v \times v - (\Lambda - \cM^2) k \times k \right) \right\}\,,
\end{equation}
is the Fierz-Pauli Lagrangian (see also \eqref{SFP}).

The two Fierz-Pauli masses, $\cM_1$ and $\cM_2$, belonging to the massive modes $k_{(\cM_1)}$ and $k_{(\cM_2)} $ respectively, are given by:
\begin{equation}
\begin{split}\label{DDGmass}
\cM_1^2 & = \frac{m^2 \gamma_{\rm crit} \gamma_2 \left( \beta _{12} + \gamma_2 \gamma_3 \beta _{23} A_+^2  \right)}{  \gamma_2\sigma_2 ( \sigma_1 + \sigma_3\gamma_3 A_+^2) + \sigma_1 \sigma_3 \gamma_3  \zeta (1 + A_+)^2   }\,, \\
\cM_2^2 & = \frac{m^2 \gamma_{\rm crit} \gamma_2 \left( \beta _{12} + \gamma_2 \gamma_3 \beta _{23} A_-^2  \right)}{  \gamma_2\sigma_2 ( \sigma_1 + \sigma_3\gamma_3 A_-^2) + \sigma_1 \sigma_3 \gamma_3  \zeta (1 + A_-)^2   }\,.
\end{split}
\end{equation}
The coefficients in front of the kinetic terms of the Fierz-Pauli Lagrangian are:
\begin{equation}
\begin{split}\label{DDGcs}
C_1 = \frac{  \gamma_2\sigma_2 ( \sigma_1 + \sigma_3\gamma_3 A_+^2) + \sigma_1 \sigma_3 \gamma_3  \zeta (1+A_+)^2 }{\gamma_{\rm crit}}\,, \\
C_2 =  \frac{  \gamma_2\sigma_2 ( \sigma_1 + \sigma_3\gamma_3 A_-^2) + \sigma_1 \sigma_3 \gamma_3  \zeta (1+A_-)^2 }{\gamma_{\rm crit}}\,.
\end{split}
\end{equation}
Finally, the dual central charge can be computed by analogy to the computation for ZDG \cite{Bergshoeff:2013xma} performed in chapter \ref{sec:AS_ZDG}. The result is
\begin{equation}\label{DDGcc}
c_{L/R} = 12\pi \ell M_P \gamma_{\rm crit} = \frac{3\ell}{2G}\gamma_{\rm crit} \,.
\end{equation}

\subsubsection{Critical lines and points}
The linear DDG theory contains critical lines and tricritical points where the central charge vanishes and the linearized Lagrangian is not diagonalizable. Consider the linear theory with $\sigma_3 =1$ and we tune
\begin{equation}
\gamma_{\rm crit} = 0\,,
\end{equation}
by parametrizing
\begin{equation}
\sigma_2 = - \frac{1}{\gamma_2}(\sigma_1 + \gamma_3)\,.
\end{equation}
This defines a set of critical lines for which (one of) the Fierz-Pauli masses \eqref{DDGmass} vanishes. Depending on the value of the following combination of parameters
\begin{equation}
A = \beta_{12} \gamma_3 + \beta_{23} \gamma_2 \sigma_1^2\,,
\end{equation} 
we find that
\begin{equation}
\begin{split}
\text{if} \quad A > 0 \,, & \qquad M_{+}^2 = \frac{A m^2 \gamma_2}{\sigma^2_1 + \gamma_3 \sigma_1} \,, \quad M_-^2 = 0\,, \\
\text{if} \quad A < 0 \,, & \qquad M_{+}^2 = 0 \,, \quad M_-^2 = \frac{A m^2 \gamma_2}{\sigma^2_1 + \gamma_3 \sigma_1}\,, \\
\text{if} \quad A = 0 \,, & \qquad M_+^2 = M_-^2 = 0\,. \\
\end{split}
\end{equation}
The last of these three equations defines a set of tricritical points. The full parameter space of ghost-free DDG is seven dimensional $(\sigma_1, \sigma_2, \alpha_1 ,\alpha_2, \alpha_3, \beta_{12}, \beta_{23} ) $ and moving to the tricritical point fixes two of the seven parameters, leaving five free parameters. Hence, the tricritical points found here are a generalization of the ENMG tricritical point found in chapter \ref{sec:ENMG_crit}.

\subsection{The Extended NMG Limit}
The Extended NMG theory obtained in section \ref{ExtendedNMG} has the same linear spectrum as ghost-free Drei-Dreibein Gravity, although the parameter space of the latter is larger. In fact, there exists a scaling limit, or a flow, from DDG to the ENMG theory presented in chapter \ref{ExtendedNMG}. Consider the DDG Lagrangian with the same ghost-free parameter choice as was discussed in the linear theory above:
\begin{equation}\label{LDDGlim}
\begin{split}
L_{\text{\tiny{DDG}}}= - M_P  \bigg\{ & \sigma_1 e_{1} \cdot R(\omega_1) + \sigma_2  e_{2} \cdot R(\omega_2) + \sigma_3 e_{3} \cdot R(\omega_3) \\ 
& + \frac{1}{6} m^2 \big( \alpha_1 e_1 \cdot e_1 \times e_1 +\alpha_2 e_2 \cdot e_2 \times e_2 + \alpha_3e_3 \cdot e_3 \times e_3 \\
&  -3\beta_{12} e_1 \cdot e_1\times e_2 - 3\beta_{23} e_2 \cdot e_2 \times e_3 \big) \bigg\}\,.
\end{split}
\end{equation}
We introduce the following parametrization as an expansion in $\lambda$ for the three dreibeine and the spin connections,
\begin{eqnarray}\label{limit}
  \begin{split}
    e_1 &= e\,,\\
    e_2 &= e +\tfrac{\lambda}{m^2} f_1\,,\\
    e_3 &= a_{31}\,e+a_{32}\tfrac{\lambda}{m^2}  f_1 +\tfrac{\lambda^2}{m^4}  f_2\,,
  \end{split}
\qquad\qquad
  \begin{split}
    \omega_1 &= \omega\,,\\
    \omega_2 &= \omega +\tfrac{\lambda}{m^2}  h_1\,,\\
    \omega_3 &= \omega +\tfrac{a_{32}}{a_{31}}\tfrac{\lambda}{m^2}  h_1+ \tfrac{1}{a_{31}}\tfrac{\lambda^2}{m^4}  h_2\,,
  \end{split}
\end{eqnarray}
where
\begin{equation}
a_{31}=-\frac{(a-b)^2}{(a-2b)}\,, \qquad \text{and} \qquad a_{32}= -\frac{(a-b)b}{(a-2b)}\,.
\end{equation}
We take the Planck mass and the $\sigma_I$ parameters as
\begin{equation}
M_P= \frac{M}{\lambda^2}\,, \qquad \sigma_1 = a - b + \lambda^2 \sigma \,,\qquad \sigma_2 = \frac{(a-b)b}{(a-2b)}\,,\qquad \sigma_3 = 1\,,
\end{equation}
where $\sigma=\pm1$ is a new sign parameter. The cosmological parameters and the two coupling constants are expanded as
\begin{align}
& \alpha_1 = -2 - \frac{2(a-b)}{\lambda}\,, \qquad
\alpha_2 = - 1 + \frac{(a-b)(a^2-2b^2)}{(a-2b)^2\lambda} \,,\qquad 
\alpha_3 = - \frac{(a-2b)}{(a-b)^3\lambda}\,, \nonumber \\
&\beta_{12} = \frac{\Lambda_0\lambda^2}{3 m^2 }- 1 -\frac{a-b}{\lambda} \,, \qquad 
\beta_{23} = -\frac{a-b}{(a - 2b)\lambda}\,.
\end{align}
After plugging this into the Lagrangian \eqref{LDDGlim} and taking the limit  $\lambda\rightarrow0$ we arrive at the Extended New Massive Gravity action given in \eqref{LENMG}.

In order to take the limit in the DDG central charge to find the central charge of the Extended NMG (ENMG) theory we need to know how $\gamma_2$ and $\gamma_3$ scale with $\lambda$. We can deduce this from the parameter relations \eqref{gammas} which in this case read
\begin{align}\label{gammas2}
\sigma_1 \frac{\Lambda}{m^2}  & =   2\beta_{12}\gamma_2 - \alpha_1  \,, \nonumber \\
 \sigma_2 \frac{\Lambda}{m^2}  &=   \beta_{12} + 2\beta_{23}\gamma_2\gamma_3 - \gamma_2^2 \alpha_2\,, \\
 \sigma_3 \frac{\Lambda}{m^2} &=  \beta_{23}\gamma_2^2 - \alpha_3\gamma_3^2 \,. \nonumber
\end{align}
If we expand the cosmological constant as $\Lambda = - 1/\ell^2 + \Lambda^{(1)} \lambda + \cO (\lambda^2)$, then it is possible to solve these equations order by order in $\lambda$. The result is
\begin{equation}
\gamma_2 =1 + \frac{1}{2\ell^2 m^2 } \lambda - \left( \frac{1}{2(a - b)m^2\ell^2} + \frac{\Lambda^{(1)}}{2m^2}\right) \lambda^2  +\mathcal{O}(\lambda^3)\,,
\end{equation}
\begin{equation}
\begin{split}
\gamma_3 & = -\frac{( a- b)^2}{(a - 2b)} - \frac{(a - b) b}{2 (a -2 b) \ell^2 m^2} \lambda \\
& \qquad - \left( \frac{a}{8\ell^4m^4} - \frac{(a-b)}{2(a-2b)\ell^2m^2} - \frac{(a-b)b\Lambda^{(1)}}{2(a-2b)}\right)\lambda^2 +\mathcal{O}(\lambda^3)\,.
\end{split}
\end{equation}
Taking the $\lambda \to 0$ limit in the DDG central charge \eqref{DDGcc}, we find that
\begin{equation}
c_{L/R} = \frac{3 \ell}{2G} \left( \sigma +\frac{1}{2\ell^2m^2}-\frac{a}{8\ell^4m^4} \right) \,.
\end{equation}
Which agrees with the result \eqref{ENMGcc}.

\subsection{Positive Central Charge and the Absence of Ghosts in DDG}
A problem for the higher-derivative extensions of New Massive Gravity was the presence of massive spin-2 ghosts. The linear theory of DDG contains the same spectrum as ENMG, however, the original theory does not contain higher derivatives and the parameter space is larger. This gives the additional freedom to achieve positive energy for the massive modes consistent with a positive central charge.

For DDG with the parameter choice $\beta_{12} \neq 0$ and $\beta_{23} \neq 0$ the bulk requirements are:
\begin{equation}\label{restrict1}
C_1, C_2, \cM_1^2, \cM_2^2 > 0\,.
\end{equation}
Positivity of the dual central charge restricts the DDG parameters as
\begin{equation}\label{restrict2}
\gamma_{\rm crit} = \sigma_1 + \gamma_2 \sigma_2 + \gamma_3 \sigma_3 > 0\,.
\end{equation}
The theory has seven free parameters (two of the $\sigma$'s, three $\alpha$'s and two $\beta$'s) which have to satisfy the five inequalities \eqref{restrict1} and \eqref{restrict2}. This is possible and an explicit choice satisfying all inequalities is:
\begin{align}
 & \beta_{12}  = A > 0\,, && \beta_{23}  = B > 0\,, && \sigma_I  = 1\,, \quad \text{for: }   I = 1,2,3\,,\nonumber \\
& \alpha_1 m^2  = \zeta + 2 m^2 A \,, &&  \alpha_2 m^2  = \zeta + m^2 (A + 2 B) \,, & &\alpha_3 m^2  = \zeta + m^2 B\,.
\end{align}
Where here $\zeta >0 $. For this choice of parameters there is an AdS vacuum satisfying \eqref{gammas2} with $\gamma_2 = \gamma_3 =1 $ and an AdS length $\ell = 1/\sqrt{\zeta}$. Moreover, like in ZDG, any nearby DDG model also satisfies the restrictions \eqref{restrict1}-\eqref{restrict2}, since these inequalities are satisfied, but not saturated. Hence there is a continuous range of parameters defining a class of DDG models with good bulk and boundary properties.

\section{Viel-Dreibein Gravity}
\label{sec:VDG}
The constraint analysis of Drei-Dreibein Gravity can be extended to a theory with an arbitrary number of interacting dreibeine. From the DDG analysis, we know that interaction terms mixing more than two dreibeine do not lead to secondary constraints. To investigate the secondary constraints in a Viel-Dreibein theory with $N$ interacting dreibeine, we only consider interaction terms involving two dreibeine and define the Lagrangian as:
\begin{equation}\label{VDG}
L = - M_P \sum_{I=1}^{N}  \left( \sigma_I \, e_{I} \cdot R (\omega_I) + \frac16 m^2 \alpha_I e_I \cdot e_I \times e_I \right) + L_{\rm int}\,,
\end{equation}
where
\begin{equation}
L_{\rm int} = \,\, \frac12 m^2 M_P \sum_{I=1}^N \sum_{J \neq I}^{N} \beta_{IJ} e_{I} \cdot e_{I} \times e_{J}\,.
\end{equation}
From the Bianchi and Cartan identities ( $\cD_I R_I =0$, and $\cD_I T_{I} = R_{I} \times e_{I}$) we can derive secondary constraints if we can use the inverse of some of the dreibeine to construct two-form equations of them. But let us first investigate the number of secondary constraints which are needed to remove the additional degrees of freedom.

The theory should describe $N-1$ massive spin-2 modes and one massless mode. The desired number of degrees of freedom is thus $2(N-1)$; two for every massive mode, or a physical phase space of dimension $4(N-1)$. After a space-time decomposition of the fields, the dynamical phase space consists out of the $12N$ components of the spatial parts of the dreibeine and the spin connections. The time components of the fields act as Lagrange multipliers for $6N$ primary constraints, out of which 6 are first class, corresponding to the diagonal gauge symmetries of the theory. In the absence of secondary constraints a counting of the physical phase-space gives:
\begin{equation}\label{VDGcount}
12N  - 6N - 6 = 6(N-1)\,.
\end{equation}
To arrive at $4(N-1)$, we need to derive $2(N-1)$ secondary constraints from the Bianchi and Cartan identities. Let us first consider the Cartan identities for this model. They are $N$ equations, each containing $N-1$ terms
\begin{equation}
\begin{split} \label{VDGCartan}
 (\beta_{12} e_1 + \beta_{21} e_2)e_1 \cdot e_2 + (\beta_{13} e_1 + \beta_{31} e_3) & e_1 \cdot e_3  + \ldots  \\
\ldots + & (\beta_{1N}e_{1} + \beta_{N1}e_{N})e_1 \cdot e_N  = 0\,,  \\[.2truecm]
(\beta_{12} e_1 + \beta_{21} e_2)e_2 \cdot e_1 +  (\beta_{23} e_2 + \beta_{32} e_3)& e_2 \cdot e_3  + \ldots  \\
\ldots  + & (\beta_{2N}e_{2} + \beta_{N2}e_{N})e_2 \cdot e_N = 0\,,  \\
&\;\; \vdots \\  
(\beta_{1N}e_1 + \beta_{N1}e_{N}) e_N \cdot e_1   +
\ldots + (\beta_{NN-1}  e_{N}  & +  \beta_{N-1N} e_{N-1}) e_N \cdot e_{N-1}  = 0\,.
\end{split}
\end{equation}
where the uncontracted form fields have a free Lorentz index.
To derive $N-1$ secondary constraints from \eqref{VDGCartan}, we need to constrain the parameters such that $N-1$ of these involve solely an invertible combination of dreibeine carrying a free Lorentz index (the terms in the parenthesis). Then we can use the inverse of this combination of dreibeine to construct a two-form equation, whose spatial projection is a secondary constraint. In other words, out of the $\frac12 N(N-1)$ combinations
\begin{equation}\label{VDGlincom}
\beta_{IJ} e_I + \beta_{JI} e_J\, \qquad \text{with } \qquad I \neq J\,,
\end{equation}
we need that $N-1$ have an inverse, while the others all vanish. 

We now turn our attention to the $N$ Bianchi identities $\cD_I R_I = 0$. They can be written, in full generality, as
\begin{align}\label{VDGBianchi} \nonumber
& \beta_{12} \omega_{12}\, e_1 \cdot e_2 + \beta_{13} \omega_{13} \, e_1 \cdot e_3 + \ldots + \beta_{1N} \omega_{1N}\, e_1 \cdot e_N + e_2\, (\beta_{12} e_1 + \beta_{21} e_2) \cdot \omega_{12}  \\ \nonumber
&  + e_3\, (\beta_{13} e_1 + \beta_{31} e_3) \cdot \omega_{13} +  \ldots  + e_N\, ( \beta_{1N} e_1 + \beta_{N1} e_N) \cdot \omega_{1N} = 0 \,, \\[.3truecm] \nonumber
& \beta_{21} \omega_{21}\, e_2 \cdot e_1 + \beta_{23} \omega_{23}\, e_2 \cdot e_3 + \ldots + \beta_{2N} \omega_{2N}\, e_2 \cdot e_N  + e_1\, (\beta_{12} e_1 + \beta_{21} e_2) \cdot \omega_{21}  \\ \nonumber
& + e_3\, (\beta_{32} e_3 + \beta_{23} e_2) \cdot \omega_{23} + \ldots  + e_N\, ( \beta_{2N} e_2 + \beta_{N2} e_N) \cdot \omega_{2N} = 0 \,, \\
 & \qquad \qquad \qquad \qquad \qquad \qquad \qquad \qquad \qquad 
 \vdots \\ \nonumber
& \beta_{N1} \omega_{N1}\, e_N \cdot e_1 + \ldots + \beta_{NN-1} \omega_{NN-1}\, e_N \cdot e_{N-1}  \\ \nonumber
& + e_1\, (\beta_{1N} e_1 + \beta_{N1} e_N) \cdot \omega_{N1} + \ldots  + e_{N-1}\, ( \beta_{N-1N} e_{N-1} + \beta_{NN-1} e_N) \cdot \omega_{NN-1} = 0 \,.
\end{align}
For simplicity, we will restrict our attention here to a special case of the assumption \eqref{VDGlincom} where only a single dreibein is assumed to be invertible. We may pick the invertible dreibein to be $e_1$ without loss of generality. There is then a single parameter choice for which eqn.~\eqref{VDGCartan} leads to secondary constraints. This is 
\begin{equation}\label{VDGparam}
\beta_{1J} \neq 0 \, \qquad \text{for } \qquad J = 2, \ldots, N\,,
\end{equation}
and all other $\beta$-parameters vanish. The theory graph for this model is a tree where all edges are arrows pointing away from $e_1$.  After this parameter restriction, $N-1$ secondary constraints can be derived by acting with $e_1^{-1}$ on \eqref{VDGCartan} and taking the spatial part. They are
\begin{equation}
\label{VDGconstraints1}
\varepsilon^{ij} e_{1\,i} \cdot e_{J\,j} = \Delta^{e_1e_J} = 0\, \qquad \text{for } \quad J = 2, \ldots, N\,.
\end{equation}
The parameter constraint \eqref{VDGparam} and the identities $e_1 \cdot e_J = 0$ reduce the set of Bianchi identities \eqref{VDGBianchi} to:
\begin{equation}
\begin{split}
\beta_{12} e_2\, e_1 \cdot \omega_{12} + \ldots + \beta_{1N} e_N e_1 \cdot \omega_{1N} & = 0\,, \\[.2truecm]
\beta_{12}  e_1\, e_1 \cdot \omega_{21} & = 0\,, \\[.2truecm]
\beta_{13}  e_1\, e_1 \cdot \omega_{31} & = 0 \,, \\
 & \vdots \\
\beta_{1N} e_1\, e_1 \cdot \omega_{N1} & = 0 \,.
\end{split}
\end{equation} 
The assumed invertibility of $e_1$ is sufficient to derive another $N-1$ secondary constraints. They are:
\begin{equation}\label{VDGconstraints2}
\varepsilon^{ij} e_{1\,i} \cdot (\omega_{1\,j} - \omega_{J\,j}) = 0 \qquad \text{for} \quad J = 2, \ldots, N\,.
\end{equation}
Provided that the secondary constraints \eqref{VDGconstraints1} and \eqref{VDGconstraints2} are second-class, these remove an additional $2(N-1)$ components from the counting performed in \eqref{VDGcount}, leading to a physical phase space of dimension $4(N-1)$, corresponding to $2(N-1)$ degrees of freedom, the correct number of degrees of freedom to account for the two helicity states of $N-1$ massive spin-2 modes. 

In a similar fashion, secondary constraints can be derived when assuming more or all dreibeine to be invertible, or when assuming $N-1$ of the linear combinations in \eqref{VDGlincom} to be invertible. In the latter case, to derive the necessary constraints from the Bianchi identities \eqref{VDGBianchi}, we need that any dreibein which couples to more than one other dreibein should have an inverse. In other words, any internal node in the theory graph should correspond to an invertible dreibein.

\begin{figure}
\centering
\subfigure[$N=4$]{
\includegraphics[height=0.14\textwidth]{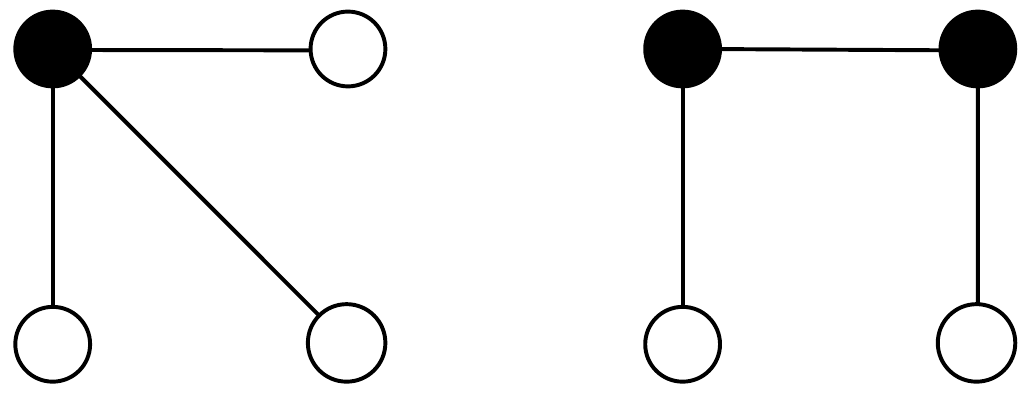}\label{fig:vdgtrees} 
} \\
\subfigure[$N=5$]{\includegraphics[height=0.14\textwidth]{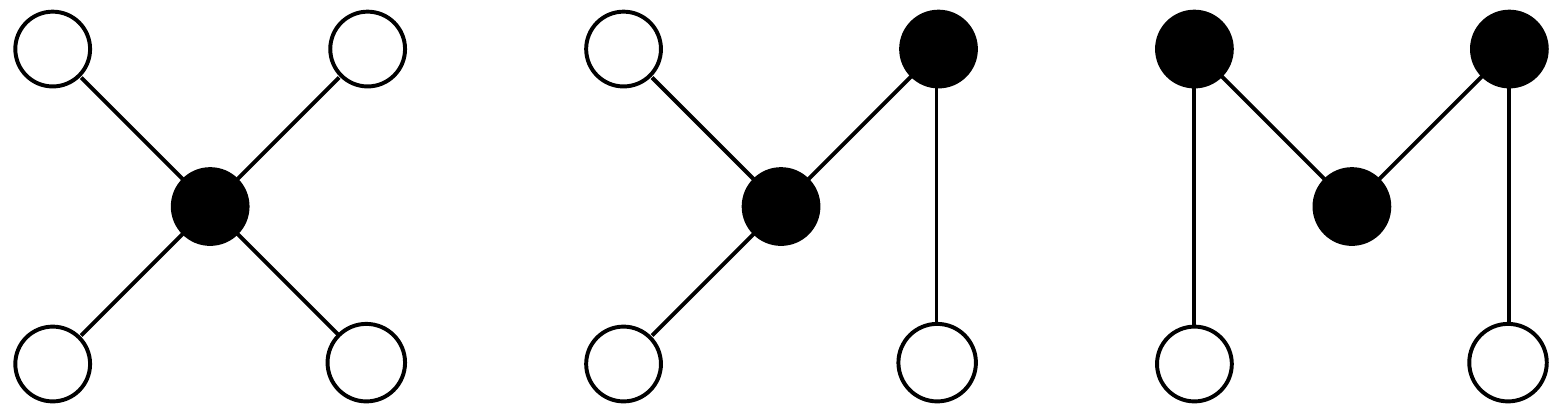}\label{fig:fdgtrees}}
\caption{The ghost-free theory diagrams for $N=4$ (a) and $N=5$ (b). In general the ghost-free VDG theory diagram is a tree where all the edges correspond to invertible linear combinations of frame fields and nodes with degree $\geq 2$ (internal nodes) correspond to invertible frame fields.} \label{fig:dreibein3}
\end{figure}

To conclude, in all of the ghost-free viel-dreibein gravity models, the parameters of the theory must be restricted such that in a theory diagram similar to figure \ref{fig:dreibein2}, but now containing $N$ nodes, the following rules hold
\begin{itemize}
\item Interaction terms with more than two dreibeine are not allowed (the theory diagram is a graph).
\item There must be exactly $N-1$ lines connecting the nodes and no node should be disconnected. (The theory graph is a tree)
\item Any internal dreibein (coupling to more than one other dreibein) should have an inverse (Nodes of degree $\geq 2$ must correspond to invertible frame fields).
\end{itemize}
The second point excludes the possibility to form loops in the diagrams. The mathematical tool to display these allowed diagrams is the notion of unoriented unlabeled trees; the diagrams for $N=4,5$ are shown in figure \ref{fig:dreibein3}.

Whenever the original dreibeine are assumed to be invertible, and not necessarily their linear combinations, then the interaction terms are restricted even further and the number of ghost-free VDG models equals  the number of different oriented trees with $N$ unlabeled nodes.\footnote{The number of unoriented and oriented trees with $N=1,2,3,\cdots$ unlabeled nodes, is
$1, 1, 1, 2, 3, 6, \cdots$ and $1,1,3,8,27,91,\cdots$ respectively --- see \href{http://oeis.org/A000055}{A000055} and \href{http://oeis.org/A000238}{A000238} at ``The On-Line Encyclopedia of Integer Sequences''.}.

Any multi-gravity theory constructed in this way will have the correct number of degrees of freedom to describe $N-1$ massive gravitons interacting with each other, provided that the secondary constraints derived here are also second-class and that the consistency conditions of the secondary constraints do not introduce any further tertiary constraints \cite{Bergshoeff:2014bia}. This has been checked explicitly for VDG with $N=2$ (ZDG) in chapter \ref{sec:ZDG_Hamil} and $N=3$ (DDG) in the appendix \ref{app:Hamil}.

\section{Discussion}

In this chapter we have discussed several Chern-Simons--like models with more than four fields which propagate multiple massive spin-2 modes in three dimensions. The theories discussed here can be divided into two classes. The first class of models we have discussed are CS--like models for higher-derivative extensions of general relativity containing a finite number of higher-derivative terms. We have shown how to derive higher-derivative terms which give a scalar ghost-free extension of GR, however, the theory does contain unphysical massive spin-2 modes. These unphysical modes disappear at special critical lines and points in the parameter space, a subject we will revisit in the next chapter. 

The scalar ghost-free terms found here are unique at each order in $1/m^2$, however, it is possible to find other theories which extend GR by (scalar ghost-free) higher-derivative terms. For instance, if we supplement the NMG Lagrangian in a CS--like form with a $\tfrac{1}{m^4} f \cdot f \times f$ term, then the solution of the auxiliary field $f$ becomes an infinite expansion in $1/m^2$, each term containing a higher power of $S^n$ contributions (with $S$ we mean the Schouten tensor). The resulting theory, while maintaining the two degrees of freedom of NMG, involves an infinite series of $R^n$ terms and is scalar ghost-free by construction. At a specific value for the free parameter of the new interaction term, the theory becomes equivalent to a Born-Infeld extension of NMG \cite{Gullu:2010pc,Gullu:2010st}, as we will discuss in chapter \ref{sec:discussions}.

On the other hand, we discussed extensions of Zwei-Dreibein Gravity to models with three or more interacting dreibeine. For these models it is possible to describe physical spin-2 modes, together with a positive central charge for the boundary CFT. This fact, together with the existence of AdS$_3$ vacua, suggests that the BTZ black hole solutions in these models will also have positive mass. In contrast to the higher-derivative extensions of GR, the DDG and VDG models discussed here fall into the class of theories which do not have an action in terms of a single metric. This is much like MMG and ZDG, as was discussed in previous chapters. In fact, in DDG, it is possible to solve the equations of motion for two of the three dreibeine in terms of the third, invertible, dreibein in a procedure much like in ZDG (see chapter \ref{section:ZDGHD}). This will result in a field equation involving a single dreibein and an infinite series of higher-derivative contributions. At the level of the action, this procedure is hardly possible, which is a sign that these theories are fundamentally different from the `conventional' gravitational theories; even though their field equation may depend on a single metric, the action requires the use of auxiliary fields. This novel property poses new difficulties, such as the question of how to couple them to matter, but it also introduces new possibilities, like the resolution of the `bulk-boundary clash'.

%% file: chapter_8/chapter_8.tex
\pagestyle{empty}
\setcounter{chapter}{7}

\chapter[The AdS/LCFT correspondence]{The AdS/LCFT correspondence}
\label{chapter:AdS_LCFT}

\pagestyle{headings}

\begin{quote}\em
Most of the Chern-Simons--like models that we have encountered so far all share a special property which we have not yet discussed. At special points in the parameter space, the mass of the spin-2 mode vanishes and the linearized Lagrangian cannot be diagonalized. In this chapter we will investigate these special points in more detail. We will show that in critical ZDG the massive mode is replaced by a logarithmic mode, which, together with the pure gauge mode, forms a Jordan cell of rank 2 with respect to the Hamiltonian. This leads to the conjecture that ZDG, at certain critical points, is dual to a logarithmic CFT (LCFT). We lend further support for this conjecture by showing that the non-linear theory contains logarithmic pp-wave solutions at the critical points, using the higher-derivative formulation of ZDG discussed in \ref{section:ZDGHD}. After this, we study the sixth-order extended NMG (ENMG) model of chapter \ref{ExtendedNMG} at a tricritical point where both massive modes become massless. The corresponding log and log$^2$ modes form a Jordan cell of rank 3 with the pure gauge modes and the conjectured dual LCFT is of rank 3. The results with respect to critical ZDG were obtained in \textsc{[vii]} and the section dealing with tricritical ENMG is adapted from \textsc{[iv]} and \textsc{[x]}. 
\end{quote}

\newpage
\section{Introduction}

Most of the Chern-Simons--like theories discussed in this thesis share a common feature. At a specific point or range of points in the parameter space the mass of the spin-2 mode vanishes. At this point the massive mode degenerates with the pure gauge modes in the theory. The field equations, which are typically higher-order differential equations, still have the same number of solutions and hence new solutions to the linearized equations of motion arise. These new solutions have different asymptotics towards the AdS$_3$ boundary; in Poincar\'e coordinates they fall of like $\log r$ and hence they are called logarithmic modes. They were first found in TMG at the chiral point in \cite{Grumiller:2008qz}. As the mass of the spin-2 mode goes to zero, the weights characterizing the conformal dimension of the operators dual to the massive modes degenerates with those of the stress-energy tensor of the dual conformal field theory (CFT).  This led to the conjecture that critical TMG (sometimes called Log Gravity \cite{Maloney:2009ck}) is dual to a logarithmic CFT\footnote{Gravitational duals for logarithmic CFTs\cite{Gurarie:1993xq} were studied before in the contexts of singletons in AdS backgrounds \cite{Kogan:1999bn,Kogan:2002mg} and in theories with a set of degenerate scalar fields on AdS \cite{Ghezelbash:1998rj}.} (LCFT) \cite{Grumiller:2008es,Skenderis:2009nt,Grumiller:2009mw}. Soon afterwards, logarithmic modes were also found in NMG at a critical point and in its higher-dimensional analogues, see for instance \cite{Liu:2009kc,Liu:2009pha,Grumiller:2009sn,Lu:2011zk,Deser:2011xc,Alishahiha:2011yb,Bergshoeff:2011ri,Porrati:2011ku,Lu:2011ks}. See \cite{Grumiller:2013at} for a recent review on the AdS/LCFT correspondence in critical gravity theories. 

In the next section we will show explicitly how these logarithmic modes arise in critical gravity theories, but let us first comment briefly on the nature of LCFTs and their difference from `ordinary' CFTs. For a more detailed review on LCFTs and references we refer to \cite{Gurarie:1993xq,Flohr:2001zs,Gaberdiel:2001tr}.

Conformal Field Theories are invariant under scale transformations that preserve angles. The two-point correlation functions are restricted by conformal symmetry to behave like:
  \begin{equation}\nonumber
  \vev{\cO_i(x) \cO_j(y)} = \delta_{ij} \frac{c}{|x-y|^{2\Delta}}\,,
  \end{equation}
 and there exists a Hamiltonian operator that is diagonalizable, i.e. the operators of the CFT are eigenstates of the energy operator $H$:
  \begin{equation}\nonumber
  [H,\cO_i] = E_0 \cO_i\,.
  \end{equation}
Logarithmic CFT's arise when two operators degenerate in all quantum numbers. The two operators will form an indecomposable, but non-irreducible representation of the CFT symmetry algebra (Jordan cells). The Hamiltonian no longer acts diagonally on the operators, but instead:
  \begin{equation}
    [H,\cO^{\log}_i] = E_0 \cO^{\log}_i + \cO_i \,, \quad
    [H,\cO_i] = E_0 \cO_i\,,
  \end{equation}
where $\cO^{\log}_i$ is called the logarithmic partner of $\cO_i$. The two-point correlation functions will now involve logarithmic terms:
  \begin{subequations}
  \begin{align}
 & \vev{\cO_i(x) \cO_j(y)}  = 0 \,,  \label{lcft1} \\
 & \vev{\cO^{\log}_i(x) \cO_j(y)}   = \delta_{ij} \frac{b}{|x-y|^{2\Delta}}\,, \label{lcft2} \\
 & \vev{\cO^{\log}_i(x) \cO^{\log}_j(y)} = \delta_{ij} \frac{1}{|x-y|^{2\Delta}}(- 2 b \log |x-y| + \lambda) \,. \label{lcft3}
 \end{align}
  \end{subequations}
The rank of the LCFT determines the amount of logarithmic partner operators. Typically, this can be any integer $n+1$ and the corresponding correlation functions will then contain log$^n$-terms. The two-point functions above are again fixed by conformal symmetry, up to the constant $b$, which is called the new anomaly. 

In this chapter, we will argue that the existence of critical points in the parameter space at which logarithmic modes appear in the linearized spectrum, is not only a feature of the higher-derivative gravity models, but also of the recently introduced ZDG model.  We will show this first at the linearized level. We will in particular show that there exist regions in the ZDG parameter space, where the massive gravitons become massless and thus coincide with pure gauge modes. We will show that instead logarithmic modes appear in the linearized spectrum, that behave similarly to the analogous modes found in higher-derivative critical gravities. The existence of these modes can be seen as a serious hint that  ZDG can be added to the class of gravity theories that are, at specific parameter values, dual to LCFTs. We will argue that the dual LCFTs have zero central charges and we will calculate the value of the new anomalies on the gravity side via a procedure outlined in \cite{Grumiller:2010tj}.

We will also confirm that the existence of these logarithmic modes is not an artefact of the linearized approximation, but that there exist solutions of the full theory, that exhibit this logarithmic fall-off behavior. This lends further credibility to the idea that the ZDG parameter space features critical points at which the theory can be dual to LCFTs. The non-linear solutions we will discuss are AdS waves, that are the ZDG analogue of the solutions discussed in \cite{AyonBeato:2004fq,Carlip:2008eq,AyonBeato:2009yq}. In order to find these solutions, we will use the fact that bimetric theories, such as ZDG, can alternatively be thought of as higher-derivative theories for a single metric field \cite{Hassan:2013pca}, as was discussed in detail in chapter \ref{section:ZDGHD}. This allows us to explicitly find AdS waves that exhibit logarithmic asymptotic behavior at critical points. 

After treating critical ZDG, we will turn to the sixth-order extension of general relativity discussed in chapter \ref{ExtendedNMG} at its critical line and tricritical point. This theory contains two massive gravitons, which both become massless at a tricritical point in the parameter space. At this point, the theory is conjectured to be dual to a rank-3 LCFT, much like the sixth-order derivative theories discussed in \cite{Bergshoeff:2012ev} and we compute the values of the new anomalies.

\section{Critical ZDG}

In chapter \ref{sec:linZDG} we have discussed Zwei-Dreibein Gravity linearized around a maximally symmetric background. For general values of the parameters, we found that the linear theory contains two pure gauge modes and two massive spin-2 modes. We did not, however, analyse the theory at special, critical points in its parameter space, for which the diagonalization procedure fails. At these critical points the massive modes become massless and are replaced by logarithmic solutions as we will show below.

Let us first recap some useful formulae obtained in earlier chapters. In chapter \ref{sec:ZDG_Hamil} we analysed the absence of additional, non-linear degrees of freedom in ZDG and found that assuming the invertibility of the linear combination $\beta_1 e_1{}^a + \beta_2 e_2{}^a$ was a sufficient and necessary condition to achieve this. In this chapter we will focus on a special case of this assumption, where we take $\beta_2=0$ (while $\beta_1 \neq 0$) and require that $e_1{}^a$ has an inverse. 

The quadratic Lagrangian for fluctuations around an AdS$_3$ background \eqref{lin_L} is then given by:
\begin{align} \label{lin_Lc8}
\cL^{(2)} =  - \sigma M_P & \left[k_{1} \cdot \bar{\cD}v_{1} + \frac12 \bar{e} \cdot \left(v_{1} \times v_{1} - \Lambda k_{1} \times k_{1}\right) \right] \nonumber \\
 - \gamma M_P & \left[k_{2} \cdot \bar{\cD}v_{2} + \frac12 \bar{e} \cdot \left(v_{2} \times v_{2} - \Lambda k_{2} \times k_{2}\right) \right] \\
\nonumber  & - \frac12 m^2 \gamma \beta_1M_P \bar{e} \cdot \left( k_{1}-  k_{2}\right)\times \left(  k_{1} -  k_{2}\right) \,.
\end{align}
Provided $\sigma + \gamma \neq 0$, this Lagrangian can be diagonalized by performing the linear field redefinition:
\begin{equation} 
\begin{split}
(\sigma + \gamma)  k_{+}{} &  = \sigma  k_{1} +  \gamma  k_{2}\,, \qquad \qquad
k_{-}   =  k_{1} -  k_{2}\,, \\
(\sigma + \gamma) v_{+}  & = \sigma v_{1} +  \gamma  v_{2} \,, \qquad \qquad
v_{-}  =  v_{1} -  v_{2} \,.
  \label{lin_redefc8}
\end{split}
\end{equation}
In terms of these fields the linearized Lagrangian becomes:
\begin{align} \label{linear_bi_lagrangianc8}
\cL^{(2)}   =  - (\sigma + \gamma) M_P & \left[ k_{+} \cdot \bar{\cD} v_{+} + \frac12  \bar{e} \cdot \left( v_{+} \times v_{+} - \Lambda k_{+} \times k_{+} \right) \right]  \\[.2truecm]
  - \frac{\sigma \gamma }{(\sigma + \gamma)} M_P & \bigg[ k_{-} \cdot \bar{\cD} v_{-} + \frac12  \bar{e} \cdot \left( v_{-} \times v_{-} -( \Lambda - \cM^2) k_{-} \times k_{-} \right)   \bigg]\,,
\nonumber\end{align}
where the Fierz-Pauli mass $\cM$ is given in terms of the ZDG parameters as:
\begin{equation}\label{MFPc8}
\cM^2 =  m^2 \beta_1\frac{\sigma + \gamma}{\sigma} \,.
\end{equation}

\subsection{Critical Points}
The diagonalization described above fails when $ \sigma + \gamma = 0$ and the Fierz-Pauli mass in eq.~\eqref{MFPc8} vanishes, along with the central charge \eqref{ccZDG}. This corresponds to a critical point\footnote{In fact, there is a line of critical points. Indeed, for $\sigma + \gamma = 0,$ the parameter relations \eqref{cc12} reduce to $\alpha_1 = -\sigma \left(2 \beta_1 + \Lambda/m^2\right)$ and $\alpha_2 = \beta_1 - \Lambda/m^2$. For a given value of the cosmological constant, there is thus a free parameter $\beta_1$ left. In the following, we will however keep on using the terminology `critical point', often using the plural form to emphasize that there is a continuous family of critical points in ZDG.} in the ZDG parameter space, where logarithmic modes appear, as we will now show. 

In terms of the fields
\begin{equation}
\begin{split}
k_-  & = m^2 \beta_1 \left( k_{1} -  k_{2} \right) \,, \qquad \qquad
 k_{+}  =  k_{1} +  k_{2}\,, \\ 
v_{-}  & = m^2 \beta_1  \left( v_{1} - v_{2} \right) \,, \qquad \qquad
v_{+}  = v_{1} +  v_{2} \,,
\end{split}
\end{equation}
the Lagrangian \eqref{lin_Lc8} becomes:
\begin{align} \label{crit_bi_lagrangian}
\cL^{(2)}   =  \frac{M_P}{2m^2 \beta_1 } \big(  & k_{+} \cdot \bar{\cD} v_{-} +  k_{-} \cdot \bar{\cD} v_{+}  + \bar{e} \cdot \left( v_{+} \times v_{-} - \Lambda k_{+} \times k_{-} \right)
\\ \nonumber & -  \bar{e} \cdot k_- \times k_{-} \big)\,.
\end{align}
This Lagrangian corresponds to the first-order form of the Lagrangian for linearized critical NMG, where the massive modes degenerate with the massless ones and logarithmic solutions appear \cite{Liu:2009kc,Bergshoeff:2011ri}. The only difference with the critical NMG case is the appearance of the coupling constant $\beta_1$ as an overall factor.

The equations of motion derived from the Lagrangian density \eqref{crit_bi_lagrangian} are given by:
\begin{align}\label{crit_eom}
& \bar{\cD}v_{-} - \Lambda  \bar{e} \times k_{-}  = 0\,, \nonumber \\
& \bar{\cD}v_{+} - \Lambda  \bar{e} \times k_{+}  = 2 \bar{e} \times k_{-}\,, \\
& \bar{\cD}k_{\pm} + \bar{e} \times v_{\pm}  = 0\,. \nonumber
\end{align}
The last of these equations can be used to express $v_{\pm}$ in terms of $k_{\pm}$ in the usual way. One finds (see also \eqref{vsol}):
\begin{equation}\label{vpm}
v_{\pm\,\mu}{}^a(k_{\pm}) = -\det(\bar{e})^{-1}\varepsilon^{\nu\rho\sigma} \left( \bar{e}_{\sigma}{}^a \bar{e}_{\mu\,b} - \frac12 \bar{e}_{\mu}{}^a \bar{e}_{\sigma\, b} \right) \bar{\cD}_{\nu} k_{\pm\,\rho}{}^b \,.
\end{equation}
Furthermore, by acting on the equations \eqref{crit_eom} with $\bar{\cD}$ and using the identity $\bar{\cD} \bar{\cD} f^a = (\bar{R} \times f)^a =  \Lambda \bar{e}^a\, \bar{e} \cdot f $ one can derive the constraints:
\begin{equation}
\bar{e}^a \bar{e} \cdot k_{-} = 0\,, 
\end{equation}
which imply that the field $k_{-\,\mu\nu}= k_{-\,\mu\,a}\bar{e}_{\nu}{}^a$ is symmetric:
\begin{equation}
	 k_{-\,[\mu\nu]}=0\,. \label{lin_symm_constr}
\end{equation}
Plugging \eqref{vpm} into the first two equations of \eqref{crit_eom} and writing them with free space-time indices we obtain:
\begin{align}\label{lineom1}
& \mathcal{G}_{\mu\nu}(k_-) = 0\,, \\
& \mathcal{G}_{\mu\nu}(k_+) = k_{-\mu\nu} - \bar{g}_{\mu\nu}k_-\,, \label{lineom2}
\end{align}
where $\mathcal{G}_{\mu\nu}(k)$ is the linearized Einstein tensor which is invariant under linearized diffeomorphisms by construction. It is defined in \eqref{linEinstein}.

One can see that these linearized equations of motion are equivalent to the NMG linearized equations of motion at the critical point, see ref.~\cite{Bergshoeff:2011ri}. Explicitly, by making use of $\bnabla^{\mu} \cG_{\mu\nu}(k_+)=0$, we can derive that
\begin{equation}
\bnabla_{\nu}k_- = \bnabla^{\mu}k_{-\,\mu\nu}\,,
\end{equation}
and together with the trace of \eqref{lineom1}, we find that
\begin{equation}
k_- = 0\,,
\end{equation}
provided that $\Lambda \neq 0$. We may then combine \eqref{lineom1} and \eqref{lineom2} into a single, fourth-order differential equation for the transverse traceless fields $k_{+\mu\nu}$
\begin{equation}\label{lineomcrit}
\cG_{\mu\nu} (\cG(k_+)) = \frac14 \left(\bBox - 2 \Lambda\right)^2 k_{+\mu\nu} = 0\,.
\end{equation}

\subsection{Modes at the Critical Point}
The linearized field equation \eqref{lineomcrit} is solved by the primary state of the massless mode $\psi_{\mu\nu}^0$, given by \eqref{psisol} with $\cM = 0$ (or equivalently, with the weights $(h,\bar{h}) = (2,0)$ and $(0,2)$ for the left- and right-moving modes respectively). But there is another solution, which does not solve the massless equation, but instead satisfies
\begin{equation}
\left( \bBox - 2 \Lambda \right)\psi_{\mu\nu}^{\rm log} = \psi_{\mu\nu}^0\,.
\end{equation}
As was shown in \cite{Grumiller:2008qz}, this logarithmic mode can be obtained by differentiation of the massive mode \eqref{psisol} with respect to $\cM^2 \ell^2$ and by setting $\cM^2 = 0$ afterwards:
 \begin{equation}\label{logsc}
   \psi^{\log}_{\mu\nu}  =  \frac{\partial \psi_{\mu\nu}(\cM^2)}{\partial(\cM^2\ell^2)}  \bigg|_{\cM^2 = 0} \,.
 \end{equation}
Here $\psi_{\mu\nu}(\cM^2)$ is the explicit solution obtained by filling in the weights $(h,\bar{h})$ corresponding to a massive graviton, in \eqref{psisol}. Explicitly, these weights are given by (see also \eqref{hleft} and \eqref{hright}):
\begin{align}
& {\rm Left:} \;\; (h_L, \bar{h}_L) = (2,0)\,, \qquad  {\rm Right:} \;\; (h_R, \bar{h}_R) = (0,2)\,, \label{hm0}\\
& \text{Massive L:} \; (h_{M,L},\bar{h}_{M,L}) = \left(\frac{3}{2} + \frac12 \sqrt{1+\ell^2 \cM^2}, -\frac12 + \frac12 \sqrt{1+\ell^2 \cM^2} \right)\,, \label{hleftc8} \\
& \text{Massive R:} \; (h_{M,R},\bar{h}_{M,R}) = \left( -\frac12 + \frac12 \sqrt{1+\ell^2 \cM^2}, \frac{3}{2} + \frac12 \sqrt{1+\ell^2 \cM^2} \right)\,, \label{hrightc8}
\end{align}
The resulting logarithmic modes are given by
 \begin{equation}\label{logmode}
\psi_{\mu\nu}^{\log} = f(u,v,\rho)\, \psi_{\mu\nu}^0 \,,
\end{equation}
where
\begin{equation}\label{fuvr}
f(u,v,\rho) =  - \frac{i}{2}(u+v)  -\log(\cosh\rho) \,.
\end{equation}
The log modes are not eigenstates of the AdS energy operator $H = L_0 + \bar{L}_0$. Instead they form a rank-2 Jordan cell with respect to this operator (or similarly, with respect to the Virasoro algebra). The normalization of the log modes has been chosen such that when acting on the modes
$k_{\mu\nu}= \{\psi_{\mu\nu}^0 , \psi^{\log}_{\mu\nu}\} $ with $H$, the
off-diagonal element in the Jordan cell is 1:
\begin{equation} \label{nondecompHr2}
H \, k_{\mu\nu} = \left( \begin{array}{cc}
(h+\bar{h}) 	&	0			\\
	1		& (h+\bar{h})			
\end{array}
\right) k_{\mu\nu} \,.
\end{equation}
The presence of the Jordan cell shows that the states form indecomposable but non-irreducible representations
of the Virasoro algebra. Furthermore, we have that
\begin{equation} \label{qprimr2}
L_1 \psi_{\mu\nu}^{\log} = 0 = \bar{L}_1 \psi_{\mu\nu}^{\log} \,.
\end{equation}
These properties form the basis for the conjecture that critical ZDG is dual to a rank-2 LCFT. The modes correspond
to states in the LCFT and \eqref{nondecompHr2} translates to the statement that the LCFT Hamiltonian is non-diagonalizable
and that the states form a rank-2 Jordan cell. The conditions \eqref{nondecompHr2} and \eqref{qprimr2} indicate that the states associated to $\psi^{\log}_{\mu\nu}$ are quasi-primary. The only proper primary state is the one associated to $\psi^0_{\mu \nu}$.

Note that the critical points in ZDG constitute a generalization of the NMG critical point. NMG can be retrieved from ZDG by performing a limiting procedure, outlined in chapter \ref{sec_NMGLim}, but this limit requires starting from the part of the ZDG parameter space where the sign parameter $\sigma = -1$. In contrast, in the above discussion, we have not assumed this and it is possible to find regions in the ZDG parameter space where $\sigma + \gamma = 0$ for positive $\sigma$. The critical points found here are thus indeed more general than the NMG one.

\subsection{New anomaly}

In the above section, we have confirmed the existence of critical points in linearized ZDG, where logarithmic modes appear that have the same properties as the logarithmic modes that appear in critical higher-derivative gravity theories, such as critical NMG. In the NMG case, the appearance of these modes led to the conjecture that the field theory dual to critical NMG is an LCFT with zero central charges, once appropriate boundary conditions are taken into account. The NMG logarithmic modes can be seen to be dual to the logarithmic partners of the stress-energy tensor components in the dual LCFT. Even though the central charges are zero, the two-point functions of the stress-energy tensor modes and their logarithmic partners are non-trivial and determined by new quantities, called the `new anomalies'. A simple way to calculate these new anomalies on the gravity side was given in \cite{Grumiller:2010tj}.

Similar conclusions hold at the ZDG critical points. Again, the central charges \eqref{ccZDG} \cite{Bergshoeff:2013xma}
\begin{equation}
c_{L/R} = 12 \pi \ell M_{P}(\sigma + \gamma)\,,
\end{equation} 
where $\Lambda = -1/\ell^2$, vanish at the critical points. The two-point functions of the $c_{L/R}=0$ LCFT should instead be characterized by the new anomalies $b_{L/R}$. These can be calculated via the limiting procedure of ref.~\cite{Grumiller:2010tj}. 
\begin{equation}\label{blimit}
b_{L/R} = \lim\limits_{\sigma + \gamma \to 0} \frac{c_{L/R}}{h_{L/R}-h_{M,L/R}}\,,
\end{equation}
where the weights of the massless and massive modes are given in \eqref{hm0}-\eqref{hrightc8}. Evaluating this limit explicitly, we find the critical ZDG new anomalies:
\begin{equation}\label{newanom}
b_{L/R} = - \frac{48 \pi \sigma M_P }{\ell m^2 \beta_1}\,.
\end{equation}
Equality of the new anomalies is due to the fact that ZDG is a parity even theory. In the NMG limit of ZDG (see appendix \ref{sec_NMGLim}) the new anomalies reduce to $b_{L/R}^{\rm NMG} = - 12 \sigma' \ell/ G$, where $\sigma'$ is the NMG sign parameter. This result agrees with the known expression obtained in \cite{Grumiller:2009sn,Alishahiha:2010bw} at the NMG critical point defined by $m^2 = - (2 \sigma' \ell^2)^{-1} $. 
The CFT dual to critical ZDG is conjectured to be a rank-2 LCFT with vanishing central charges. The two-point functions for such an LCFT are of the form
\begin{subequations}\begin{align}
 \langle \mathcal{O}^L(z) \,\mathcal{O}^L(0) \rangle &= 0 \,, \\
 \langle \mathcal{O}^L(z) \,\mathcal{O}^{\rm log}(0) \rangle &= \frac{b_L}{2 z^4} \,, \\
 \langle \mathcal{O}^{\rm log}(z,\bar{z}) \,\mathcal{O}^{\rm log}(0) \rangle &= -\frac{b_L\,\log|z|^2}{ z^4} \,,
\end{align}\end{subequations}
where $\mathcal{O}^\mathrm{log}(z,\bar{z})$ denotes the logarithmic partner of $\mathcal{O}^L(z)$. The two-point function for the right-moving sector are equivalent, since the theory is even under parity.

The difference with respect to critical NMG is that here the new anomaly \eqref{newanom} is a function of the coupling constant $\beta_1$, instead of a fixed combination of $\ell/G$. This again makes clear that the ZDG critical points are a generalization of the NMG critical point.

\section{Logarithmic AdS waves in ZDG}

To study critical behaviour in non-linear ZDG we look for propagating waves on an AdS$_3$ background with logarithmic decay, analogously to the situation in NMG, as studied in \cite{AyonBeato:2009yq}. As an Ansatz we consider  a Kerr-Schild deformation of AdS$_3$:
\begin{equation}
    g_{\mu \nu} = \bar{g}_{\mu \nu} - f(u,y) k_{\mu}k_{\nu} \,,
\end{equation}
where $\bar{g}_{\mu \nu}$ is the AdS background and $k^{\mu}$ is a light-like vector\footnote{We take $k^{\mu} \partial_{\mu} = (y/\ell)\partial_{v}$.}.
The function $f(u,y)$ is the wave profile. Using Poincare coordinates this leads to the following expression for the  AdS$_3$ wave space-time  ansatz:
\begin{equation}
    ds^2 = \dfrac{\ell^2}{y^2} (-f(u,y)du^2-2 du dv + dy^2) \,.
\end{equation}
We will choose the following dreibeine for this metric:
\begin{equation}
e^{0}  = \dfrac{\ell}{y} \left( \sqrt{f(u,y)} du + \dfrac{1}{\sqrt{f(u,y)}} dv \right) \,, \quad
e^{1}  = \dfrac{\ell}{y} \frac{1}{\sqrt{f(u,y)}} dv \,, \quad  e^{2} = \dfrac{\ell}{y} dy \,.
\label{e1AdSwave}
\end{equation}
In order to find the AdS wave solutions, we will follow the same procedure that was outlined in chapter \ref{section:ZDGHD}, in which ZDG was rewritten as a higher-derivative theory. We recall that the equation of motion in this formulation is given by eq. \eqref{R2eom} with $\beta_2=0$, and $e_2{}^a$ is understood to be written in terms of $e_1{}^a$ as in \eqref{e2}. We thus use \eqref{e1AdSwave} as an ansatz for the dreibein $e_1{}^a$ and find that $e_2{}^a$, as determined by \eqref{e2}, is  given by:
\begin{equation}
\begin{split}
& e_{2}{}^{0}  = g(u,y) du +h (u,y) dv \,, \\
& e_{2}{}^{1} = p(u,y)du + q(u,y) dv \,, \\
& e_{2}{}^{2} =s(u,y) dy \,, 
\label{e2sol}
\end{split}
\end{equation}
where:
\begin{align}
	g(u,y) &= \dfrac{1}{2m^2 \ell \beta_1 y \sqrt{f(u,y)}} \left( 2m^2 \ell^2 \gamma \beta_1 f(u,y) + \sigma y \left( \dfrac{\partial}{\partial y } - y \dfrac{\partial^2}{\partial y^2 } \right) f(u,y) \right) \,,  \nonumber \\
	h(u,y) &= \dfrac{\gamma \ell}{y \sqrt{f(u,y)}} = q(u,y) \,, \\
	p(u,y) &= \dfrac{\sigma}{2m^2 \ell \beta_1 \sqrt{f(u,y)}} \left( \dfrac{\partial}{\partial y } - y \dfrac{\partial^2}{\partial y^2 } \right) f(u,y) \,, \quad s(u,y) = \dfrac{\gamma \ell}{y} \,. \nonumber
	\label{e2AdSwave}
\end{align}
The parameter $\gamma$ appearing in these functions can be determined as a function of the ZDG parameters from \eqref{cc12} with $\beta_2=0$. In order to write down the equation of motion \eqref{R2eom}, we also need to evaluate the series expansion \eqref{seriesomega2} for $\omega_2{}^a(e_1)$. Explicitly calculating \eqref{Omega} for this solution, we see that all contributions to $\Omega^{(2n)}_{\mu}{}^a$ with $n>0$ have the same form:
\begin{equation} \label{OmAdSwave}
\Omega^{(2n)}_{\mu}{}^a = \left( \frac{\sigma}{\ell^2 \alpha_1} \right)^n \ell y \frac{\partial^3 f(u,y)}{\partial y^3} k_{\mu} k_{\nu} e_1^{\nu\,a}\,.
\end{equation}
It might seem strange that this expression contains only three derivatives for every value of $n$, in view of the fact that $\Omega_\mu^{(2n)a}$ contains more derivatives for larger $n$, in order to balance the mass dimensions of the corresponding $m^{-2n}$ in the series \eqref{seriesomega2}. For this particular ansatz however, the $\ell$-parameter features as an extra dimensionfull parameter that can be used to balance dimensions and this explains why it is possible that all $\Omega^{(2n)}_{\mu}{}^a$ feature the same number of derivatives.

We can now sum all orders of $1/m^2$ into a closed expression for $\omega_2{}^a$. We find:
\begin{equation} \label{summedupom}
\omega_{2\,\mu}{}^a = \omega_{1\,\mu}{}^a - \frac{\sigma \ell y}{\sigma - \alpha_1 \ell^2 m^2} \frac{\partial^3 f(u,y)}{\partial y^3} k_{\mu} k_{\nu} e_1^{\nu\,a}\,.
\end{equation} 
Replacing this into the equation of motion (\ref{R2eom}), we see that the latter reduces to the following fourth-order differential equation for the wave profile:
\begin{align} \label{wavediffeq}
	\frac{1}{y^2 \sqrt{f(u,y)}} \left[ y^4 \frac{\partial^4 f(u,y)}{\partial y^4} + 2y^3 \frac{\partial^3 f(u,y)}{\partial y^3 } \right. &- \notag \\
	\left( 1 + \cM^2\ell^2  \right) \left( y^2 \frac{\partial^2 f(u,y)}{\partial y^2} \right.
	&- \left. \left. y \frac{\partial f(u,y)}{\partial y } \right) \right] = 0 \,.
\end{align}
Here $\cM^2$ is the Fierz-Pauli mass \eqref{MFPc8} and we have used the parameter relations \eqref{cc12}. 

The equation \eqref{wavediffeq} can be solved by separation of variables and proposing that the solutions behave polynomially in $y$ : $f(u,y) = \tilde{f}(u) y^n$.
The power $n$ is determined as a solution of the indicial equation:
\begin{equation}
	n(n-2)\left( n(n-2) - \cM^2 \ell^2 \right) = 0 \,.
	\label{ChPolynom}
\end{equation}
In general, this equation has four roots $n = \left\lbrace 0, 2, n_+, n_- \right\rbrace$, with $n_{\pm} = 1 \pm \sqrt{1 + \cM^2 \ell^2}$.

The generic solution for the wave profile is then:
\begin{equation}
	f(u,y) = f_0(u) + f_2(u) \left(\dfrac{y}{\ell}\right)^2 + f_+(u) \left(\dfrac{y}{\ell}\right)^{n_+} + f_-(u) \left(\dfrac{y}{\ell}\right)^{n_-} \,.
	\label{GeneralSol}
\end{equation}
The constant and the quadratic terms can always be removed by local transformations \cite{AyonBeato:2005qq}. The relevant parts are then given by the terms involving $y^{n_{\pm}}$. At special points in the parameter space the roots $n_{\pm}$ become degenerate, as we will discuss in the next subsection.

Since the expressions for $\Omega^{(2n)\,a}$ are all proportional to each other, the AdS wave solution \eqref{GeneralSol} is not only a solution to the full theory, but it will solve the equations of motion at every order of $\frac{1}{m^2}$, provided that the parameters appearing in $n_\pm$ are properly adjusted.

\subsection{Special points}

At the ZDG critical points, one has that $\sigma + \gamma = 0$ and thus $\cM^2 =0$. At such points $n_+ = 2$ and $n_- = 0$ and the indicial equation \eqref{ChPolynom} thus has two degenerate solutions, instead of four distinct ones. The order of the differential equation \eqref{wavediffeq} at the critical points is still four however, and four distinct, albeit potentially non-polynomial solutions should still exist. Ignoring the constant and quadratic solutions that can be removed by local transformations, one finds the following solutions:
\begin{equation}
	f_c(u,y) = f_L(u) \ln\left(\dfrac{y}{\ell}\right) + f_{2L}(u) \left(\dfrac{y}{\ell}\right)^2 \ln\left(\dfrac{y}{\ell}\right) \,.
	\label{SpecialSol2}
\end{equation}
One thus finds AdS waves with logarithmic decay at the critical point  and this is a clear sign that the existence of logarithmic modes in critical ZDG persists at the non-linear level and is not merely an artifact of the linearized approximation.

There is another class of special points, where a degeneracy in the indicial equation \eqref{ChPolynom} takes place. At these points, $\cM^2 = -1 / \ell^2$ and $n_{\pm} = 1$. The indicial equation \eqref{ChPolynom} thus only has three roots, one of which is degenerate. The equation \eqref{wavediffeq} is again still of order four and thus one non-polynomial solution should exist. The following solutions (again ignoring the ones that can be removed by local transformations) are found:
\begin{equation}
	f_s(u,y) =\left(\dfrac{y}{\ell}\right) \left(  f_1(u)  + f_{1L}(u) \ln\left(\dfrac{y}{\ell}\right) \right) \,.
	\label{SpecialSol}
\end{equation}
One thus finds one AdS wave with logarithmic decay at these special points. This solution was also found in the Isham, Salam, Strathdee $f-g$ theory \cite{Isham:1971gm} in \cite{Afshar:2009rg}. 

This point already appears as a special point in NMG and it is known that for this special point, NMG has black hole solutions that are not locally isometric to AdS$_3$ \cite{Bergshoeff:2009aq,Oliva:2009ip}. At this point, the linearized Fierz-Pauli action in AdS$_3$ features an extra gauge invariance, with scalar parameter. Linearized NMG, being a sum of a linearized Einstein-Hilbert and Fierz-Pauli action for two different fluctuations, inherits this linearized gauge invariance. The same holds for ZDG as can be seen from eq. \eqref{linear_bi_lagrangian} and the discussion in chapter \ref{sec:PMpoint}. At the linearized level, NMG and ZDG at those critical points thus only propagate one degree of freedom. This however is no longer true at the non-linear level and the extra linearized gauge invariance is an accidental one.

NMG at this point has been dubbed `Partially Massless Gravity' (PMG), as the massive mode becomes partially massless \cite{Deser:1983mm,Deser:2001pe}. In \cite{Grumiller:2010tj}, it was argued that solutions with logarithmic decay appear in PMG, of the type given in \eqref{SpecialSol} and that this can be taken as a sign that the dual field theory is an LCFT. Interestingly, we have found above that also PMG, originally found as a special version of NMG, can be generalized to a class of special points in the ZDG parameter space.

\section{ENMG and Tricritical Gravity}
\label{sec:ENMG_log}

In this section we investigate the critical lines and points in the parity even sixth-order Extended New Massive Gravity (ENMG) model introduced in chapter \ref{ExtendedNMG}. Away from the tricritical point, the theory propagates one massless and two massive gravitons. At the critical point, the two massive gravitons degenerate with the massless one and are replaced by new solutions. In contrast to the massless graviton modes, that obey Brown--Henneaux boundary conditions, these new solutions exhibit $\log$ and $\log^2$ behaviour towards the $\mathrm{AdS}_3$ boundary
and are referred to as $\log$ and $\log^2$ modes. The existence of these various logarithmic modes naturally leads to the conjecture that ``tricritical'' ENMG is dual to a rank-3 logarithmic CFT. In this section, we will discuss these modes and their AdS/CFT consequences in more detail. 

The linearized equations of motion for ENMG, derived from the quadratic Lagrangian \eqref{linShk} can be written as
\begin{equation}
\left(\bBox + \frac{2}{\ell^2} - \cM_+^2 \right)\left(\bBox + \frac{2}{\ell^2} - \cM_-^2 \right)\left(\bBox + \frac{2}{\ell^2} \right)k_{\mu\nu} = 0\,,
\end{equation}
for a transverse traceless field $k_{\mu\nu}$. The Fierz-Pauli masses $\cM_{\pm}^2$ are given in terms of the ENMG parameters as
\begin{equation}
\cM^2_{\pm} =  \frac{a-2b-2m^2 \ell^2  \pm \sqrt{f(a,b,m^2)}}{4b\ell^2}\,,
\end{equation}
with:
\begin{equation}\label{fabm}
f(a,b,m^2) = a^2 - 2 a\left(b+2\ell^2 m^2\right) + 4\left(b^2 + \ell^4m^4 - 4b \ell^4 m^4 \sigma \right)\,.
\end{equation}
From the analysis of chapter \ref{linearth} we know that the presence of massive spin-2 ghosts or tachyons cannot be avoided for non-zero FP masses, however, there is a critical line and a tricritical point where, respectively, one and both FP masses vanish. At this point the central charge also vanishes and the theory is a gravitational dual for Logarithmic Conformal Field Theories (LCFT) of rank 2 and 3 respectively. Even though the parameter $b$ does not appear in the expression for the central charge, it does play a role in the analysis of the critical lines and points. If we tune $a$ to its critical value $a_{\rm crit}$, defined such that the central charge \eqref{ENMGcc} vanishes
\begin{equation}\label{acrit}
a = a_{\rm crit} = 4 \ell^2 m^2(1 + 2 \ell^2 m^2 \sigma)\,,
\end{equation} 
the FP masses become
\begin{equation}
\cM_{\pm}^2 = \frac{1}{2b\ell^2} \left( B(b,m^2) \pm \sqrt{ \left(B(b,m^2)\right)^2 } \right)\,,
\end{equation}
with
\begin{equation}
B(b,m^2) = -b + \ell^2 m^2 + 4 \ell^4 m^4 \sigma \,.
\end{equation}
This implies that on the critical line defined by \eqref{acrit}
\begin{equation} \label{bcrit}
 \text{if:} \quad B(b,m^2) < 0\,, \qquad \text{then:} \quad   \cM^2_+ = 0 \,, \quad \cM_-^2 = \frac{B}{b\ell^2}\,, 
 \end{equation}
 \begin{equation}
 \begin{split}\label{bcrit1}
 \text{if:} \quad & B(b,m^2) > 0\,, \qquad \text{then:} \quad  \cM^2_- = 0 \,, \quad \cM_+^2 = \frac{B}{b\ell^2}\,, \\
\text{if:} \quad & B(b,m^2) = 0\,, \qquad \text{then:} \quad  \cM^2_+ = 0 \,, \quad \cM_-^2 = 0 \,.   
\end{split}
\end{equation}
The last scenario, where $b = b_{\rm crit} = \ell^2 m^2(1+4\ell^2 m^2 \sigma)$, defines the tricritical point where both FP masses vanish and the dual LCFT has rank 3.

\subsection{Modes at the tricritical point}
At the critical line defined by \eqref{acrit}, one of the two Fierz-Pauli masses vanishes, while the other is given in \eqref{bcrit}-\eqref{bcrit1}. Since the expression for the non-vanishing mass is the same, regardless of the sign of $B(b,m^2)$, we will assume that $B(b,m^2) \geq 0$ and $\cM_-^2 = 0$ on the critical line. In the case that $B<0$, we only have to switch $\cM_+ \leftrightarrow \cM_-$. At this point the spectrum contains a massive mode \eqref{psisol} with mass $\cM_+$, a massless mode $\psi_{\mu\nu}^0$ and a logarithmic mode \eqref{logmode}.

At the tricritical point $\cM_\pm^2 = 0$ and the weights (and therefore the solutions) of the massive modes degenerate with those of the massless modes. There are now two new solutions, called $\mathrm{log}$ and $\mathrm{log}^2$ modes. Denoting these modes by $\psi^\mathrm{log}$ and $\psi^{\mathrm{log}^2}$ resp., they satisfy
\begin{align}
\cG_{\mu\nu}(\cG(\psi^{\log})) & = 0\,, & \textrm{but} & \; \; \; \cG_{\mu\nu}(\psi^{\log}) \neq 0\,, \\
\cG_{\mu\nu}(\cG(\cG(\psi^{\log^2}))) & = 0\,, & \textrm{but} & \; \; \; \cG_{\mu\nu}(\cG(\psi^{\log^2})) \neq 0\,.
\end{align}
The log mode can be obtained, as before, by differentiating the massive mode with respect to its mass and setting the mass to zero afterwards. The $\mathrm{log}^2$ mode can be obtained in a similar way, by differentiating
twice with respect to $\cM_\pm ^2 \ell^2$. The resulting modes are given by
 \begin{align}
\psi_{\mu\nu}^{\log} & = f(u,v,\rho)\, \psi_{\mu\nu}^0 \,, \\
\psi_{\mu\nu}^{\log^2} & = \frac12 f(u,v,\rho)^2 \,\psi_{\mu\nu}^0  \,,
\end{align}
where $f(u,v,\rho)$ is given in \eqref{fuvr}.

Note that the massless, log and $\mathrm{log}^2$ modes all behave differently when approaching the boundary $\rho \rightarrow \infty$. The massless mode obeys Brown--Henneaux boundary conditions. In contrast, the log mode shows a linear behaviour in $\rho$ when taking the $\rho \to \infty$ limit, whereas the $\log^2$ mode shows $\rho^2$ behaviour in this limit. The three kinds of modes therefore all show different boundary behaviour in $\mathrm{AdS}_3$ and the boundary conditions obeyed by log and $\mathrm{log}^2$ modes are correspondingly referred to as log and $\mathrm{log}^2$ boundary conditions. 

The log and $\log^2$ modes now form a rank-3 Jordan cell with respect to the AdS energy operator $H$. With the proper normalisation, the Jordan cell for the modes
$k_{\mu\nu}= \{\psi_{\mu\nu}^0 , \psi^{\log}_{\mu\nu} ,\psi^{\log^2}_{\mu\nu} \} $ is now:
\begin{equation} \label{nondecompH}
H \, k_{\mu\nu} = \left( \begin{array}{ccc}
(h+\bar{h}) 	&	0		&	0	\\
	1		& (h+\bar{h})	&	0	\\
	0		&	1		&	(h+\bar{h}) 	
\end{array}
\right) k_{\mu\nu} \,.
\end{equation}
The states form indecomposable but non-irreducible representations of the Virasoro algebra and like in \eqref{qprimr2} we now have that
\begin{equation} \label{qprim}
L_1 \psi_{\mu\nu}^{\log} = 0 = \bar{L}_1 \psi_{\mu\nu}^{\log} \,, \qquad 
L_1\psi_{\mu\nu}^{\log^2} = 0 = \bar{L}_1 \psi_{\mu\nu}^{\log^2} \,.
\end{equation}
This indicates that also the states corresponding to the $\log^2$ modes are quasi-primary. All these properties combined are the basis for the conjecture that tricritical ENMG is dual to a rank-3 LCFT.

\subsection{The structure of the dual CFT}\label{stresstenssec}
In chapter \ref{sec:cc} we computed the central charges for ENMG. They are
\begin{equation}\label{ccENMG8}
c_{L/R}^{\text{\tiny ENMG}} = \frac{3\ell}{2G} \left( \sigma + \frac{1}{\ell^2m^2} - \frac{a}{8\ell^4 m^4} \right)\,.
\end{equation}
The central charges vanish at the tricritical point and this lends further support for the conjecture that the dual CFT is logarithmic. Indeed, as  unitary $c=0$ CFTs have no non-trivial representations, CFTs with central charge $c=0$ are typically non-unitary and thus possibly logarithmic.

The central charges also vanish on the rest of the critical line \eqref{acrit} where just one of the massive modes becomes massless. On this critical line, the dual CFTs are still expected to be logarithmic, but the rank must decrease by one with respect to the tricritical point.  The dual theory on the critical line is thus expected to be an LCFT of rank 2. As a consistency check, we note that (non-critical) NMG is contained in our model in the limit $a\to0$ and $b\to0$. Substituting these values in \eqref{ccENMG8}, we see that the central charge agrees with the central charge found for NMG in \cite{Bergshoeff:2009aq}. 

The dual CFT of tricritical ENMG is thus conjectured to be a rank-3 LCFT with $c_L = c_R = 0$. In that case the general structure of the two-point correlators is known. The two-point functions are determined by quantities called new anomalies. If one knows the central charges, one can employ a short-cut \cite{Grumiller:2010tj} to derive these new anomalies. We do this for the left-moving sector. Similar results hold for the right-moving sector as the two sectors are related to each other by a parity transformation.

Let us start from the non-critical case, where the correlators of the left-moving components $\mathcal{O}^L(z)$ of the boundary stress tensor are given by
\begin{align}
 \langle \mathcal{O}^L(z) \,\mathcal{O}^L(0) \rangle = \frac{c_L}{2 z^4} \,,
\end{align}
where $c_L$ is given by \eqref{ccENMG8}. Let us first consider the case where only one of the two masses vanishes, e.g.~when $B(b,m^2)>0$ and $M_-^2 \rightarrow 0$. In this case, we are on the critical line \eqref{acrit}. The CFT dual is conjectured to be a rank-2 LCFT with vanishing central charges.
The two-point functions for such an LCFT are of the form
\begin{subequations}\begin{align}
& \langle \mathcal{O}^L(z) \,\mathcal{O}^L(0) \rangle = 0 \,, \\
& \langle \mathcal{O}^L(z) \,\mathcal{O}^{\rm log}(0) \rangle = \frac{b_L}{2 z^4} \,, \\
& \langle \mathcal{O}^{\rm log}(z,\bar{z}) \,\mathcal{O}^{\rm log}(0) \rangle = -\frac{b_L\,\log|z|^2}{ z^4} \,,
\end{align}\end{subequations}
where $\mathcal{O}^\mathrm{log}(z,\bar{z})$ denotes the logarithmic partner of $\mathcal{O}^L(z)$. The parameter $b_L$ is the new anomaly. It can be calculated with knowledge of the weights \eqref{hm0}-\eqref{hrightc8} and the limiting procedure of \cite{Grumiller:2010tj}.
The new anomaly is given by
\begin{align} \label{bL}
	b_L&= \lim_{a \to a_{\rm crit}} \frac{c_L}{h_L - h_{M_-,L}} = - \frac{6 B(b,m^2)}{G\ell^3m^4} = - \frac{6 b \cM_+^2}{\ell G m^4}\,.
\end{align}
Note that in the limit $b\to0$ and $a = a_{\rm crit} \to 0$ we recover critical NMG and we find that the result \eqref{bL} agrees with the new anomaly of NMG \cite{Grumiller:2009sn}.

At the tricritical point, the correlators are conjectured to be the ones of a rank-3 LCFT with vanishing central charges:
\begin{subequations}\label{2ptlcft}\begin{align}
& \langle \mathcal{O}^L(z) \,\mathcal{O}^L(0) \rangle = \langle \mathcal{O}^L(z) \,\mathcal{O}^{\rm log}(0) \rangle =0 \,, \\
& \langle \mathcal{O}^{L}(z) \,\mathcal{O}^{\rm log^2}(0) \rangle = \langle \mathcal{O}^{\rm log}(z) \,\mathcal{O}^{\rm log}(0) \rangle = \frac{a_L}{2 z^4}  \,, \label{loglog} \\
& \langle \mathcal{O}^{\rm log}(z,\bar{z}) \,\mathcal{O}^{\rm log^2}(0) \rangle = -\frac{a_L\,\log|z|^2}{z^4} \,, \\
& \langle \mathcal{O}^{\rm log^2}(z,\bar{z}) \,\mathcal{O}^{\rm log^2}(0) \rangle = \frac{a_L\,\log^2|z|^2}{z^4}\,.
\end{align}\end{subequations}
Here $\mathcal{O}^{\rm log}(z,\bar{z})$, $\mathcal{O}^{\rm log^2}(z,\bar{z})$ are the two logarithmic partners of $\mathcal{O}^L(z)$.
The new anomaly $a_L$ at the tricritical point is obtained via another limit:
\begin{equation}\label{aL}
	a_L=\lim_{b \to b_{\rm crit}} \frac{b_L}{h_{L}- h_{M_+,L} }    =\frac{96\ell}{G} \left( \sigma + \frac{1}{4m^2 \ell^2} \right)\,.
\end{equation}
Knowledge of the central charges thus allows one to obtain the new anomalies and hence fix the structure of the
two-point correlators, via the limit procedure of \cite{Grumiller:2010tj}.

Note that after fixing $a$ and $b$ to their critical values, the free parameter $m^2$ is undetermined in the expression for $a_L$. This implies that tricritical ENMG in fact has a continuous line of tricritical points dual to a family of rank-3 LCFTs with different values for the new anomaly \eqref{aL}.

\section{Discussion} 

In this chapter, we have shown that the parameter space of ZDG around AdS$_3$ has critical points, where solutions with logarithmic fall-off behavior appear, both at the linearized and non-linear level. These critical points and logarithmic solutions are similar to the ones that appear in critical NMG. Although NMG can be retrieved from ZDG in a particular limit, the critical points found here do however not simply correspond to the NMG one, but can rather be seen as a generalization of the NMG critical point.  

Note that the existence of the ZDG logarithmic solutions found here is non-trivial. In both NMG and ZDG, criticality is signalled when massive modes become massless and degenerate with pure gauge modes. In both cases, one can argue via continuity that new solutions, that are not massive nor massless modes, should appear at a critical point. In NMG, that is naturally formulated as a four-derivative theory, this argument is based on the fact that the equations of motion remain fourth order at the critical point, and hence there should still be four distinct linearized modes. Since the massive modes coincide with the pure gauge modes, new solutions that are not massive, nor pure gauge should appear in the spectrum. They turn out to be solutions of a particular fourth-order differential equation, featuring a fourth-order differential operator that is the square of a second-order one \cite{Bergshoeff:2011ri} and such equations typically feature the logarithmic modes. In ZDG, the equations of motion at critical points are still a system of coupled second-order differential equations for two metric fields and there should again still be four distinct linearized modes. What is non-trivial in ZDG is that the new modes that appear instead of the massive modes at the critical point are logarithmic. Indeed, the logarithmic behavior is typical for solutions of particular differential equations of order higher than two and ZDG is naturally formulated in terms of coupled second-order equations. In this chapter, we have however seen that the new solutions at the ZDG critical points are logarithmic. At the linearized level, this stems from the fact that the linearized critical ZDG equations of motion are the same as the linearized NMG ones. At the non-linear level, we have used the fact that ZDG can alternatively be rewritten as a higher-derivative theory (involving an infinite number of higher-derivatives) for a single dreibein and it is this higher-derivative character of ZDG that is ultimately responsible for the existence of AdS waves with logarithmic fall-off behavior at the critical points.

As in the critical higher-derivative massive gravity cases, the existence of logarithmic modes  can be seen as a hint that critical ZDG theories are dual to logarithmic conformal field theories, once appropriate boundary conditions are imposed. In-order to show this in more detail, more checks need to be performed however. In particular, precise calculations of two- and three-point functions, as was done for TMG and NMG in \cite{Skenderis:2009nt,Grumiller:2009mw,Grumiller:2009sn,Alishahiha:2010bw}, via e.g. holographic renormalization should be performed. The conjecture can also be shown by calculations of the classical and one-loop partition functions on the gravity side (see \cite{Gaberdiel:2010xv,Bertin:2011jk} for examples in higher-derivative theories) and checking that the results conform with the structure expected for an LCFT. Performing these checks will require an extension of the AdS/CFT holographic dictionary and methods to Chern-Simons--like theories, such as ZDG.

In addition to critical ZDG, we have also considered three-dimensional, tricritical higher-derivative gravity theories around $\mathrm{AdS}_3$.
These tricritical theories are obtained by considering higher-derivative gravities constructed from CS--like models with four auxiliary fields, as outlined in chapter \ref{ExtendedNMG}. Ordinarily, these theories propagate one massless and two massive graviton states, but at a special point in their parameter space all massive gravitons become massless. The massive graviton solutions, that ordinarily obey Brown--Henneaux boundary conditions, are in tricritical theories replaced by new solutions that obey $\log$ and $\log^2$ boundary conditions towards the AdS boundary: the so-called $\log$ and $\log^2$ modes.

GMG at the tricritical point constitutes a parity odd example of such a tricritical gravity theory and was studied in \cite{Liu:2009pha,Bertin:2011jk}. It was also shown that this theory is dual to a parity violating, rank-3 LCFT. In this chapter, we considered a parity even example, that is of sixth order in derivatives. The results obtained here can be put in the context of the findings of \cite{Bergshoeff:2012sc}, where a scalar field model was studied, that (in the six-derivative case) can be seen as a toy model for tricritical ENMG. There it was argued that odd rank LCFTs allow for a non-trivial truncation, that on the gravity side can be seen as restricting oneself to Brown--Henneaux and $\log$ boundary conditions. In \cite{Bergshoeff:2012ev} it was found that similar conclusions hold for tricritical gravity, at the linearized level. Indeed, upon applying this truncation to the two-point correlators of the dual LCFT,  the truncated theory still has one non-trivial correlator; the log-log two-point function \eqref{loglog}. Similar truncations were considered in even higher-derivative extensions of general relativity in \cite{Nutma:2012ss,Kleinschmidt:2012rs}.

In order to go beyond the linearized level, one should first address the issue of the consistency of the truncation, in the presence of interactions. In \cite{Bergshoeff:2012ev} a step in this direction was made by rephrasing the truncation for tricritical gravity as restricting oneself to a zero charge sub-sector of the theory, with respect to the Abbott--Deser--Tekin charges associated to (asymptotic) symmetries. Similar
conclusions can be made for tricritical GMG using the results for the conserved charges in \cite{Liu:2009pha}. However, it was shown in \cite{Apolo:2012vv} that the consistency of the truncation is an artefact of the linearized approximation and it is no longer valid at the next order in metric perturbations.

%% file: chapter_9/chapter_9.tex
\pagestyle{empty}
\setcounter{chapter}{8}

\chapter[Conclusions and Discussions]{Conclusions and Discussions}
\label{chapter:conclusions}

\pagestyle{headings}

\begin{quote}\em
\end{quote}
\newpage

\section{Conclusions}\label{sec:conclusions}
In this thesis we have discussed a wide variety of three dimensional massive gravity models which can be characterized as Chern-Simons--like; all of the discussed models fit a general form defined by the Lagrangian three-form
\begin{equation}\label{LCSlikec9}
L = \frac{1}{2} g_{rs} a^r \cdot \ed a^s + \frac{1}{6} f_{rst} a^r \cdot (a^s \times a^t )\,.
\end{equation}
Here $a^{r\,a}$ are a set of $N$ Lorentz vector-valued one-form fields, labelled by flavor indices $r,s,t,\ldots$. The symmetric matrix $g_{rs}$ is an invertible metric on the flavor space and $f_{rst}$ a totally symmetric flavor tensor which contains information on the interaction terms in the theory. The Latin Lorentz indices $a,b,c,\ldots$ are made implicit by denoting contractions with $\eta_{ab}$ and $\ve_{abc}$ by dots and  crosses respectively. In addition, wedge products between the form fields are implicit. 

Whenever $f^{r}{}_{st}\ve^a{}_{bc}$ are the structure constants of some Lie algebra, and $g_{rs}\eta_{ab}$ a group invariant symmetric tensor on this Lie algebra, then the Lagrangian three-form \eqref{LCSlikec9} defines a pure Chern-Simons theory. Two examples of pure CS theories were discussed in this thesis, summarized in the table below. In all other cases, the theory is only CS--like and the individual fields are no longer Lie algebra-valued connections. Note that the full class of CS theories is larger that the one which can be obtained from \eqref{LCSlikec9}, as not all of the fields need to be vector-valued under the Lorentz-group. 
\begin{table}[b]
\centering
{\bf Chern-Simons theories of gravity} \\
\begin{tabular}{|c|c|c|c|} \hline & & & \\
\# of fields & name  & gauge group & chapter \\ \hline  & & & \\
$N= 2$		& Einstein-Cartan & $ \begin{array}{cc} ISO(2,1) & \Lambda = 0 \\ 
SO(2,2) & \Lambda < 0  \\ 
SO(3,1) & \Lambda > 0 \end{array}$ & \ref{chapter:GRin3D} \\ & & & \\
\hline & & & \\
$ N=3$ & Conformal Gravity & $SO(3,2)$ &  \ref{sec:CSG} \\ & & & \\ \hline
\end{tabular}
\end{table}

The advantage of the formulation \eqref{LCSlikec9} is that it allows for theories with local degrees of freedom, while retaining a relatively simple Hamiltonian formulation. In chapter \ref{chapter:three_crowd} we have analyzed this Hamiltonian form and found that the presence of secondary constraints is related to the presence of invertible fields in the theory. This is especially important if we wish to view these CS--like theories as models for three dimensional gravity, since this link can only be made when the dreibein is invertible. 

The CS--like theories of gravity considered in this thesis can be characterized in two classes: those with an alternative action principle in terms of a single metric and higher-derivative terms and those without. The specific models which were discussed in this thesis are summarized in the tables on the following pages. 

\begin{table}[t]
\centering
{\bf CS--like theories: Class I } \\
Models with a higher-derivative action \\
\begin{tabular}{|c|c|c|c|} \hline & & & \\
\# of fields & name  & parity & chapter \\ \hline  & & & \\
$N= 3$		& Topologically Massive Gravity & violating & \ref{sec:TMG} \\ & & & \\
\hline & & & \\
\multirow{2}{*}{$N=4$} & New Massive Gravity & even & \ref{sec:NMG} \\
& General Massive Gravity & violating & \ref{sec:GMG}, \ref{sec:GMG_Hamil}, \ref{sec:AS_GMG}
\\ & & & \\ \hline & & & \\
$N= 5 $ & Extended gravitational CS-term & odd & \ref{ExtendedCSG} \\ & & & \\ \hline & & & \\
$N =6$ & Extended New Massive Gravity & even & \ref{ExtendedNMG}\\ & & & \\ \hline & & & \\
\multirow{2}{*}{$N > 6$} & \multirow{2}{*}{Extended Massive Gravity} 
& even for $N$ = even & \multirow{2}{*}{\ref{Extended}} 
\\
& & odd for $N$ = odd & \\ & & & \\ \hline 
\end{tabular}
\end{table}

The models which can be written in terms of a higher-derivative theory share a number of properties. Besides the dreibein $e$ and spin connection $\omega$, they contain a number of auxiliary fields. The order of the derivatives after integrating out the auxiliary fields equals to the number of fields $N$ denoted in the table above. Furthermore, the number of propagating degrees of freedom equals to $N-2$ and all local degrees of freedom are states with helicity $\pm 2$. The CS--like construction of the theory, together with the assumption of an invertible dreibein, guarantees the absence of additional scalar-ghost degrees of freedom, as was verified in this thesis by the Hamiltonian analysis outlined in chapter \ref{chapter:three_crowd}. This is a big advantage of the CS--like formulation; since it is very clear how secondary constraints arise in the Hamiltonian form, it is also clear how to define theories which posses the necessary number of constraints needed to remove the Boulware-Deser ghost which typically plague theories of massive gravity. Hence, by using the Chern-Simons--like formulation it becomes possible to derive the scalar-ghost free combinations of higher-derivative terms.

Unfortunately, the Boulware-Deser ghost is not the only pathology which affects these type of models. It seems that all of the higher-derivative models suffer from a ``bulk-boundary clash''\footnote{Technically, the term ``bulk-boundary clash'' is not a very good one, since the BTZ black hole, which is a bulk solution, would have negative mass whenever the central charge is negative. This implies that even without considering the boundary central charge, the bulk has unphysical properties. Furthermore, in light of the AdS/CFT-correspondence, a non-unitary boundary theory should correspond to a non-unitary bulk theory and vice versa. This implies that in regions with perturbative unitarity in the bulk, the non-linear theory may still suffer from pathologies, since there the dual CFT is non-unitary.}: there are no regions in parameter space where a boundary central charge is consistent with perturbative unitary in the bulk, at least not for perturbations around AdS$_3$ spacetimes. For theories with $N \geq6$, when multiple massive spin-2 modes are propagated, we have shown there is always an unphysical (tachyonic or ghost) massive mode in the spectrum. These considerations seem generic for higher-derivative theories of gravity and have their origin in the Ostrogradski instability. 

\begin{table}[t]
\centering
{\bf CS--like theories: Class II } \\
Models without a higher-derivative action \\
\begin{tabular}{|c|c|c|c|} \hline & & & \\
\# of fields & name  & parity & chapter \\ \hline  & & & \\
$N= 3$		& Minimal Massive Gravity & violating & \ref{chapter:MMG} \\ & & & \\
\hline & & & \\
\multirow{2}{*}{$N=4$} & Zwei-Dreibein Gravity & even & \ref{section:ZDG}, \ref{sec:ZDG_Hamil}, \ref{sec:AS_ZDG} \\
& General Zwei-Dreibein Gravity & violating & \ref{sec:GZDG} 
\\ & & & \\ \hline & & & \\
$N =6$ & Drei-Dreibein Gravity & even & \ref{sec:DDG}\\ & & & \\ \hline & & & \\
$N>6$ & \multirow{2}{*}{Viel-Dreibein Gravity} 
&\multirow{2}{*}{even}& \multirow{2}{*}{\ref{sec:VDG}} 
\\
$N =$ even & & & \\ & & & \\ \hline 
\end{tabular}
\end{table}

The general CS--like model \eqref{LCSlikec9} allows for more possibilities than the class of theories discussed above. Another class can be loosely defined as CS--like models without an action in terms of a single metric. The various examples which we have discussed in this thesis are summarized in the above table. These models also share a number of properties. They all admit an AdS$_3$ background and linearization around this background shows they propagate $N-2$ massive states of helicity $+2$ and/or $-2$. A background independent Hamiltonian analysis confirms that the same number of degrees of freedom are propagated non-linearly. An analysis of the asymptotic symmetry group when restricting to Brown-Henneaux boundary conditions shows that these models do contain regions in their parameter space where positive central charge is consistent with perturbative unitarity in the bulk. 

The improved behavior with respect to the AdS/CFT-correspondence in these models stems from the inclusion of additional interaction terms, and hence additional parameters, with respect to the CS--like theories with an equal number of fields belonging to the first class. However, this entails by no means an obvious modification of the higher-derivative gravity theories which follow from the models of the first class. Even though the field equations can still be written in terms of a metric and its derivatives, the models of the second class cannot be obtained by the inclusion of new higher-derivative terms to the action of the models of the first class. In this sense, they constitute novel modifications of gravity in 3 dimensions.

The absence of an action formulation in terms of a single metric implies that a minimal coupling to matter in the metric formulation is not possible. We have explicitly seen this for Minimal Massive Gravity (MMG) in chapter \ref{chapter:MMG}; the inclusion of a symmetric stress-tensor to the modified field equations of MMG do not lead to the conservation of this stress-tensor. In a sense this should not be a big surprise, since there is no action from which these equations can be derived. It would therefore be more logical to investigate matter couplings directly in the CS--like formulation of the theory, where there is a simple and manifestly gauge invariant action. Here it is clear, from the Hamiltonian analysis, which field should have an inverse\footnote{The invertible field is usually the dreibein, but it may also be some linear combination of fields as in Zwei-Dreibein Gravity, see the discussion of chapter \ref{chapter:three_crowd}.} and from this field we may construct a metric. It would be interesting to study the coupling of matter to this invertible field. In the second class of theories one usually needs the field equation obtained by varying with respect to the dreibein to solve for the auxiliary fields, therefore the coupling to matter could change the solution for the auxiliary fields and modify the final higher-derivative equation for the metric (now plus matter) in a non-trivial (and non-minimal) way which may be consistent. 

The latter class of models generalizes and extends the former class of CS--like theories. In the explicit examples treated in this thesis, we have shown how (some of) the models belonging to the first class can be obtained from the theories in the second class as scaling limits. This can also be seen at a range of critical points in the parameter space where the masses of the massive modes vanish. At this point, as we have shown explicitly for ZDG, a new, logarithmic solution appears and the dual theory (with modified boundary conditions) is conjectured to be a logarithmic CFT. We expect this feature to be a general one for all CS--like theories constructed in the manner outlined in this thesis. 

\section{Discussions}\label{sec:discussions}
The Chern-Simons--like formulation seems well adapted to describe ghost-free theories of massive gravity in three dimensions and in the preceding section we have made a distinction into two classes; theories with and without a higher-derivative action. The models discussed in this work, however, do not saturate all possibilities; there are several ways to extend the work presented in this thesis. Let us comment on a number of ways to do so here.

First of all, the CS--like description is also capable of describing theories which contain an infinite sum of higher-derivative terms in a very compact and powerful way. Consider, for example, extending the Lagrangian three-form of NMG with a $f \cdot f \times f$-term in its CS-like formulation:
\begin{equation}\label{ESinha}
L = - \sigma e \cdot R + \frac{\Lambda_0}{6} e \cdot  e \times e + h \cdot \cD e - \frac{1}{m^2} f \cdot \left( R + \frac12  e \times f + \frac{a}{6m^2} f \times f \right) \,,
\end{equation}
where $a$ is a free dimensionless parameter. The field equations obtained by varying with respect to $f$ are given by
\begin{equation}\label{feom}
R + e \times f+ \frac{a}{2m^2}  f \times f = 0\,.
\end{equation}
This equation can be solved for $f_{\mu\nu} \equiv f_{\mu}{}^a e_{\nu}{}^b \eta_{ab}$ in terms of an infinite expansion in $m^{-2}$
\begin{equation}
f_{\mu\nu} = \sum_{n=0}^{\infty} \frac{1}{m^{2n}} f^{(n)}_{\mu\nu}\,,
\end{equation}
 as follows:
\begin{equation}\label{fiterative}
f^{(n+1)}_{\mu\nu} = - \frac{a}{2} \left(g_{\mu\rho}g_{\nu\sigma} - \frac12 g_{\mu\nu} g_{\rho\sigma} \right) \epsilon^{\rho\alpha\beta} \epsilon^{\sigma \gamma\delta} \sum_{k=0}^{n} f^{(k)}_{\alpha\gamma} f^{(n-k)}_{\beta\delta} \,.
\end{equation}
The starting value at order $m^0$ is $f^{(0)}_{\mu\nu} = - S_{\mu\nu}$. Having found the solution for $f_{\mu\nu}$, we can go to the metric formulation by plugging the solution of \eqref{fiterative} back into the Lagrangian. The result can be written as
\begin{equation}\label{LESinha}
\cL = \frac{1}{2} \left\{ \sigma R - 2 \Lambda_0 - \frac{2}{m^2} \left[ \frac{2}{3} f_{\mu\nu} G^{\mu\nu} + \frac16 \left( f_{\mu\nu} f^{\mu\nu} - f^2 \right) \right] \right\}\,.
\end{equation}
Here $ f = f_{\mu\nu} g^{\mu\nu}$ and $f_{\mu\nu}$ is given in terms of the coefficients $f^{(n)}_{\mu\nu}$ in eq.~\eqref{fiterative}. Explicitly, up to order $1/m^6$ we have checked that the scalar ghost free $R^n$ combinations  of \eqref{EENMG} are recovered. These terms are the same as those that follow from the expansion of the Born-Infeld extension of NMG \cite{Gullu:2010pc,Gullu:2010st,Gullu:2010em,Paulos:2012xe,Bergshoeff:2014ida} when $a=\sigma$.
The linearized spectrum of this model includes only one massive graviton. Using the prescription explained in chapter \ref{chapter:asymptotic_symm} we obtain the following expression for the central charge of the model:
\begin{equation}
 c=\frac{3\ell}{2G}\left[\sigma+\frac{1}{a}\left(-1+\sqrt{1+\frac{a}{m^2\ell^2}}\,\right)\right]\,.
\end{equation}
This coincides with the central charge computed in \cite{Gullu:2010st}, when $a=\sigma$. 
So besides the advantages already listed in the previous section, the CS--like formulation is capable of describing theories with actions containing infinite expansions in a compact and closed form.

A second possible extension of the models discussed here would be to investigate the most general theories with an odd number of fields. We have discussed $N=3$ in chapter \ref{chapter:MMG}, and the Viel-Dreibein Gravity models of chapter \ref{chapter:ENMG_VDG} represent models with even $N \geq 4$. It may be possible to generalize the Viel-Dreibein models to parity violating theories by including Lorentz Chern-Simons terms for the spin connections. 
The resulting ``General Viel-Dreibein'' model may then have a limit where the mass of one of the helicity states goes to infinity, much like the limit from General Zwei-Dreibein Gravity to Minimal Massive Gravity. The resulting theory would be a parity violating, odd $N$ CS--like model. In this sense the models discussed in chapter \ref{sec:DDG} and \ref{sec:VDG} are not the most general ones.

A further possible extension may be to expand the definition of CS--like to include fields which are Lorentz scalar-valued one-forms or Lorentz tensor-valued one-forms. The latter may be interesting from the point of view of bosonic higher-spin theories in three dimensions. Three dimensional massless higher spin fields can be described by a CS-theory of $SL(N,\bR) \times SL(N,\bR)$ \cite{Blencowe:1988gj,Campoleoni:2010zq}. The $SL(N,\bR)$ generators can then be written to contain an $SL(2,\bR)$ generator, plus some higher-spin generators which carry multiple Lorentz indices. The gauge fields corresponding to these higher-spin generators are irreducible Lorentz tensors and they are responsible for the higher-spin fields in a metric formulation. It may be possible, like in the spin-2 case discussed in this thesis, to give up the requirement that all fields are Lie algebra valued connections and to describe in a similar fashion massive higher-spin fields in three dimensions. This would require a more serious modification of the general CS--like formalism and its Hamiltonian form.

The final possible extension which we will comment on here deals with the inclusion of half-integer spin fields. It would be interesting to investigate whether the CS--like models can be made supersymmetric. The formulation in terms of one-forms suggests it could be compatible with local $\mathcal{N}=1$ supersymmetry and a step in this direction was initiated in \cite{Routh:2013uc} where the CS--like form of $\mathcal{N}= 1$ Topological Massive Supergravity was given and analyzed in a Hamiltonian form. Especially for theories which lead to novel field equations in terms of the metric alone, a possible supersymmetric extension could lead to matter couplings which are consistent in a similar novel manner. Perhaps an adaptation of the general Hamiltonian analysis of chapter \ref{chapter:three_crowd} to include spin-3/2 fields could aid in these type of constructions.

In addition to possible extensions of the work presented in this thesis, much progress can still be made in setting up and refining the holographic dictionary for CS--like theories. In this work we have considered the first steps in this direction, by imposing Brown-Henneaux boundary conditions for the CS--like models. In general, the theories with $N\geq3$ fields also contain massive modes which should be represented by a corresponding operator in the dual CFT. In the cases we have discussed, the additional fields are auxiliary and can be solved for in terms of the dreibein. Imposing Brown-Henneaux boundary conditions may then be too restrictive, as one would expect that it may be possible to include sources for the massive modes in the boundary conditions. For higher-derivative gravity theories, it is known how to include the sources for massive modes in a Fefferman-Graham expansion and how to holographically renormalize the metric form of the action and compute the boundary correlation functions (see for instance \cite{Bianchi:2001kw,Bianchi:2001de,Skenderis:2002wp,Skenderis:2009nt,Alishahiha:2010bw,Hohm:2010jc}). For models which do not have an action in terms of a single metric, this procedure is less clear. It would be very interesting to investigate how to include these massive modes and how to extract information on the correlation functions of their corresponding CFT operators from the CS--like formulation directly. 

In addition, the richer dynamics in these theories, as compared to Einstein-Cartan theory, may allow for solutions with different asymptotic behavior and hence different asymptotic symmetry algebras. The sector of parity odd theories discussed in \ref{sec:2odd} and appendix \ref{ExtendedCSG} would be interesting in the light of flat space holography (see for instance \cite{Barnich:2006av,Afshar:2011yh,Afshar:2011qw,Bagchi:2012yk,Afshar:2013bla,Afshar:2014rwa}) since the unitarity of the dual CFT requires the left and right central charges to be opposite. Furthermore, it was shown in \cite{Anninos:2008fx} that TMG admits warped AdS$_3$ solutions with global $SL(2,\bR) \times U(1)$ symmetries and appropriate boundary conditions can be found for these solutions \cite{Compere:2013bya}. It would be interesting to investigate these and novel types of solutions in the second class of CS--like theories, which show improved bulk-boundary behavior and hence do not have to be restricted to 'chiral' points for bulk unitarity.

To conclude, in this thesis we have presented novel ways to couple massive spin-2 modes to gravity in three dimensions and investigated some of the consequences with regard to the asymptotic symmetry algebra. The results so far are promising, but a lot of further work needs to be performed in order to thoroughly understand the dual field theories and to set-up a holographic dictionary for these models. Besides these issues, one may wonder about the lessons we can learn from these models for higher-dimensional theories of massive gravity. The Chern-Simons formulation is native to three dimensional physics, however, the departure from pure Chern-Simons theory discussed in this thesis implies that perhaps we may also depart from three dimensional spacetimes. After all, some of the desirable properties of the CS--like formulation (first order actions, simple Hamiltonian form, clear relation to gauge theory, relatively easy analysis of asymptotic symmetries) need not be restricted to three dimensional spacetimes. A generalization of CS--like models to four dimensions could lead to novel theories of massive spin-2 particles coupled to gravity and this may lead to interesting new possibilities for cosmological models.

%% file: appendix_a/appendix_a.tex
\chapter{Extended Gravitational Chern-Simons Term}
\label{ExtendedCSG}

\pagestyle{headings}

In this appendix we analyze the first parity odd higher-derivative extension of conformal gravity, defined by $S_3$ in \eqref{S3}. Its Lagrangian 3-form is given explicitly as
\begin{equation}\label{LECSG}
\begin{split}
 L_3 =   \frac{1}{2\mu}\omega \cdot \left(d \omega + \frac{1}{3}  \omega \times \omega\right) +\frac{1}{\mu^3}\Big[ &e \cdot \cD f_2 + h_{1} \cdot \left( R + e \times f_{1} \right)  
 \\  & + \frac{\alpha}{2}\,  f_{1}\cdot \cD f_1  \Big]  \,,
\end{split}
\end{equation}
where $\alpha $ is an arbitrary dimensionless parameters.
The equations of motion for this Lagrangian, obtained by varying with respect to $f_2{}^a$, $h_1{}^a$, $f_1{}^a$, $\omega^a$ and $e^a$ respectively, are,
\begin{equation}\label{ECSGeom2}
\begin{split}
& \cD e = 0 \,, \\
& R +  e \times f_1 = 0\,, \\
& \alpha \cD f_1 +  e \times h_1 = 0 \,, \\
&\cD h_1 + \tfrac{1}{2} \left(\alpha \,f_1 \times f_1 + 2e \times f_2 - 2\mu^2 \sigma  e \times f_1\right) = 0\,, \\
& \cD f_2 + f_1 \times h_1 = 0 \,.
\end{split}
\end{equation}
After acting on the equations of motion with an exterior derivative and doing some algebra, we can derive that the auxiliary fields are symmetric
\begin{equation}
f_{1\,[\mu\nu]} = h_{1\,[\mu\nu]} = f_{2\,[\mu\nu]} = 0\,.
\end{equation}
We can solve them in turns of derivatives on the dreibein. Explicitly:
\begin{align}
 & f_{1\,\mu\nu} = - S_{\mu\nu}\,, \qquad h_{1\,\mu\nu} = \alpha C_{\mu\nu} \,, \\
 & f_{2\, \mu\nu} = - \alpha \,D_{\mu\nu} + \alpha\left( P_{\mu\nu}-\tfrac{1}{4}P\,g_{\mu\nu}  \right) - \mu^2 S_{\mu\nu}  \,.
\end{align}
Here $S_{\mu\nu}$ is the Schouten tensor defined in \eqref{Schouten} and $D_{\mu\nu}$ and $P_{\mu\nu}$ are defined in \eqref{DP}. They are the covariant exterior derivative of the Cotton tensor and the contraction of the Einstein and Schouten tensor respectively. Back substitution in the action leads to the following higher-derivative extension of the gravitational Chern-Simons term:
\begin{equation} \label{ECSG}
\mathcal{L}_3=\frac{1}{\mu}\epsilon^{\mu\nu\lambda}\left\{\Gamma_{\mu\sigma}^\rho\partial_\nu\Gamma^\sigma_{\lambda\rho}+\tfrac{2}{3}\Gamma^\rho_{\mu\sigma}\Gamma^\sigma_{\nu\tau}\Gamma^\tau_{\lambda\rho}+\frac{\alpha}{\mu^2}\, R_\mu{}^\sigma\nabla_\nu R_{\sigma\lambda} \right\}\,.
\end{equation}
In a similar manner, we can construct the seventh order derivative theory $S_5$, defined by taking $N=2$ in \eqref{SNodd}. It is, in a metric form
\begin{align}\label{EECSG}
\mathcal{L}_5=\mathcal{L}_{3}+\frac{e}{\mu^5}\bigg\{\beta_1P_{\mu\nu}C^{\mu\nu}+\beta_2D_{\mu\nu} C^{\mu\nu}\bigg\}\,,
\end{align}
where $P_{\mu\nu}$ and $D_{\mu\nu}$ are defined in \eqref{DP}.

\section{Linearization}
We now turn to the quadratic Lagrangian for the parity odd model $S_3$. 
After plugging \eqref{fluctuations} with $m \to \mu$ and $a \to \alpha$ into \eqref{LECSG} and shifting the fluctuations as
\begin{equation}
k_2\rightarrow k_2+(\mu^2+\tfrac{\alpha\Lambda}{2})k_1-\tfrac{\alpha\Lambda^2}{8}k\,,\qquad\text{and}\qquad k_1\rightarrow k_1-\tfrac{\Lambda}{2}k\,,
\end{equation}
we find the following quadratic Lagrangian for $S_3$:
\begin{align}\label{quadL3}
L_{3}^{(2)} =  L_{1}^{(2)}+\frac{1}{\mu^3} \bigg\{& k_{2} \cdot \left(  \bar{\cD} k + \bar{e} \times v \right)   +\frac{\alpha}{2}k_1 \cdot \bar{\cD}k_1 \nonumber\\ 
&+ v_{1} \cdot \left( \bar{\cD} v - \Lambda  \bar{e} \times k +   \bar{e} \times k_1 \right)\bigg\}\,.
\end{align}
Here $L_{1}^{(2)}$ is the linearized Lagrangian three-form of conformal gravity \eqref{LinCSG}. Upon eliminating the auxiliary fields by their equations of motion, we find the fifth-order derivative Lagrangian density
\begin{equation}
\cL_3^{(2)} =  - \frac{1}{\mu}\epsilon^{\mu\alpha}{}_{\rho} \bar{\nabla}_{\alpha} k_0^{\rho\nu} \cG_{\mu\nu}(k_0) + \frac{\alpha}{\mu^3} \epsilon^{\mu\alpha}{}_{\rho} \bar{\nabla}_{\alpha} \cG^{\rho\nu}(k_0) \cG_{\mu\nu}(k_0)\,.
\end{equation}
The linearised equations of motion for this action can be written as:
\begin{equation}
(\cD^0 \cD^M \tilde{\cD}^M \cD^L \cD^R k_0)_{\mu\nu} = 0\,,
\end{equation}
where the differential operators $\cD^{M}$ and $\tilde{\cD}^M$ are defined as:
\begin{align}
(\cD^M)_{\mu}^{\rho} = \delta_{\mu}^{\rho} + \frac{1}{M} \ve_{\mu}{}^{\alpha\rho} \bar{\nabla}_{\alpha} \,, &&
(\tilde{\cD}^M)_{\mu}^{\rho} = \delta_{\mu}^{\rho} - \frac{1}{M} \ve_{\mu}{}^{\alpha\rho} \bar{\nabla}_{\alpha} \,, 
\end{align}
and the mass parameter $M$ is given as:
\begin{equation}
M =  \sqrt{ \cM^2 -  \Lambda } = 
\sqrt{\frac{1}{\ell^2} -\frac{\mu^2}{\alpha}}\,.
\end{equation}
The linear theory hence describes a partially massless mode, and two helicity $\pm 2$ massive modes, with a Fierz-Pauli mass $\cM^2 = - \mu^2/ \alpha$. Now the conformal symmetry is broken, due to the additional interaction term. The theory hence propagates three degrees of freedom, corresponding to the two helicity states of the massive mode and the partially massless mode.

In accordance with what is expected from the linear spectrum, it is possible to diagonalise the quadratic Lagrangian. After the appropriate field redefinitions \eqref{quadL3} can be written as:
\begin{align}\label{linESCG}
L_3{}^{(2)} = & \; \frac{1}{\mu} \left( k_{L} \cdot \bar{\cD} k_{L} + \frac{1}{\ell}  \bar{e} \cdot k_{L} \times k_{L} \right) 
+ \frac{1}{\mu} \left( k_{R} \cdot \bar{\cD} k_{R} - \frac{1}{\ell}  \bar{e} \cdot k_{R} \times k_{R} \right) \nonumber \\
& + \frac{(\alpha - \ell^2 \mu^2)}{2 \mu} k_{0} \cdot \bar{\cD} k_{0}  - \frac{1}{\mu}\left( k_{M_+} \cdot \bar{\cD} k_{M_+} + M \bar{e} \cdot k_{M_+} \times k_{M_+} \right)  \\
& - \frac{1}{\mu} \left( k_{M_-} \cdot \bar{\cD} k_{M_-} - M \bar{e} \cdot k_{M_-} \times k_{M_-} \right)\,, \nonumber 
\end{align}
where we have furthermore assumed that $\mu^2\ell^2 \neq \alpha$. This point correspond to a special case in the linear spectrum. At $\alpha = \ell^2 \mu^2$, the massive modes become partially massless and degenerate with the partially massless mode $k_0$. Note that in this case, there is no (finite) parameter choice possible where the massive mode degenerates with the massless mode.

From inspection of \eqref{linESCG} it is immediately obvious that the kinetic terms for the massive modes come with the opposite sign as the massless modes. Moreover, following the analysis for positive energy of the Fierz-Pauli modes outlined in chapter \ref{sec:pertunit}, we must conclude that the left and right moving sectors have an opposite sign for the energy. This implies that even in the absence of the massive mode, either the left or right moving massless mode has negative energy. This is reflected in the central charge of the right- and left-moving sectors, which have opposite signs. 

\section{Central charge}
We will now compute the central charges of the model defined by \eqref{LECSG}. It fits the general Chern-Simons--like model \eqref{Lgeneral} with flavor space metric and structure constants:
\begin{align}
g_{\omega\omega} = \frac{1}{\mu}\,, && g_{ef_2} = g_{\omega h_1} = \frac{1}{\mu^3}\,, && g_{f_1f_1} = \frac{\alpha}{\mu^3} \,, \\
f_{\omega \omega \omega} = \frac{1}{\mu}\,, && f_{\omega e f_2} = f_{\omega\omega h_1}= f_{ef_1h_1} = \frac{1}{\mu^3}\,, && f_{\omega f_1 f_1} = \frac{\alpha}{\mu^3}\,.
\end{align}
The matrix of Poisson brackets \eqref{Pmat_def} in the flavor space basis $(\omega, e, f_1,h_1,f_2)$ gives
\begin{equation}\label{PmatECSG}
\cP = \left( \begin{array}{cc}
0 & 0 \\
0 & Q
\end{array} \right)\,,
\end{equation}
where
\begin{equation}
Q = \frac{1}{\mu^3} \left( \begin{array}{cccc}
\mu^2 V_{ab}^{f_1f_1} - \frac{1}{\alpha} V_{ab}^{h_1h_1} - 2 V_{[ab]}^{f_1f_2} & - \mu^2 V_{ab}^{f_1 e} + V_{ab}^{f_2e} & \frac{1}{\alpha} V_{ab}^{h_1e} & V_{ab}^{f_1e} \\[.2truecm]
- \mu^2 V_{ab}^{ef_1} + V_{ab}^{ef_2} &\mu^2 V_{ab}^{ee} & 0 & - V_{ab}^{ee} \\[.2truecm]
\frac{1}{\alpha} V_{ab}^{eh_1} & 0 & - \frac{1}{\alpha} V_{ab}^{ee} & 0 \\[.2truecm]
V_{ab}^{ef_1} & - V_{ab}^{ee} & 0 & 0
\end{array} \right)\,.
\end{equation}
From \eqref{PmatECSG} it is immediately obvious that $\phi_{\rm LL}[\chi]$ defined as in \eqref{phiLL} is indeed first-class. To show that the brackets of $\phi_{\rm diff}[\zeta]$ vanish, we may use that, by virtue of
\begin{equation}
e_{[\mu}{}^a f_{1\,\nu]\,a} = e_{[\mu}{}^a h_{1\,\nu]\,a} = e_{[\mu}{}^a f_{2\,\nu]\,a} = 0\,,
\end{equation}
the gauge parameters $\xi^r_a = a_{\mu\,a}^r \zeta^{\mu}$ satisfy
\begin{equation}
e_i{}^a \xi^{f_1}_a = f_{1\,i}{}^a \xi^{e}_{a}\,, \qquad e_i{}^a \xi^{h_1}_a = h_{1\,i}{}^a \xi^{e}_{a}\,, \qquad e_i{}^a \xi^{f_2}_a = f_{2\,i}{}^a \xi^{e}_{a}\,.
\end{equation}
Using these identities, explicit computation shows that $\phi_{\rm diff}[\zeta]$ as defined in \eqref{phidiff} has weakly vanishing brackets with all other primary constraint functions. It is also possible to show that the Poisson brackets of $\phi_{\rm LL}[\chi]$ and $\phi_{\rm diff}[\zeta]$ with the secondary constraints vanish on the AdS vacuum. This is sufficient to identify them as the generators of gauge symmetries at the AdS boundary, since close to the AdS boundary, we may use the background values for the fields. Then, it becomes possible to split the first-class constraint functions into a set of mutually commuting constraints $L_{\pm}$ defined by \eqref{Lpm}. The background values for the fields \eqref{fluctuations} give that:
\begin{equation}\label{ECSGparam}
\xi^{f_1}_a = \frac{1}{2 \ell^2} \xi^{e}_a\,, \qquad \xi^{h_1}_a = 0 \,, \qquad \xi^{f_2}_a = \frac{1}{2\ell^2}\left( \mu^2 - \frac{\alpha}{4 \ell^2}\right) \xi^{e}_a \,.
\end{equation}
We define the constraint functions $L_{\pm}[\xi]$ as
\begin{equation}
L_{\pm}[\xi] = \phi_{\rm diff}'[\zeta] \pm \frac{1}{\ell} \phi_{\rm LL}[\xi] \,,
\end{equation}
where $\phi_{\rm diff}'[\zeta] = \phi_{\rm diff}[\zeta] - \phi_{\rm LL}[\omega_{\mu} \zeta^{\mu}]$. Upon using the AdS background identities \eqref{ECSGparam} in the expression for the boundary charges \eqref{Qpm}, we find that
\begin{equation}
\delta Q_{\pm}^{\text{\tiny ECSG}}[\xi^{\pm}] = \pm \frac{1}{8\pi\mu G} \int_{\partial \Sigma} dx^i \; \xi^{\pm} \cdot \left( \delta \omega_i \pm \frac{1}{\ell} \delta e_i \right)\,,
\end{equation}
where we have reinstated the overall factor of $(8 \pi G)^{-1}$. Following the analysis of section \ref{sec:cc}, this leads to a central charge given by.
\begin{equation}
c_{L/R} = \pm \frac{3}{2\mu G} \,.
\end{equation}
The result does not depend on the new coupling constant $\alpha$, which may be expected since the $\alpha$ term in \eqref{ECSG} vanishes on the AdS background. 

%% file: appendix_b/appendix_b.tex
\chapter{Hamiltonian Analysis of Drei-Dreibein Gravity}
\label{app:Hamil}

\pagestyle{headings}

In this appendix we give the details on the Hamiltonian analysis of the Drei-Dreibein Gravity model. The calculation is performed along the lines of the Hamiltonian analysis of general Chern-Simons--like models involving only Lorentz-vector valued one-forms as presented in chapter \ref{chapter:three_crowd}. For more details on the general model we refer to there and \cite{Bergshoeff:2014bia}. 

The DDG model is described by a Lagrangian:
\begin{equation}\label{LgeneralB}
\cL = \frac12 g_{rs}  a^r \cdot da^s + \frac16 f_{rst} a^r \cdot (a^s \times a^t)\,,
\end{equation}
where $a^{r\,a}$ describe the six fields $(\omega_1{}^a, \omega_2{}^a, \omega_3{}^a ,e_1{}^a ,e_2{}^a ,e_3{}^a)$. After omitting an overall factor $M_P$, the field symmetric space metric $g_{rs}$ has non-zero entries for:
\begin{align}
g_{e_1\omega_1} = -\sigma_1 \,, && g_{e_2\omega_2} = - \sigma_2 \,, && g_{e_3\omega_3} = -\sigma_3 \,.
\end{align}
The non-zero entries for the symmetric field space matrix $f_{rst}$ are
\begin{align}
& f_{e_1 \omega_1 \omega_1} = -\sigma_1 \,, && f_{e_2 \omega_2 \omega_2} = -\sigma_2 \,, && f_{e_3 \omega_3 \omega_3} = - \sigma_3 \,, \nonumber \\
& f_{e_1e_1e_1} = -\alpha_1 m^2\,, &&  f_{e_2e_2e_2} = -\alpha_2 m^2\,, && f_{e_3e_3e_3} = -\alpha_3 m^2\,,   \nonumber \\ 
& f_{e_1e_1e_2} = \beta_{12}m^2\,, && f_{e_1e_2e_2} = \beta_{21}m^2\,, && f_{e_1e_1e_3} = \beta_{13} m^2\,, \\  \nonumber
& f_{e_1e_3e_3} = \beta_{31}m^2\,, && f_{e_2e_2e_3} = \beta_{23}m^2\,, && f_{e_2e_3e_3} = \beta_{32} m^2\,, \\ \nonumber
&  && f_{e_1e_2e_3} = \beta_{123} m^2\,.
\end{align}
The Lagrangian \eqref{LgeneralB} is a first order Lagrangian and after a space-time decomposition of the fields, the corresponding Hamiltonian is solely a function Lagrange multipliers (the time-components of $a^{r\,a}$) and a set of constraints $\phi_r{}^a$. Since the time-components do not propagate, only the spatial parts of the fields contribute to the dynamical phase-space.

From the equations of motion of \eqref{LgeneralB} a set of conditions can be derived which must hold on-shell. They are
\begin{equation}\label{IntconB}
f^t{}_{q[r}f_{s]pt}a^{r\,a} a^p \cdot a^q = 0\,.
\end{equation}
Where the index of $f^t{}_{qr}$ is raised with the inverse of $g_{rs}$. Eqn.~\eqref{IntconB} are six three-form equations from which we can derive the secondary constraints if they are a function of solely an invertible field. Three of these equations give the Cartan identities \eqref{Cartan1}-\eqref{Cartan3} which we analyse in section \ref{sec:constraints}. Assuming only invertibility of $e_1{}^a$ led to a unique choice where two secondary constraints on the spatial components of $a^{r\,a}$ could be derived from the identities \eqref{Cartan1}-\eqref{Cartan3}. This parameter restriction was to take only $\beta_{12}$ and $\beta_{13}$ non-zero.\footnote{For two or three invertible fields there are three inequivalent choices of two non-zero coupling constants leading to secondary constraints, as was shown in section \ref{sec:constraints}. One of these three possibilities is worked out here. The Hamiltonian analysis for the other two choices is similar to the analysis presented here and yields the same results.} The corresponding secondary constraints are given in \eqref{seccon1}. For this choice of parameters, the other three equations in \eqref{IntconB} reduce to:
\begin{equation}
\begin{split}
\beta_{12} e_2{}^a (\omega_1 - \omega_2)\cdot e_1 + \beta_{13} e_3{}^a (\omega_1 - \omega_3)\cdot e_1 = 0 \,, \\
 e_1{}^a (\omega_1 - \omega_2) \cdot e_1 = 0 \,, \qquad 
 e_1{}^a (\omega_1 - \omega_3) \cdot e_1 = 0\,.
 \end{split}
\end{equation}
These equations lead to another two secondary constraints, given in \eqref{seccon2}. To check the consistency of the primary constraints $\phi_r^a$ under time evolutions, we calculate $d\phi [\xi]/dt$, where $\phi[\xi]$ is a smeared operator defined by integrating $\phi_r^a$ against a vector field $\xi_a^r$ (see \eqref{phi}). This amounts to calculating the matrix of Poisson brackets \cite{Hohm:2012vh}\footnote{We are focusing on the bulk degrees of freedom, hence we omit boundary terms here.}
\begin{equation} \label{gen_poissonbrB}
\left\{  \phi[\xi] , \phi[\eta]  \right\}_{\rm P.B.} =  \phi[[\xi, \eta]] + \int_{\Sigma} \xi^r_a \eta^s_b \, \cP_{rs}^{ab} \,,  
\end{equation}
with
\begin{align}
& \cP_{rs}^{ab}  = f^t{}_{q[r} f_{s] pt} \eta^{ab} \Delta^{pq}  +  2f^t{}_{r[s} f_{q]pt} (V^{ab})^{pq}\,, \label{Pmat_defB} \\
 & V_{ab}^{pq} =  \varepsilon^{ij} a^p_{i\, a} a^q_{j\, b}\,, \text{  and }\;\; \Delta^{pq} = \varepsilon^{ij} a_i^p \cdot a_j^q\,, \\
& [\xi ,\eta]^t_c = f_{rs}{}^{t} \epsilon^{ab}{}_{c} \xi^r_a \eta^s_b \,.
\end{align}
By virtue of our parameter choice and the secondary constraints, the first term in \eqref{Pmat_defB} is identically zero. The remaining term gives a $18 \times 18$ matrix, $\cP'{}_{rs}^{ab}$ whose entries are given by
\begin{align}\label{Pmat} \hspace{-.5cm}
\nonumber (\cP'_{ab})_{rs} &=  m^2 \beta_{12}
\left(
\begin{array}{cccccc}
 0 & 0 & 0 & V_{ab}^{e_1e_2} & - V_{ab}^{e_1e_1} & 0 \\
 0 & 0 & 0 & -V_{ab}^{e_1e_2} & V_{ab}^{e_1e_1} & 0 \\
 0 & 0 & 0 & 0 & 0 & 0 \\
  V_{ab}^{e_2e_1} & - V_{ab}^{e_2e_1} & 0 & - (V_{[ab]}^{\omega_1e_2}-V_{[ab]}^{\omega_2e_2}) &
   V_{ab}^{\omega_1e_1} - V_{ab}^{\omega_2e_1} & 0 \\
 - V_{ab}^{e_1e_1} &  V_{ab}^{e_1e_1} & 0 & V_{ab}^{e_1\omega_1}-V_{ab}^{e_1\omega_2} & 0 & 0 \\
  0 & 0 & 0 & 0 & 0 & 0
\end{array}
\right) \\ 
& + m^2 \beta_{13}
\left(
\begin{array}{cccccc}
 0 & 0 & 0 & V_{ab}^{e_1e_3} & 0 &  - V_{ab}^{e_1e_1} \\
  0 & 0 & 0 & 0 & 0 & 0 \\
 0 & 0 & 0 & - V_{ab}^{e_1e_3} & 0 & V_{ab}^{e_1e_1} \\
 V_{ab}^{e_3e_1} & 0 & - V_{ab}^{e_3e_1} & - (V_{[ab]}^{\omega_1e_3}-V_{[ab]}^{\omega_3e_3}) & 0 & V_{ab}^{\omega_1 e_1}-V_{ab}^{\omega_3 e_1} \\
   0 & 0 & 0 & 0 & 0 & 0 \\
- V_{ab}^{e_1e_1} & 0 & V_{ab}^{e_1e_1} &  V_{ab}^{e_1 \omega_1} -  V_{ab}^{e_1 \omega_3} &
0 & 0
\end{array}
\right) \,. \nonumber
\end{align}
We can determine the rank of this matrix at any point in space-time by an arbitrary parametrisation of the fields and plugging it into Mathematica. We find that this matrix has rank 8. To complete the analysis we must add to this the matrix the Poisson brackets of primary constraints with the secondary ones. We define
\begin{equation}
\psi_1 = \Delta^{e_1e_2}\,, \quad \psi_2 = \Delta^{e_1e_3} \,, \quad \psi_3 = \Delta^{\omega_1e_1}- \Delta^{\omega_2 e_1}\,, \quad \psi_4 = \Delta^{\omega_1e_1} - \Delta^{\omega_3e_1}\,.
\end{equation}
The Poisson brackets of the secondary constraints among themselves vanish on the constraint surface and the brackets with the primary constraints are given by:
\begin{align}
\{\psi_1, \phi[\xi]\} = & \; \varepsilon^{ij}\left(e_{1\,i} \cdot \partial_{j} \xi^{e_2} - e_{2\,i} \cdot \partial_{j} \xi^{e_1} - (\xi^{\omega_1} - \xi^{\omega_2}) \cdot  e_{1\,i} \times e_{2\,j} - \xi^{e_1} \cdot \omega_{1\,i} \times e_{2\,j} \right. \nonumber  \\
& \left. +\xi^{e_2} \cdot \omega_{2\,i} \times e_{1\,j} \right) \,, 
\\
\{\psi_2, \phi[\xi]\} = & \; \varepsilon^{ij}\left(e_{1\,i} \cdot \partial_{j} \xi^{e_3} - e_{3\,i} \cdot \partial_{j} \xi^{e_1} - (\xi^{\omega_1} - \xi^{\omega_3}) \cdot e_{1\,i} \times e_{3\,j} - \xi^{e_1} \cdot \omega_{1\,i}\times  e_{3\,j} \right. \nonumber  \\
& \left. +\xi^{e_3}\cdot \omega_{3\,i} \times e_{1\,j}  \right) \,, 
\\
\{ \psi_3, \phi[\xi] \} = & \; \varepsilon^{ij} \Big( - e_{1\,i} \cdot \partial_j( \xi^{\omega_1} - \xi^{\omega_2})  + (\omega_{1\,i} - \omega_{2\,i}) \cdot \partial_j \xi^{e_1} 
 \nonumber \\ \nonumber
& + \xi^{e_1} \cdot (\omega_{1\,i} - \omega_{2\,i}) \times \omega_{1\,j} + m^2 \left(\sigma_1^{-1} \beta_{12} \xi^{e_1} + \sigma_2^{-1} \alpha_2 \xi^{e_2} \right) \cdot e_{1\,i} \times e_{2\,j}   \\
& - m^2\left( ( \beta_{12} \sigma_2^{-1} + \alpha_1 \sigma_1^{-1}) \xi^{e_1} - \beta_{12} \sigma_1^{-1} \xi^{e_2} - \beta_{13} \sigma_1^{-1} \xi^{e_3} \right) \cdot e_{1\,i} \times e_{1\,j} \nonumber \\
&  + \beta_{13} \sigma_1^{-1} m^2 \xi^{e_1} \cdot e_{1\,i} \times e_{3\,j} - (\xi^{\omega_1} - \xi^{\omega_2}) \cdot e_{1\,i} \times \omega_{2\,j} \Big)\,,  
\end{align}
\begin{align}
\{ \psi_4, \phi[\xi] \} = & \; \varepsilon^{ij} \Big( - e_{1\,i} \cdot \partial_j( \xi^{\omega_1} - \xi^{\omega_3})  + (\omega_{1\,i} - \omega_{3\,i}) \cdot \partial_j \xi^{e_1} 
 \nonumber \\ \nonumber
& + \xi^{e_1} \cdot (\omega_{1\,i} - \omega_{3\,i}) \times \omega_{1\,j} + m^2 \left(\beta_{13} \sigma_1^{-1} \xi^{e_1} +  \alpha_3 \sigma_3^{-1} \xi^{e_3} \right) \cdot e_{1\,i} \times e_{3\,j}   \\
&  - m^2\left( ( \beta_{13} \sigma_3^{-1} + \alpha_1 \sigma_1^{-1}) \xi^{e_1} - \beta_{12} \sigma_1^{-1}  \xi^{e_2} - \beta_{13} \sigma_1^{-1} \xi^{e_3} \right) \cdot e_{1\,i} \times e_{1\,j} \nonumber\\
& + \beta_{12} \sigma_1^{-1} m^2 \xi^{e_1} \cdot e_{1\,i} \times e_{2\,j} - (\xi^{\omega_1} - \xi^{\omega_3}) \cdot e_{1\,i} \times \omega_{3\,j} \Big)\,.
\end{align}
For general values of the coupling constants adding these brackets to the total matrix of Poisson brackets will increase the rank of that matrix by 8, making a $22 \times 22$ matrix of rank 16. This implies that there are $22-16= 6 $ first class constraints, while the remaining 16 constraints are second class. This leads to the degree of freedom count as
\begin{eqnarray}
\# \text{ d.o.f} = \frac12 \left( 6 \times 3 \times 2 - 16 - 2 \times 6\right) = 4\,.
\end{eqnarray}
This result is consistent with the linear analysis of the DDG model, which propagates two massive spin--2 particles, each with two helicity states. 

%% file: publications.tex
\chapter[List of publications]{List of publications} 
\label{chapter:publications}

\pagestyle{headings}

\begin{itemize}

\item[{[}{\sc i}{]}] Eric A. Bergshoeff, Sjoerd de Haan, Wout Merbis and Jan Rosseel, {\it A Non-relativistic Logarithmic Conformal Field Theory from a Holographic Point of View}, JHEP {\bf 1109} 038, (2011),  \href{http://www.arXiv.org/abs/1106.6277}{\tt [arXiv:1106.6277]} \newline

\item[{[}{\sc ii}{]}] Wout Merbis, { \it A Holographic Approach to Logarithmic CFTs},  Proceedings of the Corfu Summer Institute 2011, PoS(\textsc{corfu2011}) 039, (2011) \\

\item[{[}{\sc iii}{]}]  Eric A. Bergshoeff, Sjoerd de Haan, Wout Merbis, Massimo Porrati and Jan Rosseel, {\it Unitary Truncations and Critical Gravity: A Toy Model},  JHEP {\bf 1204} 134, (2012), \href{http://www.arXiv.org/abs/1201.0449}{\tt [arXiv:1201.0449]} \\

\item[{[}{\sc iv}{]}] Eric A. Bergshoeff, Sjoerd de Haan, Wout Merbis, Jan Rosseel and Thomas Zojer, {\it On Three-Dimensional Tricritical Gravity},  Phys. Rev. {\bf D86} 064037, (2012), \href{http://www.arXiv.org/abs/1206.3089}{\tt [arXiv:1206.3089]} \\

\item[{[}{\sc v}{]}] Wout Merbis, {\it Massive Gravity}, Proceedings of the Barcelona Postgrad Encounters on Fundamental Physics 55-65, (2013). \\

\item[{[}{\sc vi}{]}] Eric A. Bergshoeff, Sjoerd de Haan, Olaf Hohm, Wout Merbis and Paul K. Townsend, {\it Zwei-Dreibein Gravity: A Two-Frame-Field Model of 3D Massive Gravity},  Phys. Rev. Lett. {\bf 111} 111102, (2013), \href{http://www.arXiv.org/abs/1307.2774}{\tt [arXiv:1307.2774]} \\

\item[{[}{\sc vii}{]}] Eric A. Bergshoeff, Andr\'es F. Goya, Wout Merbis and Jan Rosseel, {\it Logarithmic AdS Waves and Zwei-Dreibein Gravity},  JHEP {\bf 1404} 012, (2014), \href{http://www.arXiv.org/abs/1401.5386}{\tt [arXiv:1401.5386]}  \\

\item[{[}{\sc viii}{]}] Eric A. Bergshoeff, Olaf Hohm, Wout Merbis, Alasdair J. Routh and Paul K. Townsend, { \it The Hamiltonian Form of Three-Dimensional Chern-Simon-like Gravity Models}, Proceedings of the Seventh Aegean Summer School {\it Beyond Einstein's theory of gravity}, Lect.Notes Phys. {\bf 892} 181-201, (2015), \href{http://www.arXiv.org/abs/1402.1688}{\tt [arXiv:1402.1688]} \\

\item[{[}{\sc ix}{]}] Eric A. Bergshoeff, Olaf Hohm, Wout Merbis, Alasdair J. Routh and Paul K. Townsend, {\it Minimal Massive 3D Gravity}, Class. Quant. Grav. {\bf 31} 145008, (2014), \href{http://www.arXiv.org/abs/1404.2867}{\tt [arXiv:1404.2867]}\\

\item[{[}{\sc x}{]}] Hamid R. Afshar, Eric A. Bergshoeff, Wout Merbis, { \it Extended Massive Gravity in Three Dimensions}, JHEP {\bf 1408} 115, (2014), \href{http://www.arXiv.org/abs/1405.6213}{\tt [arXiv:1405.6213]} \\

\item[{[}{\sc xi}{]}] Hamid R. Afshar, Eric A. Bergshoeff, Wout Merbis, {\it 
Interacting spin-2 fields in three dimensions}, \href{http://www.arXiv.org/abs/1410.6164}{\tt [arXiv:1410.6164]}

\end{itemize}

%% file: dankwoord/dankwoord.tex
\chapter*{Acknowledgements}
\label{chapter:acknowledgements}
\addcontentsline{toc}{chapter}{Acknowledgements}

Contrary to popular belief, a PhD-project is far from a lonely venture. As such there are many people to whom I owe gratitude, not always related to my scientific output, but rather to my general happiness in work and my peace of mind. I would like to take the opportunity here to express my heartfelt appreciation towards them.

First of all, I wish to thank my supervisor Eric Bergshoeff. Your experience and expertise in the field has thought me many valuable lessons. Your enduring questions on ongoing research topics made me sharp and forced me to think critically about our projects. Your motivational speeches greatly simplified any tough time and your warm and friendly appearance is one of the main reasons why the institute in Groningen is such a nice place to work.

I would also like to express my gratitude to the reading committee. Mees de Roo, whom I know from the institute in Groningen, thank you for your careful reading and for your wise remarks and calming presence during the first years of my PhD. Maximo Ba\~nados, thank you for important discussions and communications on one of my papers in the fall of 2013 and also for the thorough reading of my manuscript. And Daniel Grumiller, thank you for the many comments on the thesis, as well as your overall and continued interest in my work and your ability to immediately spot new ideas and follow-up projects.

In the course of my four years in Groningen, I have collaborated with many people, both in Groningen and elsewhere, whose many contributions to the shaping of this thesis were essential. Jan Rosseel, thank you for your all your help in my first few years in Groningen, for your explanations and your sense of humor. Sjoerd de Haan, we shared many of the same burdens in our starting years and it was a pleasure to collaborate and cope with them together. Massimo Porrati, even though we never had the opportunity to meet in person, your contribution to one of my papers was clear, powerful and to the point. Thomas Zojer, we shared many discussions during our time in Groningen and many adventures into the academic world outside. Thank you for all the questions and comments made on the various topics discussed here and in my previous work. Thank you also for reading parts of my thesis but more importantly, thanks for being a consistent provider of the institute coffee during the last year. Olaf Hohm, thank you for discussions and your critical look. Your ability to pinpoint any weak spot in an argument is greatly appreciated. Also many thanks for your reference letters which I am sure have benefited me more than I could know. Paul Townsend, it was a true pleasure to collaborate with you on various projects in the last year of my PhD. Your thorough approach and immaculate rewriting of the many drafts have been an essential ingredient in the successful completion of our papers. It was also a great pleasure to discuss with you in Paros. Andr\'es Goya, thank you for a very pleasant collaboration, your enthusiasm and cheerful character. Alasdair Routh, thank you for the many discussions and for having patience with my stupid questions. I'm glad that you accepted the invitation to visit Groningen, which became a memorable and productive working week. Finally, thank you, Hamid Afshar, for your many explanations and for some great discussions; many of which are still not concluded at this moment. I look forward to continuing them.

The institute in Groningen is filled with great people who made the work environment more pleasant than I could have imagined. Thanks especially to Niels and Roel for being excellent officemates. It seemed many times that the law of conservation of misery applied within our office as whenever someones problems were solved, someone else would get stuck or have some unexplainable anomalous results. Fortunately our office humor would always lighten the day. I wish you both the best of luck in your future careers at G\&S. 

I also wish to express many thanks for enlightening conversations on various aspects of physics and beyond to the staff of the CTN, especially Diederik Roest, Elisabetta Palante and Kyriakos Papadodimas. Thanks also to all the PhDs and postdocs during my time in Groningen; Ana, Andrea, Avihay, Dries, Giuseppe, G\"ohkan, Ivonne, Jacob, Jan-Willem, Jasper, Joyce, Jordy, Keri, Lorena, Luca, Marco, Marija, Mario, Mehmet, Roel, Siebren, Souvik, Tiago, Victor, Wouter and Yihao, thank you for being great colleagues. And Annelien and Iris, thanks for being the hub in the wheels of the institute.

Also outside Groningen there are many people which always make me look forward to any national or international high-energy physics meeting. In particular, I would like to thank Ivano Lodato for our many adventures and our shared loves and interests. You always know how to make every meeting we had unforgettable. Thanks to Sander Mooij for your enthusiasm; your inspiring personality and optimism always made me happy to be a physicist. Many thanks to all the people I have shared a good time with on physics conferences all around the world, especially Daniel, Drazan, Jan, Kristiaan, Nico, Pablo, Paul and Val.

Ook de verbintenis met de wereld buiten de natuurkunde was essentieel voor mij en mijn gemoedsrust. Heel veel mensen hebben hier positief aan bijgedragen en het is onmogelijk om iedereen hier te bedanken. Toch zijn er enkelen die hier een vermelding verdienen. Ward, jouw invloed op mij kan niet worden onderschat. Jouw unieke perspectief is altijd een grote relativerende factor geweest en ik kijk altijd uit naar onze ontmoetingen. Joeri, ik ben super blij dat jij mijn paranimf bent en ik denk dat onze vriendschap alleen maar sterker kan worden. Andere mensen die niet onbelangrijk zijn, zijn al mijn vrienden uit mijn Amsterdamse dagen en daarvoor, vooral Alice, Anne, Carmen, Chandar, Chris, Ellen, Julia, Kim, Mathijs, Paul, Rian, Sander en Thomas. Many thanks also to HH, GW and ZA for the more practical interpretation of string theory and for our combined redefinitions of space and time.

Tot slot richt ik me tot mijn dierbare familie en bovenal mijn lieve ouders, Max en Jeanette. Jullie onvoorwaardelijke steun in al mijn ondernemingen geven me het vertrouwen en doorzettingsvermogen dat ik nodig heb. Ik kan me geen betere ouders voorstellen. Vincent, jij ook bedankt dat je mijn paranimf wil zijn en dat je mijn lieve broer bent. Je vrolijke en avontuurlijke karakter is inspirerend.

Y por \'ultimo, mi querida Irene. Gracias por creer siempre en m\'i, incluso (o especialmente) cuando yo no lo hice. Estoy muy agradecido por tu amor incondicional y tu paciencia y por hacer que mis d\'ias terminen con una sonrisa. Te amo y espero que sigas tu coraz\'on y tus sue\~nos. Veo un futuro brillante para nosotros.

\begin{flushright}
Wout Merbis\\
Groningen\\
July 2014
\end{flushright}